\documentclass[twocolumn,appendixfloats]{aastex6}

\newcommand{\Lsun}{$L_{\odot}$}
\newcommand{\Msun}{$M_{\odot}$}
\newcommand{\Rsun}{$R_{\odot}$}

\shorttitle{HOPS: Protostellar SEDs and Model Fits}
\shortauthors{Furlan et al.}
\submitted{ApJS, in press}

\begin{document}

\title{The Herschel Orion Protostar Survey: Spectral Energy Distributions and
Fits Using a Grid of Protostellar Models}


\author{E. Furlan\altaffilmark{1}, W. J. Fischer\altaffilmark{2}, B. Ali\altaffilmark{3},
A. M. Stutz\altaffilmark{4}, T. Stanke\altaffilmark{5}, J. J. Tobin\altaffilmark{6,7}, \\
S. T. Megeath\altaffilmark{8}, M. Osorio\altaffilmark{9}, 
L. Hartmann\altaffilmark{10}, N. Calvet\altaffilmark{10},
C. A. Poteet\altaffilmark{11}, J. Booker\altaffilmark{8}, \\
P. Manoj\altaffilmark{12}, D. M. Watson\altaffilmark{13}, L. Allen\altaffilmark{14}}
\altaffiltext{1}{Infrared Processing and Analysis Center, California
Institute of Technology, 770 S. Wilson Ave., Pasadena, CA 91125, USA; 
furlan@ipac.caltech.edu}
\altaffiltext{2}{{\it NASA Postdoctoral Program Fellow}; Goddard Space Flight 
Center, 8800 Greenbelt Road, Greenbelt, MD 20771, USA}
\altaffiltext{3}{Space Science Institute, 4750 Walnut Street, Suite 205, Boulder, 
CO 80301, USA}
\altaffiltext{4}{Max-Planck-Institut f\"ur Astronomie, K\"onigstuhl 17, D-69117
Heidelberg, Germany}
\altaffiltext{5}{ESO, Karl-Schwarzschild-Strasse 2, D-85748 Garching bei 
M\"unchen, Germany}
\altaffiltext{6}{{\it Hubble Fellow}; National Radio Astronomy Observatory,
Charlottesville, VA 22903, USA}
\altaffiltext{7}{current address: Leiden Observatory, Leiden University, 
P.O. Box 9513, 2300-RA Leiden, The Netherlands}
\altaffiltext{8}{Ritter Astrophysical Research Center, Department of Physics and
Astronomy, University of Toledo, 2801 W. Bancroft Street, Toledo, OH 43606, USA}
\altaffiltext{9}{Instituto de Astrof\'isica de Andaluc\'ia, CSIC, Camino Bajo de 
Hu\'etor 50, E-18008 Granada, Spain}
\altaffiltext{10}{Department of Astronomy, University of Michigan, 500 Church
Street, Ann Arbor, MI 48109, USA}
\altaffiltext{11}{New York Center for Astrobiology, Rensselaer Polytechnic Institute,
110 Eighth Street, Troy, NY 12180, USA}
\altaffiltext{12}{Department of Astronomy and Astrophysics, Tata Institute
of Fundamental Research, Homi Bhabha Road, Colaba, Mumbai 400005, India}
\altaffiltext{13}{Department of Physics and Astronomy, University of Rochester, Rochester, 
NY 14627, USA}
\altaffiltext{14}{National Optical Astronomy Observatory, 
950 N. Cherry Avenue, Tucson, AZ 85719, USA}

\begin{abstract}
We present key results from the Herschel Orion Protostar Survey (HOPS):
spectral energy distributions (SEDs) and model fits of 330 young stellar objects, 
predominantly protostars, in the Orion molecular clouds. This is the largest sample of 
protostars studied in a single, nearby star formation complex. With near-infrared 
photometry from 2MASS, mid- and far-infrared data from {\it Spitzer} and {\it Herschel}, 
and submillimeter photometry from APEX, our SEDs cover 1.2 -- 870 $\mu$m 
and sample the peak of the protostellar envelope emission at $\sim$~100~$\mu$m. 
Using mid-IR spectral indices and bolometric temperatures, we classify our sample 
into 92 Class 0 protostars, 125 Class I protostars, 102 flat-spectrum sources, and 
11 Class II pre-main-sequence stars. We implement a simple protostellar model
(including a disk in an infalling envelope with outflow cavities) to generate a grid of 
30,400 model SEDs and use it to determine the best-fit model parameters for each 
protostar. We argue that far-IR data are essential for accurate constraints on 
protostellar envelope properties. We find that most protostars, and in particular the 
flat-spectrum sources, are well fit. The median envelope density and median inclination 
angle decrease from Class 0 to Class I to flat-spectrum protostars, despite the broad 
range in best-fit parameters in each of the three categories. We also discuss 
degeneracies in our model parameters. Our results confirm that the different 
protostellar classes generally correspond to an evolutionary sequence with 
a decreasing envelope infall rate, but the inclination angle also plays a role in the 
appearance, and thus interpretation, of the SEDs.
\end{abstract}


\section{Introduction}
\label{intro}

The formation process of low- to intermediate-mass stars is divided into several 
stages, ranging from the deeply embedded protostellar stage to the period when 
a young star is dispersing its protoplanetary disk in which planets may have formed.
During the protostellar phase, which is estimated to last $\sim$ 0.5 Myr 
\citep{evans09,dunham14}, the growing central source accretes dust and gas 
from a collapsing envelope. The material from the envelope is most likely accreted 
through a disk, feeding the growing star. A fraction of the mass is ejected in outflows, 
which carve openings into the envelope along the outflow axis. 
Despite our understanding of the basic processes operating in low-mass protostars, 
fundamental questions remain \citep[e.g.,][]{dunham14}. In particular, it is not 
understood how the processes of infall, feedback from outflows, disk accretion, 
as well as the surrounding birth environment, affect mass accretion and determine the 
ultimate stellar mass. The luminosity of protostars, which can be dominated by 
accretion, is observed to span more than three orders of magnitude, yet the underlying 
physics of this luminosity range is also not understood \citep{dunham10,offner11}. 
It is in this protostellar phase that disks are formed, setting the stage for planet 
formation, yet how infall, feedback, accretion, and environment influence the 
properties of disks and of planets that eventually form from them is unknown. 
The large samples of well-characterized protostars identified from surveys with 
{\it Spitzer} and {\it Herschel} now provide the means to systematically study the 
processes controlling the formation of stars and disks; the goal of this work is to
provide such a characterization for the protostars found in the Orion A and B 
clouds, the largest population of protostars for any of the molecular clouds within 
500 pc of the Sun \citep{kryukova12,dunham13,dunham15}.
  
In protostars, dust in the disk and envelope reprocesses the shorter-wavelength 
radiation emitted by the central protostar and the accretion shock on the stellar surface 
and reemits it prominently at mid- to far-infrared wavelengths. As a result, the combined 
emission of most protostellar systems (consisting of protostar, disk, and envelope) peaks 
in the far-IR. 
Young, deeply embedded protostars have spectral energy distributions (SEDs) 
with steeply rising slopes in the infrared, peaking around 100 \micron, and large 
fractional submillimeter luminosities \citep[e.g.,][]{enoch09,stutz13}. Near 10 
\micron\ and 18 \micron, absorption by sub-micron-sized silicate grains causes broad 
absorption features; in addition, there are several ice absorption features across 
the infrared spectral range \citep{boogert08, pontoppidan08}. These absorption 
features are indicative of the amount of material along the line of sight, with the 
deepest features found for the most embedded objects. In addition, due to the
asymmetric radiation field, the orientation of a protostellar system to the line of 
sight, whether through a dense disk or a low-density cavity, plays a role in the
appearance of the SED. It influences the near- to far-IR slope, the depth of the 
silicate feature, the emission peak, and the fraction of light emitted at the longest 
wavelengths \citep[see, e.g.,][]{whitney03a}. 
 
To classify young stellar objects (YSOs) into observational classes, the near- to 
mid-infrared spectral index $n$ ($\lambda F_{\lambda} \propto \lambda^n$) 
from about 2 to 20 $\mu$m has traditionally been used \citep{lada87,adams87,
andre94,evans09,dunham14}. This index is positive for a Class 0/I protostar, 
between $-0.3$ and 0.3 for a flat-spectrum source, and between $-1.6$ and $
-0.3$ for a Class II pre-main-sequence star. Class 0 protostars are distinguished 
from Class I protostars as having $L_{submm}/L_{bol}$ ratios larger than 0.5\%, 
according to the original definition by \citet{andre93}. Other values for this threshold 
that have recently been used are 1\% \citep{sadavoy14} and even 3\% \citep{maury11}.
Another measure for the evolution of a young star is the bolometric temperature 
($T_{bol}$), which is the temperature of a blackbody with the same flux-weighted 
mean frequency as the observed SED \citep{myers93}. A Class 0 protostar has 
$T_{bol}<70$ K, a Class I protostar 70 K $<T_{bol}<$ 650 K, and a Class II 
pre-main-sequence star 650 K $<T_{bol}<$ 2800 K \citep{chen95}. 
These observational classes are inferred to reflect evolutionary stages, with the 
inclination angle to the line of sight being the major source of uncertainty in 
translating classes to ``stages'' \citep{robitaille06,evans09}. Also the accretion
history, which likely includes episodic accretion events and thus temporary 
increases in luminosity, adds to this uncertainty \citep{dunham10,dunham12}.
Protostars with infalling envelopes of gas and dust correspond to Stages 0 and I, 
with the transition from Stage 0 to I occurring when the stellar mass becomes 
larger than the envelope mass \citep{dunham14}. Young stars that have dispersed 
their envelopes and are surrounded by circumstellar disks correspond to Stage II. 

By modeling the SEDs of protostars, properties of their envelopes, and to some
extent of their disks, can be constrained. The near-IR is particularly sensitive to 
extinction and thus constrains the inclination angle and cavity opening angle, 
as well as the envelope density.  Mid-IR spectroscopy reveals the detailed emission 
around the silicate absorption feature and thus provides additional constraints for 
both disk and envelope properties \citep[see, e.g.,][]{furlan08}. At longer wavelengths, 
envelope emission starts to dominate. Thus, photometry in the far-IR is necessary 
to determine the peak of the SED and constrain the total luminosity and envelope 
properties.

Here we present 1.2--870 $\mu$m SEDs and radiative transfer model fits
of 330 YSOs, most of them protostars, in the Orion star formation complex. 
This is the largest sample of protostars studied in a single, nearby star-forming 
region (distance of 420 pc; \citealt{menten07,kim08}) and therefore significant 
for advancing our understanding of protostellar structure and evolution.
These protostars were identified in {\it Spitzer Space Telescope} \citep{werner04} 
data by \citet{megeath12} and were observed at 70 and 160 \micron\ with 
the Photodetector Array Camera and Spectrometer \citep[PACS;][]{poglitsch10} 
on the {\it Herschel Space Observatory}\footnote{{\it Herschel} is an ESA 
space observatory with science instruments provided by European-led 
Principal Investigator consortia and with important participation from NASA.}
\citep{pilbratt10} as part of the Herschel Orion Protostar Survey (HOPS), a 
{\it Herschel} open-time key program \citep[e.g.,][W. J. Fischer et al. 2016, 
in preparation; B. Ali et al. 2016, in preparation]{fischer10, stanke10, 
manoj13, stutz13}. To extend the SEDs into the sub-mm, most of the
YSOs were also observed in the continuum at 350 and 870~$\mu$m with 
the Atacama Pathfinder Experiment (APEX) telescope \citep{stutz13}. 
Our sample also includes 16 new protostars identified in PACS data obtained 
by the HOPS program (\citealt{stutz13,tobin15}; see section \ref{sample}).
We use a grid of 30,400 protostellar model SEDs to find the best fit to the SED for each
object and constrain its protostellar properties. As mentioned above, the far-infrared 
data add crucial constraints for the model fits, given that for most protostars the 
SED peaks in this wavelength region, and therefore, within the framework of the 
model grid, our SED fits yield the most reliable protostellar parameters to date 
for these sources.

\section{Sample Description}
\label{sample}

The 488 protostars identified in {\it Spitzer} data by \citet{megeath12}
represent the basis for the HOPS sample\footnote{The selection of HOPS
targets is based on an earlier version of the {\it Spitzer} Orion Survey,
and in addition some objects likely in transition between Stages I and II
were included; thus, not all protostars in the HOPS sample are classified
as protostars with envelopes in \citet{megeath12}.} \citep[see][]{fischer13,
manoj13,stutz13}. They have 3.6-24 $\mu$m spectral indices $\geq$ 
$-0.3$ and thus encompass flat-spectrum sources.  
To be included in the target list for the PACS observations, the predicted flux 
of a protostar in the 70 $\mu$m PACS band had to be at least 42 mJy as
extrapolated from the {\it Spitzer} SED. Since targets were required to have 
a 24 $\mu$m detection, protostars in the Orion Nebula -- where the {\it Spitzer} 
24 $\mu$m data are saturated -- are excluded. In addition, after the PACS data 
were obtained, several new point sources that were very faint or undetected 
in the {\it Spitzer} bands were discovered in the {\it Herschel} data 
\citep{stutz13}. Fifteen of them were found to be reliable new protostars.
One more protostar, which was not included in the sample of \citet{stutz13} due 
to its more spatially extended appearance at 70 $\mu$m, was recently confirmed 
by \citet{tobin15}. We have added these 16 protostars to the HOPS sample for 
this work (see Table D\ref{New_proto} in the Appendix). Most of these new 
protostars have very red colors and are thus potentially the youngest protostars 
identified in Orion \citep[see][]{stutz13}. 

Each object in the target list was assigned a ``HOPS'' identification number, resulting
in 410 objects with such numbers; HOPS 394 to 408 are the new protostars
identified by \citet{stutz13}, and HOPS 409 is the new protostar from \citet{tobin15}.
Four of the 410 HOPS targets turned out to be duplicates, and 31 are likely 
extragalactic contaminants (see Appendix \ref{exgal_not_modeled} for details). 
Some objects in the HOPS target list were not observed by PACS; of these 
33 objects, 16 are likely contaminants, while the remaining objects were 
originally proposed but were not observed since they were too faint to have 
been detected with PACS in the awarded observing time. In addition, 
35 HOPS targets were not detected at 70 \micron\ (see Appendices
\ref{YSOs_not_modeled} and \ref{exgal_not_modeled}); eight of these are 
considered extragalactic contaminants, while two of them (HOPS 349 and 381) 
have only two measured flux values each, making their nature more uncertain. 
One more target, HOPS 350, also has just two measured flux values (at 24 and 
70 $\mu$m) and is therefore also excluded from the analysis of this paper.
Similarly, we excluded HOPS 352, since it was only tentatively detected at 24 
\micron\ (it lies on the Airy ring of HOPS 84) and in none of the other data sets.

To summarize, starting from the sample of 410 HOPS targets, but excluding likely 
contaminants and objects not observed or detected by PACS, there are 330 
remaining objects that have {\it Spitzer} and {\it Herschel} data and are considered 
protostars (based on their {\it Spitzer} classification from \citealt{megeath12}). 
They form the sample studied in this work. Their SEDs are presented in the 
next section, and in later sections we show and discuss the results of SED fits 
for these targets. Their coordinates, SED properties, and classification, as well 
as their best-fit model parameter values, are listed in Table A\ref{bestfit}. 
The 41 likely protostars that lack PACS data (either not observed or not detected) 
are presented in Appendix \ref{YSOs_not_modeled}.

\section{Spectral Energy Distributions}
\label{SEDs}

\subsection{Data}

\begin{splitdeluxetable*}{lcccccccccccBlcccccc}
\tablewidth{0.9\linewidth}
\tabletypesize{\scriptsize}  
\tablecaption{SED Data for the HOPS targets
\label{SED_data}}
\tablehead{
\colhead{Object} & \colhead{$J$ Flux} & \colhead{$J$ Unc.} & \colhead{$J$ Flag} &
\colhead{\nodata} &
\colhead{[70] Flux} & \colhead{[70] Unc.} & \colhead{[70] Flag} &
\colhead{\nodata} &
\colhead{[870] Flux} & \colhead{[870] Unc.} & \colhead{[870] Flag} &
\colhead{Object} & \colhead{[5.4] Flux} & \colhead{[5.4] Unc.} & \colhead{\nodata} & 
\colhead{[35] Flux} & \colhead{[35] Unc.} & \colhead{IRS scaling}}
\startdata
 HOPS 1 &    0.000E+00 &    0.000E+00 &   3 & \nodata &    3.697E+00 &    1.850E-01 &   1 & \nodata &    6.354E-01 &    1.271E-01 &   2 &   HOPS 1 &  8.631E-03 &    6.069E-04 & \nodata &    1.185E+00 &    6.460E-02 &  1.17 \\
 HOPS 2 &    2.770E-04 &    5.000E-05 &   1 & \nodata &    5.188E-01 &    2.617E-02 &   1 & \nodata &    3.865E-01 &    7.730E-02 &   2 &    HOPS 2 &  4.360E-02 &    3.757E-03 & \nodata &    3.704E-01 &    3.035E-02 &  1.00 \\
 HOPS 3 &    2.198E-03 &    8.900E-05 &   1 & \nodata &    3.187E-01 &    1.622E-02 &   1 & \nodata &    1.201E-01 &    2.402E-02 &   1 &    HOPS 3 &  4.460E-02 &    7.925E-03 & \nodata &    4.050E-01 &    2.443E-02 &  1.66 \\
 HOPS 4 &    3.820E-04 &    5.300E-05 &   1 & \nodata &    6.116E-01 &    3.083E-02 &   1 & \nodata &    1.840E-01 &    3.680E-02 &   2 &    HOPS 4 &  2.055E-02 &    1.240E-03 & \nodata &    4.943E-01 &    3.484E-02 &  1.00 \\
 HOPS 5 &    0.000E+00 &    0.000E+00 &   3 & \nodata &    7.103E-01 &    3.573E-02 &   1 & \nodata &    6.973E-02 &    1.395E-02 &   2 &    HOPS 5 &  1.475E-02 &    2.695E-03 & \nodata &    3.077E-01 &    4.484E-03 &  1.00 \\
 HOPS 6 &    0.000E+00 &    0.000E+00 &   3 & \nodata &    9.110E-02 &    5.523E-03 &   1 & \nodata &    2.311E-01 &    4.622E-02 &   2 &    HOPS 6 &  1.271E-03 &    3.935E-04 & \nodata &    5.350E-02 &    5.244E-03 &  1.00 \\
 HOPS 7 &    0.000E+00 &    0.000E+00 &   3 & \nodata &    1.342E+00 &    6.728E-02 &   1 & \nodata &    3.577E-01 &    7.154E-02 &   2 &    HOPS 7 &  6.459E-04 &    3.090E-04 & \nodata &    3.258E-01 &    2.114E-02 &  1.00 \\
 \enddata
\tablecomments{
Each object has up to 13 photometric data points and 16 IRS data points
(see Section \ref{method}). Here we only show some of the data points for
a few HOPS targets. For each measurement, we provide the measured flux
in Jy, its uncertainty (also in Jy) and, for the photometry only, a flag value 
(0---not observed, 1---measured, 2---upper limit, 3---not detected). For those 
HOPS targets with IRS spectra, we also provide the scaling factor that was 
applied to all IRS fluxes in each spectrum to bring them in agreement 
with the IRAC and MIPS fluxes (see Section \ref{method} for details). \\
To convert the 2MASS magnitudes and the {\it Spizter} magnitudes from
\citet{megeath12} to fluxes, we used the following zero points: 1594 Jy for
$J$, 1024 Jy for $H$, 666.7 Jy for $K_s$, 280.9 Jy for [3.6], 179.7 Jy for
[4.5], 115.0 Jy for [5.8], 64.1 Jy for [8], and 7.17 Jy for [24] (2MASS: 
\citealt{cohen03}; IRAC: \citealt{reach05}; MIPS: \citealt{engelbracht07}). \\
{\it This table is published in its entirety in the machine-readable format on 
the journal website. A portion is shown here for guidance regarding its 
form and content.}}
\end{splitdeluxetable*}

In order to construct SEDs for our sample of 330 YSOs, we combined 
our own observations with data from the literature and existing catalogs. 
For the near-infrared photometry, we used $J$, $H$, and $K_s$ data from 
the Two Micron All Sky Survey \citep[2MASS;][]{skrutskie06}. For the 
mid-infrared spectral region, we used {\it Spitzer} data from \citet{kryukova12}
and \citet{megeath12}: the Infrared Array Camera \citep[IRAC;][]{fazio04} 
provided 3.6, 4.5, 5.8, and 8.0 \micron\ photometry, while the Multiband 
Imaging Photometer for Spitzer \citep[MIPS;][]{rieke04} provided 24 \micron\ 
photometry. In addition, most of the YSOs in the HOPS sample 
were also observed with the Infrared Spectrograph \citep[IRS;][]{houck04} 
on {\it Spitzer} using the Short-Low (SL; 5.2-14 \micron) and Long-Low 
(LL; 14-38 \micron) modules, both with a spectral resolution of about 90
(see, e.g., \citet{kim16} for a description of IRS data reduction). 
{\it Herschel} PACS data at 70, 100, and 160 \micron\ yielded far-infrared 
photometric data points (B. Ali et al. 2016, in preparation; the 100 $\mu$m 
data are from the Gould Belt Survey; e.g., \citealt{andre10}). Most YSOs 
were also observed at 350 and 870 \micron\ (see \citealt{stutz13}) by the 
APEX telescope using the SABOCA and LABOCA instruments 
\citep[][respectively]{siringo10, siringo09}. Thus, our SEDs have well-sampled
wavelength coverage from 1.2 to 870 $\mu$m; we did not include additional 
data from the literature in order to preserve a homogeneous data set for 
all the objects in our sample.

The aperture radius used for the photometry varies depending on the
instrument and wave band. The photometry in the 2MASS catalog was
derived from point-spread function (PSF) fits using data from 4\arcsec\ 
apertures around each object (see the Explanatory Supplement to the 2MASS 
All Sky Data Release and Extended Mission Products). \citet{megeath12} used 
an aperture radius of 2{\farcs}4 for IRAC and PSF photometry for MIPS
24 $\mu$m data. We used aperture radii of 9{\farcs}6 and sky annuli 
of 9{\farcs}6-19{\farcs}2 for PACS 70 and 100 \micron\ images; we then applied 
aperture correction factors of 0.7331 and 0.6944 to the 70 and 100 \micron\ fluxes,
respectively. For  PACS 160 \micron, we used an aperture radius of 
12{\farcs}8, a sky annulus of 12{\farcs}8-25{\farcs}6, and an aperture correction
factor of 0.6602. In some cases (background contamination, close companions)
we used PSF photometry at 70 and 160 $\mu$m instead (see B. Ali et al. 2016, 
in preparation, for details). Finally, we adopted beam fluxes at 350 and 870 
$\mu$m (with FWHMs of 7{\farcs}34 and 19\arcsec, respectively).
The IRS SL module has a slit width of 3{\farcs}6, while the LL module
is wider, with a slit width of 10{\farcs}5. Sometimes the flux level of the
two segments did not match at 14 $\mu$m (due to slight mispointings or 
more extended emission from surrounding material measured in LL), 
and in these cases usually the SL spectrum was scaled by at most a 
factor of $\sim$ 1.4 (typically 1.1-1.2). In a few cases, especially when 
the LL spectrum included substantial amounts of extended emission or 
flux from a nearby object, the LL spectrum was scaled down to match the 
flux level of the SL spectrum at 14 $\mu$m, typically by a factor of 0.8-0.9.
We discuss how the different aperture sizes are accounted for in the model
fluxes in section \ref{model_ap}.

The SEDs of our HOPS sample are shown in Figure A\ref{bestSEDs} together
with their best-fit models from our model grid (see sections below); the data
are listed in Table \ref{SED_data}.
Many objects display a deep silicate absorption feature at 10 $\mu$m and
ice features in the 5-8 $\mu$m region, as expected for protostars. Those 
objects with very deep 10 $\mu$m features and steeply rising SEDs are likely 
deeply embedded protostars, often seen at high inclination angles.

\subsection{Multiplicity and Variability}

A large fraction (203 out of 330) of the young stars in our sample have at least one
{\it Spitzer}-detected source within a radius of 15\arcsec; in most cases, this
``companion'' is faint in the infrared and likely a background star or galaxy.
Thus, the emission at far-IR and sub-mm wavelengths is expected to be 
dominated by the protostar or pre-main-sequence star, and we can assume that 
the SEDs are representative of the YSOs even if the nearby sources cannot be 
separated at these wavelengths. There are a few YSOs that have objects 
separated by just 1\arcsec-3\arcsec\ and are only resolved in one or two IRAC 
bands (HOPS 22, 78, 108, 184, 203, 247, 293, 364); in these cases we used 
the flux at the IRAC position that most closely matched those at longer wavelengths. 
We note that some of these very close ``companions'' are likely outflow knots.
There are also unresolved binaries, which appear as single sources even in the 
IRAC observations \citep{kounkel16}; in these cases our SEDs 
show the combined flux in all wave bands. If two point sources are not fully resolved 
and the resulting blended source is elongated, no IRAC photometry was extracted.  
In such cases, a protostar may not have IRAC fluxes even though it was detected 
in the {\it Spitzer} images.

There are also several protostars that lie close to other protostars:
HOPS 66 and 370 ($d$=14.9\arcsec), HOPS 76 and 78 ($d$=14.1\arcsec),
HOPS 86 and 87 ($d$=12.1\arcsec), HOPS 117 and 118 ($d$=13.7\arcsec),
HOPS 121 and 123 ($d$=7.6\arcsec), HOPS 124 and 125 ($d$=9.8\arcsec),
HOPS 165 and 203 ($d$=13.3\arcsec), HOPS 175 and 176 ($d$=8.0\arcsec), 
HOPS 181 and 182 ($d$=10.2\arcsec), HOPS 225 and 226 ($d$=9.2\arcsec), 
HOPS 239 and 241 ($d$=12.4\arcsec), HOPS 262 and 263 ($d$=6.3\arcsec), 
HOPS 316 and 358 ($d$=6.9\arcsec), HOPS 332 and 390 ($d$=11.2\arcsec), 
HOPS 340 and 341 ($d$=4.7\arcsec), and HOPS 386 and 387 ($d$=9.9\arcsec).
HOPS 105 lies 8.7\arcsec\ to the north of an infrared-bright source, identified
by \citet{megeath12} as a young star with a protoplanetary disk. This source
is brighter than HOPS 105 in all {\it Spitzer} bands and at 70 $\mu$m, but it is
well separated at all wavelengths. A similar situation applies to HOPS 128, which
has a disk-dominated source 6.3\arcsec\ to the southeast. 
HOPS 108 is 6.6\arcsec\ from HOPS 64, which is brighter than HOPS 108 out
to 8 $\mu$m, but not detected in the far-IR and sub-mm. HOPS 108 also lies
16.6\arcsec\ from HOPS 369 and 28.2\arcsec\ from HOPS 370.
HOPS 140 has two neighboring sources, at 9.6\arcsec\ and 13.9\arcsec, that
are likely surrounded by protoplanetary disks; they are both brighter than 
HOPS 140 out to 8 $\mu$m, but at 70 $\mu$m and beyond HOPS 140
dominates.
HOPS 144 lies 7.9\arcsec\ from HOPS 377; there is also a somewhat fainter,
red source 11.7\arcsec\ to the northeast, which is not detected beyond 24 
$\mu$m. This source also lies 9.7\arcsec\ to the southwest of HOPS 143.
HOPS 173 forms a small cluster with HOPS 174 (at 7.1\arcsec) and HOPS 380
(at 11.4\arcsec); HOPS 174 is the brightest source out to 24 $\mu$m, but at
70 $\mu$m HOPS 173 takes on this role. 
Also HOPS 322, 323, and 389 form a group of protostars; HOPS 322 lies
13.4\arcsec\ from HOPS 389 and 20.1\arcsec\ from HOPS 323, while
HOPS 323 and 389 are 10.2\arcsec\ apart. HOPS 323 is the brightest source.

Thus, there are 45 targets in our sample that have an object within 15\arcsec\
that is bright in the mid- or far-IR and that is resolved with IRAC and MIPS.
Given that \citet{megeath12} used PSF photometry for the MIPS 24 $\mu$m 
observations, they obtained reliable fluxes even for companions separated by 
less than 6\arcsec, the typical PSF FWHM. For fluxes at 70 and 160 $\mu$m, 
we also used PSF photometry for objects that were point sources, but too close 
for aperture photometry. In cases where the fluxes could not be determined 
even with PSF photometry, we had to adopt upper limits instead. Similarly, 
we performed PSF photometry on protostars without companions, but 
contaminated by extended or filamentary emission; if the PSF photometry
did not return a good fit, we used the flux value from aperture photometry 
as an upper limit.

Since most of our targets have an IRS spectrum, in addition to data points from
IRAC at 5.8 and 8 \micron\ and from MIPS at 24 \micron, we can detect
discrepancies if flux values at similar wavelengths, but from different instruments,
do not agree. They might be due to calibration or extraction problems in the 
IRS spectrum (for example, some extended emission around the target or a 
close companion), but also to variability. We assumed the former scenario if 
the flux deviations between IRS and IRAC and between IRS and MIPS were 
similar (and more than 10\%, a conservative estimate for the typical calibration 
uncertainty), and in such cases scaled the IRS spectrum to the MIPS 24 
\micron\ flux. Even though this scaling could mask actual variability, it 
creates a representative SED for the YSO and yields an estimate of 
the protostellar parameters from model fits of the SED.

\begin{figure*}[!t]
\centering
\includegraphics[angle=90, scale=0.68]{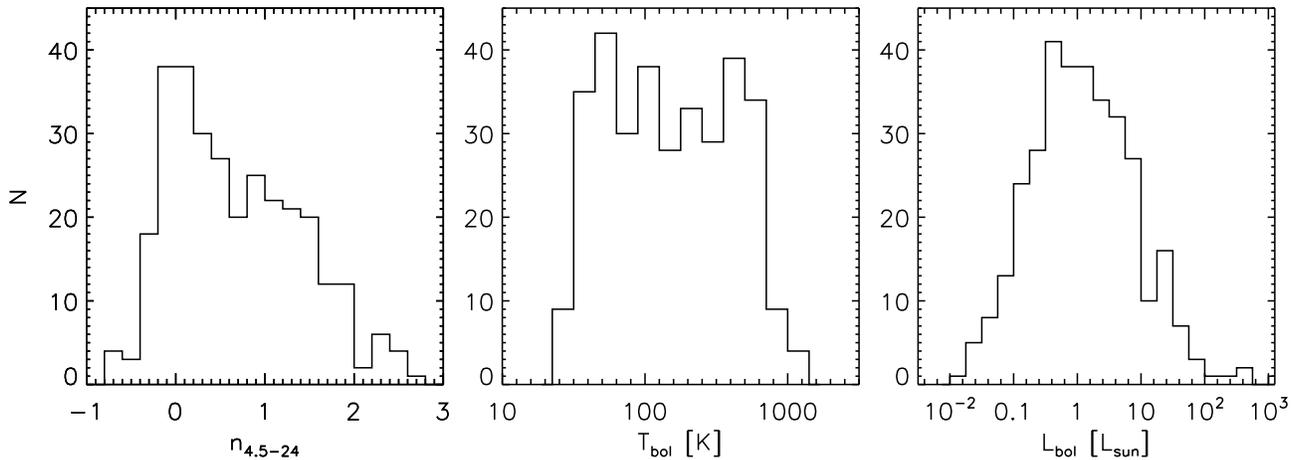}
\caption{Histograms of the 4.5-24 $\mu$m spectral indices ({\it left}), bolometric 
temperatures ({\it middle}), and bolometric luminosities ({\it right}) for the 330 
YSOs in our sample.  
\label{HOPS_n_Tbol_Lbol_histo}}
\end{figure*}

In Appendix \ref{variability} we identify potentially variable HOPS targets 
based on their mid-IR fluxes and find that about 5\% of the protostars 
with IRS, IRAC, and MIPS data could be variable. The Young Stellar Object 
Variability (YSOVAR) program, which monitored large samples of protostars 
and pre-main-sequence stars in nearby star-forming regions with {\it Spitzer} 
at 3.6 and 4.5 $\mu$m \citep{morales11, cody14, rebull14, guenther14, 
poppenhaeger15, wolk15, rebull15}, found that up to $\sim$ 90\% of flat-spectrum 
and Class I YSOs are variable on a timescale of days, with typical changes in 
brightness of 10\%-20\%. On longer timescales (years as opposed to days), 
20\%-40\% of members of young clusters show long-term variability, with the 
highest fraction for those clusters with more Class I protostars \citep{rebull14}. 
In Orion, the fraction of variable Class I protostars is $\sim$ 85\% 
\citep{morales11}. Using a larger sample of protostars in Orion and IRAC 
data at 3.6, 4.5, 5.8, and 8.0 $\mu$m, \citet{megeath12} found that, on a 
timescale of about 6 months, 60\%-70\% of Orion protostars show brightness 
variabilities of $\sim$ 20\%, with some as high as a factor of four.
Thus, given that our SEDs consist of noncontemporaneous data sets, small 
flux discrepancies should be common, but we also expect some protostars 
with large mismatches.

One protostar with a large discrepancy between various data sets is HOPS 223.
It is an outbursting protostar (also known as V2775 Ori; \citealt{caratti11}), 
and for its SED we had 2MASS, IRAC, and MIPS data from the pre-outburst 
phase available, while the IRS spectrum, PACS, and APEX data are from the 
post-outburst period. Thus, its SED does not represent an actual state of the 
object, and the derived $T_{bol}$ and $L_{bol}$ values are unreliable. 
Pre- and post-outburst SEDs and model fits for this protostar can be found
in \citet{fischer12}.
HOPS 223 is the only protostar with an SED affected by extreme variability.
A few more protostars, HOPS 71, 132, 143, 228, 274, and 299, show notable 
discrepancy between the IRAC and IRS fluxes, and to a minor extent between 
MIPS 24 $\mu$m and IRS, and thus have somewhat unreliable SEDs and 
SED-derived parameters. 
HOPS 383, which was identified as an outbursting Class 0 protostar by
\citet{safron15}, does not appear variable in the SED presented here,
since we adopted post-outburst IRAC 3.6 and 4.5 $\mu$m fluxes obtained 
by the YSOVAR program \citep{morales11,rebull14} to construct a 
representative post-outburst SED for this object.

\vspace{6ex}

\subsection{Determination of $L_{bol}$, $T_{bol}$, Spectral Index, and
SED Classification}

The SEDs provide the means to determine $L_{bol}$, $T_{bol}$, and
the 4.5-24 $\mu$m spectral indices for our sample of protostars. For measuring 
the near- to mid-IR SED slope ($n= d \log(\lambda F_{\lambda})/d \log(\lambda)$), 
we chose a spectral index between 4.5 and 24 $\mu$m to minimize the effect of 
extinction on the short-wavelength data point; also, the IRAC 4.5 $\mu$m fluxes 
for our HOPS targets are more complete than the IRAC 3.6 $\mu$m fluxes due 
to the lower extinction at this wavelength. 
For calculating $L_{bol}$ and $T_{bol}$, we used all available fluxes for each 
object, including the IRS spectrum, assumed a distance of 420 pc, and used trapezoidal 
summation; for $T_{bol}$, we applied the equation from \citet{myers93}.
Figure \ref{HOPS_n_Tbol_Lbol_histo} shows the distribution of $n_{4.5-24}$, 
$T_{bol}$, and $L_{bol}$ values for our targets. There is a peak in the distribution 
of spectral indices around 0, while the distribution of $T_{bol}$ values is relatively 
uniform from about 30 K to 800 K. The bolometric luminosities cover a wide range, 
with a broad peak around 1 \Lsun. The median $L_{bol}$, $T_{bol}$, and
$n_{4.5-24}$ values are 1.1 \Lsun, 146 K, and 0.68, respectively.

Our distribution of $L_{bol}$ values is very similar to the observed luminosity function 
of Orion protostars presented in \citet{kryukova12}; both distributions peak around
1 \Lsun\ and include values from $\sim$ 0.02 \Lsun\ up to several hundred \Lsun.
Some differences between the two distributions are expected, given that
\citet{kryukova12} only had {\it Spitzer} 3.6-24 $\mu$m data available and thus
had to extrapolate $L_{bol}$ from the measured near- to mid-infrared luminosity.
The main difference is a somewhat larger number of protostars with $L_{bol} 
\lesssim$ 0.5 \Lsun\ for the \citet{kryukova12} Orion sample; our median $L_{bol}$ 
value amounts to 1.1 \Lsun, while their value is 0.8 \Lsun. However, with the 
contaminating sources removed from their sample (which tend to have lower 
luminosities; see \citealt{kryukova12} for details), their median bolometric luminosity 
and our value match. 
Overall, given that Orion is considered a region of high-mass star formation, its 
luminosity function is similar to that of other regions where massive star forms 
\citep{kryukova12,kryukova14}, and it is different from that of low-mass star-forming 
regions such as Taurus and Ophiuchus \citep{kryukova12, dunham13}. Compared 
to the sample of 230 protostars in 18 different molecular clouds studied by 
\citet{dunham13}, the observed (i.e., not extinction-corrected) $L_{bol}$ values 
of those protostars span from 0.01 to 69 \Lsun, with a median value of 0.9 \Lsun. 
However, given that almost half the protostars in the \citet{dunham13} sample lack 
far-IR and sub-mm data, the true luminosities are likely higher, which would bring the 
median closer to the Orion value. Finally, we note that the distribution of observed 
$T_{bol}$ values from \citet{dunham13} is similar to our distribution for Orion protostars;  
the median $T_{bol}$ of their and our sample is 160 K and 146 K, respectively, and the 
bulk of their protostars also has $T_{bol}$ values between 30 K and 1000 K, with
a tail down to temperatures of $\sim$ 10 K and another tail up to $T_{bol}$=2700~K.

\begin{figure}[!t]
\centering
\includegraphics[angle=90, scale=0.41]{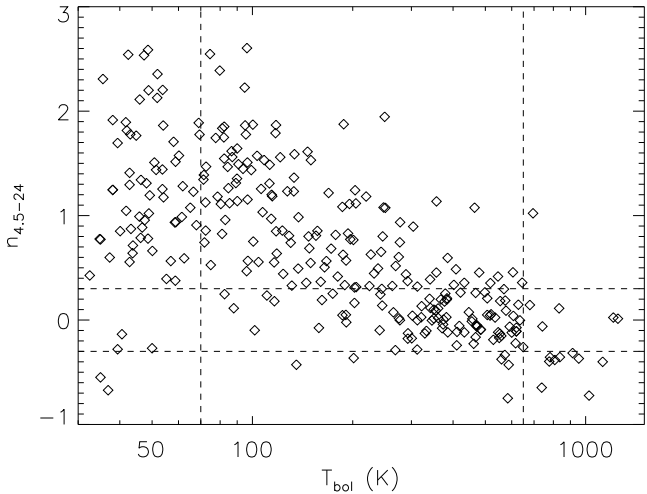}
\caption{The 4.5-24 $\mu$m spectral index versus the bolometric temperature
for the 330 YSOs in our sample. The dashed lines delineate 
the regions that define the various SED classes (see text for details).
\label{HOPS_n_Tbol}}
\end{figure}

To separate our targets into Class 0, Class I, Class II, and flat-spectrum sources, 
we used the 4.5 to 24 \micron\ spectral index ($n_{4.5-24}$) and/or bolometric 
temperature ($T_{bol}$): Class 0 protostars have $n_{4.5-24}>0.3$ and 
$T_{bol}<70$~K, Class I protostars have $n_{4.5-24}>0.3$ and $T_{bol}>70$~K, 
flat-spectrum sources have $-0.3 < n_{4.5-24} < 0.3$, and Class II pre-main-sequence 
stars have $n_{4.5-24}<-0.3$.
Based on this, we identify 92 targets as Class 0 protostars, 125 as Class I protostars, 
102 as flat-spectrum sources, and 11 as Class II pre-main-sequence stars (see Table
A\ref{bestfit} and Figure \ref{HOPS_n_Tbol}). 
There are nine protostars with $T_{bol}$ values between 66.5 and 73.5 K (which 
corresponds to a $\pm$ 5\% range around the Class 0--I boundary of 70 K); 
six of them have $T_{bol}$ $>$ 70 K (HOPS 1, 18, 186, 256, 322, 370), and 
the other three have $T_{bol}$ values just below 70 K (HOPS 75, 250, 361). 
These protostars' classification is less firm than for the other HOPS targets. 
There are also a few flat-spectrum sources whose classification is more 
uncertain: HOPS 45, 183, 192, 194, 210, 264, and 281 should be Class I 
protostars based on their 4.5-24 $\mu$m spectral index, but when considering 
the IRS spectrum (specifically, the 5-25 $\mu$m spectral index), they fall 
into the flat-spectrum regime ($n_{5-25} < 0.3$). Also, for HOPS 45 and 194 
the $T_{bol}$ values are relatively high ($>$ 500 K).
Similarly, HOPS 33, 134, 242, 255, and 284 should be Class II pre-main-sequence
stars based on their 4.5-24 $\mu$m spectral index, but the spectral slope over 
the IRS wavelength range suggests that they are flat-spectrum sources.
In these cases where the $n_{4.5-24}$ and $n_{5-25}$ spectral indices were 
somewhat discrepant, we adopted the latter, and thus these objects were 
classified as flat-spectrum sources. 

There are five objects with $T_{bol}$ $<$ 70 K and $n_{4.5-24}$ $<$ 0
(HOPS 164, 340, 341, 373, 405); despite their negative 4.5-24 $\mu$m SED 
slopes, their SEDs either show or imply a deep silicate absorption feature at 
10 $\mu$m, rise steeply in the mid- to far-IR, and their long-wavelength emission 
is strong. Thus, their $T_{bol}$ values are low, and we identify them as Class 0 
protostars, even though they have 4.5-24 $\mu$m spectral indices not typical 
of embedded protostars. In particular, HOPS 341, 373, and 405 are likely young 
protostars with dense envelopes (\citealt{stutz13}; see also section \ref{Class0}). 
In the case of HOPS 373, the 4.5 $\mu$m flux may be contaminated by bright
H$_2$ emission from an outflow shock, rendering the $n_{4.5-24}$ value more 
unreliable. This might also explain the negative 4.5-24 $\mu$m spectral index 
for the other four protostars.

Finally, the few Class II objects in our sample were thought to be potential
protostars prior to their observations with {\it Herschel}. Their 4.5-24 \micron\ 
SED slopes are usually just slightly more negative than the cutoff for a 
flat-spectrum source ($-0.3$); three Class II pre-main-sequence stars 
(HOPS 22, 184, 201) have SEDs that are typical of disks with inner holes, 
displaying a 10 $\mu$m silicate emission feature and a rising SED from 
12 to about 20 \micron\ \citep[e.g.,][]{kim13}. The SEDs of the other
Class II objects are similar to those of flat-spectrum sources; thus, they could
have (remnant) envelopes that contribute to their long-wavelength emission.

Our HOPS sample is mostly complete in the number of Class 0, Class I, and 
flat-spectrum sources in the areas of Orion surveyed by {\it Spitzer} excluding
the Orion Nebula \citep[see][]{megeath12,stutz13}. Of the 357 unique YSOs 
originally identified in {\it Spitzer} data that were included in the HOPS sample and 
observed with PACS, 322 were detected at least at 70 $\mu$m, which amounts 
to a fraction of 90\%. We removed likely contaminants and added 16 new
protostars discovered in PACS data to get to our sample of 330 YSOs,
most of which are protostars.
Our lowest $L_{bol}$ source is HOPS 208, with $L_{bol}$= 0.017 \Lsun. This 
protostar also has the lowest PACS 70 $\mu$m flux in our sample (8.2 mJy). 
Overall, our sample has 27 protostars with $L_{bol}<$ 0.1 \Lsun, which places 
them in the luminosity range of very low luminosity objects 
\citep[VeLLOs;][]{diFrancesco07,dunham08}. The number of VeLLOs in our 
sample is likely larger, given that VeLLOs are defined as having internal 
luminosities less than 0.1 \Lsun, and the bolometric luminosity has contributions 
from both the internal luminosity and that due to external heating 
\citep[see][]{dunham08}. In addition, our sample could miss fainter flat-spectrum 
sources and Class 0 and Class I protostars. In fact, there are several faint YSOs 
without PACS data that were excluded from our sample, but do have {\it Spitzer} 
detections (see Appendix section \ref{YSOs_not_modeled}).

\vspace{3ex}

\section{Model Grid}
\label{grid}

To characterize the SEDs of our HOPS sample in a uniform manner,
we fit the data to simple but physically plausible models. In this way we
can assess how well such simple models can fit the data, and how the
quality of the fits changes with evolutionary class. We can also determine
the full range of physical parameters implied by the fits and the range of
parameters for each protostellar class. There are degeneracies and biases 
in the fits, and the uncertainties in model parameters will vary from object 
to object, but our results represent a first step in estimating physical 
parameters that describe the protostars in our sample.

We use a large model grid calculated using the 2008 version of the
\citet{whitney03a,whitney03b} Monte Carlo radiative transfer code 
\citep[see][]{stutz13}; an early version of the grid was presented in 
\citet{ali10}. 
Each model consists of a central protostar, a circumstellar disk, and an envelope;
the radiation released by the star and the accretion is reprocessed by the
disk and envelope. The density in the disk is described by power laws in the 
radial and vertical directions, while the density distribution in the envelope 
corresponds to that of a rotating, collapsing cloud core with constant infall
rate (the so-called TSC model, after \citealt{terebey84}; see also 
\citealt{ulrich76,cassen81}). The envelope also contains an outflow cavity, 
whose walls are assumed to follow a polynomial shape. At favorable inclination 
angles, this evacuated cavity allows radiation from the inner envelope and 
disk regions to reach the observer directly. Also, radiation is scattered off the 
cavity walls and can increase the near-IR emission from a protostellar system.

\begin{figure}[!t]
\centering
\includegraphics[scale=0.49]{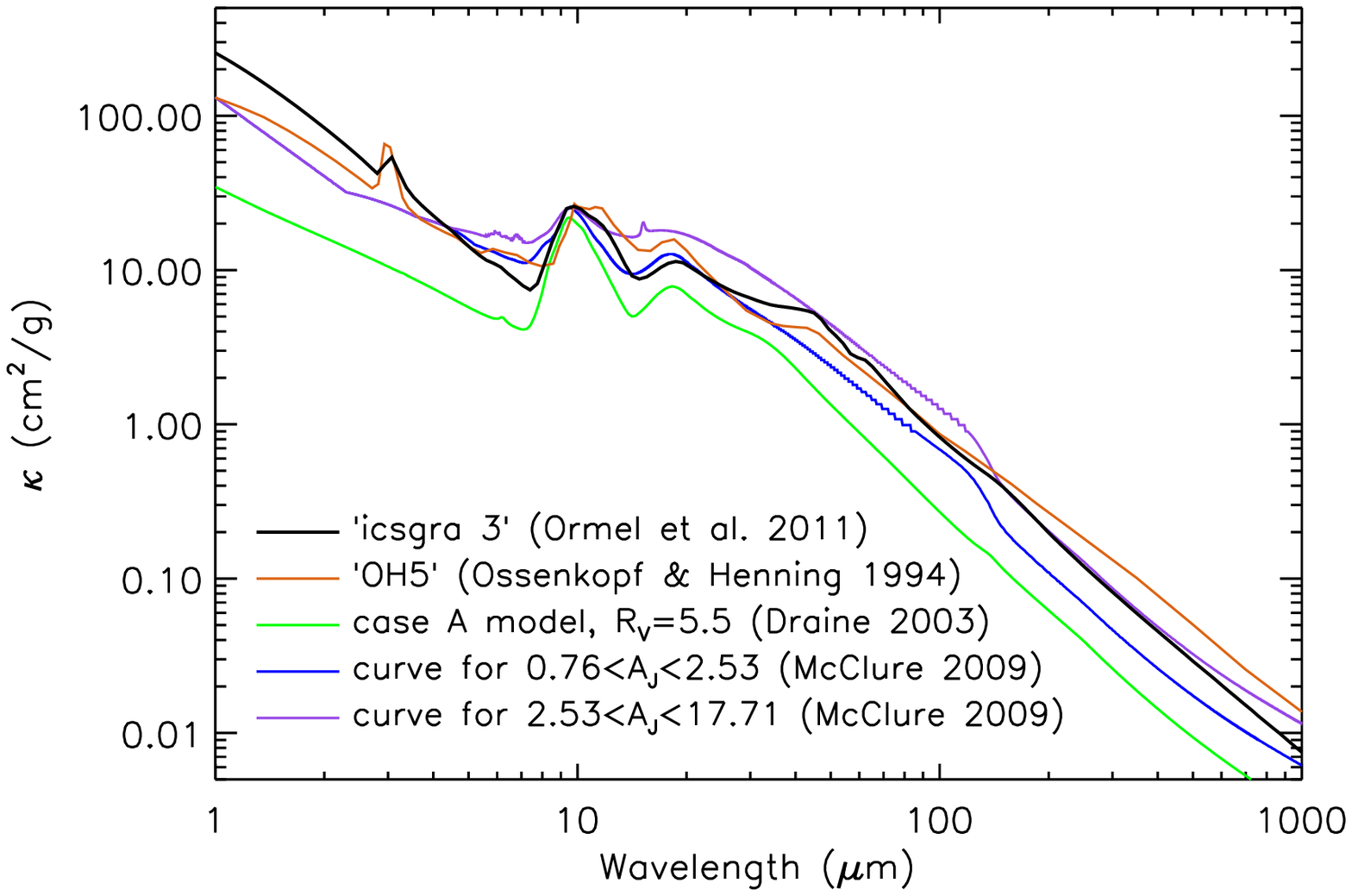}
\caption{Extinction opacities of the \citet{ormel11} dust model ``icsgra3''
({\it black}) compared to other dust opacities from the literature: grains with
thin ice mantles after 10$^5$ years of coagulation with a gas density of
10$^6$ cm$^{-3}$ from \citet{ossenkopf94} ({\it orange});
case A model of carbon and silicate dust for R$_V$=5.5 from \citet{draine03}
({\it green}); two extinction curves derived for star-forming regions by
\citet{mcclure09}, one for $0.76 < A_J < 2.53$ ({\it blue}), and one
for $2.53 < A_J < 17.71$ ({\it purple}).
\label{opacities}}
\end{figure}

We used dust opacities from \citet{ormel11} to account for larger, icy grains
(as opposed to the small grains made of amorphous silicates typically found in
the interstellar medium). We adopted their dust model that includes graphite 
grains without ice coating and ice-coated silicates, with a size distribution that 
assumes growth of aggregates for $3 \times 10^5$ years, when grains have 
grown up to 3 $\mu$m in size (``icsgra3''). Particle sizes range from 0.1 to 
3 $\mu$m, with a number density that is roughly proportional to $a^{-2.3}$ 
(where $a$ is the particle radius).
Figure \ref{opacities} shows our adopted opacities compared to different 
ones found in the literature. The opacities from \citet{draine03} assume a 
mixture of small carbonaceous and amorphous silicate grains. Including larger 
and icy grains broadens the 10 $\mu$m silicate feature (which is mostly due to 
the libration mode of water ice) and causes additional absorption at 3 $\mu$m 
and in the 40-60 $\mu$m range (all mostly due to the presence of water ice). 
The mid-IR opacities of the ``icsgra3'' dust model are similar to the ones 
determined by \citet{mcclure09} for star-forming regions and also to those 
used by \citet{tobin08} to model an edge-on Class 0 protostar; in the mid- to 
far-IR, they resemble the opacities of \citet{ossenkopf94}, which are often 
used to model embedded sources. In Figure \ref{opacities}, we show model 
`OH5' from \citet{ossenkopf94}, which is listed as the fifth model in their 
Table 1 and corresponds to grains with thin ice mantles after 10$^5$ years 
of coagulation and a gas density of 10$^6$ cm$^{-3}$. We could not use the 
`OH5' opacities for our model grid, since that opacity law does not include 
scattering properties (which are required by the Whitney Monte Carlo 
radiative transfer code). Other authors have modified the `OH5' dust  
to include the scattering cross section and extend the opacities to shorter
and longer wavelengths \citep{young05,dunham10}.

\subsection{Model Parameters}
\label{model_parameters}

There are 3040 models in the grid; they cover 8 values for the total (i.e.,
intrinsic) luminosity, 4 disk radii, 19 envelope infall rates (which correspond 
to envelope densities), and 5 cavity opening angles. Each model is calculated 
for 10 different inclination angles, from 18.2\degr\ to 87.2\degr, in equal steps 
in $\cos(i)$ (starting at 0.95 and ending at 0.05), resulting in 30,400 different 
model SEDs. The values for the various model parameters are listed in Table 
\ref{modelpars}. Since there are a large number of parameters that can be set 
in the Whitney radiative transfer models, we focused on varying those 
parameters that affect the SED the most, leaving the other parameters 
at some typical values. For example, we assumed a stellar mass of 0.5 \Msun, 
a disk mass of 0.05 \Msun, and an envelope outer radius of 10,000 AU.
The stellar mass enters the code in two ways. First, it is needed to relate the 
density of the envelope to the infall rate (see Equation 1 below). Since we fit the 
density of the envelope, the infall rate plays no role in the best-fit envelope 
parameters; any stellar mass can be chosen to determine the infall rate for a 
given best-fit envelope density. Second, the stellar mass is combined with the 
stellar radius and disk accretion rate to set the disk accretion luminosity. Given 
that the accretion luminosity is the actual parameter that influences the SED, 
it does not matter which of the three factors is varied. For simplicity and reasons 
described below, we varied the disk accretion rate and the stellar radius, but
left the stellar mass constant, to achieve different values for this component 
of the luminosity.
 
The total luminosity for each system consists of the stellar luminosity 
(derived from a 4000 K stellar atmosphere model), the accretion luminosity 
resulting from material accreting through the disk down to the disk truncation 
radius, and the accretion luminosity from the hot spots on the stellar surface, 
where the accretion columns, which start at the magnetospheric truncation 
radius, land (these columns are not included in the modeled density distribution, 
since they do not contain dust and do not have a source of opacity in the radiative 
transfer models). Typically, the accretion luminosity from the hot spots is much 
larger than the disk accretion luminosity; in our models, the former is about a factor 
of 9 larger than the latter.
We chose three different stellar radii, 0.67, 2.09, and 6.61 \Rsun\ (with the same 
stellar temperature), resulting in three different stellar luminosities. Since both 
components of the accretion luminosity depend on the disk accretion rate, 
choosing a total of eight different disk accretion rates (three for the 0.67 \Rsun\ 
star, two for the 2.09 \Rsun\ star, and three for the 6.61 \Rsun\ star) results in 
eight values for the total luminosity used in the grid (see Table \ref{modelpars}). 
The input spectrum produced by the central protostar depends on the relative 
contributions from the intrinsic stellar luminosity (which peaks at 0.7~$\mu$m) 
and the accretion luminosity (which is radiated primarily in the UV). In the 
models, it can be altered to some degree by choosing different combinations of 
the disk accretion rate and stellar radius (the former affects only the accretion 
luminosity, while the latter affects both the stellar and accretion luminosity). 
However, the effect of the input spectrum on the output SED is negligible. 
Consequently, we cannot reliably measure the relative contributions of stellar 
and accretion luminosity through our SED fits. Instead, we adjusted the particular 
values for the stellar radius and disk accretion rate to set the values of the
total luminosity.

For our model grid, we chose four values for the disk outer radius, which we
set equal to the centrifugal radius ($R_c$). In a TSC model, the centrifugal 
radius is the position in the disk where material falling in from the envelope 
accumulates; due to envelope rotation, material from the envelope's
equatorial plane lands at $R_c$, while material from higher latitudes falls
closer to the star. The disk could extend beyond $R_c$, but in our models 
it ends at $R_c$. In this work, we use the terms ``disk (outer) radius'' and 
``centrifugal radius'' interchangeably. The primary effect of $R_c$ is to set 
the rotation rate of the infalling gas and thereby determine the density 
structure of the envelope \citep{kenyon93}.  
 
\begin{deluxetable*}{clcc}
\tablecaption{Model Parameters
\label{modelpars}}
\tablehead{
\colhead{Parameter} & \colhead{Description} & \colhead{Values} & 
\colhead{Units}}
\startdata
\multicolumn{4}{c}{\it{\bf Stellar Properties}} \\ 
$M_{\ast}$ &  Stellar mass & 0.5 & \Msun \\
$T_{\ast}$ &  Stellar effective temperature & 4000 & K \\
$R_{\ast}$ &  Stellar radius & 0.67, 2.09, 6.61 & \Rsun \\ 
\hline
\multicolumn{4}{c}{\it{\bf Disk Properties}} \\ 
$M_{disk}$ &  Disk mass & 0.05 & \Msun \\
$R_{disk}$ &  Disk outer radius & 5, 50, 100, 500 & AU \\
A & Radial exponent in disk density law & 2.25 & \nodata \\
B & Vertical exponent in disk density law  & 1.25 & \nodata \\
$\dot{M}_{disk,1}$ & Disk-to-star accretion rate for $R_{star}$=0.67 \Rsun\ & 
0, $1.14 \times 10^{-8}$, $5.17 \times 10^{-8}$  & \Msun\ yr$^{-1}$ \\
$\dot{M}_{disk,2}$ & Disk-to-star accretion rate for $R_{star}$=2.09 \Rsun\ & 
$3.67 \times 10^{-7}$, $1.63 \times 10^{-6}$ & \Msun\ yr$^{-1}$ \\
$\dot{M}_{disk,3}$ & Disk-to-star accretion rate for $R_{star}$=6.61 \Rsun\ & 
$1.14 \times 10^{-5}$, $5.15 \times 10^{-5}$,  $1.66 \times 10^{-4}$ & 
\Msun\ yr$^{-1}$ \\
$R_{trunc}$ & Magnetospheric truncation radius$^a$ & 3 & $R_{\ast}$ \\
$f_{spot}$ & Fractional area of the hot spots on the star$^b$ & 0.01 & \nodata \\
\hline
\multicolumn{4}{c}{\it{\bf Envelope Properties}} \\ 
$R_{env}$ &  Envelope outer radius$^c$ & 10,000 & AU \\
$\rho_{1000}$ & Envelope density at 1000 AU$^d$ & 0.0, 1.19 $\times 10^{-20}$, 
1.78 $\times 10^{-20}$, 2.38 $\times 10^{-20}$, &  g cm$^{-3}$ \\
& & 5.95 $\times 10^{-20}$, 1.19 $\times 10^{-19}$, 1.78 $\times 10^{-19}$, 
&  g cm$^{-3}$ \\
& & 2.38 $\times 10^{-19}$, 5.95 $\times 10^{-19}$, 1.19 $\times 10^{-18}$,
&  g cm$^{-3}$ \\
& & 1.78 $\times 10^{-18}$, 2.38 $\times 10^{-18}$, 5.95 $\times 10^{-18}$,
&  g cm$^{-3}$ \\
& & 1.19 $\times 10^{-17}$, 1.78 $\times 10^{-17}$, 2.38 $\times 10^{-17}$,
&  g cm$^{-3}$ \\
& & 5.95 $\times 10^{-17}$, 1.19 $\times 10^{-16}$, 1.78 $\times 10^{-16}$
&  g cm$^{-3}$ \\
$R_c$ & Centrifugal radius of TSC envelope & $= R_{disk}$ & AU \\ 
$\theta$ & Cavity opening angle & 5, 15, 25, 35, 45 & degrees \\
$b_{cav}$ & Exponent for cavity shape$^e$ (polynomial) & 1.5 & \nodata \\
$z_{cav}$ & Vertical offset of cavity wall & 0 & AU \\
 \hline
\multicolumn{4}{c}{\it{\bf Derived Parameters}} \\ 
$L_{\ast}$ &  Stellar luminosity$^f$ &  0.1, 1, 10 & \Lsun \\
$L_{tot}$ & Total luminosity (star + accretion)$^g$ & 0.1, 0.3, 1.0, 3.1, 
         10.1, 30.2, 101, 303 & \Lsun \\
\hline
\multicolumn{4}{c}{\it{\bf Parameters for Model SEDs}} \\ 
$i$ & Inclination angle & 18.2, 31.8, 41.4, 49.5, 56.7, & degrees \\
  &   &  63.3, 69.5, 75.6, 81.4, 87.2 & degrees \\
  &  Aperture radii for model fluxes$^h$ &  420, 840, 1260, 1680, ..., 10080 & AU \\
\enddata
\tablecomments{
The dust opacities used for these models are those called ``icsgra3'' from
\citet{ormel11}.\\
$^a$ This radius applies to the gas. The inner disk radius for the dust is equal to the 
dust destruction radius. The scale height of the disk at the dust sublimation radius is set 
to the hydrostatic equilibrium solution. \\
$^b$ The hot spots are caused by the accretion columns that reach from the 
magnetospheric truncation radius to the star. \\
$^c$ The inner envelope radius is set to the dust destruction radius. \\
$^d$ The actual input parameter for the Whitney code is the envelope infall
rate, which can be derived from $\rho_{1000}$ using Equation (2). The first six 
$\rho_{1000}$ values correspond to envelope infall rates of 0, $5.0 \times 10^{-8}$, 
$7.5 \times 10^{-8}$, $1.0 \times 10^{-7}$, $2.5 \times 10^{-7}$, and $5.0 \times 10^{-7}$  
\Msun\ yr$^{-1}$; the other values can be similarly deduced. \\
$^e$ The cavity walls are assumed to have a polynomial shape; no material is assumed
to lie inside the cavity. Also, the ambient density (outside the envelope) is 0. \\
$^f$ The three values of $L_{\ast}$ correspond to the three different stellar radii.\\
$^g$ The total luminosities combine the stellar luminosities and the accretion luminosities
(which depend on  $\dot{M}_{disk}$). \\
$^h$ For each model, the emitted fluxes are calculated for 24 apertures ranging from 
420 to 10080 AU, in steps of 420 AU. }
\end{deluxetable*}

\begin{figure*}[!t]
\centering
\includegraphics[scale=0.37,angle=90]{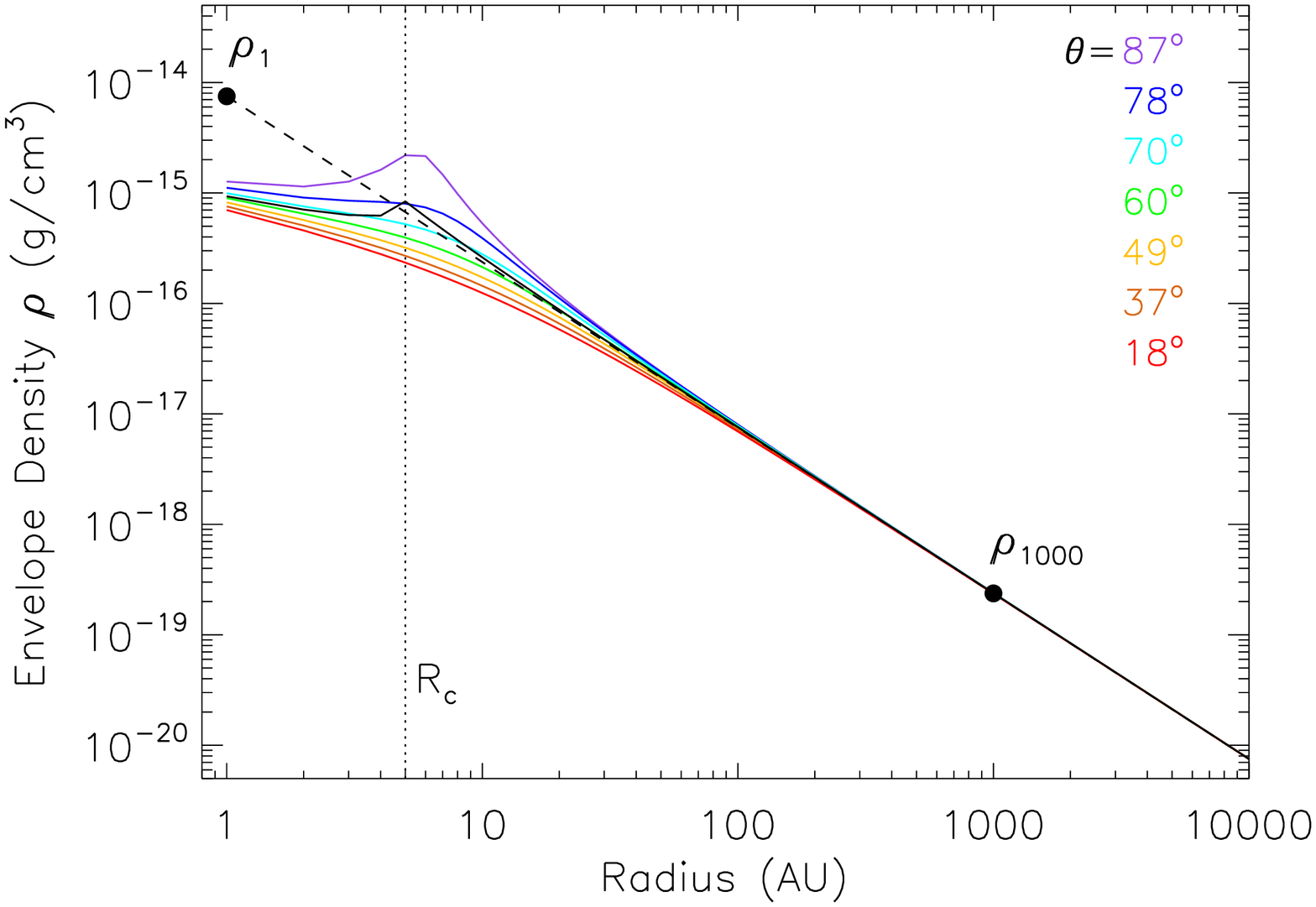}
\includegraphics[scale=0.37,angle=90]{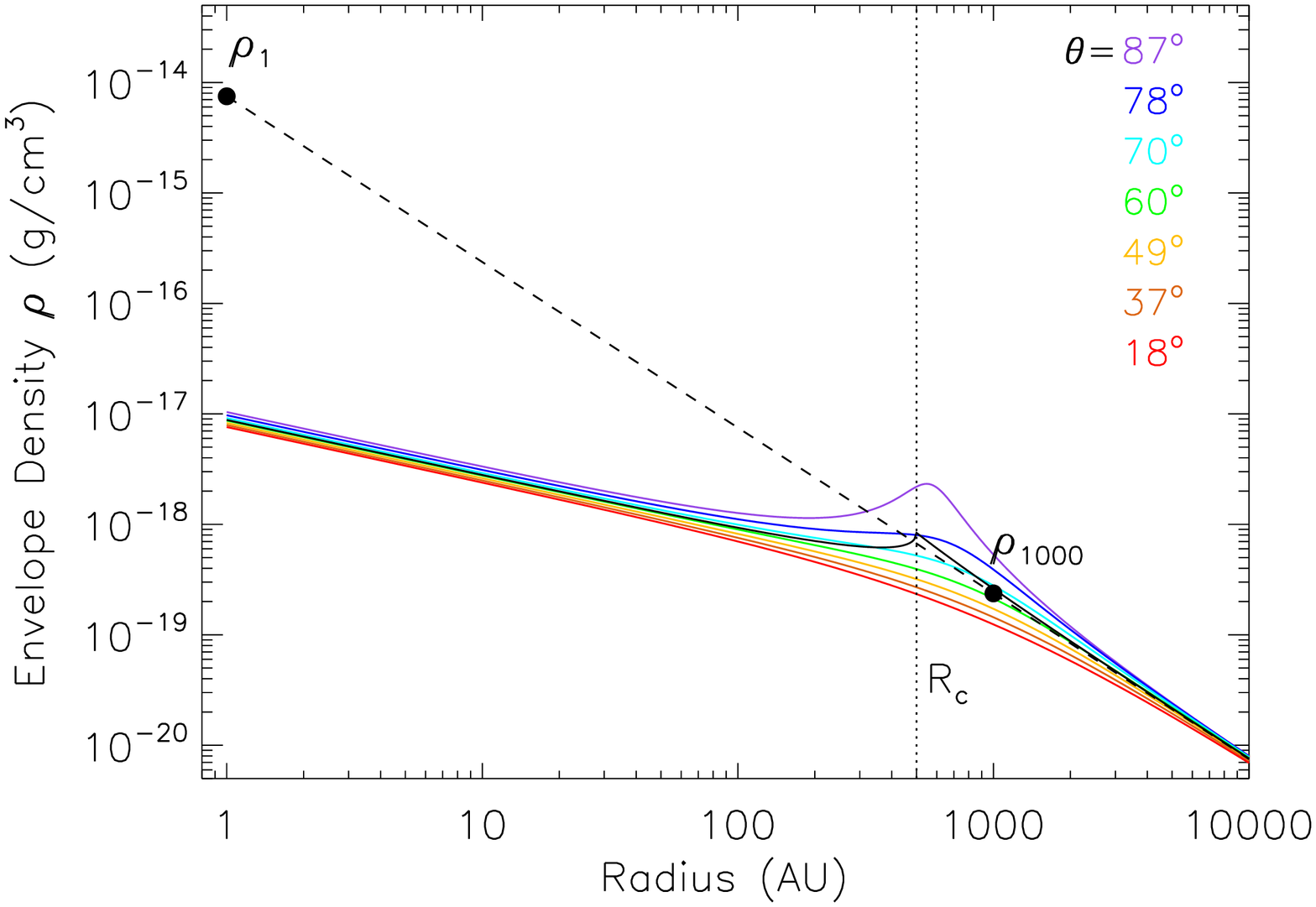}
\caption{Envelope density versus radius for a model protostar with
$\dot{M}_{env}=1.0 \times 10^{-6}$ \Msun\ yr$^{-1}$,
$M_{\ast}$=0.5 \Msun, and $R_c$=5 AU ({\it left}) and 500
AU ({\it right}) to show the difference between the reference densities
$\rho_1$ and $\rho_{1000}$. The lines with different colors represent 
radial density profiles for different polar angles $\theta$; the black line represents
the angle-averaged density profile (for equations see \citealt{whitney03a,
adams86}). The dashed line represents an $r^{-3/2}$ power law. The vertical
dotted line marks the location of the centrifugal radius.
\label{Rho_env_profiles}}
\end{figure*}

The largest number of parameter values in our grid is for the envelope 
infall rate. The envelope infall rate used as an input in the Whitney code 
sets the density of the envelope for a given mass of the protostar.  
Since the SED depends on the density of the envelope (and not directly 
on the infall rate, which is only inferred from the density and the acceleration 
due to gravity from the central protostar), in this work we report a  
reference envelope density instead of the envelope infall rate as one of our 
model parameters. For the TSC model, the envelope infall rate $\dot{M}_{env}$ 
and the reference density at 1 AU in the limit of no rotation ($R_c$=0) are 
related as follows \citep[see][]{kenyon93}:
\begin{equation}
\rho_1 = 5.318 \times 10^{-14} \left( \frac{\dot{M}_{env}}{10^{-5} 
M_{\odot} \, \mathrm{yr}^{-1}} \right) \left(\frac{M_{\ast}}
{1\, M_{\odot}}\right)^{-1/2} \mathrm{g}\, \mathrm{cm}^{-3},
\end{equation} 
where $M_{\ast}$ is the mass of the central protostar, which is assumed 
to be 0.5 \Msun\ in our model grid. The density distribution in the
envelope follows a power law, $\rho \propto r^{-3/2}$, at radii
larger than the centrifugal radius, $R_c$, but then flattens as a result
of the rotation of the envelope. The density reported by $\rho_1$ 
assumes a spherically symmetric envelope with a $-3/2$ power-law 
exponent valid down to the smallest radii, and it is higher than the 
angle-averaged density of a rotating envelope at 1 AU. To quote 
densities that are closer to actual values found in the modeled rotating 
envelopes (which have $R_c$ values ranging from 5 to 500 AU), we 
report $\rho_{1000}$, the density at 1000 AU for a $\rho \propto r^{-3/2}$ 
envelope with a 0.5 \Msun\ protostar:
\begin{eqnarray}
\rho_{1000} & = & \rho_1 \left(\frac{1}{1000}\right)^{3/2} \nonumber \\
& = & 2.378 \times 10^{-18} \left( 
\frac{\dot{M}_{env}}{10^{-5} M_{\odot} \, yr^{-1}} \right) 
\mathrm{g}\, \mathrm{cm}^{-3}.
\end{eqnarray} 
Thus, the range of reference densities probed in our model grid,
from $1.2 \times 10^{-20}$ to $1.8 \times 10^{-16}$ g cm$^{-3}$ 
(see Table \ref{modelpars}), would correspond to envelope
infall rates from $5.0 \times 10^{-8}$ to $7.5 \times 10^{-4}$ \Msun\ yr$^{-1}$,
assuming $M_{\ast}$=0.5 \Msun\ (this does not account for a reduction 
of the infalling mass due to clearing by outflow cavities).
In Figure \ref{Rho_env_profiles}, we show the radial density profiles
for two TSC models with 5 AU and 500 AU centrifugal radii. The density 
profiles are azimuthally symmetric and show the flattening of the density 
distribution inside $R_c$ due to envelope rotation. These plots demonstrate
that the density $\rho_1$ is much higher than the angle-averaged 
density at 1 AU; $\rho_{1000}$ seems to yield more physical values for 
the density in the envelope at 1000 AU, even for $R_c$ values of 500 AU.
 
As can be seen from the values of the envelope density in Table \ref{modelpars}, 
there is one set of models with an envelope density of 0. These are models 
that do not contain an envelope component; the entire excess emission 
is caused by the circumstellar disk. If an object is best fit by such a model, 
it would indicate that it is more evolved, having already dispersed its envelope.

\begin{figure}[!t]
\centering
\includegraphics[scale=0.55]{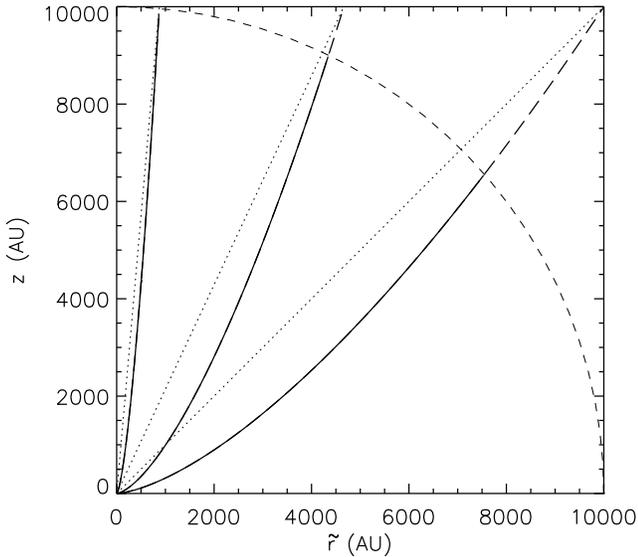}
\caption{Schematic showing the shape of the cavity assumed in our models
for three cavity opening angles $\theta$: 5\degr, 25\degr, and 45\degr\ 
(from left to right). The cavity walls are defined as a polynomial with exponent 
1.5 ($z \propto \tilde r^{1.5}$), with $\tilde r_{max} = z_{max}\, \tan\theta$, and 
are shown as solid lines. The outer envelope radius ($R_{env}$) at 10,000 AU 
is shown with a short-dashed line. The dotted lines show a different definition 
of the cavity size, where $\tilde r_{max} = R_{env} \sin\theta$ 
and $z_{max} = R_{env} \cos\theta$.
\label{cavity_shape}}
\end{figure}

The cavities in our models range from 5\degr\ to 45\degr\ and are defined
such that $z \propto \tilde r^{1.5}$, where $\tilde r$ and $z$ are the cylindrical
coordinates for the radial and vertical direction, respectively, and $\tilde r_{max} 
= z_{max}\, \tan\theta$, with $\theta$ defined as the cavity opening angle that 
is specified in the parameter file of the Whitney radiative transfer code. In this 
code, $z_{max}$ is set to the envelope outer radius. Thus, a polynomial-shaped 
cavity, which is wider at smaller $\tilde r$ values and then converges toward the 
specified opening angle, is somewhat larger than this opening angle at the outer 
envelope radius (see Figure \ref{cavity_shape}). This effect is most noticeable at 
larger cavity opening angles, but negligible for small cavities. A different definition 
of the cavity size, where $\tilde r_{max} = R_{env} \sin\theta$ and 
$z_{max} = R_{env} \cos\theta$ (with $R_{env}$ as the envelope outer radius), 
results in $z$ values that are a factor of $1/\cos\theta$ larger, and thus the cavity 
reaches the specified opening angle at the outer envelope radius. For this 
work, the adopted definition of the cavity opening angle is inconsequential, 
but it becomes relevant when comparing the results of SED modeling to 
scattered light images that reveal the actual cavity shape and size. 
We also note that in our models the cavities are evacuated of material,
so there is no dust and gas inside the cavity; in reality, there might be
some low-density material left that would add to the scattered light
\citep[see][]{fischer14}.

\begin{figure}[!t]
\centering
\includegraphics[scale=0.52]{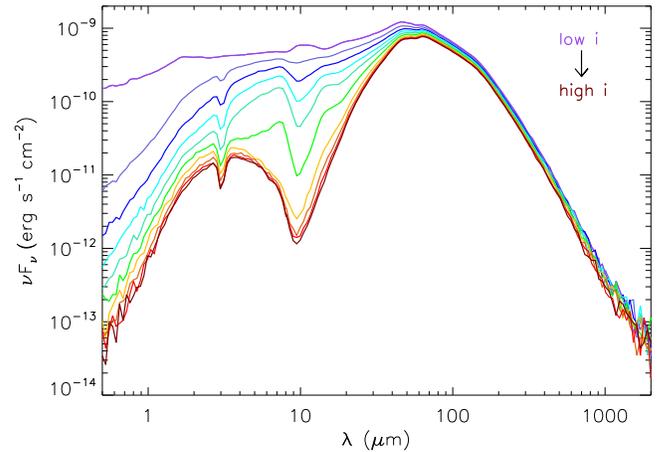}
\caption{A model from the grid seen at 10 different inclination
angles to illustrate the effect of viewing angle on the SED. 
The model has $L_{tot}$=10.1 \Lsun, $R_c$=50 AU, 
$\rho_{1000}$=$1.2 \times 10^{-18}$ g cm$^{-3}$, $\theta$=15\degr, 
and is seen at inclination angles 18\degr, 32\degr, 41\degr, 49\degr, 
57\degr, 63\degr, 69\degr, 76\degr, 81\degr, and 87\degr\ (from top to bottom). 
\label{Models_inc}}
\end{figure}

\begin{figure}[!t]
\centering
\includegraphics[scale=0.52]{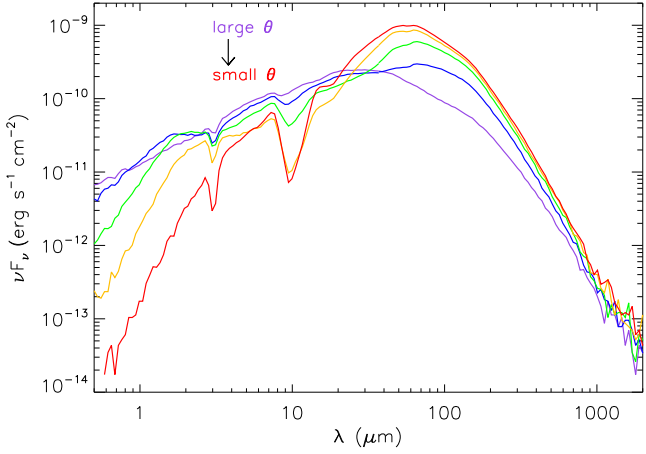}
\caption{Models from the grid to illustrate the effect of cavity opening
angle on the SED. The models have $L_{tot}$=10.1 \Lsun, 
$R_c$=50 AU, $\rho_{1000}$=$1.2 \times 10^{-18}$ g cm$^{-3}$, i=63\degr, 
but each has a different cavity opening angle: 5\degr\ ({\it red}),
15\degr\ ({\it yellow}), 25\degr\ ({\it green}), 35\degr\ 
({\it blue}), 45\degr\ ({\it purple}).
\label{Models_cavity}}
\end{figure}

\begin{figure}[!t]
\centering
\includegraphics[scale=0.52]{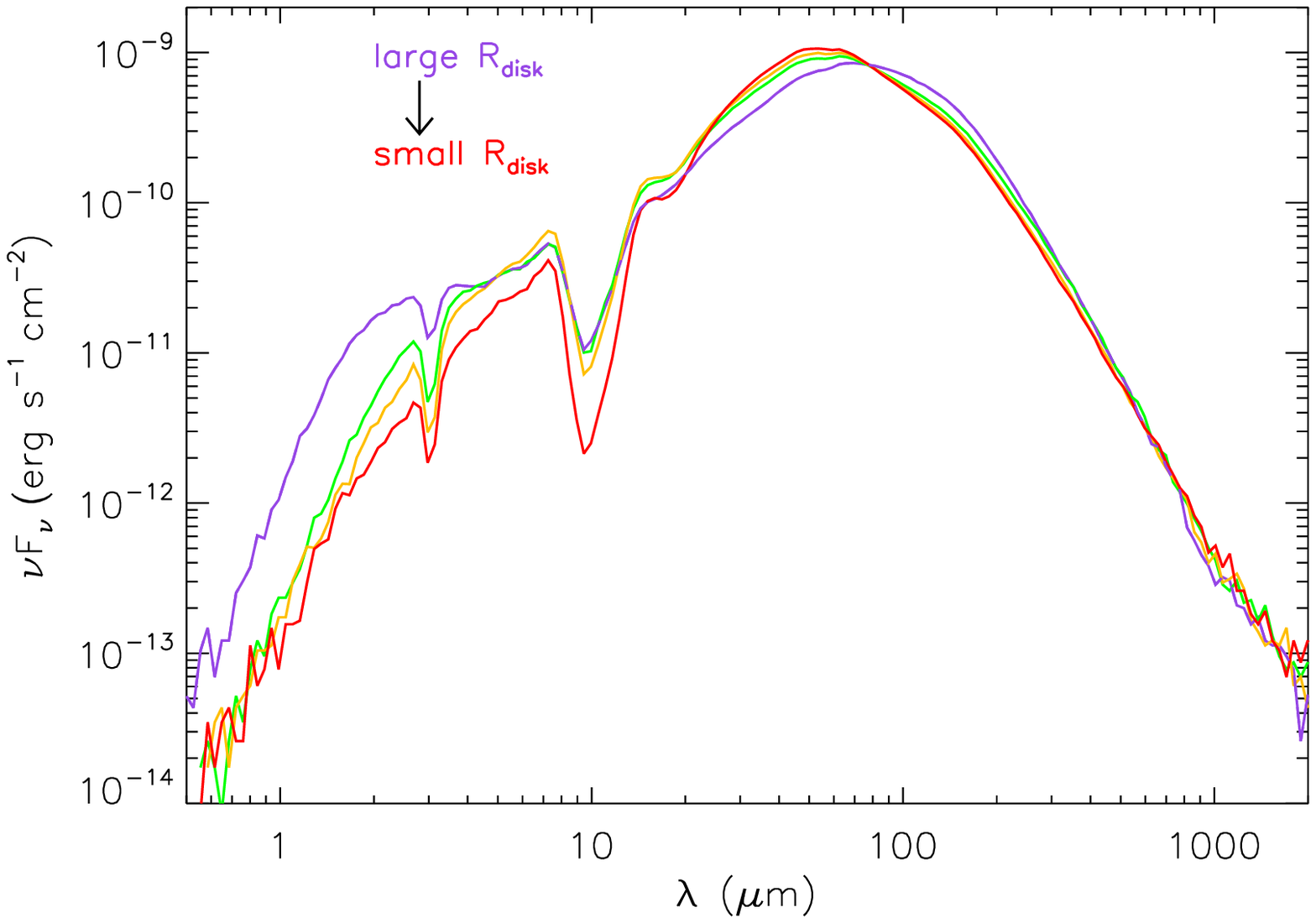}
\caption{Models from the grid to illustrate the effect of the centrifugal radius
($=R_{disk}$) on the SED. The models have $L_{tot}$=10.1 \Lsun, 
$\rho_{1000}$=$1.2 \times 10^{-18}$ g cm$^{-3}$,  $\theta$=5\degr, i=63\degr, 
but different disk radii: 5 AU ({\it red}), 50 AU ({\it yellow}), 100 AU ({\it green}), 
500 AU ({\it purple}).
\label{Models_Rdisk}}
\end{figure}

\begin{figure}[!t]
\centering
\includegraphics[scale=0.52]{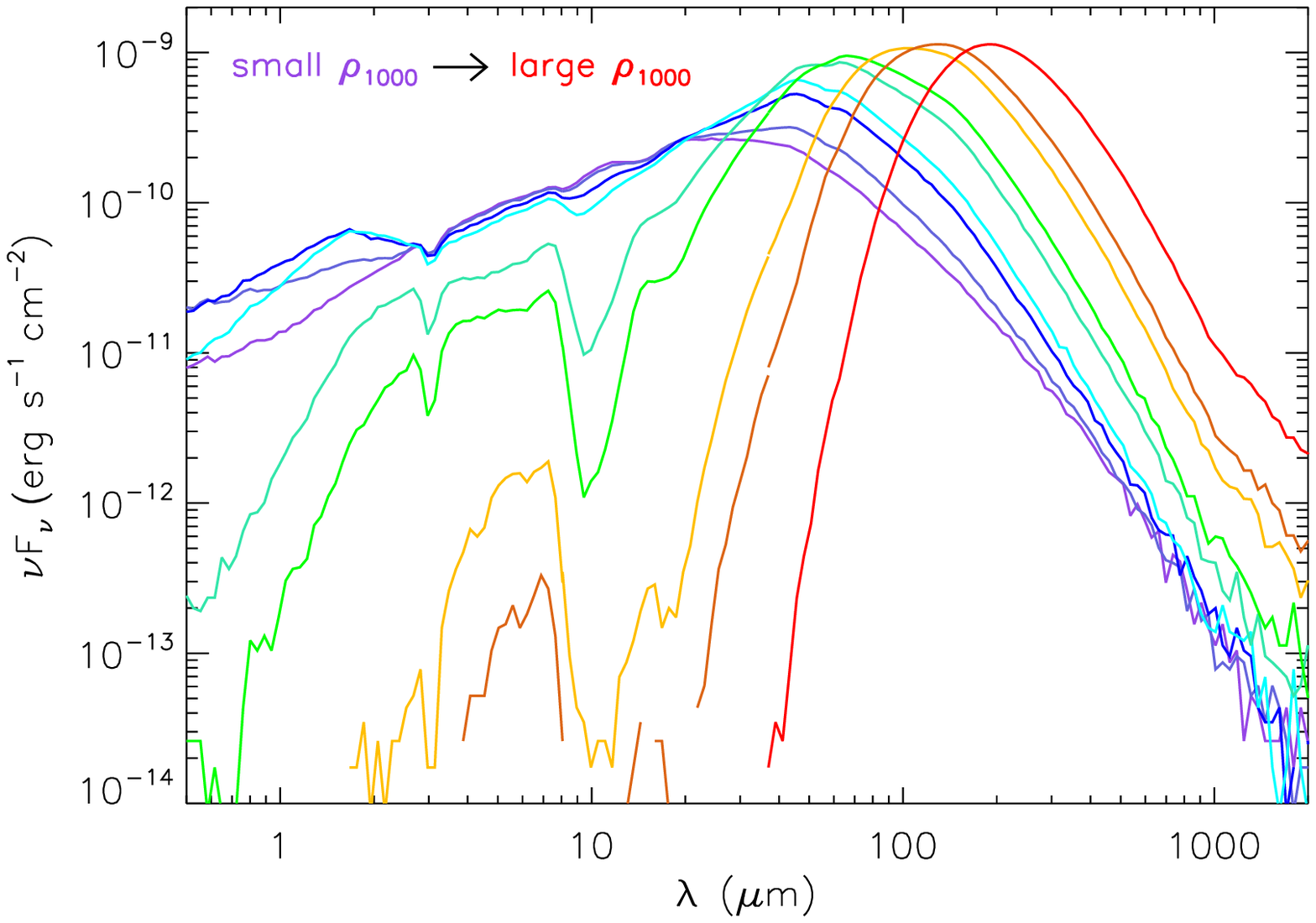}
\caption{Models from the grid to illustrate the effect of envelope
density on the SED. The models have $L_{tot}$=10.1 \Lsun, 
$R_c$=50 AU, $\theta$=15\degr, i=63\degr, but different reference 
densities $\rho_{1000}$:
0, $2.4 \times 10^{-20}$, $1.2 \times 10^{-19}$, $2.4 \times 10^{-19}$, 
$1.2 \times 10^{-18}$, $2.4 \times 10^{-18}$, $1.2 \times 10^{-17}$, 
$2.4 \times 10^{-17}$,  and $1.2 \times 10^{-16}$ g cm$^{-3}$ (the 
peak of the SED moves to longer wavelengths as $\rho_{1000}$ 
increases).
\label{Models_density}}
\end{figure}

\begin{figure}[!t]
\centering
\includegraphics[scale=0.52]{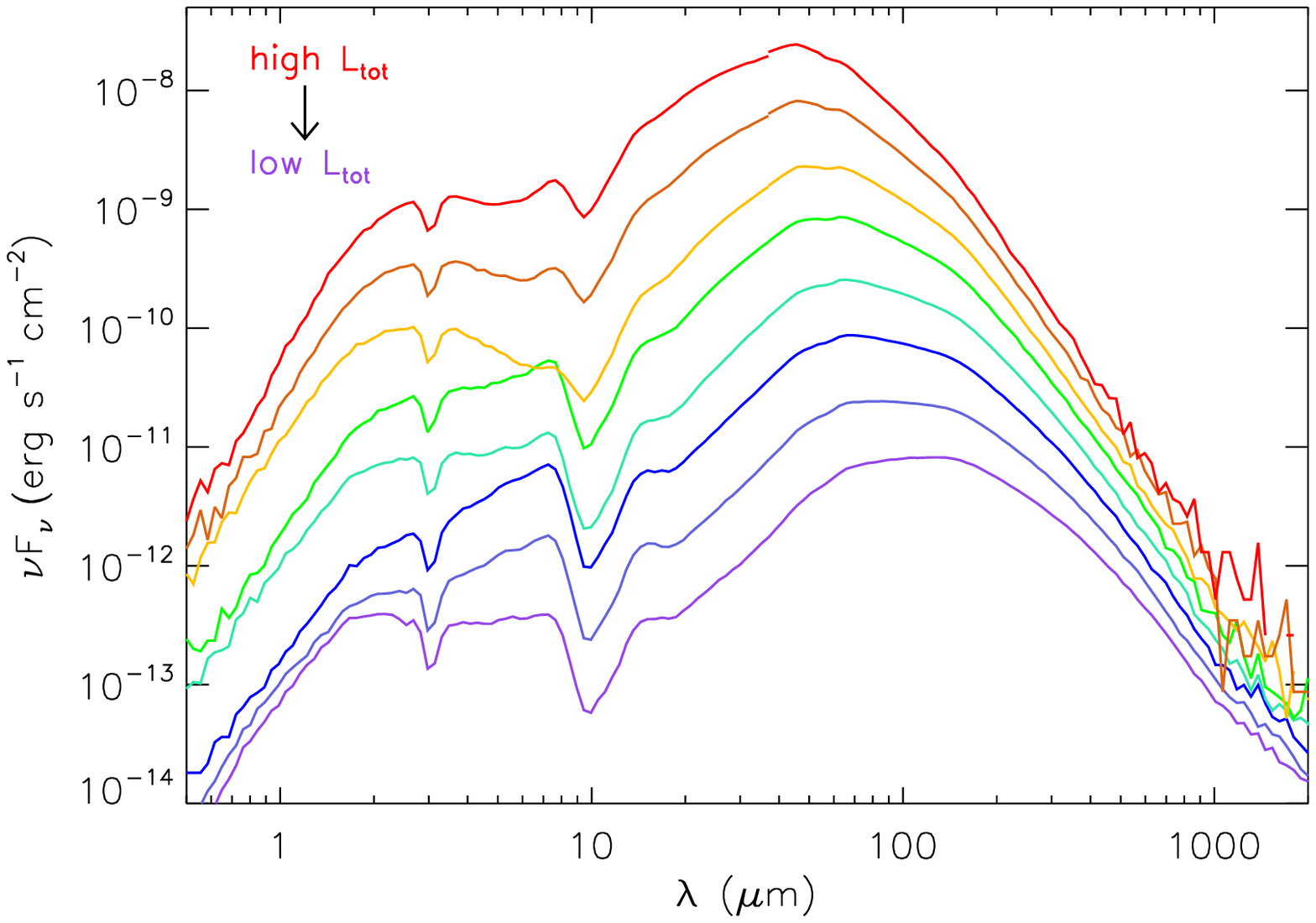}
\caption{Models from the grid to illustrate the effect of the total
luminosity on the SED. The models have $R_c$=50 AU, $\rho_{1000}$=
$1.2 \times 10^{-18}$ g cm$^{-3}$, $\theta$=15\degr, i=63\degr, but 
different values for the total luminosity: 0.1, 0.3, 1.0, 3.1, 10.1, 30.2, 101, 
and 303 \Lsun\ (from bottom to top).
\label{Models_Ltot}}
\end{figure}

Figures \ref{Models_inc} to \ref{Models_Ltot} display a few examples
of model SEDs from our grid to show the effect of changing those model 
parameters that influence the resulting SED the most. 
The inclination angle has a strong effect on the near- and mid-infrared SED
(Figure \ref{Models_inc}). While a low inclination angle results in an overall 
flat SED in this wavelength region, increasing the inclination angle causes 
a deeper silicate absorption feature at 10 $\mu$m and a steep slope 
beyond it. The far-infrared to millimeter SED is not affected by the 
inclination angle, since emission at these wavelengths does not suffer 
from extinction through the envelope.

The cavity opening angle affects the SED shape at all wavelengths (Figure
\ref{Models_cavity}). A small cavity only minimally alters the SED compared
to a case without a cavity; there is still a deep silicate absorption at 10 $\mu$m 
and steep SED slope, but the cavity allows some scattered light to escape in
the near-IR. A larger cavity results in higher emission at near- and mid-infrared 
wavelengths and reduced emission in the far-infrared. 
The effect of the cavity on the SED would change if a different shape
for the cavity walls were adopted. For example, cavities where the outer
wall follows the streamlines of the infalling gas and dust evacuate less inner 
envelope material than our polynomial-shaped cavities, resulting in deeper 
silicate absorption features and steeper mid-infrared SED slopes for the 
same cavity opening angle \citep[see][]{furlan14}. Thus, our cavity 
opening angles are tied to our assumed cavity shape.

The effect of the centrifugal radius is somewhat similar to those of the cavity
opening angle and inclination angle, but less pronounced (Figure 
\ref{Models_Rdisk}). Small disk radii imply more slowly rotating, less flattened 
envelopes and depress the near- and mid-infrared fluxes more than larger 
disk radii, but even with large disk radii (and more flattened envelopes) there 
is still sufficient envelope material along the line of sight to cause a pronounced 
10 $\mu$m absorption feature. Overall, our models do not directly constrain the 
size of the disk; the opacity is dominated by the envelope. Furthermore, the 
flattening of the envelope that is determined by $R_c$ has a similar effect on 
the SED as changing the outflow cavity opening angle.  

Changing the envelope density causes shifts in the SED in terms of both
wavelength and flux level: the higher the envelope density, the less 
flux is emitted at shorter wavelengths, and the more the peak of the 
SED shifts to longer wavelengths (Figure \ref{Models_density}). Deeply 
embedded protostars have SEDs that peak at $\lambda >$ 100 $\mu$m, 
steep mid-IR SED slopes, and deep silicate absorption features.
The effect of the envelope density on the SED is different from that of the
inclination angle, especially in the far-IR: while the SED is not very
sensitive to the inclination angle in this wavelength region, the ratio of, 
e.g., 70 and 160 $\mu$m fluxes changes considerably depending on 
the envelope density. 

The total luminosity of the source has an effect on the overall emission
level of the protostar, but does not strongly affect the SED shape. The
main effect is that the peak of the SED shifts to longer wavelengths as 
the luminosity decreases ($\lambda_{peak} \propto L^{-1/12}$; 
\citealt{kenyon93}). Especially when comparing models with $L_{tot}$ 
values that differ by a factor of a few, the SED shapes are similar 
(Figure \ref{Models_Ltot}). Thus, one could scale a particular model by 
a factor between $\sim$ 0.5 and 2 and get a good representation of a 
protostar that is somewhat fainter or brighter, without having to rerun 
the model calculation with the different input luminosity.

\subsection{Model Apertures}
\label{model_ap}

The model fluxes are computed for 24 different apertures, ranging from 
420 to 10,080 AU in steps of 420 AU (which corresponds to 1\arcsec\ at
the assumed distance of 420 pc to the Orion star-forming complex). 
For these SED fluxes, no convolution with a PSF is done, and therefore
the spatial distribution of the flux is solely due to the extended nature 
of protostars. Since the envelope outer radius is chosen to be 10,000 AU, 
the largest aperture encompasses the entire flux emitted by each protostellar
system. However, most of the near- and mid-infrared emission comes 
from smaller spatial scales, so an aperture of about 5000 AU will already
capture most of the flux emitted at these wavelengths.  

For a more accurate comparison of observed and model fluxes, in each 
infrared photometric band where we have data available, we interpolate 
model fluxes from the two apertures that bracket the aperture used in 
measuring the observed fluxes (4\arcsec\ for 2MASS, 2{\farcs}4
for IRAC, PSF photometry for MIPS 24 \micron, with a typical FWHM of
6\arcsec, 9{\farcs}6 for PACS 70 and 100 \micron, 12{\farcs}8 
for PACS 160 \micron). For the IRS data points, we use fluxes interpolated 
for a 5{\farcs}3 aperture, since the spectra are composed of two segments,
SL (5.2-14 \micron; slit width of 3{\farcs}6) and LL (14-38 \micron, slit 
width of 10{\farcs}5), and, if any flux mismatches were present, the SL 
segment was typically scaled to match the LL flux level at 14 \micron\ 
\citep[see, e.g.,][]{furlan08}. So, fluxes measured in an aperture with 
a radius of 5{\farcs}3 roughly correspond to fluxes from a 10{\farcs}6-wide 
slit.

\begin{figure}[!t]
\centering
\includegraphics[scale=0.39,angle=90]{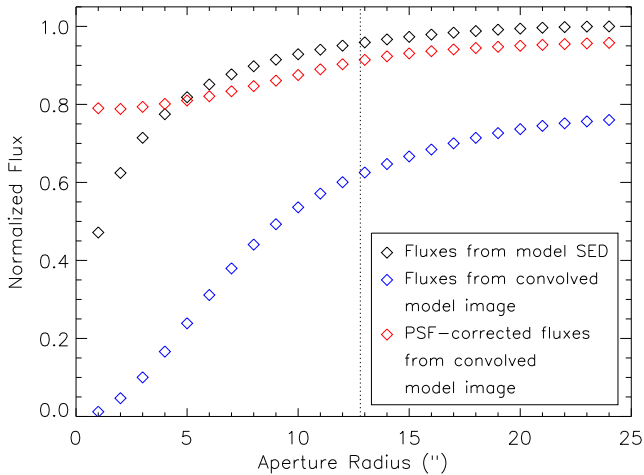}
\caption{PACS 160 $\mu$m fluxes versus aperture radius derived for 
a model ($L_{tot}=1.0$ \Lsun, $R_c=100$ AU, $\rho_{1000}=2.378 
\times 10^{-18}$ g cm$^{-3}$, $\theta$=15\degr, $i= 63$\degr) using 
different methods. 
The black symbols represent fluxes from the model SED, the blue symbols 
fluxes derived using aperture photometry on the model image convolved with 
the PACS 160 $\mu$m PSF, and the red symbols fluxes derived from the 
convolved model image and then corrected for PSF losses (see text for details). 
The maximum flux from the model SED was used to normalize all other fluxes.
The dotted line indicates an aperture radius of 12{\farcs}8.
\label{Model_fluxes_PACS160}}
\end{figure}

Given that our targets are typically extended and that the near- to mid-infrared 
data have relatively high spatial resolution, measuring fluxes in small apertures 
(a few arcseconds in radius) will truncate some of the object's flux, so it is
important to choose similar apertures for the model fluxes. 
From about 30 to 100 $\mu$m, the model fluxes calculated for smaller apertures 
are not very different from the total flux (i.e., the flux from the largest aperture),
which is a result of the emission profile in the envelope and the lower spatial 
resolution at longer wavelengths. 
To check whether extended source emission in the far-infrared might affect 
the flux we measure in our models, we calculated a small set of model
images at 160 $\mu$m, convolved them with the PACS 160 $\mu$m
PSF, and compared the fluxes from the model images to those written
out for the model SEDs (which we refer to as ``SED fluxes''; these are
the fluxes from the models in the grid).
Model images would be the most observationally consistent way to measure
the flux densities, but they are too computationally expensive and would not
represent a significant gain.

\begin{figure}[!t]
\centering
\includegraphics[scale=0.39,angle=90]{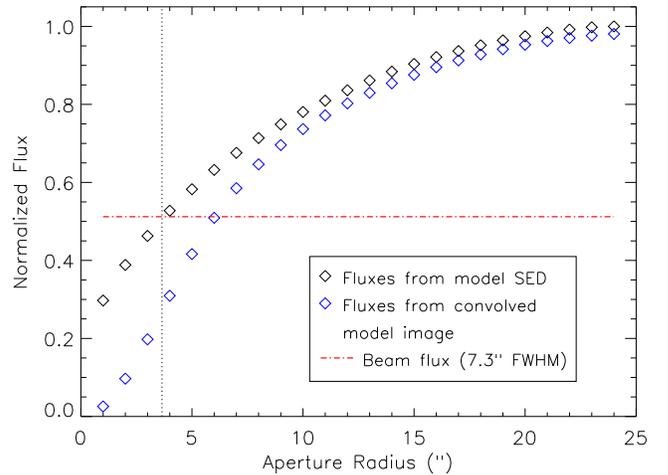}
\caption{SABOCA (350 $\mu$m) fluxes versus aperture radius derived 
for the same model as in Figure \ref{Model_fluxes_PACS160} using different
methods. The black symbols represent fluxes from the model SED, the 
blue symbols  fluxes derived using aperture photometry on the model 
image convolved with a Gaussian PSF, and the red dot-dashed line the 
beam flux (assuming a beam with a FWHM of 7{\farcs}3). The maximum 
flux from the model SED was used to normalize all other fluxes.
The dotted line indicates an aperture radius of 3{\farcs}65.
\label{Model_fluxes_SABOCA}}
\end{figure}

In Figure \ref{Model_fluxes_PACS160} we show the fluxes derived for
a particular model at 160 $\mu$m using different methods. The fluxes 
measured in the convolved model image are lower than the SED fluxes; this 
is caused by the wide PACS 160 $\mu$m PSF, which spreads flux to very
large radii. Since the shape of the PSF is known, we can correct for these 
PSF losses (assuming a point source and using standard aperture
corrections). The fluxes corrected for these PSF losses are very similar to the 
SED fluxes, typically within $\sim$ 5-10\% at apertures larger than 5\arcsec.
Since our observed fluxes correspond to these PSF-corrected fluxes (we apply 
aperture corrections to our fluxes measured in a 12{\farcs}8 aperture to 
account for PSF losses), adopting the SED fluxes from the largest aperture 
would yield model fluxes that are somewhat too high. Thus, we chose to adopt 
the SED flux measured in a 12{\farcs}8 aperture as a good approximation for 
the model flux we would get if we had model images available for all models in 
the grid and measured aperture-corrected fluxes in these images. We note that 
in our PACS data, the 160 $\mu$m sky annulus, which extends from 12{\farcs}8 
to 25{\farcs}6 (see B. Ali et al. 2016, in preparation), can include extended emission
from surrounding material and also some envelope emission. In these cases, we 
often used PSF photometry to minimize contamination from nearby sources and 
nebulosity; however, PSF fitting was not used for more isolated sources since 
the envelopes can be marginally resolved at 160 $\mu$m and thus deviate 
slightly from the adopted PSF shape.

\begin{figure}[!t]
\centering
\includegraphics[scale=0.39,angle=90]{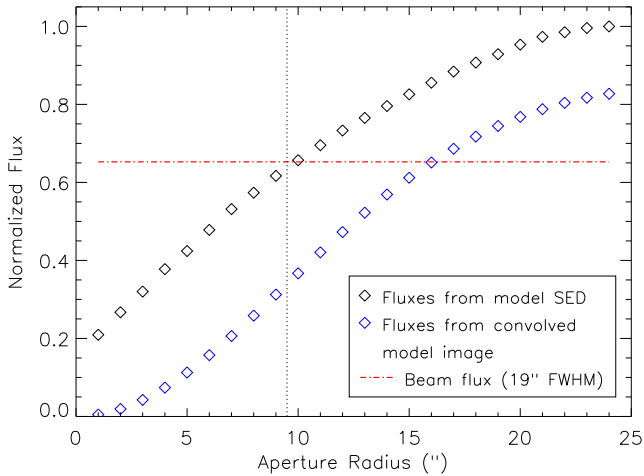}
\caption{Similar to Figure \ref{Model_fluxes_SABOCA}, but for the
LABOCA (870 $\mu$m) fluxes. The dotted line indicates an aperture 
radius of 9{\farcs}5.
\label{Model_fluxes_LABOCA}}
\end{figure}

For the SABOCA and LABOCA data, beam fluxes were adopted; the
FWHM of the SABOCA beam is 7{\farcs}3, while for the LABOCA
beam it is 19\arcsec. In order to determine which aperture radius
corresponds best to beam fluxes, we created a similar set of model 
images as above at 350 and 870 $\mu$m, convolved them with 
Gaussian PSFs, and measured fluxes in the model images using 
different apertures (see Figures \ref{Model_fluxes_SABOCA} and
\ref{Model_fluxes_LABOCA}, where we show the results for one
model). Fluxes measured in the convolved model image are smaller 
than the SED fluxes, especially at aperture radii smaller than the 
FWHM of the beam. We find that the beam fluxes for SABOCA and 
LABOCA are best matched by SED fluxes from apertures with radii 
half the size of the FWHM of the beam, i.e., 3{\farcs}65 for 
SABOCA and 9{\farcs}5 for LABOCA (thus, the aperture sizes are
the same as the beam FWHM). 
This is again an idealized situation, since the measured SABOCA
and LABOCA beam fluxes also include extended emission (if the
source lies on top of background emission), and thus they could be 
higher than those from the model.

\subsection{Effect of External Heating}
\label{model_ext_heat}

\begin{figure*}[!t]
\centering
\includegraphics[scale=0.65,angle=90]{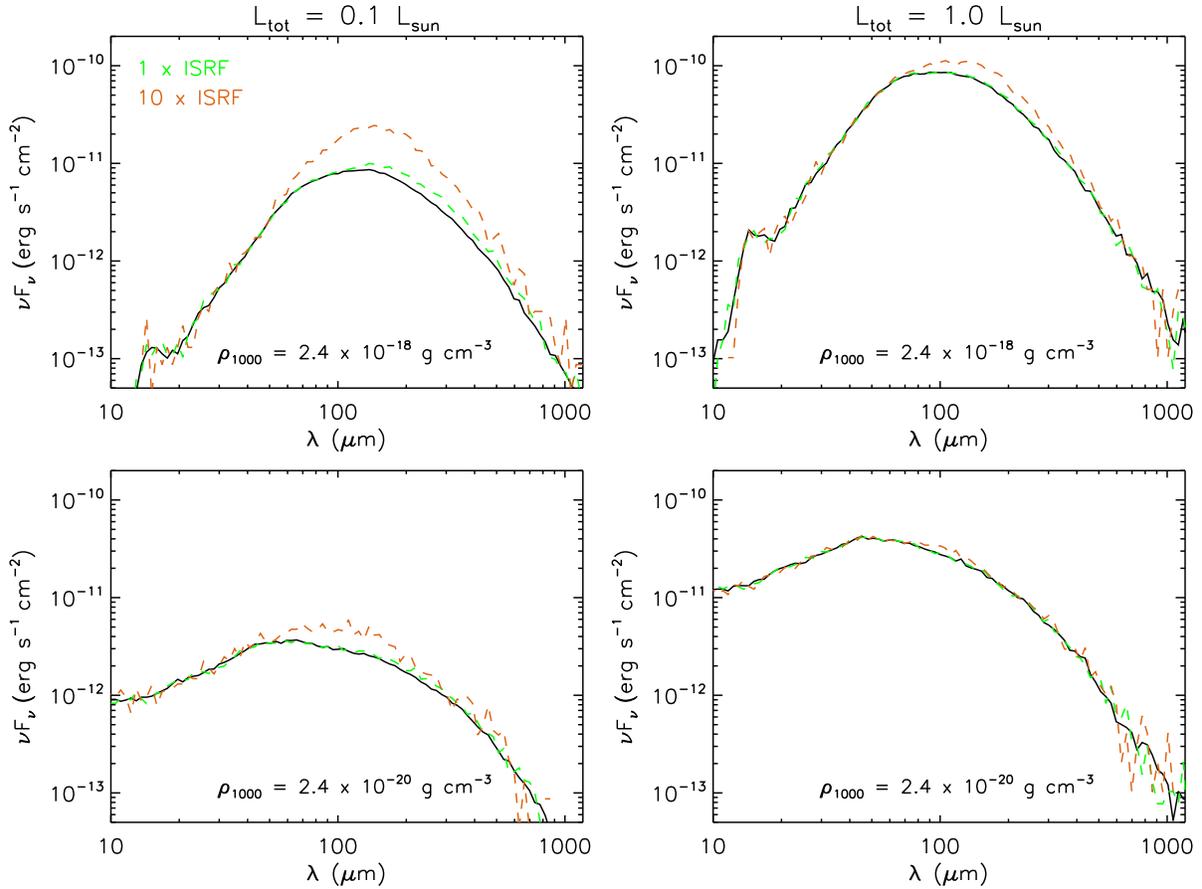}
\caption{{\it Left:} Comparison of models with $L_{tot}$=0.1 \Lsun, $R_c$=100 AU, 
$\theta$=15\degr, $\rho_{1000}$=2.4 $\times 10^{-18}$ g cm$^{-3}$ ({\it top}) or
2.4 $\times 10^{-20}$ g cm$^{-3}$ ({\it bottom}), $i$=63\degr, without external heating 
({\it black}), with external heating by an ISRF equal to that in the solar neighborhood
({\it green, dashed line}), and with heating by an ISRF 10 times stronger ({\it orange, 
dashed line}). {\it Right:} Similar to the models in the left panels, but these models have 
$L_{tot}$=1.0 \Lsun.
\label{models_ext_heat}}
\end{figure*}

In our models, the luminosity is determined by the central protostar and
the accretion; no external heating is included. The interstellar radiation
field (ISRF) could increase the temperature in the outer envelope regions,
thus causing an increase in the longer-wavelength fluxes \citep[e.g.,][]
{evans01,shirley02,young03}.
It is expected that external heating has a noticeable effect only on 
low-luminosity sources ($\lesssim$ 1 \Lsun), while objects with strong 
internal heating are not affected by the ISRF. Moreover, the strength of 
the ISRF varies spatially \citep{mathis83}, and thus its effect on each 
individual protostar is uncertain. Nonetheless, in the following we estimate
the effect of external heating on model fluxes by using a different set of
models.
 
\begin{figure*}[!t]
\centering
\includegraphics[scale=0.7,angle=90]{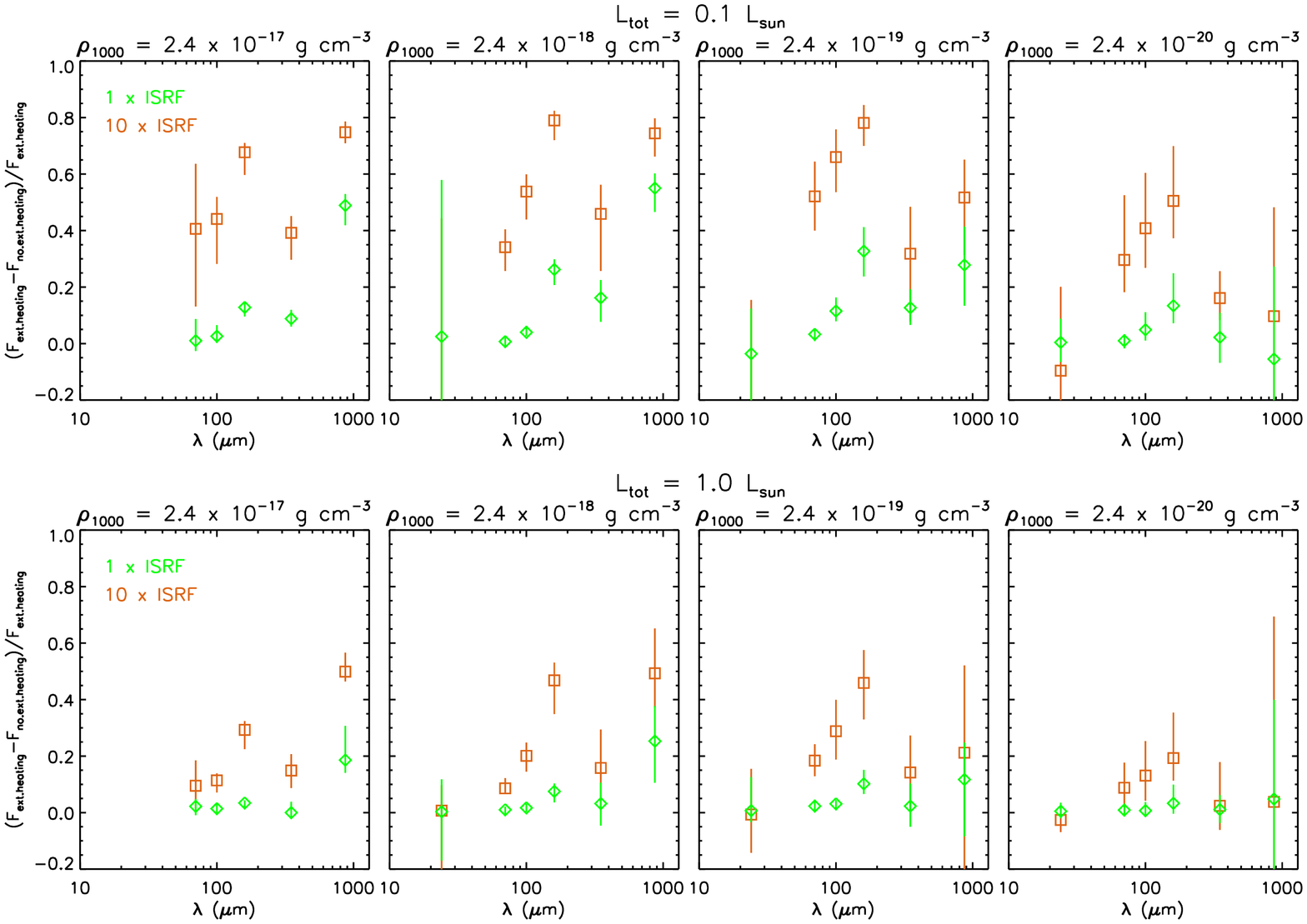}
\caption{Ratio of the excess emission due to external heating and the emission of 
the protostar with external heating in different bands, for heating by an ISRF
equal to that in the solar neighborhood ({\it green diamonds}) and by an ISRF 
10 times stronger ({\it orange squares}). The vertical lines show the range of
flux excess ratios resulting from different viewing angles (inclination angles range 
from 18\degr\ to 87\degr), while the symbols represent mean values. The top (bottom) 
panels are for models with $L_{tot}$=0.1 (1.0) \Lsun. The four columns correspond 
to the four reference densities probed.
\label{models_ext_heat_ratios}}
\end{figure*}

For this model calculation, we used the 2012 version of the Whitney radiative 
transfer code \citep{whitney13}, which allows for the inclusion of external illumination 
by using the ISRF value in the solar neighborhood from \citet{mathis83}; to vary 
the ISRF strength, the adopted value can be scaled by a multiplicative factor 
and extinguished by a certain amount of foreground extinction. 
We calculated a small number of models with and without external heating
and then compared their far-infrared and submillimeter fluxes. One set of models 
has $L_{tot}$=0.1 \Lsun, $R_c$=100 AU, $\theta$=15\degr, and four different 
reference densities $\rho_{1000}$, ranging from 2.4 $\times 10^{-17}$ g cm$^{-3}$ 
to 2.4 $\times 10^{-20}$ g cm$^{-3}$. The other set has the same parameters
except for $L_{tot}$, which is 1.0 \Lsun. We calculated models without external 
heating, with heating from an ISRF equal to that in the solar neighborhood, and 
with ISRF heating 10 times the solar neighborhood value. For these models, we 
did not include any foreground extinction for the ISRF; thus, the ISRF heating in 
these models can be considered an upper limit -- especially the 10-fold increase 
over the ISRF in the solar neighborhood represents an extreme value.
Figure \ref{models_ext_heat} shows a few examples of model SEDs with and
without external heating. External heating results in flux increases in the far-IR
and sub-mm; as expected, it affects low-luminosity sources more, and its effects 
are also more noticeable for higher-density envelopes.

For a more quantitative comparison of model fluxes in the far-IR and sub-mm,
we computed the fluxes for each model in six different bands, those of MIPS 
24 $\mu$m, PACS 70, 100, and 160 $\mu$m, and SABOCA (350 $\mu$m) 
and LABOCA (870 $\mu$m), using apertures as described in section 
\ref{model_ap}. The model fluxes are affected by poorer signal-to-noise 
ratios at the longest wavelengths, so the 870 $\mu$m fluxes are less reliable.
We subtracted the fluxes of the models without external heating ($F_{\rm no.ext.heating}$)
from those with external heating ($F_{\rm ext.heating}$) to determine the flux 
excess due to external heating. The ratios of these excess fluxes and the model fluxes 
with external heating ($(F_{\rm ext.heating}-F_{\rm no.ext.heating})/F_{\rm ext.heating}$)
are shown in Figure \ref{models_ext_heat_ratios}. Given that these ratios depend 
on the inclination angle to the line of sight, we show them as average values for all
10 inclination angles as well as the range subtended by all inclination angles.
We note overall smaller flux ratios at 350 $\mu$m due to the smaller aperture 
size chosen in this wave band (see section \ref{model_ap}).

\begin{figure*}[!t]
\centering
\includegraphics[scale=0.65,angle=90]{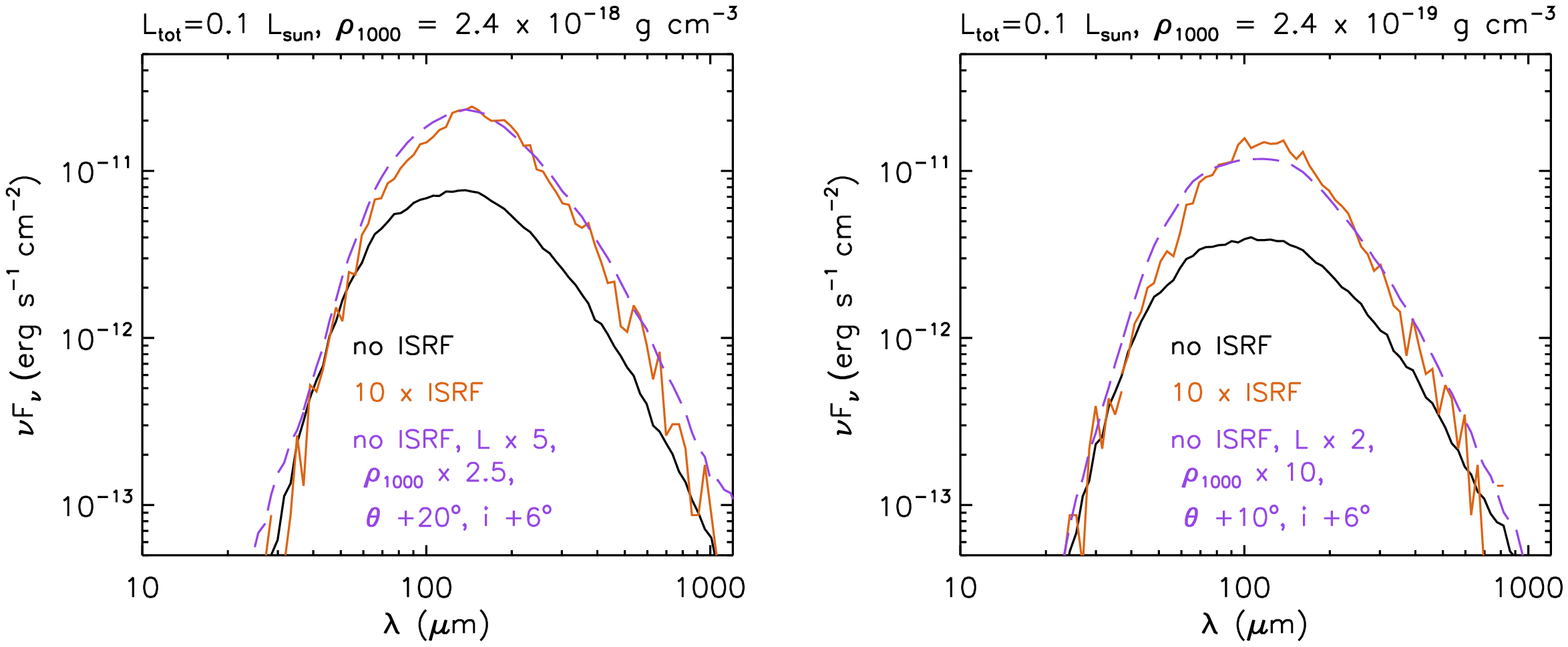}  
\caption{{\it Black and orange lines:} SEDs for models with $L_{tot}$=0.1 \Lsun, 
$R_c$=100 AU,  $\theta$=15\degr, $i$=75\degr, reference densities 
$\rho_{1000}$=2.4 $\times 10^{-18}$ g cm$^{-3}$ ({\it left}) and 2.4 $\times 10^{-19}$ 
g cm$^{-3}$ ({\it right}), without external heating ({\it black}) and with heating by an ISRF 
scaled by a factor of 10 ({\it orange}). The purple dashed lines show SEDs from our 
model grid (which does not include external heating) with model parameters
changed as indicated in the figure label; these models were chosen to closely 
match the model SEDs with external heating.
\label{ext_heat_SED1}}
\end{figure*}

\begin{figure*}[!t]
\centering
\includegraphics[scale=0.65,angle=90]{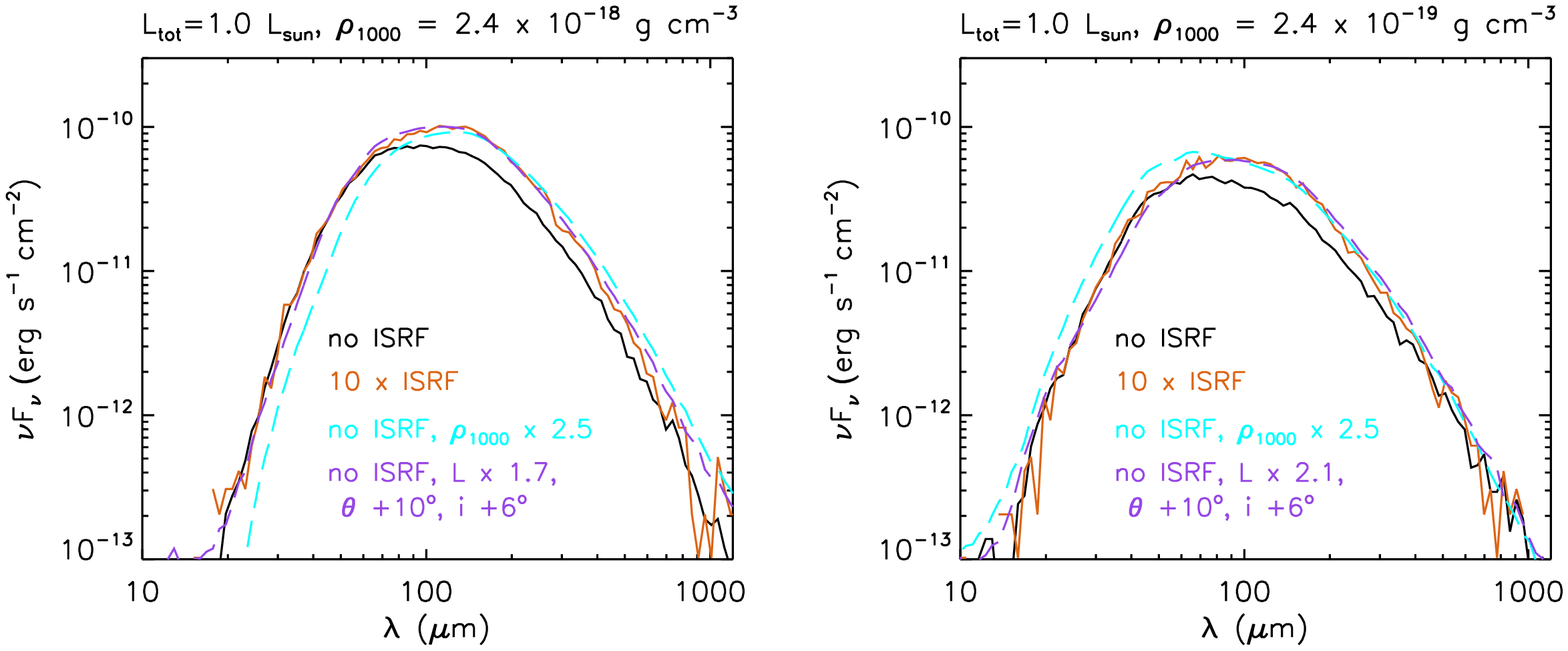}  
\caption{Similar to Figure \ref{ext_heat_SED1}, but for model SEDs with
$L_{tot}$=1.0 \Lsun\ ({\it black and orange lines}). The light blue and purple 
dashed lines show SEDs from our model grid (no external heating) with the same 
model parameters as shown except for a reference density 2.5 times higher 
({\it light blue}) and $\theta$=25\degr, $i=$81\degr, and a higher luminosity 
({\it purple}).
\label{ext_heat_SED2}}
\end{figure*}

Our analysis shows that heating by the ISRF results in flux increases in the far-IR 
and sub-mm that are about a factor of 2-3 higher for envelopes of low-luminosity 
sources ($L_{tot}$=0.1 \Lsun) than for those with higher luminosity. Also, the
effect of external heating is more noticeable at longer wavelengths (where 
apertures/beams are also larger) than at shorter ones; given our chosen apertures, 
the largest effect occurs at 160 and 870 $\mu$m. We also note that the flux increases 
due to heating by the ISRF are smallest for the lowest $\rho_{1000}$ value probed, 
2.4 $\times 10^{-20}$ g cm$^{-3}$; at 160 $\mu$m, the flux increase is largest for 
intermediate envelope densities. Finally, the flux increases in the far-IR and sub-mm 
are far larger for a solar-neighborhood ISRF scaled by factor of 10 than for an 
unscaled ISRF; for the $L_{tot}$=0.1 \Lsun\ models, an unscaled ISRF increases 
the fluxes from a few percent (at $\lesssim$ 100 $\mu$m) to 50\% (at 870 $\mu$m), 
while an ISRF scaled by a factor of 10 increases these fluxes by 30\%-75\%. Thus, 
for low-luminosity protostars, up to $\sim$ 75\% of a protostar's 870 $\mu$m flux 
could be due to external heating, if the environment is dominated by an extremely 
strong ISRF.

To estimate how the contribution of external heating would modify derived
model parameters, in Figures \ref{ext_heat_SED1} and \ref{ext_heat_SED2} 
we compare model SEDs that include external heating by an ISRF 10 times 
stronger than in the solar neighborhood and model SEDs without this additional 
heating. For the latter, we used models from our model grid and tried to reproduce
the SEDs with external heating. For the models with $L_{tot}$=0.1 \Lsun, the
effect of external heating can be reproduced by increasing the luminosity by
factors of a few, increasing $\rho_{1000}$ by up to an order of magnitude,
and increasing the cavity opening angle and inclination angle by a small
amount. For the $L_{tot}$=1.0 \Lsun\ models, just increasing the reference 
density by a factor of 2.5 results in a good match to the long-wavelength 
emission of our externally heated models; however, the shorter-wavelength 
flux is either under- or overestimated. A better match is achieved with models 
having the same reference density as the externally heated models, but with 
slightly larger cavity opening angles and inclination angles, and luminosities 
about a factor of 2 larger.
Thus, if the far-IR and sub-mm fluxes were contaminated by emission resulting 
from extremely strong external heating, a model fit using models from our grid 
(which does not include external heating) could overestimate the envelope density 
by up to an order of magnitude and the luminosity by a factor of 2-5. The cavity 
opening and inclination angles would also be more uncertain, but not by much.  
For a more realistic scenario with more modest external heating (which would 
also include the effect of local extinction), the effect on model parameters would
be smaller.

\begin{figure}[!t]
\centering
\includegraphics[scale=0.4,angle=90]{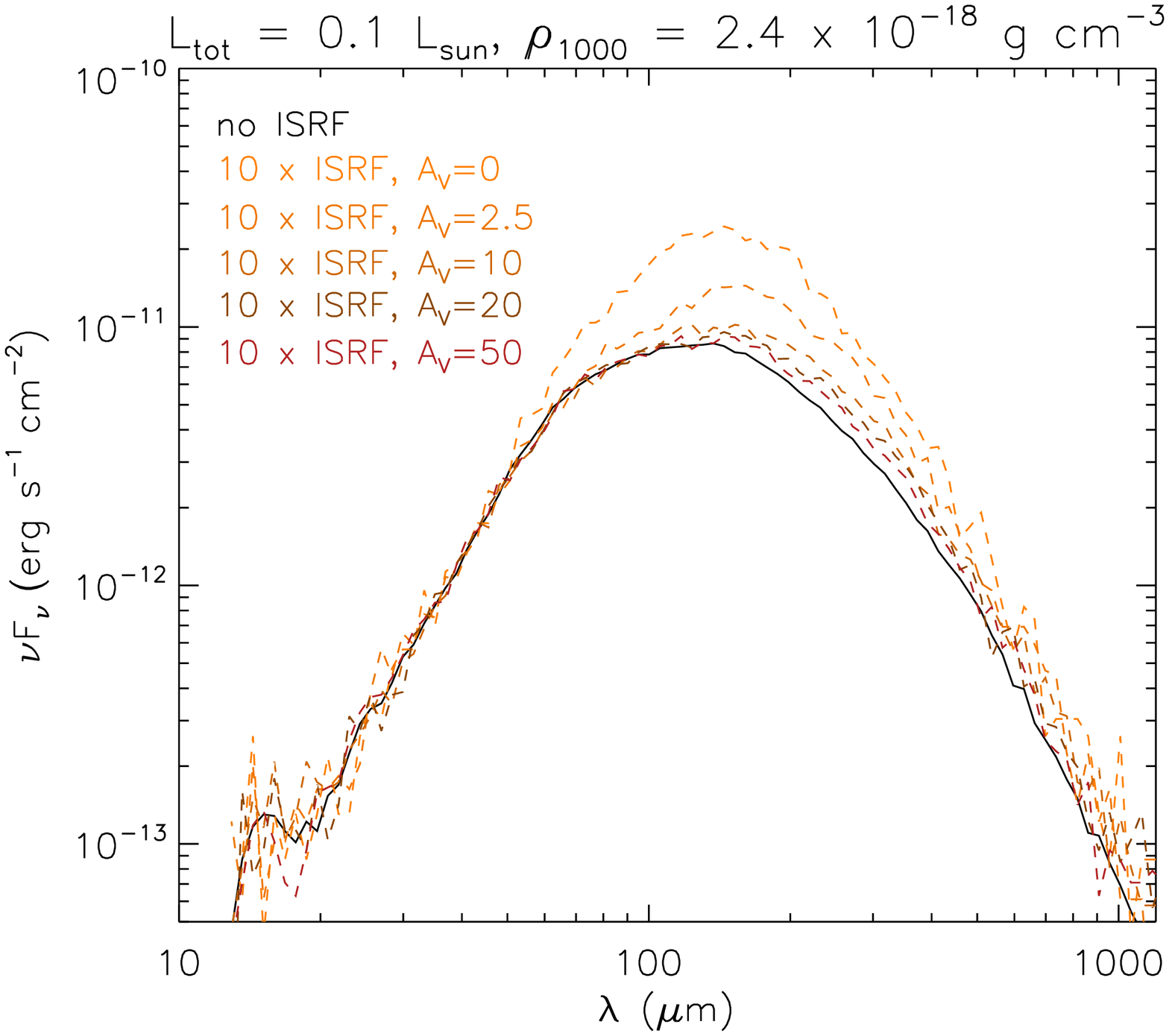}
\caption{Models with $L_{tot}$=0.1 \Lsun, $R_c$=100 AU, $\theta$=15\degr, 
$\rho_{1000}$=2.4 $\times 10^{-18}$ g cm$^{-3}$, $i$=63\degr, without external 
heating ({\it black}), with external heating by an ISRF 10 times stronger than in the 
solar neighborhood ({\it orange} to {\it brown}, {\it dashed lines}) and different 
amounts of extinction applied to the ISRF (from $A_V = 2.5$ to $A_V = 50$, 
{\it top} to {\it bottom}).
\label{models_ext_heat_Av}}
\end{figure}

For the latter point, we explored the effect of extinction on the ISRF by calculating 
a few more models with $L_{tot}$=0.1 \Lsun, $R_c$=100 AU, $\theta$=15\degr, 
$\rho_{1000} = 2.4 \times 10^{-18}$ g cm$^{-3}$, an ISRF 10 times stronger than 
that in the solar neighborhood, and $A_V$ values for the ISRF of 2.5, 10, 20, and 50. 
The model SEDs are shown in Figure \ref{models_ext_heat_Av}. Compared to ISRF
heating without any foreground extinction, already $A_V=2.5$ causes a decrease 
by a factor of 1.5-2 in the overall emission at far-IR wavelengths. With $A_V$ of
10 and 20, the far-IR emission decreases by factors of up to $\sim$ 3.5 and 4, 
respectively, compared to a strong ISRF that is not extinguished. The fraction of
excess emission due to external heating at 160 $\mu$m decreases from an
average of 0.8 for $A_V$=0 (see Figure \ref{models_ext_heat_ratios}) to 0.6,
0.3, and 0.2 for $A_V$=2.5, 10, and 20, respectively. Therefore, considering that 
typical $A_V$ values in Orion are $\sim$ 10-20 mag \citep{stutz15}, it is likely
that the effect of external heating on model parameters of low-luminosity sources
does not exceed a factor of $\sim$ 2 in luminosity and $\sim$ 5 in envelope
density.

\section{Fitting Method}
\label{method}

A customized fitting routine determines the best-fit model from the grid 
for each object in our sample of 330 YSOs (see Sections \ref{sample} 
and \ref{SEDs}) using both photometry and, where available, IRS spectroscopy.
Ideally, an object has 2MASS, IRAC, IRS, MIPS, PACS, and SABOCA and LABOCA 
data; in many cases, no submillimeter data are available, and in a few cases the 
object is too faint to be detected by 2MASS. Of the 330 modeled objects, 40 do 
not have IRS spectra. As a minimum, objects have some {\it Spitzer} photometry 
and a measured flux value in the PACS 70 $\mu$m band. No additional data 
from the literature were included in the fits to keep them homogeneous.

\begin{figure}[!]
\centering
\includegraphics[scale=0.63]{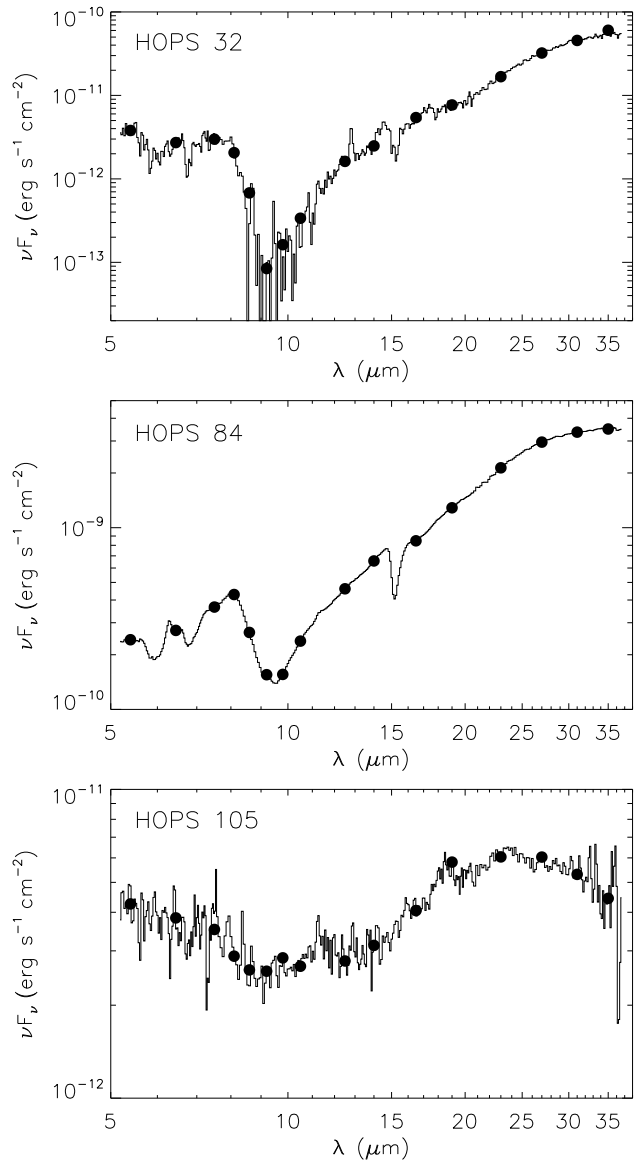}
\caption{Three IRS spectra, one for HOPS 32 (Class 0 protostar; {\it top}),
one for HOPS 84 (Class I protostar; {\it middle}), and one for HOPS 105
(flat-spectrum source; {\it bottom}), overlaid with the rebinned data points
({\it filled circles}) used by the fitting routine. Note the different flux ranges
on the y axis in the three panels and thus the big differences in slopes
among the three spectra.
\label{IRS_rebin}}
\end{figure}

In order to reduce the number of data points contained in the IRS spectral
wavelength range (such that the spectrum does not dominate over the photometry) 
and to exclude ice absorption features in the 5-8 \micron\ region and at 15.2 \micron\
that are usually observed, but not included in the model opacities, we rebin each 
IRS spectrum to fluxes at 16 wavelengths. These data points trace the continuum 
emission and the 10 and 20 \micron\ silicate features. Also, when rebinning the 
spectrum, we smooth over its noisy regions, and we scale the whole spectrum 
to match the MIPS 24 \micron\ flux if a similar deviation is also seen at the 
IRAC 5.8 and 8 \micron\ bands and is larger than 10\%. Figure \ref{IRS_rebin} 
shows three examples of our IRS spectra with the rebinned fluxes overplotted.
Our selection of 16 IRS data points in addition to at most 13 photometric points 
spread from 1.1 to 870 \micron\ puts more emphasis on the mid-IR spectral 
region in the fits. This wavelength region is better sampled by observations, 
most of the emission is thermal radiation from the protostellar envelope and
disk (as opposed to some possible inclusion of  scattered light or thermal 
emission from surrounding material at shorter and longer wavelengths, 
respectively), and it contains the 10 \micron\ silicate feature, which crucially 
constrains the SED fits. As a result, most models are expected to reproduce 
the mid-IR fluxes well and might fit more poorly in the near-IR and sub-mm.

To directly compare observed and model fluxes, we create model SEDs 
with data points that correspond to those obtained from observations,
from both photometry and IRS spectroscopy. For the former, the model 
fluxes are not only derived from the same apertures as the data (see 
section \ref{model_ap}), but also integrated over the various filter 
bandpasses, thus yielding model photometry. For the latter, the model 
fluxes are interpolated at the same 16 wavelength values as the IRS spectra.

Since the model grid contains a limited number of values for the total luminosity
(eight), but the objects we intend to fit have luminosities that likely do not
correspond precisely to these values, we include scaling factors for the luminosity 
when determining the best-fit model. As long as these scaling factors are not far
from unity, they are expected to yield SEDs that are very similar to those obtained 
from models using the scaled luminosity value as one of the input parameters.
The scaling factor can also be related to the distance of the source; for all model 
fluxes, a distance of 420 pc is assumed, but in reality the protostars in our sample
span a certain (presumably small) range of distances along the line of sight.
For example, a 10\% change in distance would result in a $\sim$ 20\% change 
in flux values (scaling factors of 0.83 or 1.23). Here we report luminosities 
assuming a distance of 420 pc.

In addition to scaling factors, each model SED can be extinguished to account
for interstellar extinction along the line of sight. We use two foreground extinction 
laws from \citet{mcclure09} that were derived for star-forming regions: one applies 
to $0.76 \leq A_J < 2.53$ (or $0.3 \leq A_K < 1$), and the other one to $A_J \geq 2.53$
(or $A_K \geq 1$). For $A_J < 0.76$, we use a spline fit to the Mathis $R_V=5$ 
curve \citep{mathis90}. Since the three laws apply to different extinction environments, 
we use a linear combination of them to achieve a smooth change in the extinction law 
from the diffuse interstellar medium to the dense regions within molecular clouds. 
Thus, to find a best-fit model for a certain observed SED, the model fluxes 
$F_{mod}(\lambda)$ are scaled and extinguished  as follows:
\begin{equation}
F_{obs}(\lambda) = s F_{mod}(\lambda) 10^{-0.4 A_{\lambda}},
\label{F_scaled_ext}
\end{equation}
where  $F_{obs}(\lambda)$ and $F_{mod}(\lambda)$ are the observed and model 
fluxes, respectively, $s$ is the luminosity scaling factor, and $A_{\lambda}$
is the extinction at wavelength $\lambda$. We use three reddening laws,
$k_{\lambda}=A_{\lambda}/A_J$; by denoting them with the subscripts 1, 2,
and 3, $A_{\lambda}$ in the above equation becomes
\begin{eqnarray}
A_{\lambda} = A_J k_{1,\lambda} \quad {\rm for}\; A_J < 0.76 \nonumber \\
A_{\lambda} = 0.76 k_{1,\lambda} + (A_J - 0.76) k_{2,\lambda} 
\nonumber \\ {\rm for}\; 0.76 < A_J < 2.53 \nonumber \\
A_{\lambda} = 0.76 k_{1,\lambda} + 2.53 k_{2,\lambda} + 
(A_J - 2.53) k_{3,\lambda} \nonumber \\ {\rm for}\; A_J > 2.53
\end{eqnarray}
Thus, equation \ref{F_scaled_ext} can be written as
\begin{eqnarray}
2.5 \log(F_{mod}(\lambda)/F_{obs}(\lambda)) 
= A_J k_{1,\lambda} -2.5 \log(s) \nonumber \\ 
{\rm for}\; A_J < 0.76 \nonumber \\
2.5 \log(F_{mod}(\lambda)/F_{obs}(\lambda)) 
- 0.76 (k_{1,\lambda}-k_{2,\lambda}) = \nonumber \\ 
A_J k_{2,\lambda} -2.5 \log(s) \quad {\rm for}\; 0.76 < A_J < 2.53 
\nonumber \\
2.5 \log(F_{mod}(\lambda)/F_{obs}(\lambda)) - 0.76 k_{1,\lambda} 
- 2.53 (k_{2,\lambda}-k_{3,\lambda}) \nonumber \\
=  A_J k_{3,\lambda} -2.5 \log(s) \quad {\rm for}\; A_J > 2.53 \qquad
\end{eqnarray}
These are linear equations in $A_J$, with the left-hand side of the 
equations as the dependent variables and $k_{\lambda}$ as the 
independent variable. For each regime of $A_J$ values, a best-fit
line can be determined that yields $A_J$ and $-2.5 \log(s)$ from the 
slope and intercept, respectively, for each model that is compared to 
the observations.

For each set of model fluxes and observed fluxes, we calculate three linear fits 
(using linear combinations of the three different extinction laws, as explained above),
thus yielding three values for scaling factors and three for the extinction value. 
If each extinction value is within the bounds of the extinction law that was 
used and smaller than a certain maximum $A_J$ value (which will be discussed
below), and the scaling factor is in the range from 0.5 to 2.0, then the result 
with the best linear fit will be used. 
However, if some of the values are not within their boundaries, then combinations
of their limiting values are explored, and the set of scaling factor and extinction
with the best fit is adopted. For example, if a model has fluxes that are
much higher than all observed fluxes, the linear fit described above will likely
yield very large extinction values and small scaling factors. In this case the fitter
would only accept the smallest possible scaling factor (0.5) and the maximum 
allowed $A_J$ value as a solution (which will still result in a poor fit). 

For each object, we allowed the model fluxes to be extinguished up to a maximum
$A_J$ value derived from column density maps of Orion (\citealt{stutz15}; see also
\citealt{stutz10,stutz13,launhardt13} for the methodology of deriving N$_H$ from 
160-500 \micron\ maps). We converted the total hydrogen column 
density from these maps to $A_V$ values ($A_V$=3.55 $A_J$)
by using a conversion factor of 
$1.0 \times 10^{21}$ cm$^{-2}$ mag$^{-1}$ \citep{winston10, pillitteri13}. 
For objects for which no column density could be derived, we set the maximum 
$A_J$ value to 8.45 (which corresponds to $A_V=30$). 

After returning a best-fit scaling factor and extinction value for each model, 
each data point is assigned a weight, and the goodness of the fit is 
estimated with 
\begin{equation}
 R = \frac{\sum_{i=1}^{N} w_i |\ln \left(\frac{F_{obs}(\lambda_i)}
{F_{mod}(\lambda_i)}\right)|}{N},
\end{equation}
where $w_i$ are the weights, $F_{obs}(\lambda_i)$ and $F_{mod}(\lambda_i)$
are the observed and the scaled and extinguished model fluxes, respectively, 
and N is the number of data points \citep[see][]{fischer12}. Thus, $R$ is a measure 
of the average, weighted, logarithmic deviation between the observed and model SED. 
It was introduced by \citet{fischer12} since the uncertainty of the fit is dominated by 
the availability of models in the grid (i.e, the spacing of the models in SED space) 
and not by the measurement uncertainty of the data, making the standard $\chi^2$ 
analysis less useful. Also, a statistic that measures deviations between models and 
data in log space more closely resembles the assessment done by eye when 
comparing models and observed SEDs in log($\lambda F_{\lambda}$) vs.\ $\lambda$ 
plots.
We set the weights $w_i$ to the inverse of the estimated fractional uncertainty 
of each data point; so, for photometry at wavelengths below 3 \micron\ they 
are equal to 1/0.1, between 3 and 60 \micron\ they are 1/0.05, at 70 and 
100 \micron\ they are 1/0.04, at 160 \micron\ the weight is 1/0.07, and 
for photometry at 350 and 870 \micron\ they are 1/0.4 and 1/0.2, respectively. 
For fluxes from IRS spectra the weights are 1/0.075 for wavelength ranges 
8-12 \micron\ and 18-38 \micron, while they are 1/0.1 for the 5-8 \micron\ 
and 12-18 \micron\ regions. These IRS weights are also multiplied by 1.5 for
 high signal-to-noise spectra and by 0.5 for noisy spectra. In this way those 
parts of the IRS spectrum that most constrain the SED, the 10 \micron\ silicate 
absorption feature and slope beyond 18 $\mu$m, are given more weight; for 
high-quality spectra, the weights in these wavelength regions are the same as 
for the 3-60 \micron\ photometry.

For small values, $R$ measures the average distance between model and data
in units of the fractional uncertainty.
In general, the smaller the $R$ value, the better the model fit, but protostars with fewer 
data points can have small $R$ values, while protostars with some noisy data can have 
larger $R$ values (but still an overall good fit). We find a best-fit model for each object, 
but we also record all those models that lie within a certain range of $R$ values from the 
best-fit $R$. These models give us an estimate on how well the various model parameters
are constrained (see Section \ref{deltaR}).

Our model grid is used to characterize the parameters that best describe the 
observed SED of each object; the $R$ values rank the models for each 
object and thus can be used to derive best-fit parameters, as well as estimates of 
parameter ranges. In several instances, better fits could be achieved if the model 
parameters were further adjusted, for example by testing more values of cavity 
opening angle or shape, or even changing the opacities (see, e.g., HOPS 68 
\citep{poteet11}, HOPS 223 \citep{fischer12}, HOPS 59, 60, 66, 108, 368, 369, 370
\citep{adams12}, HOPS 136 \citep{fischer14}, and HOPS 108 \citep{furlan14}).
However, for protostars that are well fit with one of the models from the grid or for 
which the grid yields a narrow range of parameter values, it is unlikely that a more 
extended model grid would yield much different best-fit parameters. Overall, our 
model fits yield good estimates of envelope parameters for a majority of the sample, 
and thus we can analyze the protostellar properties of our HOPS targets in a statistical 
manner.

\section{Results of the Model Fits}

The best-fit parameters resulting from our models can be found in Table 
A\ref{bestfit}, and Figure A\ref{bestSEDs} shows the SEDs and best fits for our sample.
In this section we give an overview of the quality of the fits, the distributions 
of the best-fit model parameters, both for the sample as a whole and separated 
by SED class, the parameter uncertainties, and the various degeneracies between 
model parameters.

\subsection{Quality of the Fits}
\label{fit_quality}

\begin{figure}[!t]
\centering
\includegraphics[scale=0.55]{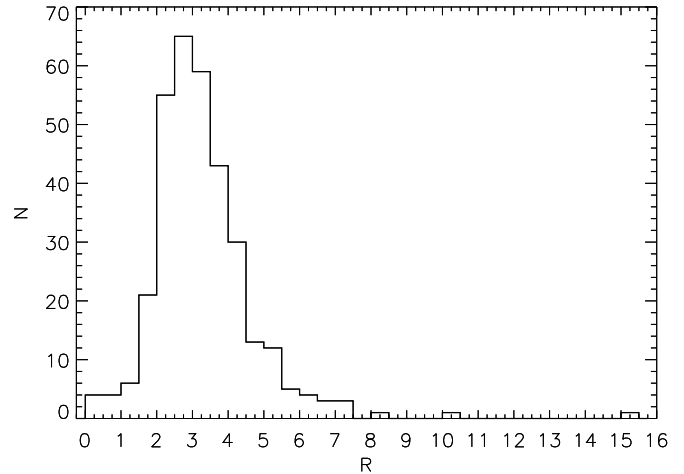}
\caption{Histogram of the $R$ values of the best fits of the 330 
YSOs in the HOPS sample that have {\it Spitzer} and {\it Herschel}
detections. \label{R_histo}}
\end{figure}

\begin{figure*}[!t]
\centering
\includegraphics[scale=0.62,angle=90]{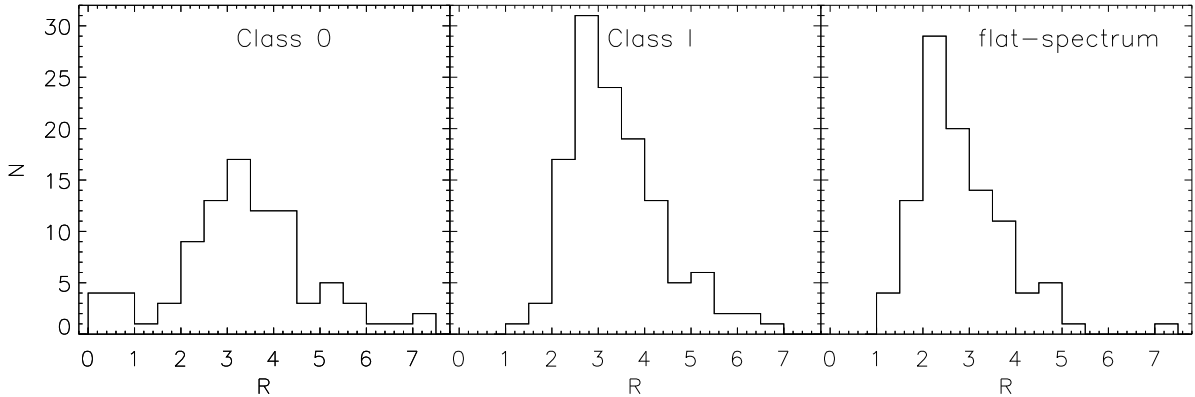}
\caption{Histograms of the $R$ values of the best fits  shown separately
for the three classes of objects (Class 0, I, and flat-spectrum). The three 
fits with $R>8$ (two Class 0 protostars, one Class I protostar) are not shown.
\label{R_histo_by_class}}
\end{figure*}

Figure \ref{R_histo} displays the histogram of $R$ values of the best model
fits for the 330 objects in our HOPS sample that have {\it Spitzer} and 
{\it Herschel} data (more than two data points at different wavelengths)
and are not contaminants (see Section \ref{sample}). The median $R$ 
value is 3.10, while the mean value is 3.29. Fitting a Gaussian to the 
histogram at $R$ $\leq$ 7 yields 3.00 and 2.24 as the center and FWHM 
of the Gaussian, respectively. 
The distribution of $R$ values implies that, on average, the model deviates
by about three times the average fractional uncertainty from the data.
This is not unexpected, given that we fit models from a grid to observed 
SEDs that span almost three orders of magnitude in wavelength range, 
with up to 29 data points. The fewer the data points, the easier it is to 
achieve a good fit; in fact, the eight protostars with $R < 1$, HOPS 371, 391, 
398, 401, 402, 404, 406, and 409, have SEDs with measured flux values at 
only 4-5 points.
Starting at $R$ values of about 1, $R$ can be used as an indicator of the 
goodness of fit. However, in some cases a noisy IRS spectrum can 
increase the $R$ value of a fit that, judged by the photometry alone, does 
not deviate much from the observed data points. In other cases, mismatches 
between different data sets, like offsets between the IRAC fluxes and the IRS
spectrum, can result in larger $R$ values. These might be interesting protostars 
affected by variability and are thus ideal candidates for follow-up observations. 

When looking at the SED fits in Figure A\ref{bestSEDs} (and the corresponding 
$R$ values in Table A\ref{bestfit}), we estimate that an $R$ value of up to 
$\sim$ 4 can identify a reliable fit (with some possible discrepancies between 
data and model in certain wavelength regions). When $R$ gets larger than 
about 5, the discrepancy between the fit and the observed data points usually
becomes noticeable; the fit might still reproduce the overall SED shape
but deviate substantially from most measured flux values. 

In Figure \ref{R_histo_by_class}, we show the histogram of $R$ values 
separately for the three main protostellar classes in our sample. The 
median $R$ value decreases from 3.27 for the Class 0 protostars to 3.18 for 
the Class I protostars to 2.58 for the flat-spectrum sources. There are 
4 Class 0 protostars and 4 Class I protostars with $R$ values between 1.0 
and 2.0, but 17 flat-spectrum sources in this $R$ range. These numbers 
translate to 17\% of the flat-spectrum sources in our sample, 4\% of the 
Class 0 protostars, and 3\% of the Class I protostars. When examining objects' 
$R$ values between 2.0 and 4.0, there are 51 Class 0 protostars (55\% of 
Class 0 protostars in the sample), 91 Class I protostars (73\% of the Class I 
sample), and 74 flat-spectrum sources (73\% of the flat-spectrum sample).

Thus, close to 90\% of flat-spectrum sources are fit reasonably well ($R$ 
values $<$ 4), representing the largest fraction among the different classes 
of objects in our sample. This could be a result of their source properties 
being well represented in our model grid, but also lack of substantial 
wavelength-dependent variability (see, e.g., \citealt{guenther14}), which, 
if present, would make their SEDs more difficult to fit.  
About three-quarters of Class I protostars also have best-fit models with $R < 4$; 
this fraction drops to about two-thirds for the Class 0 protostars. The latter 
group of objects often suffers from more uncertain SEDs due to weak
emission at shorter wavelengths (which, e.g., results in a noisy IRS
spectrum); they might also be more embedded in extended emission, 
such as filaments, which can contaminate the far-IR to submillimeter fluxes. 
Another factor that could contribute to poor fits is their presumably
high envelope density, which places them closer to the limit in parameter
space probed by the model grid.
Overall, 75\% of the best-fit models of the protostars in our sample have
$R < 4$.

When examining the SED fits of objects with $R$ values larger than 5.0,
several have very noisy IRS spectra (HOPS 19, 38, 40, 95, 164, 278, 316,
322, 335, 359). In a few cases the measured PACS 100 and 160 $\mu$m 
fluxes seem too high compared to the best-fit model (e.g., HOPS 189), 
which could be an indication of contamination by extended emission 
surrounding the protostar. 

Of particular interest are objects where variability likely plays a role in 
a poor fit. As mentioned in Section \ref{SEDs}, variability among protostars
is common; we found in Appendix \ref{variability} that about 5\% of our 
targets display noticeable ($\gtrsim$ 50\%) mismatches between 
the IRS, IRAC, and MIPS fluxes that could be due to intrinsic variability.
The SED fits of objects for which the flux mismatches between IRS and 
IRAC and between IRS and MIPS are different are particularly affected, 
since in that case we did not scale the IRS spectrum to match the 
MIPS 24 $\mu$m flux.
HOPS 228 exemplifies such a case: there is a clear discrepancy
between the IRAC and IRS fluxes (a factor of 2.1-2.7) and also
between MIPS 24 $\mu$m and IRS (a factor of 0.8); even though 
the fit gives more weight to the IRS data, they are not fit well, especially 
the silicate absorption feature. The $R$ value of 5.74 for the fit of
HOPS 228 reflects the discrepant data sets and poor fit. 
HOPS 223 is another case where the IRS  fluxes do not match the 
shorter-wavelength data (they are more than an order of magnitude 
larger); however, it is a known FU Ori source \citep[see][]{fischer12},
and the SED presented here contains both pre- and post-outburst data. 
The model fit is very poor, which can also be gauged by the $R$ value 
of 8.41. 

There are also objects with overall good fits whose SEDs show discrepancies
that may be signs of variability or contamination. 
For example, for the Class I protostar HOPS 71 the IRAC fluxes are a factor 
of 1.8-2.4 lower than the IRS fluxes in the 5-8 $\mu$m region, and also the 
MIPS flux is about 20\% lower. The best-fit model ($R=3.63$) fits the SED 
extremely well beyond about 6 $\mu$m, with some discrepancy at shorter 
wavelengths. There is a source just 11\arcsec\ from HOPS 71 that is detected
in 2MASS and {\it Spitzer} data, but not by PACS; this object, HOPS 72, is
likely an extragalactic object (see Appendix \ref{exgal_not_modeled}) that
could contaminate the IRS fluxes. Thus, in this case, wavelength-dependent
contamination by a companion could explain the discrepancies observed in
the SED.

Another example is HOPS 124, which is a deeply embedded Class 0 protostar. 
For this object, the mismatch between IRS and IRAC and MIPS fluxes 
decreases with increasing wavelength (from a factor of 2.5 to a factor of 1.4); 
for the SED fit, the IRS spectrum was scaled by 0.7 to match the MIPS 24 
$\mu$m flux. As with HOPS 71, there is a nearby source that could contaminate 
some of the fluxes, especially at shorter wavelengths: HOPS 125, a flat-spectrum 
source, lies 9.8\arcsec\ from HOPS 124 and is brighter than HOPS 124 out to 
$\sim$ 20 $\mu$m, but then much fainter at longer wavelengths. The best-fit 
model of HOPS 124 ($R=2.43$) matches the mid- to far-IR photometry and 
also most of the IRS spectrum well.

As an example of a probably variable flat-spectrum source, HOPS 132 
has IRAC fluxes that lie a factor of 1.3-1.7 above those of IRS and a 
MIPS 24 $\mu$m flux that is a factor of 0.6 lower. It does not have a close
companion; the nearest HOPS source, HOPS 133, is 27\arcsec\ away.
The IRS spectrum was not scaled, and since the SED fitter gave more 
weight to the spectrum, it is fit well, but the IRAC photometry is underestimated
and the MIPS photometry overestimated. Nonetheless, the $R$ value
of the best fit is 2.87.

Overall, the SED fits of objects that are likely variable or suffer from some
contamination are less reliable, but it is not always clear from the $R$ 
value of the best fit. The SED fitting procedures assume that the protostars 
are not variable, so when large mismatches between different data sets are 
present, the fit will appear discrepant with at least some of the observed
data points, but the $R$ value would not end up particularly high 
if, e.g., the IRS spectrum was fit exceptionally well. However, given the
data sets we have for these protostars, our SED fits will still yield the best 
possible estimate for the protostellar parameters describing these systems.

\subsection{Overview of Derived Parameters}
\label{results_overview}

\begin{figure}[!t]
\centering
\includegraphics[scale=0.55]{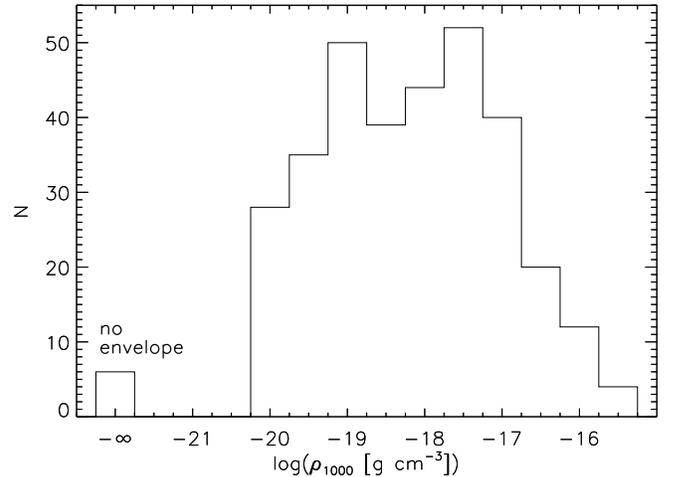}
\caption{Histogram of the envelope reference density $\rho_{1000}$ 
of the best fits for the 330 targets in our sample. 
\label{rho1000_histo}}
\end{figure}

The histogram of best-fit $\rho_{1000}$ values (which is the density of the 
envelope at 1000 AU; see Section \ref{model_parameters}) is shown in 
Figure \ref{rho1000_histo}.
The median value of the distribution amounts to $5.9 \times 10^{-19}$ 
g cm$^{-3}$; this corresponds to a $\rho_1$ value of $1.9 \times 
10^{-14}$ g cm$^{-3}$. There is a spread in values: 69 objects have 
densities $\rho_{1000}$ smaller than $5.0 \times 10^{-20}$ g cm$^{-3}$
(6 of them have actually no envelope), 89 fall in the $5.0 \times
10^{-20}$ to $5.0 \times 10^{-19}$ g cm$^{-3}$ range, 96
are between $5.0 \times 10^{-19}$ and $5.0 \times 10^{-18}$ g 
cm$^{-3}$, 60 between  $5.0 \times 10^{-18}$ and $5.0 \times 
10^{-17}$ g cm$^{-3}$, and 16 have $\rho_{1000}$ values
larger than $5.0 \times 10^{-17}$ g cm$^{-3}$.

\begin{figure}[!t]
\centering
\includegraphics[scale=0.55]{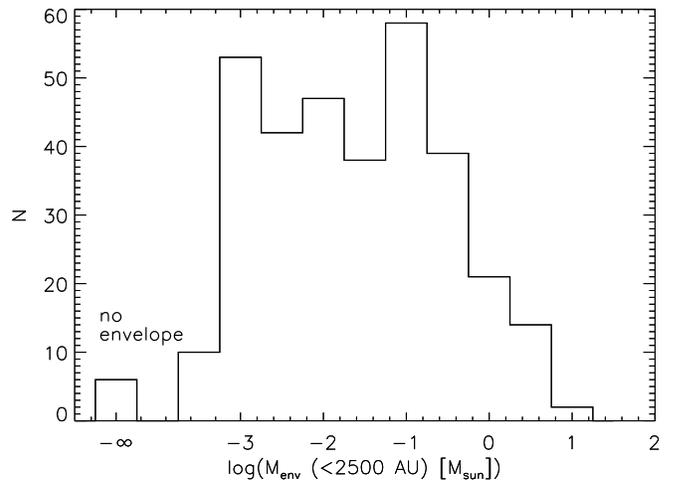}
\caption{Histogram of the envelope mass within 2500 AU derived for
the best fits for the 330 targets in our sample.
\label{Menv_histo}}
\end{figure}

We also calculated the envelope mass ($M_{env}$) within 2500 AU for 
the best-fit models (see Figure \ref{Menv_histo} for their distribution). 
The 2500 AU radius is close to half the FWHM of the PACS 160 $\mu$m 
beam at the distance of Orion (i.e., $\sim$ 6\arcsec), and thus roughly 
represents the spatial extent over which we measure the SEDs.
This envelope mass is determined from the integrated envelope density 
of our best-fit models, with allowances made for outflow cavities, and thus 
only valid in the context of our models. The median envelope mass within 
2500 AU amounts to 0.029 \Msun. The majority of protostars have 
model-derived masses in the inner 2500 AU of their envelopes around 
0.1~\Msun; just 22 objects have $M_{env}$ ($<$ 2500 AU) larger than 
1.0~\Msun. Of the 330 modeled objects, 291 have $M_{env}$ ($<$ 2500 AU)
smaller than 0.5 \Msun\ (6 of these 291 objects have no envelope).  

\begin{figure}[!t]
\centering
\includegraphics[scale=0.55]{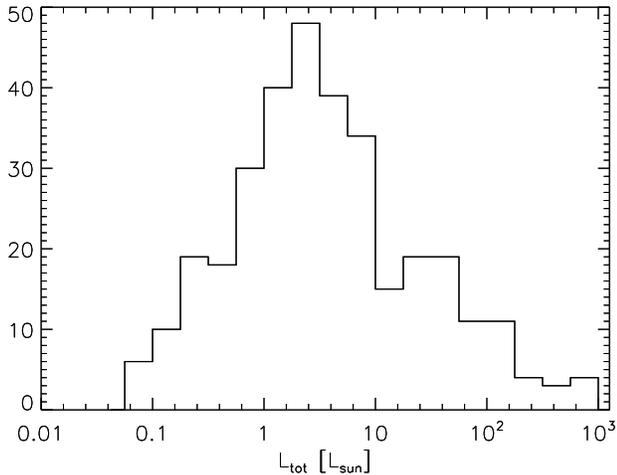}
\caption{Histogram of the total luminosities of the best fits for the 330 targets 
in our sample. 
\label{Ltot_histo}}
\end{figure}

\begin{figure}[!t]
\centering
\includegraphics[scale=0.55]{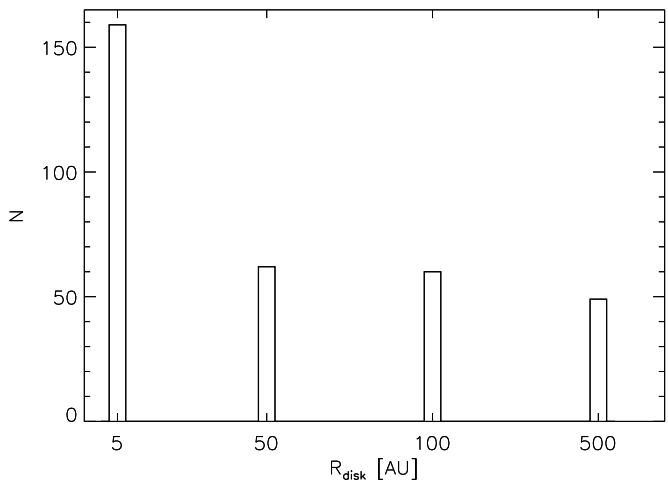}
\caption{Histogram of the disk radii of the best fits for the 330 targets 
in our sample. 
\label{Rdisk_histo}}
\end{figure}

\begin{figure}[!t]
\centering
\includegraphics[scale=0.55]{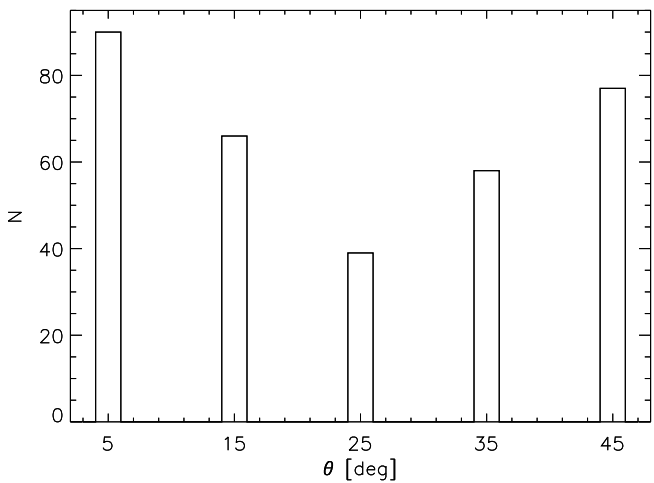}
\caption{Histogram of the cavity opening angles of the best fits for the 
330 targets in our sample. 
\label{Cav_histo}}
\end{figure}

\begin{figure*}[!]
\centering
\includegraphics[scale=0.7]{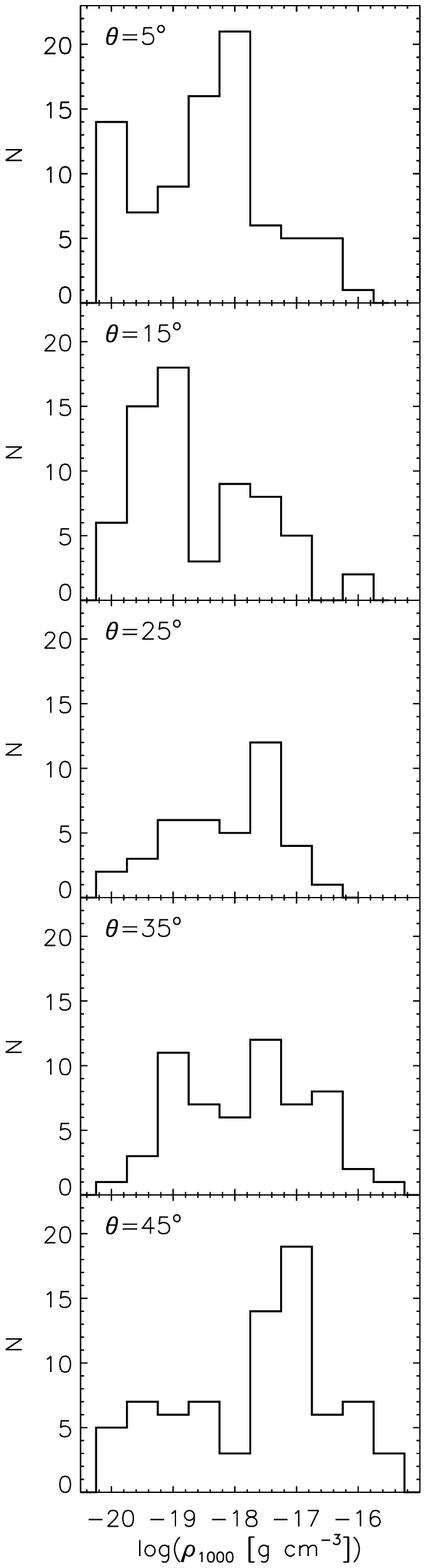}
\includegraphics[scale=0.7]{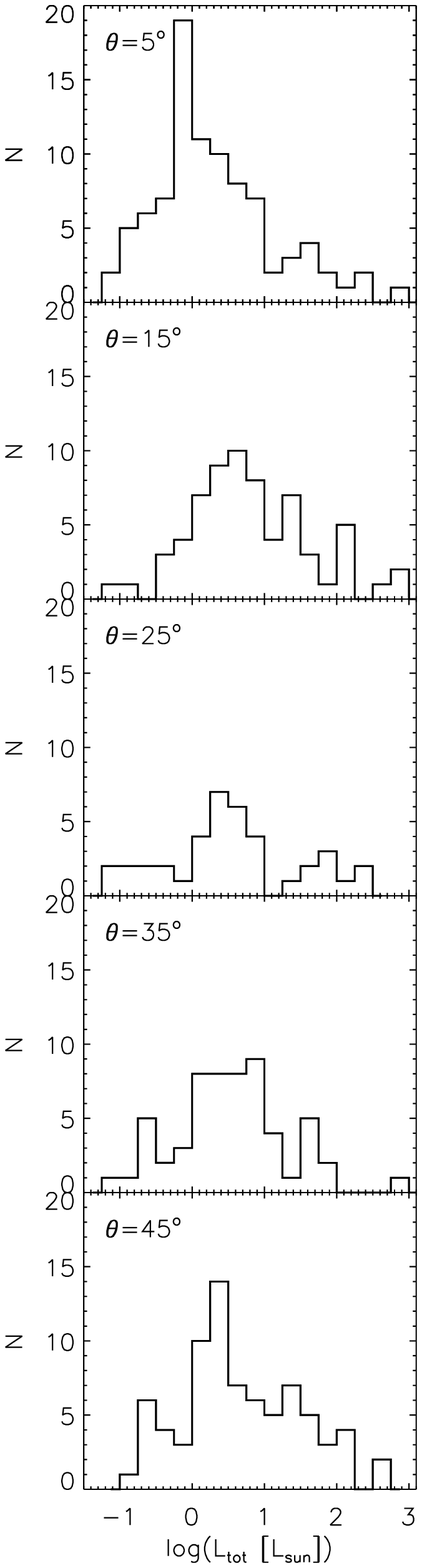}
\includegraphics[scale=0.7]{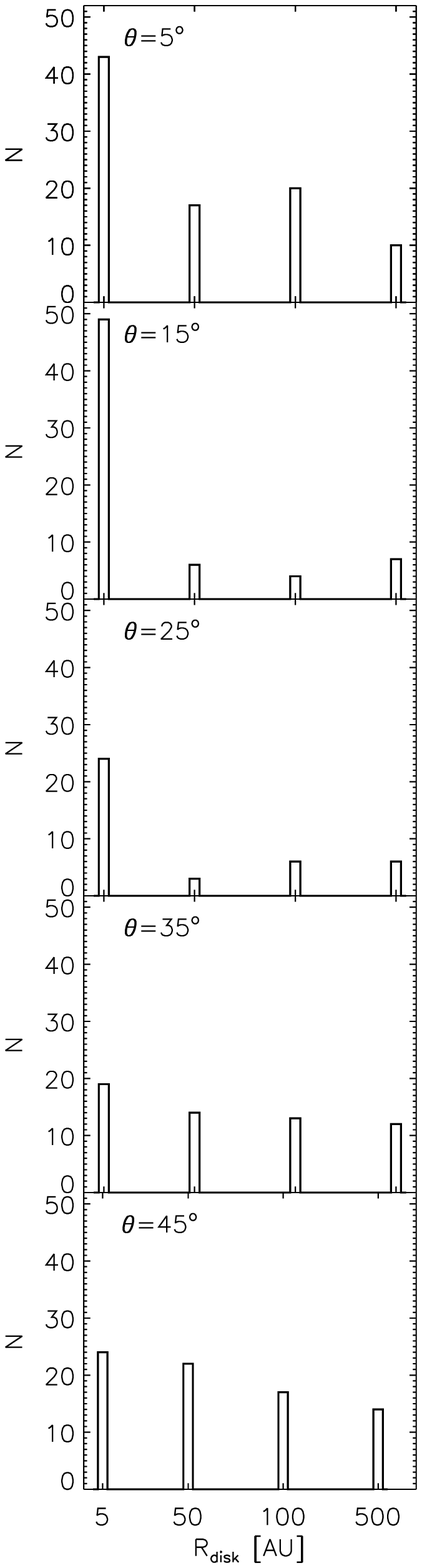}
\caption{Histograms of the envelope reference density $\rho_{1000}$ 
({\it left}), the total luminosity ({\it middle}), and the disk radius ({\it right})
of the best fits grouped by cavity opening angles. 
\label{Pars_cav_histo}}
\end{figure*}

Figure \ref{Ltot_histo} contains the histogram of the total luminosities
derived from the best-fit models. These luminosities consist of the stellar,
disk accretion, and accretion shock components. The median total luminosity 
amounts to 3.02 \Lsun, while the values cover four orders of magnitude, 
from 0.06 \Lsun\ (for HOPS 336) to 607 \Lsun\ (for HOPS 288 and 361). 
Since the minimum and maximum values for the total luminosity 
in our grid amount to 0.1 and 303.5 \Lsun, respectively, and our scaling 
factors range from 0.5 to 2.0, our fitting procedure can return best-fit
luminosities that range from 0.05 to 607 \Lsun. Thus, two protostars are
reaching the upper limit allowed for total luminosities in our grid; it is possible 
that even better fits could be achieved by increasing the luminosity further. 

From the distribution of best-fit outer disk radii in Figure \ref{Rdisk_histo}, it is 
apparent that most protostars are fit by small disks whose radius is only 5 AU. 
Since the outer disk outer radius is the centrifugal radius in our models, 
infalling material from the envelope tends to accumulate close to the star 
for most sources. Thus, the disk radius is tied to the envelope structure; 
a small centrifugal radius implies higher envelope densities at smaller radii
and a less flattened envelope structure. The median disk radius is 50 AU, 
but the number of objects with disk radii $\geq$ 50 AU is roughly evenly 
split among the values of 50, 100, and 500 AU.

The distribution of best-fit cavity opening angles is displayed in Figure 
\ref{Cav_histo}. Most protostars seem to have either very small (5\degr) 
or very large (45\degr) cavities; the median value is 25\degr.
When dividing the envelope densities by cavity opening angle (see Figure
\ref{Pars_cav_histo}, left column), differences emerge: the distributions of 
$\rho_{1000}$ values are significantly different when comparing objects 
with $\theta$=5\degr\ and $\theta \geq$35\degr, objects with 
$\theta$=15\degr\ and $\theta$ $\geq$ 25\degr, and objects with 
$\theta$=25\degr\ and $\theta$=45\degr. The Kolmogorov-Smirnov (K-S)
tests yield significance levels that these subsamples are drawn from the
same parent population of $\lesssim$ 0.015. Thus, there seems to be a 
difference in the distribution of envelope densities among the best-fit 
models with smaller cavity opening angles and those with larger cavities. 
Protostars with larger cavities ($\geq$ 35\degr) tend to have higher envelope 
densities (their median $\rho_{1000}$ values are about an order of magnitude 
larger compared to objects with cavities $\leq$ 15\degr). 

Figure \ref{Pars_cav_histo} (middle column) also shows the distribution 
of total luminosities for the different cavity opening angles. The only 
significant difference can be found for the $\theta$=5\degr\ histogram 
as compared to the histograms for larger $\theta$ values (K-S test
significance level $\lesssim$ 0.03); the luminosities of models with 
$\theta$=5\degr\ have a different distribution, and also their median 
value is 1.45 \Lsun, as compared to $\sim$ 3-5 \Lsun\ for the models
with larger cavities. So, protostars with small cavities seem to have lower
total luminosities.

\begin{figure}[!t]
\centering
\includegraphics[scale=0.55]{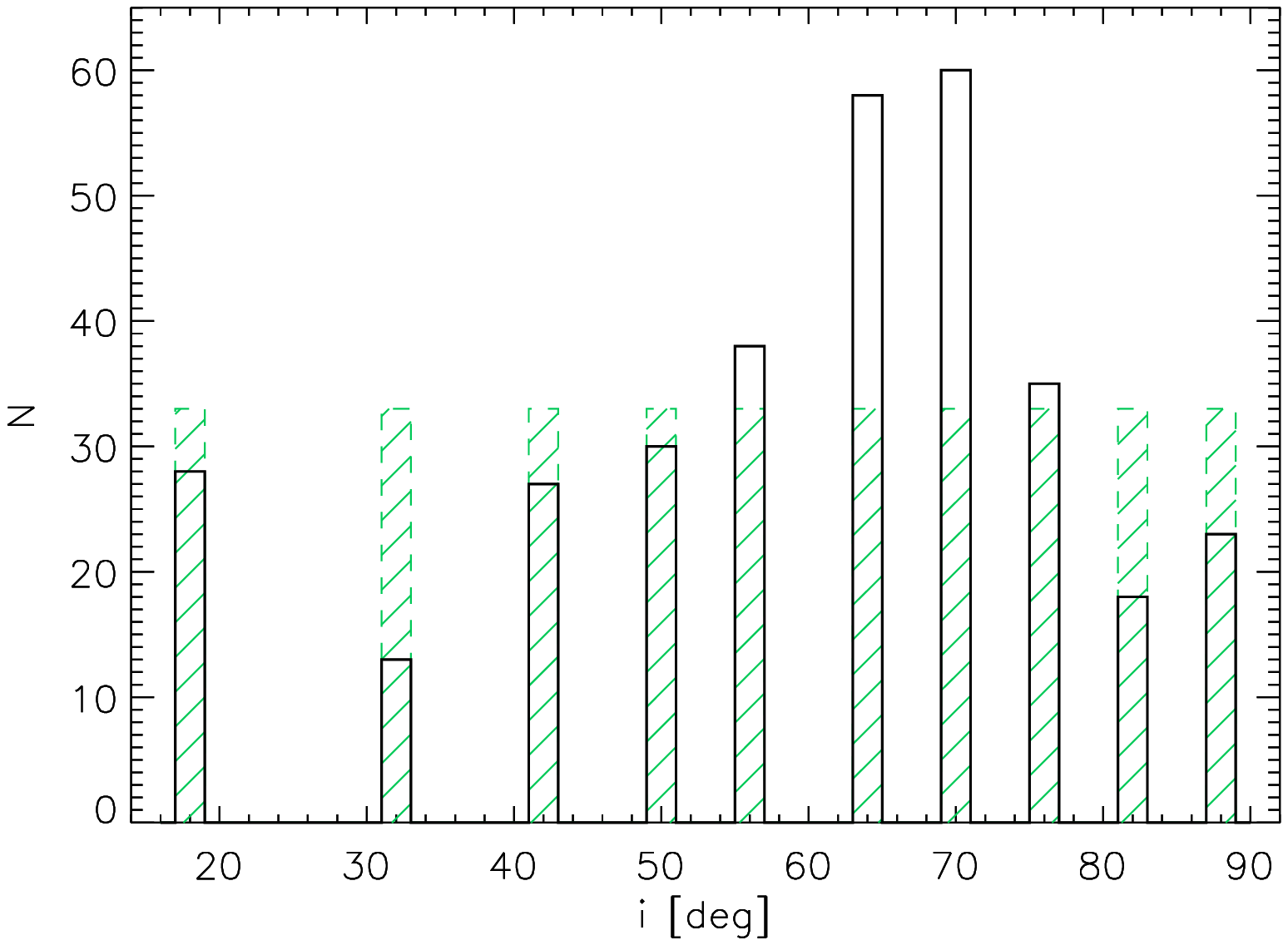}
\caption{Histogram of the inclination angles of the best fits for the 330 
targets in our sample. The green dashed histogram represents the 
distribution of uniformly (randomly) distributed inclination angles.
\label{Inc_histo}}
\end{figure}

The distribution of centrifugal radii for different cavity opening angles 
(right column in Figure \ref{Pars_cav_histo}) shows that, independent of 
cavity size, most objects have $R_{disk}$ = 5 AU. However, the
distribution among the four different disk radii becomes flatter for 
the largest cavity opening angles; the histograms for $\theta$=35\degr\ 
and $\theta$=45\degr\ are very similar (K-S test significance level of
0.98). There is also no significant difference (K-S test values $>$ 0.075)
between the $\theta$=15\degr\ and $\theta$=25\degr\ histograms and 
between the $\theta$=5\degr\ and $\theta \geq $ 35\degr\ histograms. 
The distributions of disk radii for the other cavity opening angles are all 
different from one another (K-S test significance levels $<$ 0.015). 
Overall, Figure \ref{Pars_cav_histo} shows that protostars best fit by 
models with large cavity opening angles are also fit by models with 
higher envelope densities and larger centrifugal radii.  

\begin{figure}[!t]
\centering
\includegraphics[scale=0.54]{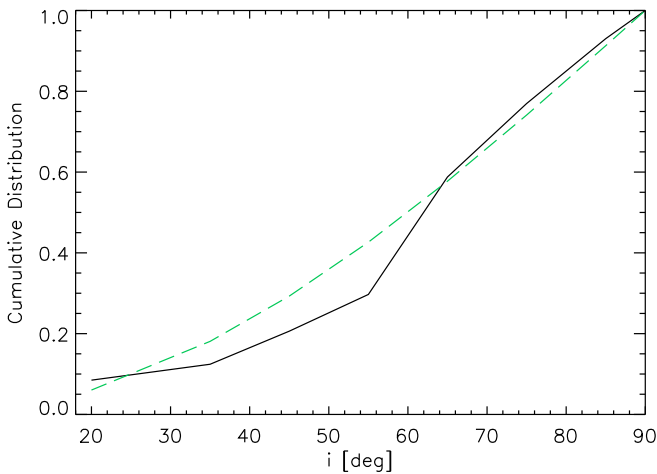}
\caption{Cumulative distribution of the inclination angles of the best fits,
normalized by the total number of fits ({\it solid line}), compared to the 
cumulative probability of finding an inclination angle below a given value
for randomly distributed inclinations ({\it green dashed line}). 
\label{Inc_CDF}}
\end{figure}

In Figure \ref{Inc_histo}, we show the distribution of the inclination angles
for the best-fit models. There is a clear concentration of models in the 
60\degr$-$70\degr\ range; the median inclination angle is 63\degr.
This median value is close to 60\degr, which is where the probability for 
isotropically distributed inclination angles reaches 50\% (i.e., the probability 
of observing an inclination angle less than 60\degr\ is the same as the 
probability of observing i$>$60\degr). However, the details of the
distributions differ.
The cumulative probability of finding an inclination angle less than a certain value, 
$i_c$, is $1-\cos(i_c)$, assuming a random distribution of inclination angles. For
inclination angles $i_1$ and $i_2$, the probability for $i_1 < i < i_2$ is
$\cos(i_1)-\cos(i_2)$. Thus, since the inclination angles in our model grid were 
chosen to be equally spaced in $\cos(i)$ (there are five values $<$60\degr\ and 
five values $>$60\degr), one would expect a flat distribution in Figure \ref{Inc_histo} 
if the best-fit inclination angles were randomly distributed (see the green dashed
histogram). However, we find a distribution peaked at 63\degr\ and 70\degr. 
This can also be seen in Figure \ref{Inc_CDF}, where we compare our observed 
cumulative distribution of inclination angles to that of randomly distributed ones. 
Our distribution shows a deficit at inclination angles below 60\degr\ and 
is just slightly higher at large inclination angles. A K-S test of the two 
distributions yields a 5.6\% chance that they are drawn from the same parent
distribution.

\begin{figure*}[!t]
\centering
\includegraphics[scale=0.59]{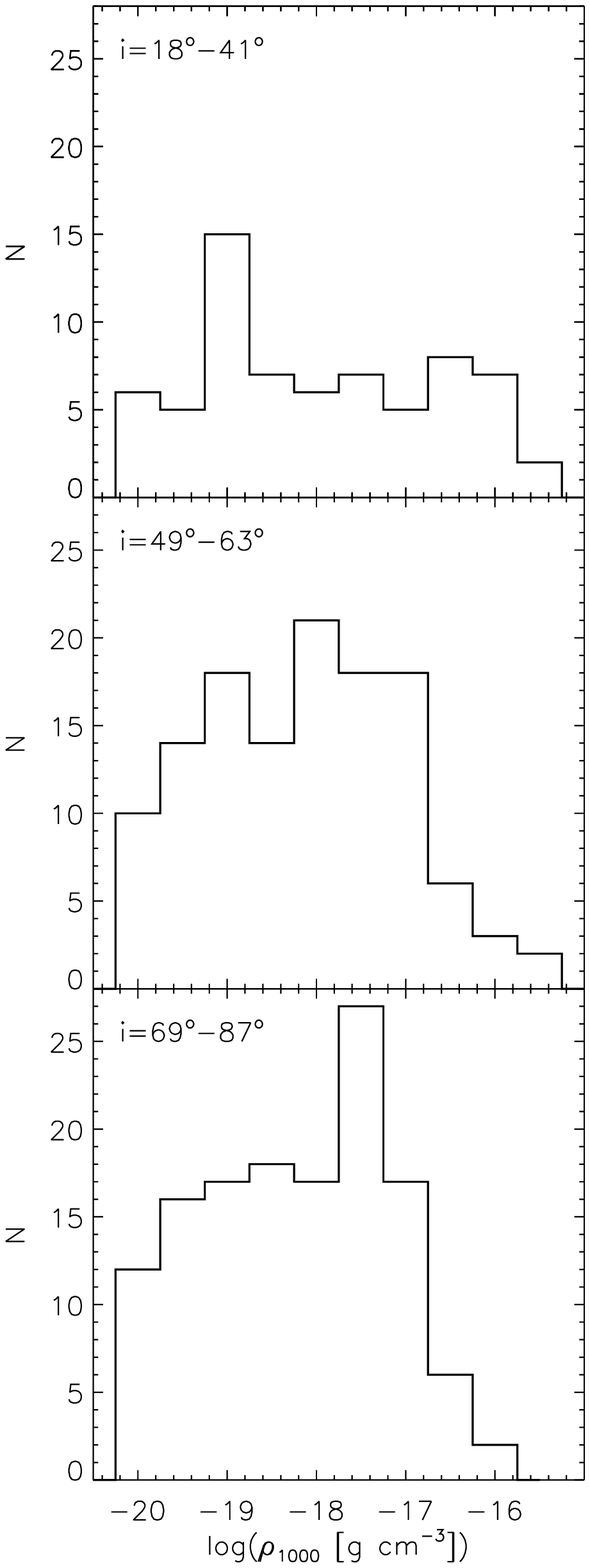} 
\includegraphics[scale=0.59]{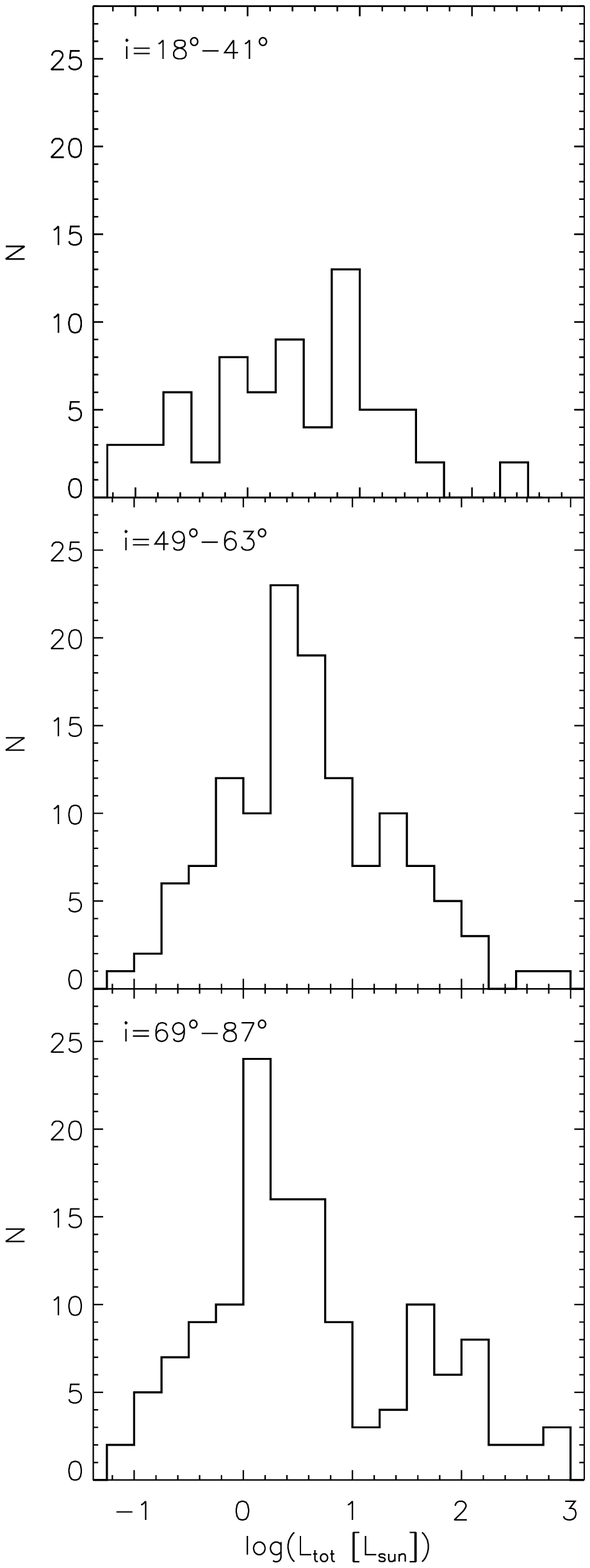}
\includegraphics[scale=0.59]{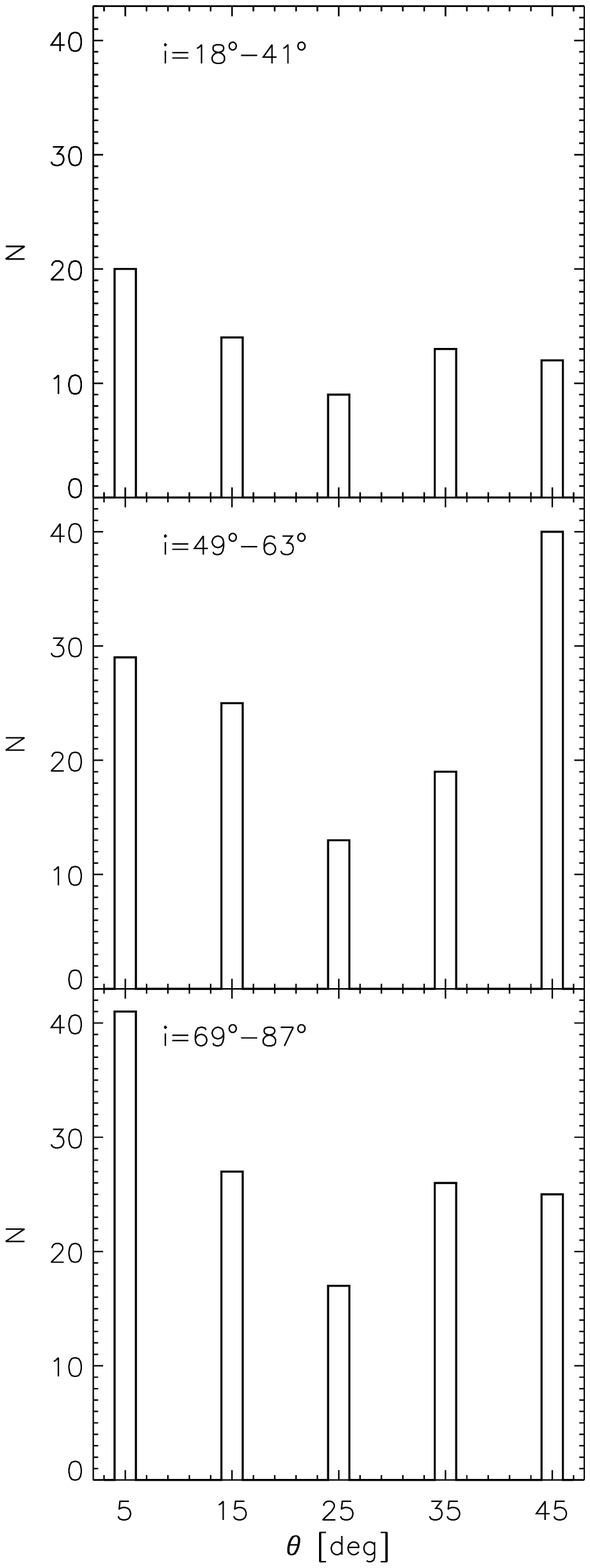}
\caption{Histograms of the envelope reference density $\rho_{1000}$
({\it left}), total luminosity ({\it middle}), and cavity opening angles 
({\it right}) of the best fits divided by bins of inclination angles. 
\label{Pars_inc_histo}}
\end{figure*}

\begin{figure*}[!t]
\centering
\includegraphics[scale=0.37,angle=90]{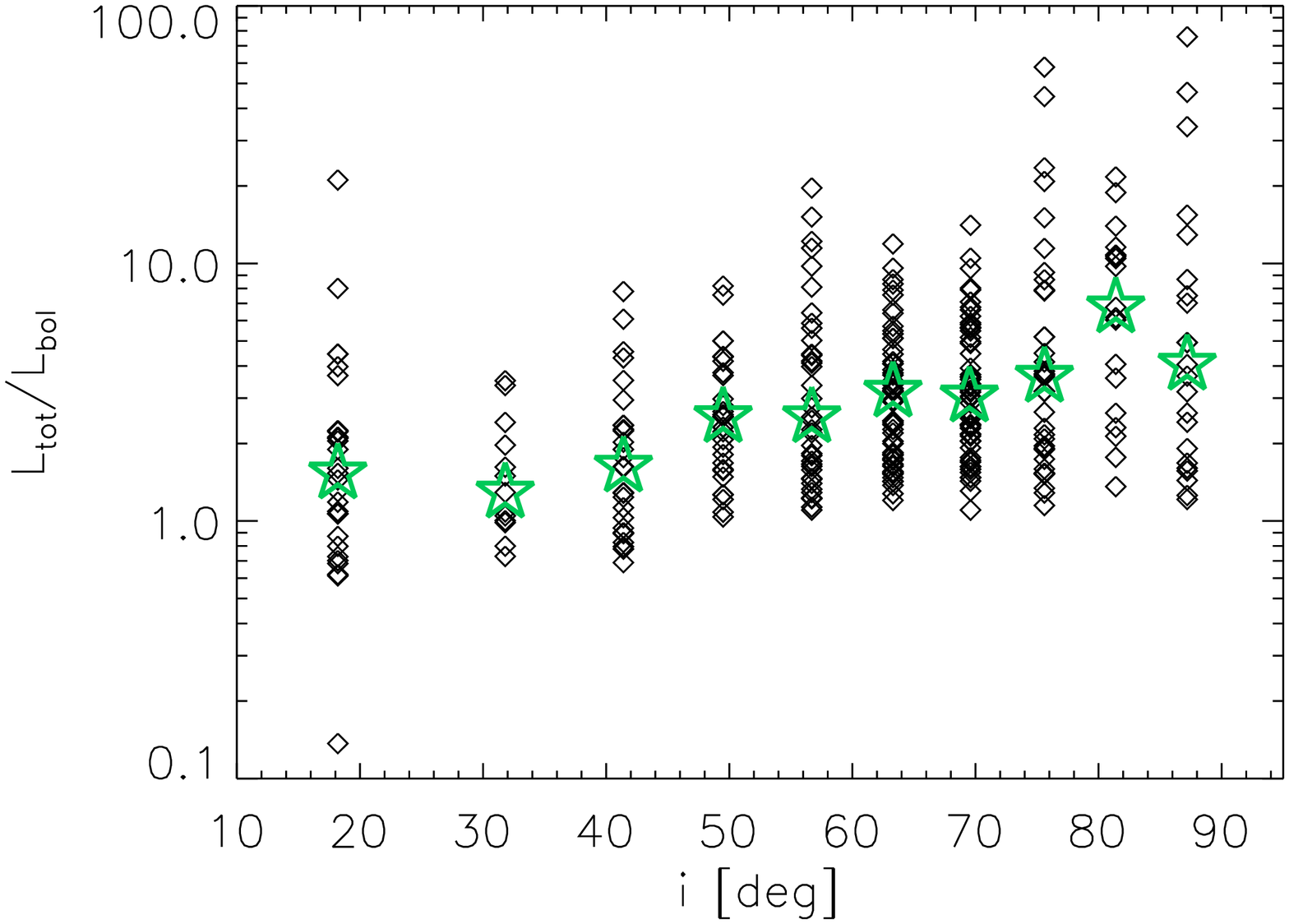}
\includegraphics[scale=0.37,angle=90]{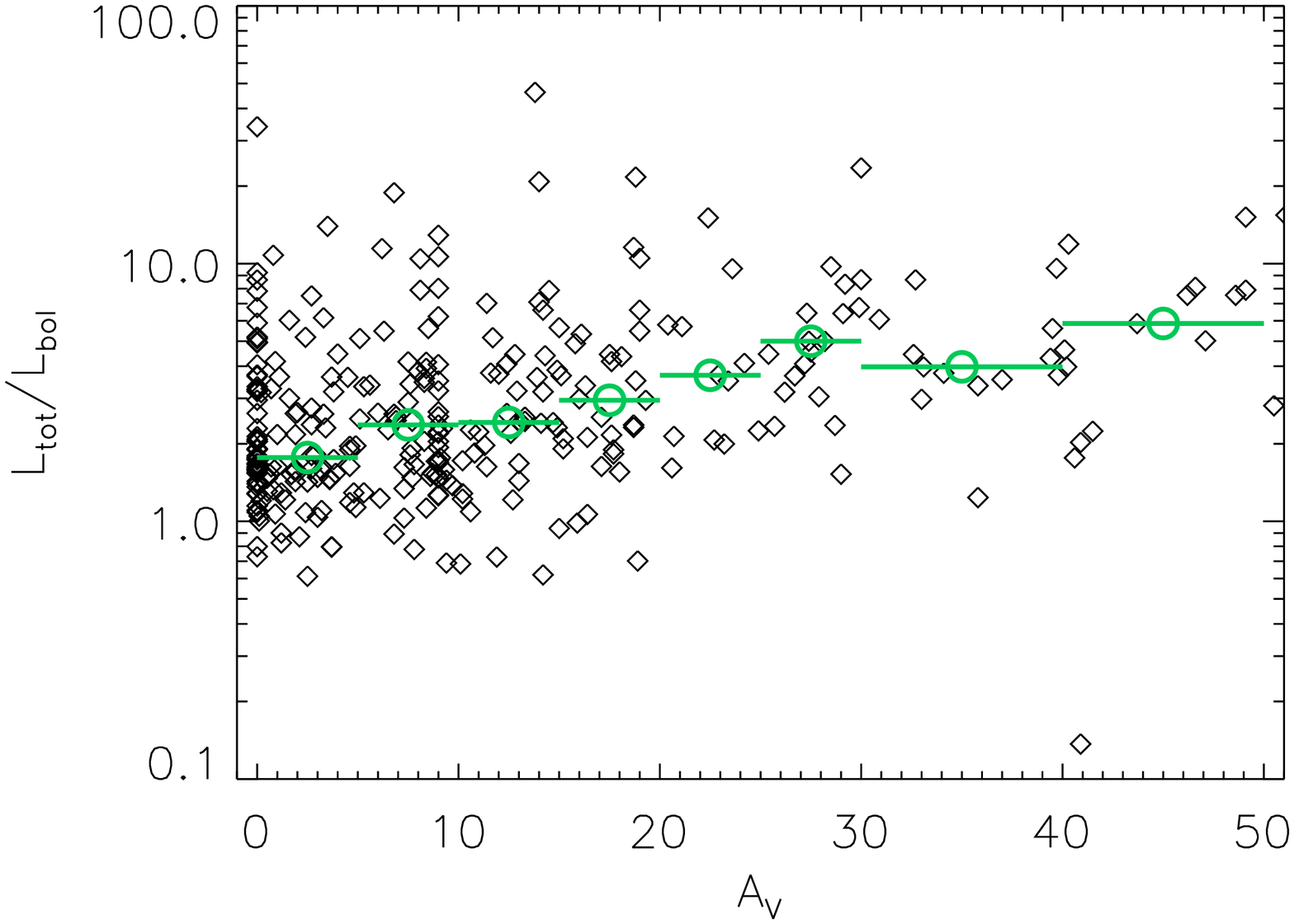}
\caption{Ratio of the total luminosity from the best fits and the bolometric 
luminosity derived from the observed SEDs versus the inclination angle ({\it 
left}) and foreground extinction ({\it right}) of the best fits. In the left
panel, the open stars represent the median ratios at each inclination angle.
In the right panel, the open circles represent the median ratios for eight
bins in $A_V$ values, represented by the horizontal lines bisecting
each circle.
\label{Ltot_Lbol}}
\end{figure*}

To examine whether the distribution of envelope parameters changes with
inclination angle (which could imply a degeneracy), Figure \ref{Pars_inc_histo}
shows the reference envelope density $\rho_{1000}$, the total luminosity, 
and the cavity opening angle binned by three ranges of inclination angles.
None of the three model parameters show a significantly different distribution 
for any of the inclination bins (K-S test significance levels are $\gtrsim$ 0.1,
except for the cavity opening angles for the lowest and middle inclination
range, for which the K-S test significance value is 0.02). 
The median $\rho_{1000}$ values for the $i=$18\degr--41\degr, 49\degr--63\degr, 
and 69\degr--87\degr\ inclination bins are all $5.9 \times 10^{-19}$ g cm$^{-3}$. 
Even though not shown in Figure \ref{Pars_inc_histo}, the objects whose best-fit
model does not include an envelope are only found at $i \geq$ 49\degr. It is
noteworthy that protostars with the highest envelope densities do not have 
inclination angles in the 69\degr--87\degr\ range; it is not clear whether this 
is an observational bias, whether our observed sample does not contain high-density, 
edge-on protostars, or whether this is due to biases in the fitting procedure and/or 
model grid.
The median values for the total luminosity do not differ by much for the 
different bins of inclination angle, increasing from 2.9 to 4.1 \Lsun\ from
the lowest to the middle inclination range and then decreasing to 2.0 \Lsun\
for the highest inclination angles. The few protostars with very high $L_{tot}$
values have large inclination angles ($i \geq$ 49\degr).
Finally, the distribution of cavity opening angles is quite similar for different
ranges in inclination, except for a somewhat larger number of $\theta =$ 45\degr\
values at intermediate inclination angles. Half the objects in the $i=$18\degr--41\degr\ 
and 69\degr--87\degr\ inclination bins have $\theta \leq $ 15\degr\ (with the most
common value 5\degr), while almost half the objects at intermediate inclination 
angles have $\theta \geq $ 35\degr\ (the most common value is 45\degr). 

In Figure \ref{Ltot_Lbol}, we show ratios of the total and bolometric luminosities
as a function of inclination angle and foreground extinction ($i$ and $A_V$ are
adopted from the best model fits). The total luminosity is the intrinsic luminosity 
from the best-fit model of each object, while the bolometric luminosity is derived 
by integrating the fluxes of the observed SED. It is expected that $L_{tot}$ is 
higher than $L_{bol}$ for objects seen at higher inclination angles, since for 
these objects a large fraction of the emitted flux is not directed toward the 
observer (and thus deriving bolometric luminosities from observed fluxes will 
underestimate the intrinsic source luminosity). Conversely, objects seen more 
face-on should have lower $L_{tot}$ values compared to $L_{bol}$. Our data 
and model fits yield $L_{tot}$ values that are usually higher than the $L_{bol}$
values measured from the SED; the discrepancy is larger for the more 
highly inclined sources. The median $L_{tot}/L_{bol}$ ratio is 1.5 for 
protostars with inclination angles in the 18\degr--41\degr\ range, 2.5
for the i=49\degr--63\degr\ range, and 3.5 for inclination angles 
$\geq$ 69\degr. 
The fact that $L_{tot}>L_{bol}$ even for $i=$18\degr--41\degr\ could
be related to the typically smaller cavity opening angles for this range of
inclination angles (see Figure \ref{Pars_inc_histo}); less flux, especially at
shorter wavelengths, is detected since the opacity along the line of sight 
is still high due to the small cavities. 

Foreground extinction also plays a role in increasing the $L_{tot}$/$L_{bol}$
ratio. The median ratio of these luminosities increases from 1.8 for the $A_V$=
0-5 mag range to 5.0 for $A_V$=25-30; it decreases somewhat for the next
$A_V$ bin, but reaches 5.9 at $A_V$=40-50 (the 23 objects with $A_V >$ 50, 
not shown in Figure \ref{Ltot_Lbol}, have a median $L_{tot}$/$L_{bol}$ ratio of 8.2). 
Among the 22 objects with best-fit $A_V$ values of 0-5 mag and inclination angles 
$\leq$ 50\degr, only four have $L_{tot}/L_{bol}$ ratios that are larger than 
1.5 (they are HOPS 57, 147, 199, and 201; in most cases the model overestimates 
the near-IR emission).

\subsection{Envelope Parameters for Different SED Classes}
\label{par_classes}

\begin{figure*}[!t]
\centering
\includegraphics[scale=0.65,angle=90]{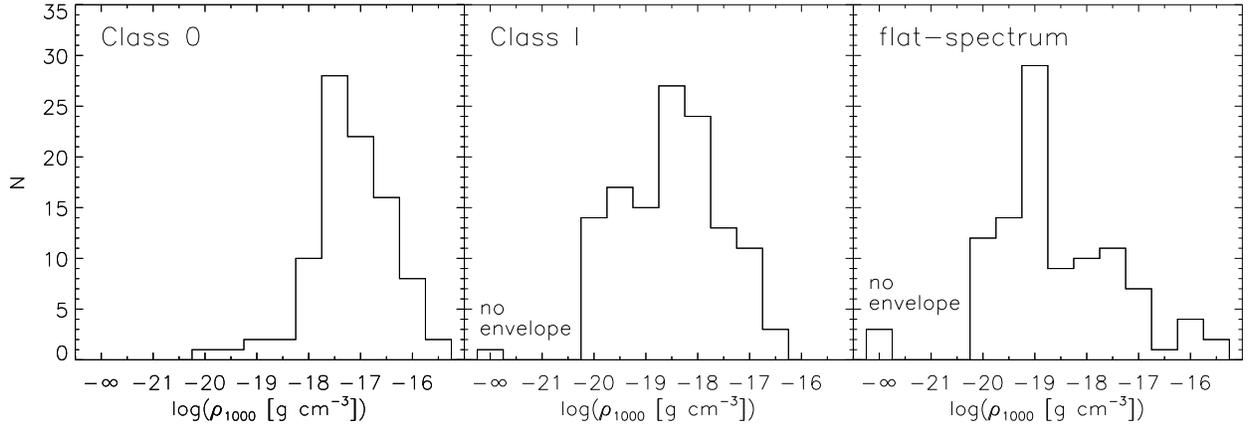}
\caption{Histograms of the envelope reference density $\rho_{1000}$ 
of the best fits for the different SED classes. 
\label{Rho_class_histo}}
\end{figure*}

\begin{figure*}[!t]
\centering
\includegraphics[scale=0.65,angle=90]{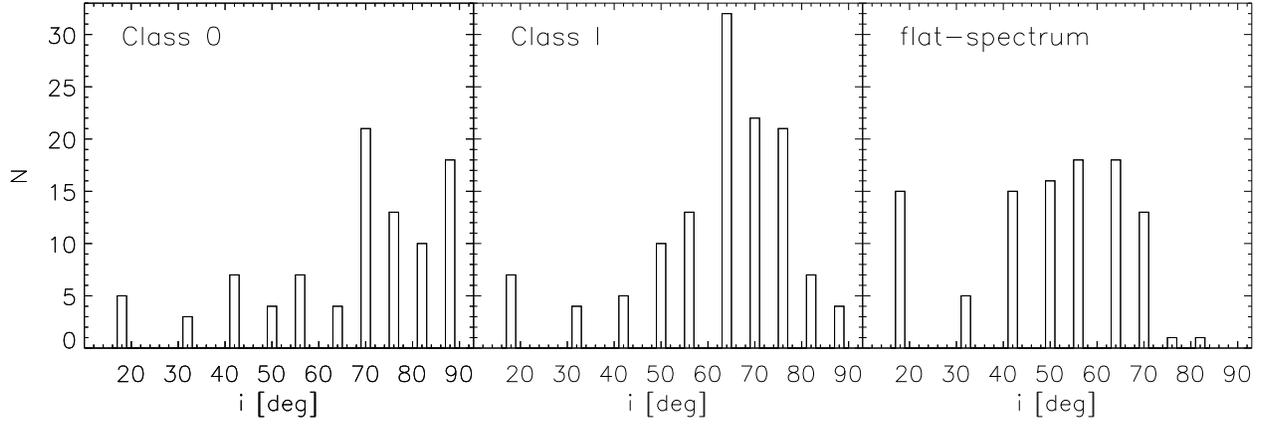}
\caption{Histograms of the inclination angles of the best fits for the different 
SED classes. 
\label{Inc_class_histo}}
\end{figure*}

\begin{figure*}[!t]
\centering
\includegraphics[scale=0.65, angle=90]{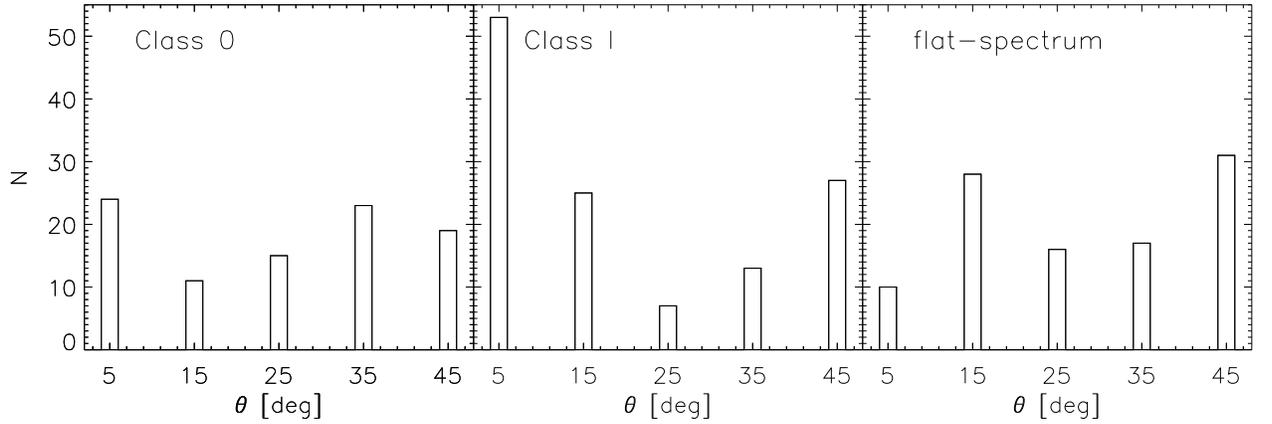}
\caption{Histograms of the cavity opening angles of the best fits for the different 
SED classes. 
\label{Cav_class_histo}}
\end{figure*}

\begin{figure*}[!t]
\centering
\includegraphics[scale=0.65, angle=90]{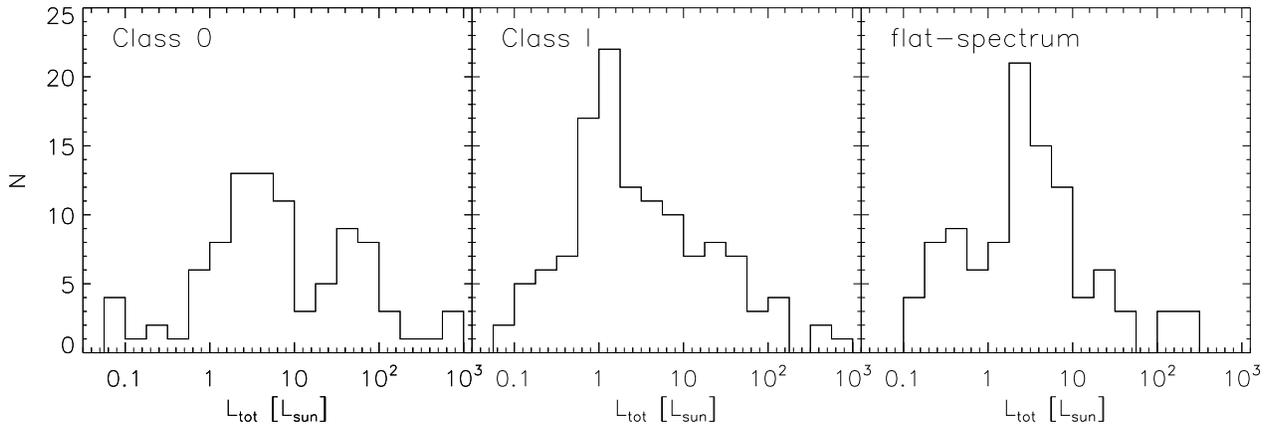}
\caption{Histograms of the total luminosity of the best fits for the different 
SED classes. 
\label{Ltot_class_histo}}
\end{figure*}

\begin{figure*}[!t]
\centering
\includegraphics[scale=0.65, angle=90]{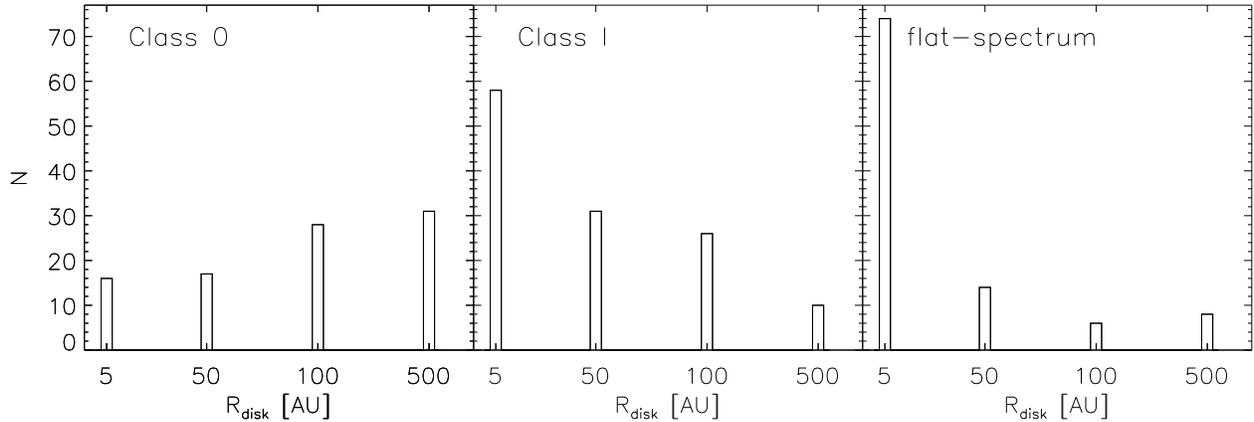}
\caption{Histograms of the disk radii of the best fits for the different 
SED classes. 
\label{Rdisk_class_histo}}
\end{figure*}

\begin{figure*}[!t]
\centering
\includegraphics[scale=0.65, angle=90]{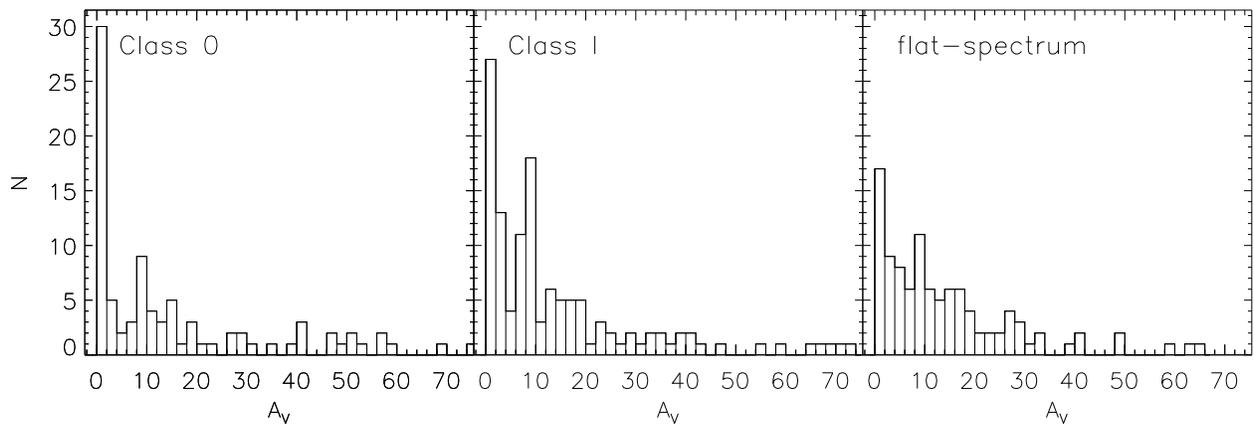}
\caption{Histograms of the foreground extinction of the best fits for the different 
SED classes. 
\label{AV_class_histo}}
\end{figure*}

Figures \ref{Rho_class_histo}--\ref{AV_class_histo} divide the histograms of the 
best-fit reference density $\rho_{1000}$, inclination angle, cavity opening angle, 
total luminosity, disk radius, and foreground extinction, respectively, by protostar 
class. As explained in Section \ref{SEDs}, we divided our targets into Class 0, 
Class I, flat-spectrum, and Class II objects based on their mid-infrared (4.5-24 \micron) 
spectral index and bolometric temperature (see also Table A\ref{bestfit}). Thus, 
Class 0 and I protostars have a spectral index $>$ 0.3, and Class 0 protostars have 
$T_{bol}$ values $<$ 70 K, but, as mentioned in Section \ref{SEDs}, there are 
a few protostars whose spectral index or $T_{bol}$ value places them very close 
to the transition region between Class 0 and I or between Class I and flat spectrum. 
Given that our sample contains just eleven Class II pre-main-sequence stars, we did 
not include them in the following histograms; they will be discussed in section 
\ref{disk_sources}.

The distributions of reference densities (Figure \ref{Rho_class_histo}) are 
different for all SED classes; none are consistent with being drawn from the 
same parent population (K-S test significance level $<$ 0.01). Overall, Class 0 
protostars have higher envelope densities than Class I and flat-spectrum sources; 
the median $\rho_{1000}$ values decrease from 5.9 $\times 10^{-18}$ g cm$^{-3}$ 
to 2.4 $\times 10^{-19}$ g cm$^{-3}$ to 1.2 $\times 10^{-19}$ g cm$^{-3}$ for 
these three groups. The lower and upper quartiles for $\rho_{1000}$ are 
1.8 $\times 10^{-18}$ and 1.8 $\times 10^{-17}$ g cm$^{-3}$ for the Class 0
protostars, and 2.4 $\times 10^{-20}$ and 1.2 $\times 10^{-18}$ g cm$^{-3}$ for the 
Class I and flat-spectrum objects.
We will discuss some implications of these differences in derived envelope
densities in section \ref{SED_class_properties}.

For the inclination angles (Figure \ref{Inc_class_histo}), the distributions 
are significantly different for all protostellar classes, too (K-S test significance 
level $\ll$ 0.01). As was shown in Figure \ref{Inc_histo}, a random distribution of 
inclination angles would result in equal numbers of protostars at each value; 
there is a deficit of Class 0 and Class I protostars at $i \lesssim$ 60\degr, and 
there are also few Class I protostars and hardly any flat-spectrum sources 
at the highest inclination angles. 
The median inclination angle is highest for Class 0 protostars (70\degr), then 
decreases somewhat for Class I protostars (63\degr) and even more for 
flat-spectrum sources (57\degr). Similar to the envelope density, 
the median inclination angle decreases as one progresses from Class 0 to 
flat-spectrum sources. 

In the distributions of cavity opening angles (Figure \ref{Cav_class_histo}), 
significant differences can be found between Class 0 and Class I protostars 
and between Class I protostars and flat-spectrum sources (K-S test significance 
level $\ll$ 0.01). The median cavity opening angle is 15\degr\ for the Class I 
protostars, but 25\degr\ for the other two classes. About 40\% of Class I protostars 
have $\theta$=5\degr, while the distribution among the different cavity opening 
angles is flatter for the other two object classes. The large fraction of Class I
protostars with small cavities could be the result of degeneracy in model parameters 
(see section \ref{SED_class_properties}) or our assumptions on envelope geometry 
(see section \ref{Model_problems}). There are notably few flat-spectrum sources 
with a 5\degr\ cavity opening angle; most of them have cavity opening angles of 
15\degr\ or 45\degr.

When comparing the total luminosities for the different SED classes
(Figure \ref{Ltot_class_histo}), the distribution of $L_{tot}$ values
is different for the Class 0 protostars when compared to the other two
classes (K-S test significance level $<$ 0.015), but similar for Class I 
protostars and flat-spectrum sources. The median total luminosity for 
Class 0 protostars is 5.5 \Lsun, compared to 2.0 \Lsun\ for Class I 
protostars and 3.0 \Lsun\ for flat-spectrum sources. Both Class 0 and I 
protostars cover close to the whole range of $L_{tot}$ values in the model grid 
($\sim$ 0.06-600 \Lsun), while flat-spectrum sources span a more limited 
range, from 0.1 to 316 \Lsun.

The distribution of centrifugal radii for the whole sample showed a preference
for 5 AU (see Figure \ref{Rdisk_histo}). When separating the best-fit disk
radii by protostellar class (Figure \ref{Rdisk_class_histo}), it is clear that the
trend for small centrifugal radii is driven by the flat-spectrum sources 
and also Class I protostars. The fraction of Class 0 protostars with $R_{disk}$=
5 AU is 17\%; it increases to 46\% and 73\% for Class I protostars and 
flat-spectrum sources, respectively. The median disk radius decreases 
from 100 AU for Class 0 protostars to 50 AU for Class I protostars to 5 AU for 
flat-spectrum sources. All three histograms are significantly different from 
one another (K-S test significance level $\ll$ 0.001). The unexpectedly small
centrifugal radii for Class I protostars and flat-spectrum sources could point
to parameter degeneracies (see section \ref{SED_class_properties}) or the
need to revise certain model assumptions (see section \ref{Model_problems}).

\begin{figure}[!t]
\centering
\includegraphics[scale=0.53]{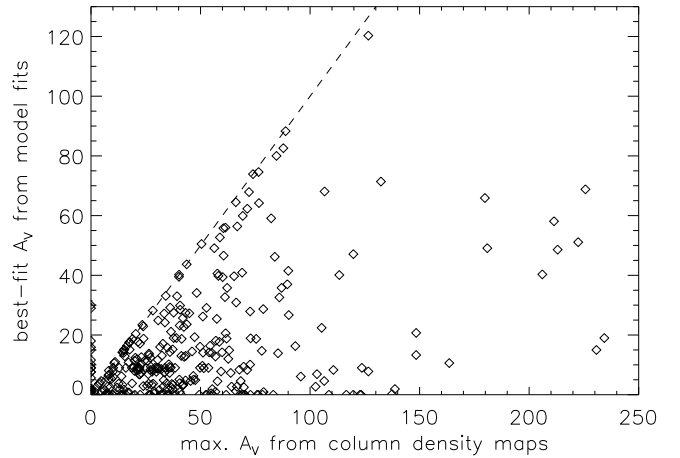}
\caption{Foreground extinction values $A_V$ from the best-fit models versus
the maximum $A_V$ value determined from column density maps of Orion.
The dashed line indicates where the two $A_V$ values are equal.
\label{AV_models_maps}}
\end{figure}

Finally, the distribution of best-fit foreground extinction values (Figure 
\ref{AV_class_histo}) is similar for all three object classes (K-S test significance 
level $>$ 0.03). Even the median values are close: $A_V$=9.2 for Class 0 
protostars, $A_V$=8.9 for Class I protostars, and $A_V$=10.1 for flat-spectrum
sources. Most objects are fit with relatively low foreground extinction values.
As can be seen from Figure \ref{AV_models_maps}, the majority of protostars
have best-fit $A_V$ values well below the maximum $A_V$ values determined
from column density maps, which were used as the largest allowed $A_V$
values for the SED fitter. The ratio of model-derived $A_V$ to observationally
constrained maximum $A_V$ is lower than 0.5 for about 60\% of the sample.

\begin{figure*}[!t]
\centering
\includegraphics[scale=0.65, angle=90]{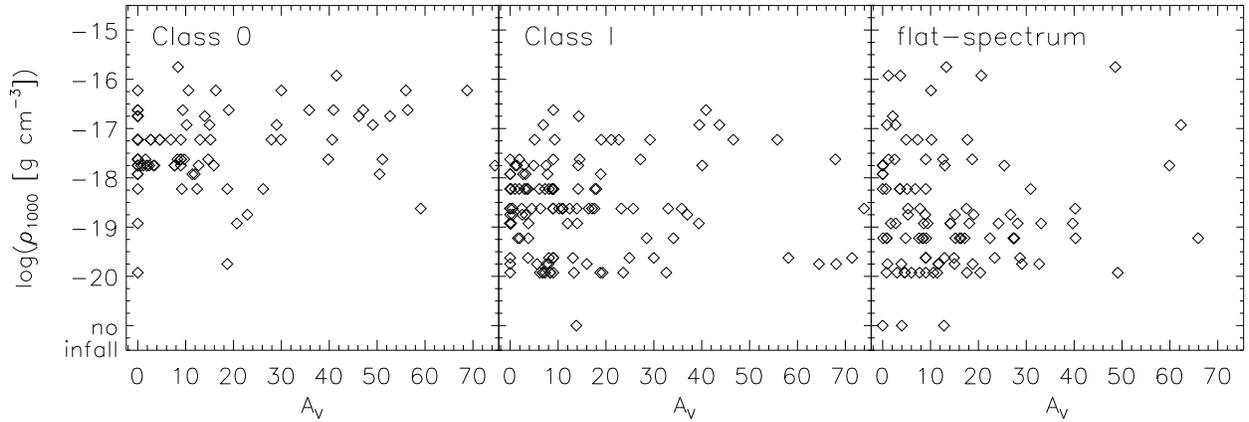}
\caption{Best-fit $\rho_{1000}$ values versus the foreground extinction for the different 
SED classes. Note that there are a few objects at $A_V > 75$, but they are not shown
for overall clarity of the figure.
\label{Rho_AV_class}}
\end{figure*}

\begin{figure*}[!t]
\centering
\includegraphics[scale=0.65, angle=90]{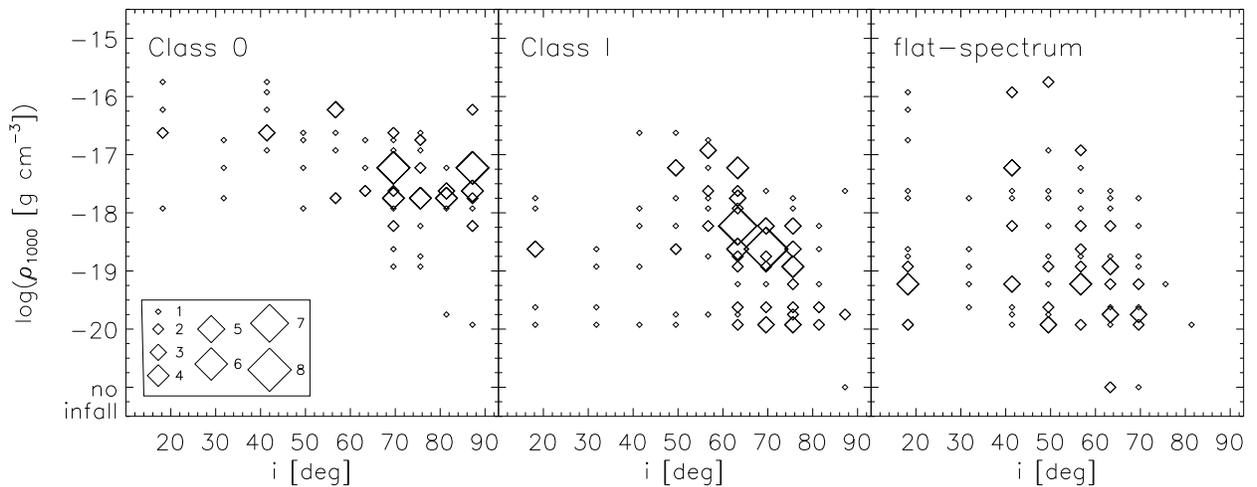}
\caption{Best-fit $\rho_{1000}$ values versus inclination angle for the different 
SED classes. The size of the plotting symbol increases with the number of objects
having the same ($i$, $\rho_{1000}$) combination; the legend in the leftmost panel
shows which symbol size corresponds to which number of objects. 
\label{Rho_inc_class}}
\end{figure*}

In Figure \ref{Rho_AV_class}, we plot the reference densities $\rho_{1000}$
versus the foreground extinction for Class 0, Class I, and flat-spectrum sources.
As was already seen in Figure \ref{AV_class_histo}, the extinction along the line
of sight is similar for all three classes, with most objects in the $A_V \sim$ 0-30
regime. Class 0 protostars, which have higher envelope densities, tend to 
have lower $A_V$ values from foreground extinction; the highest-density
envelopes are spread among a wide range of $A_V$ values. The result is
similar for Class I protostars. Flat-spectrum sources display a range in 
envelope densities at various foreground extinction values; the lowest-density
envelopes typically have $A_V < 20$.
Thus, foreground extinction does not seem to affect the classification of
protostars. This result is also supported by the statistical analysis of \citet{stutz15},
who found that, for $A_V$ values up to 35, the misclassification of a Class I
protostar as a Class 0 protostar due to foreground extinction (which results in a
lower $T_{bol}$) is low.

\begin{deluxetable*}{l|ccc}
\tablewidth{0.9\linewidth}
\tablecaption{Median Best-Fit Parameter Values for the Three Protostellar Classes
\label{Median_par}}
\tablehead{
\colhead{Parameter} & \colhead{Class 0} & \colhead{Class I} & 
\colhead{Flat-spectrum}}
\startdata
$L_{tot}$ & 5.5 \Lsun & 2.0 \Lsun & 3.0 \Lsun \\
$\rho_{1000}$ & 5.9 $\times 10^{-18}$ g cm$^{-3}$ & 
2.4 $\times 10^{-19}$ g cm$^{-3}$ & 1.2 $\times 10^{-19}$ g cm$^{-3}$ \\
$\theta$ & 25\degr & 15\degr & 25\degr \\
$R_{disk}$ & 100 AU & 50 AU & 5 AU \\
$i$ & 70\degr & 63\degr & 57\degr \\
$A_V$ & 9.2 & 8.9 & 10.1 \\
\enddata
\end{deluxetable*}

\begin{figure*}[!]
\centering
\includegraphics[scale=0.5, angle=90]{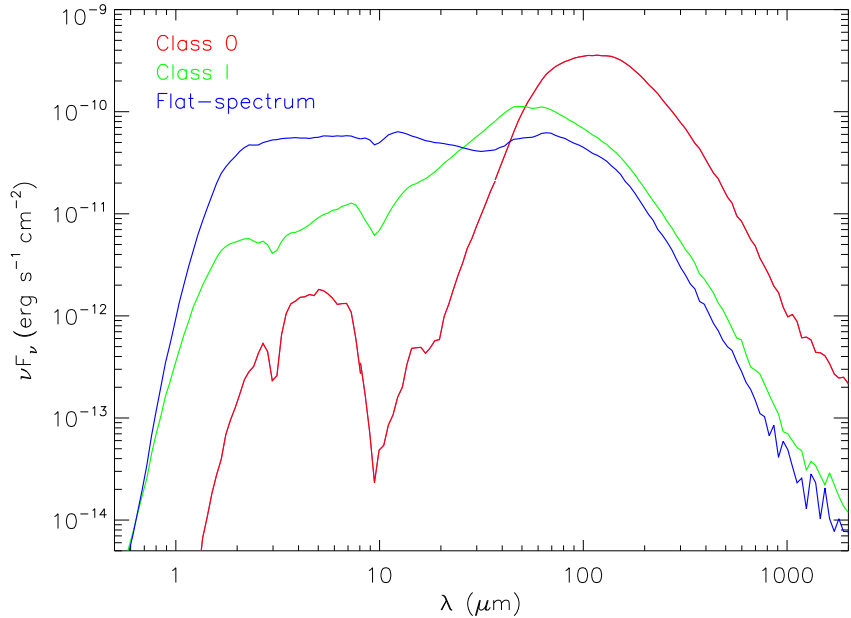}
\caption{Model SEDs for Class 0 protostars ({\it red}), Class I protostars
({\it green}), and flat-spectrum sources ({\it blue}) with parameter values 
equal to the median values for each SED class (see Table \ref{Median_par}).
\label{median_SEDs}}
\end{figure*}

We found differences in the best-fit envelope densities and inclination angles
for the various protostellar classes.
The result that Class 0 protostars tend to have larger inclination angles and 
envelope densities compared to Class I and flat-spectrum objects can also
be seen in Figure \ref{Rho_inc_class}. There are very few Class 0 protostars
with low inclination angles; most have relatively high density and $i>$60\degr. 
Class I protostars are best fit by somewhat lower inclination angles than Class 0 
protostars and also lower $\rho_{1000}$ values. The best-fit reference density
for Class I protostars decreases as the inclination angle increases; thus, higher-density
protostars are typically classified as Class I protostars only if they are not seen at
close to edge-on orientations.
Flat-spectrum sources are spread out in density--inclination space, but intermediate 
inclination angles and low envelope densities are common. There is a relatively
large number of objects at $i=$18\degr\ and a deficit of objects at high inclination
angles. The highest-density flat-spectrum sources are seen at inclination angles 
$<$ 50\degr, while the lower-density objects cover almost the full range of 
inclination angles.

The median parameter values we determined from the best fits for the 
Class 0, Class I, and flat-spectrum sources (see Table \ref{Median_par}) 
can be used to show representative median SEDs for each protostellar class. 
In Figure \ref{median_SEDs}, we show model SEDs whose parameter values 
are equal to the median values found for each of the three protostellar classes. 
It is apparent that the large envelope density and higher inclination angle for 
Class 0 protostars cause a deep absorption feature at 10 $\mu$m and a steeply 
rising SED in the mid- and far-IR, with a peak close to 100 $\mu$m. In Class I 
protostars, the SED is less steep and peaks at a shorter wavelength than 
the median SED of Class 0 protostars. Flat-spectrum sources show the 
strongest near-IR emission of the three protostellar classes; their median
SED is very flat out to 70 $\mu$m, but at longer wavelengths it is very 
similar in shape and flux level to that of Class I protostars.

\subsection{Estimating Parameter Uncertainties}
\label{deltaR}

\begin{figure*}[!t]
\centering
\includegraphics[scale=0.69, angle=90]{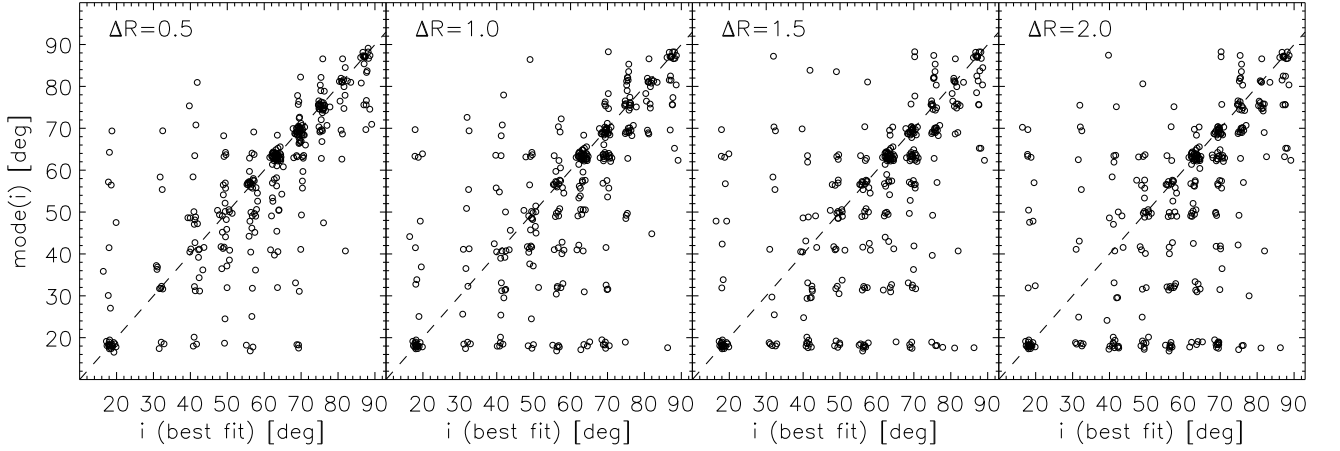}
\caption{Mode of the inclination angle of all models that lie within 
0.5, 1.0, 1.5, and 2 of the best-fit $R$ value (from left to right) versus 
the best-fit inclination angle for all 330 objects in our sample. 
Note that for each data point, small random offsets in the x and y direction 
have been applied to avoid overlap. Also, when two or more parameter 
values had the same frequency within a $\Delta R$ bin (i.e., not a unique 
mode value), we computed the average of these values and used it for 
the mode. The dashed line indicates where the mode and best-fit value 
are equal.
\label{inc_modes}}
\end{figure*}

\begin{figure*}[!t]
\centering
\includegraphics[scale=0.69, angle=90]{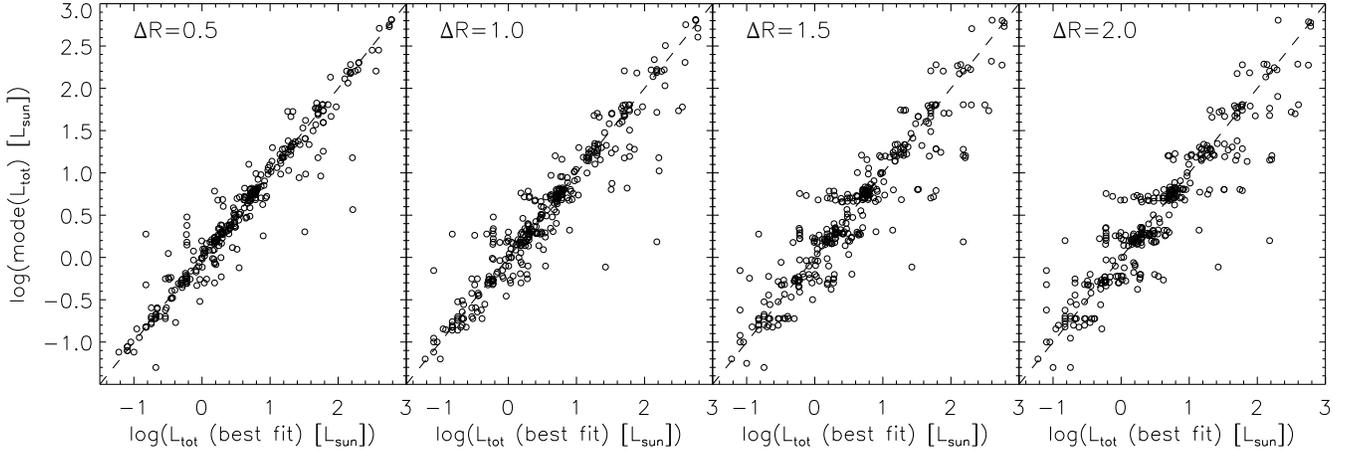}
\caption{Mode of the total luminosity of all models that lie within 
0.5, 1.0, 1.5, and 2 of the best-fit $R$ value (from left to right) versus 
the best-fit total luminosity for all 330 objects in our sample. 
\label{lum_modes}}
\end{figure*}

\begin{figure*}[!t]
\centering
\includegraphics[scale=0.69, angle=90]{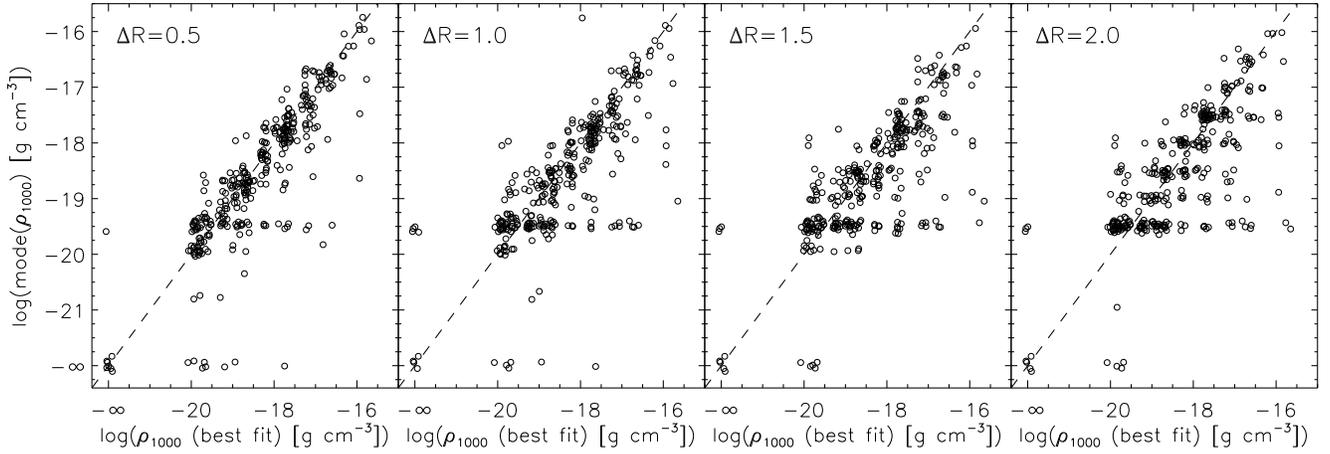}
\caption{Similar to Figure \ref{inc_modes}, but for the reference density
$\rho_{1000}$.
\label{rho1000_modes}}
\end{figure*}

\begin{figure*}[!t]
\centering
\includegraphics[scale=0.69, angle=90]{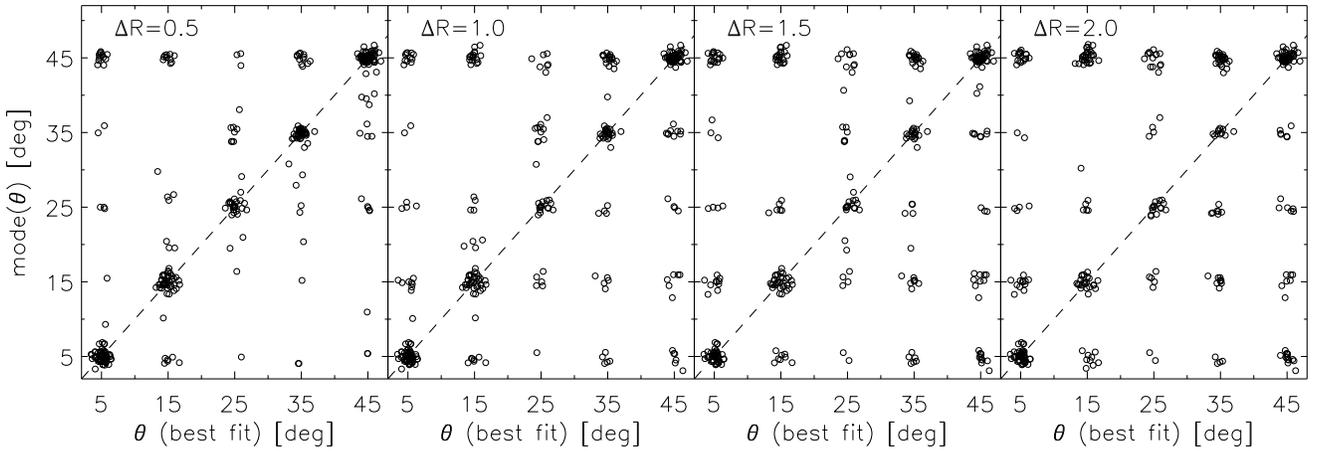}
\caption{Similar to Figure \ref{inc_modes}, but for the cavity opening
angle.
\label{cavity_modes}}
\end{figure*}

\begin{figure*}[!t]
\centering
\includegraphics[scale=0.69, angle=90]{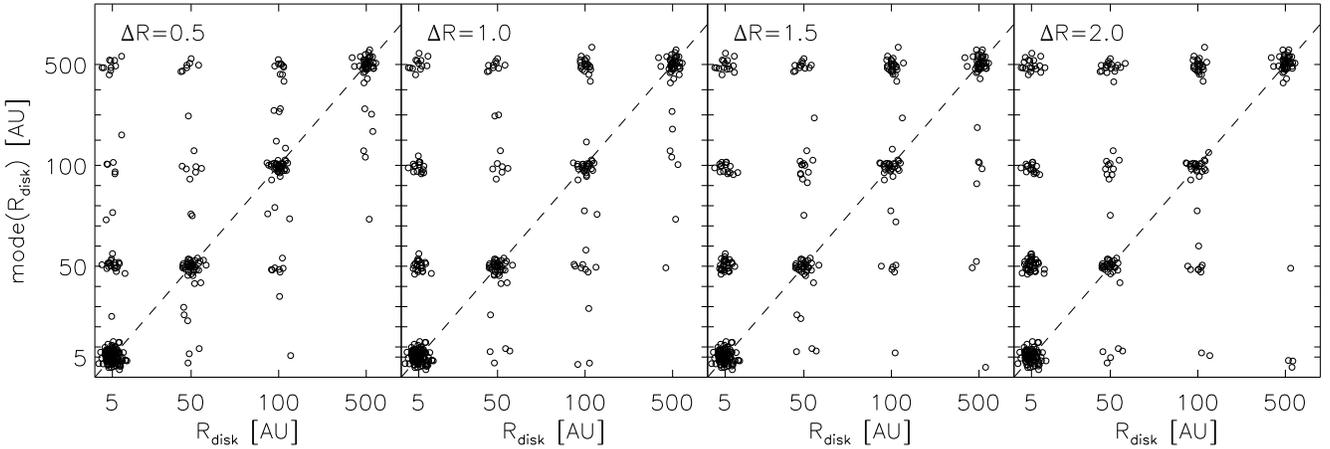}
\caption{Similar to Figure \ref{inc_modes}, but for the outer disk radius ($=R_c$).
\label{Rdisk_modes}}
\end{figure*}

\begin{figure*}[!t]
\centering
\includegraphics[scale=0.69, angle=90]{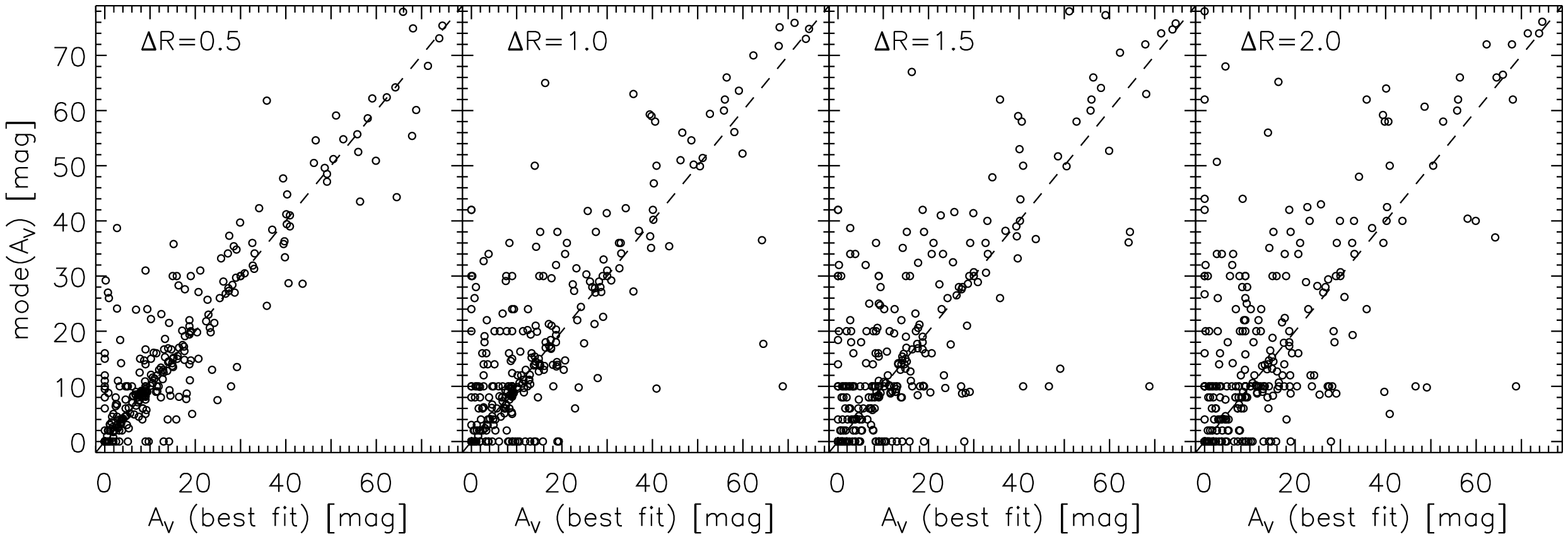}
\caption{Similar to Figure \ref{lum_modes}, but for the foreground
extinction.
\label{AV_modes}}
\end{figure*}

Given that the $R$ values are a measure of the goodness of fit in units of
the fractional uncertainty, we can use models that lie within a certain range
of the best-fit $R$ value to estimate uncertainties for the various model 
parameters. For each modeled HOPS target, we tabulated the model
parameters for all those models that lie within a difference of 0.5, 1.0, 
1.5, and 2.0 of the best-fit $R$. We then computed the mode (i.e., the 
value with the highest frequency) for the inclination angle, total luminosity, 
$\rho_{1000}$, cavity opening angle, outer disk radius, and $A_V$ in 
each of the $\Delta R$ bins for each object. 
For any given protostar, the models in each $\Delta R$ bin span certain 
ranges in parameter values; while the modes do not capture the full extent of 
these ranges, they convey the most common value within each parameter 
range. The farther away a mode is from the best-fit value, the more poorly 
constrained the model parameter. Conversely, if a mode of a certain
parameter is close to or matches the best-fit value, especially for $\Delta R=$ 
1.5 or 2, that particular model parameter is well constrained.
In Figures \ref{inc_modes} to \ref{AV_modes} we plot the mode versus the best-fit 
value for six model parameters and four $\Delta R$ bins for all 330 targets 
in our sample. The larger $\Delta R$, the larger the spread in modes is expected 
to be for each parameter value. 

For example, Figure \ref{inc_modes} shows that even when considering 
all models with an $R$ value of up to 2 larger than the best-fit $R$ 
($\Delta R=$ 2), the inclination angle for objects with a best-fit $i$ of 
18\degr\ is well constrained; most modes lie at $i=$ 18\degr, too, and 
only a few modes can be found at larger inclination angles. However, 
objects with best-fit $i$ values of 32\degr\ or 41\degr\ typically can also 
be fit by models with lower inclination angles (the majority of modes lies 
below the line where mode and best-fit value are equal). Inclination 
angles $\gtrsim$ 63\degr\ are better constrained, since their modes 
mostly lie at high inclination angle values, but there are protostars 
with modes of $i=$18\degr, too.

The modes for the total luminosity (Figure \ref{lum_modes}) show a small 
spread for models within $\Delta R$=0.5, but the spread increases as $R$ 
increases, with some objects displaying up to an order of magnitude in 
variation of $L_{tot}$.
As illustrated in Figure \ref{rho1000_modes}, the reference density 
$\rho_{1000}$ is usually well constrained; however, as $R$ increases, 
the modes of the $\rho_{1000}$ values are often lower than the best-fit 
values.
For the cavity opening angle (Figure \ref{cavity_modes}), many models up 
to $\Delta R$=2 have modes of $\theta$=45\degr, independent of the 
best-fit value. Similarly for the centrifugal radius (Figure \ref{Rdisk_modes}), 
$R_{disk}$=500 AU is a common mode. For all four disk radii, the modes 
tend to be larger than the best-fit values; in particular, objects with a best-fit 
$R_{disk}$ of 5 AU have a very uncertain disk radius. In general, it looks like 
our models do not constrain the disk radius and cavity opening angle well.
The foreground extinction (Figure \ref{AV_modes}) displays a certain range 
of modes for each best-fit value, but objects with $A_V \lesssim$ 20 typically
have more reliable $A_V$ values from their model fits. 

Figures \ref{inc_modes} to \ref{AV_modes} allow us to gauge general trends 
between best-fit values and modes for different model parameters. For results 
on individual objects, we refer to Appendix \ref{models_unc}, where we show 
plots of the difference between the modes and the best-fit values of the major 
model parameters for all modeled HOPS targets. In this way it is possible to 
estimate which models are better constrained and thus which objects have 
more reliable SED fits. 
In addition, in Appendix  \ref{models_unc} we also include contour plots of 
$R$ values for different pairs of model parameters for a few targets to 
illustrate typical parameter degeneracies, which also contribute to parameter
uncertainties.

\section{Discussion} 
  
\subsection{Deriving Envelope Parameters from a Model Grid}

We compared a grid of TSC models to each target by ranking the models 
using a statistic, $R$, which measures the deviation between observed and  
model fluxes in logarithmic space. We did not model each source by further 
adjusting the model parameters, but instead identified the best-fit SED from 
our model grid. Thus, we are bound by the range and sampling of parameters 
chosen for the grid, and while we constructed the grid with the aim of covering 
the typical parameter space for protostars, it is limited to discrete values. It is 
likely that many protostars have best-fit parameters that would fall between those 
sampled by the grid, and a few objects could have parameter values that lie 
beyond the limits set by the grid. In addition, TSC models are axisymmetric 
and have mostly smooth density and temperature profiles, and they do not
include external heating. They assume a rotating, infalling envelope with 
constant infall rate and with the gravitational force dominated by the central
protostar, but the true envelope structure is likely more complex. 
The models would not apply to the collapse of a cloud in an initial filamentary 
or sheet-like geometry or to multiple systems with, e.g., more than one 
outflow cavity \citep[e.g.,][]{hartmann96,tobin12}.

Despite the relatively simple models that we use, many of the observed SEDs are 
fit remarkably well: 75\% of the fits have $R<4$. In those cases, the continuum 
traced by the IRS spectrum, the silicate absorption feature at 10 $\mu$m, and 
the PACS fluxes are all accurately reproduced by the model. Even many 
flat-spectrum sources, which often do not display any spectral features in 
the mid-infrared and have an overall flat SED out to 30 or 70 $\mu$m, often 
have models that fit them very well. About 75\% of Class I protostars and  $\sim$ 
70\% of Class 0 protostars have $R < 4$, while close to 90\% of flat-spectrum 
sources have $R$ values in this range. This validates the choice of parameter 
values for our model grid.  
Additional constraints, like limits on foreground extinction or information on the 
inclination and cavity opening angles from scattered light images or mapping of 
outflows, would allow us to further test and refine the models.  We have used limits 
on the extinction in our analysis.  Although {\it Hubble Space Telescope (HST)} 
scattered light images have been used to constrain models for one HOPS 
protostar \citep{fischer14}, scattered light images are not available for many 
of our targets. We therefore chose to focus on fitting the SEDs of all of our 
targets in a uniform way to a well-defined set of models. Future studies will 
incorporate scattered light images and compare the results to those from 
the SED fits (J. Booker et al. 2016, in preparation).

The best-fit models from our grid for the HOPS targets both reproduce the SEDs 
and yield estimates for their protostellar parameters, mostly envelope properties.  
However, these are not necessarily unique fits to the data for three reasons. 
First, there are degeneracies in the model parameters; increasing the envelope 
density or inclination angle, or decreasing the cavity opening angle or disk radius, 
results in a steeper mid-IR SED slope and deeper silicate features. Each of these 
parameters affects the SED differently (just the general trends are the same), and 
the best fit for each object tries optimizing them. The next best fit, however, could 
be a different combination of these parameters, especially if the SED is not 
well constrained by observations (see Section \ref{deltaR}).  
Second, although the TSC models reproduce the observed SEDs, other models 
with different envelope geometries may also be able to fit the same SEDs. The 
modeling presented here is only valid in the context of the TSC models of single 
stars, and the resulting derived properties are only valid within that framework.   
Last, neglecting external heating could result in overestimated envelope densities
and luminosities, with the most noticeable effects ($\rho_{1000}$ and $L_{tot}$ too 
large by factors of a few) on low-luminosity sources exposed to strong radiation 
fields (see Section \ref{model_ext_heat}). From the distribution of best-fit $L_{tot}$ 
values, we estimate that $\sim$ 20\% of HOPS targets in our sample could be 
affected by external heating. Even though we do not know the strength of 
external heating for each protostar, it is likely that external heating would only 
result in relatively small changes in the derived envelope parameters for these 
protostars.

\subsection{Envelope Properties and SED Classes}
\label{SED_class_properties}

When comparing envelope parameters sorted by SED classes, we found that 
envelope densities and inclination angles decrease from the sample of
Class 0 protostars through that of Class I protostars to that of flat-spectrum objects. 
The former is likely an evolutionary effect, while the latter confirms the results of 
previous work \citep[e.g.,][]{evans09} that  the inclination angle has an important 
effect on the SED and that the evolutionary state of an object cannot be derived 
from SED slopes alone. Thus, there is a difference between the ``stage'' and 
``class'' of an object \citep{robitaille06}; Stage 0 and I objects are characterized 
by substantial envelopes, Stage II objects are surrounded by optically thick disks, 
with possibly some remnant infalling envelopes, and Stage III objects have 
optically thin disks. 

In general, the trends we see among model parameters are a consequence 
of the definition of a protostar based on its SED: in order to be classified 
as a Class 0 or I object, a protostar is required to have a near- to mid-infrared 
SED slope larger than 0.3. A protostellar  model with a small cavity opening 
angle, small centrifugal radius, and/or high inclination angle will generate 
such an SED, since it increases the  optical depth along the line of sight. 
Models with a large cavity will only yield a rising SED in the 2$-$40 $\mu$m 
spectral range if their  envelope density is large or the inclination angle is 
relatively high. 

We find that Class 0 protostars can be best fit not only by very high envelope 
densities but also moderately high envelope densities and large inclination 
angles. The bolometric temperature, which is used to separate 
Class~0 from Class~I protostars, is inclination dependent; some Class~I 
protostars are shifted to the Class~0 regime if they are viewed more edge-on. 
The higher-density Class~I protostars tend to have lower inclination angles (but 
still $>$ 50\degr); thus, their evolutionary stage could be similar to more 
embedded protostars that are seen edge-on and classified as Class 0 protostars. 
Conversely, some Class~0 objects with large inclination angles, but lower 
envelope densities, could be in a later evolutionary stage than typical Class~0 
protostars. Similarly, Class I protostars with large $i$ and low $\rho_{1000}$ 
values could be edge-on Stage II objects (whose infrared emission is dominated 
by a disk). Finally, low-inclination Stage 0 and I protostars can appear 
as a flat-spectrum sources \citep{calvet94}.

\begin{figure*}[!]
\centering
\includegraphics[scale=0.58, angle=90]{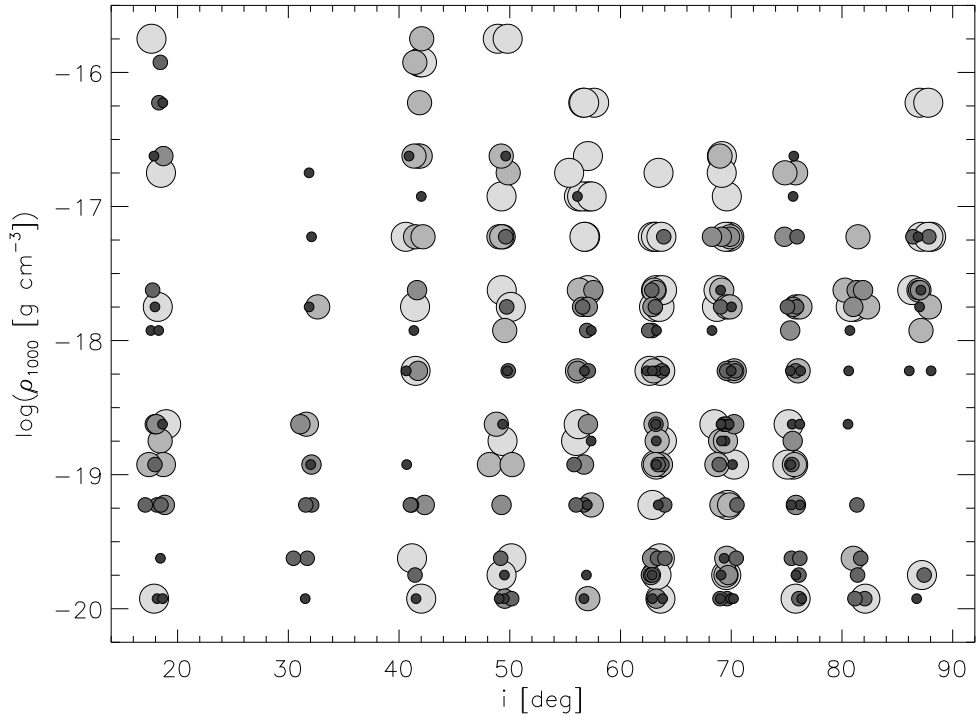}
\caption{Best-fit $\rho_{1000}$ values versus inclination angle with the cavity
size indicated by the different symbol sizes and gray shades: symbols become
larger and lighter colored with increasing cavity size (5\degr, 15\degr, 25\degr,
35\degr, 45\degr). A small random offset in the x direction has been applied to 
each data point to prevent too much overlap.
\label{Rho1000_inc_cavity}}
\end{figure*}

\begin{figure*}[!t]
\centering
\includegraphics[scale=0.58, angle=90]{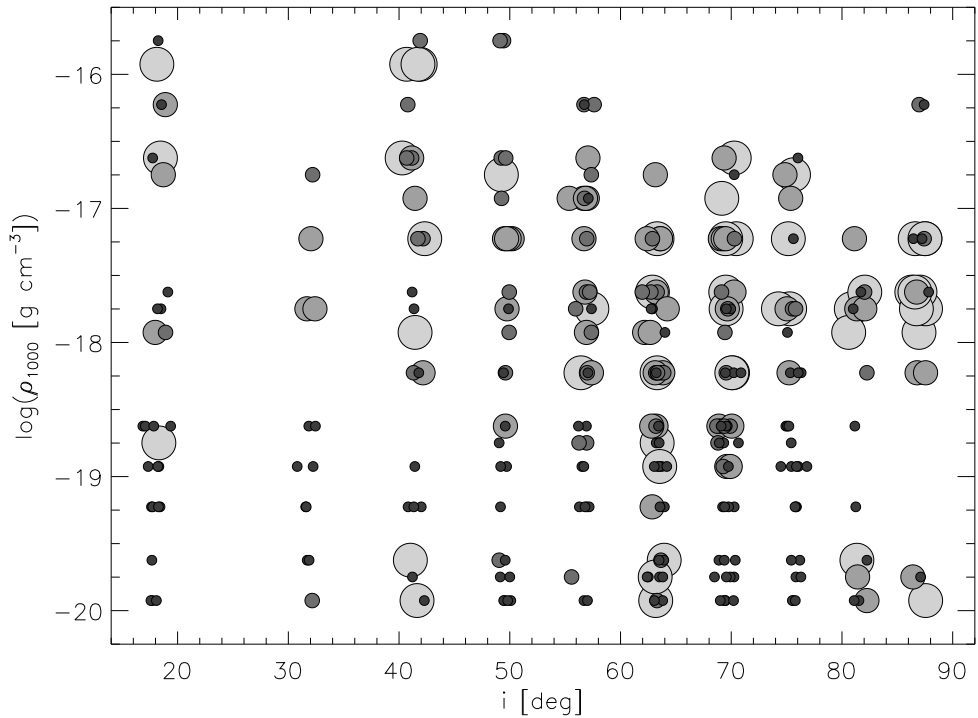}
\caption{Similar to Figure \ref{Rho1000_inc_cavity}, but with the outer disk 
radius indicated by the different symbol sizes and gray shades: symbols 
become larger and lighter colored with increasing disk radius (5, 50, 100,
500 AU).
\label{Rho1000_inc_Rdisk}}
\end{figure*}

Nevertheless, the observed trend in envelope densities suggests that the variations 
in the observed SEDs track, in great part, an evolution toward lower envelope 
densities and lower infall rates. Assuming a certain mass for the central 
star, the reference density in our models can be used to infer an envelope infall rate
($\dot{M}_{env} \propto \rho_{1000} \sqrt{M_{\ast}}$). 
As mentioned in section \ref{model_parameters}, this infall rate is model dependent 
and therefore tied to the assumptions of the models. With this in mind, the median 
$\rho_{1000}$ values for the Class 0, Class I, and 
flat-spectrum protostars in our sample correspond to envelope infall rates of 
$2.5 \times 10^{-5}$, $1.0 \times 10^{-6}$, and $5.0 \times 10^{-7}$ \Msun\ yr$^{-1}$, 
respectively, for a 0.5 \Msun\ star. 
Using a more realistic assumption of larger stellar mass for more evolved protostars, 
the infall rates for Class I and flat-spectrum protostars would be larger than these 
values by a factor of a few. However, just larger stellar masses cannot explain the
large decrease of a factor of 50 in the median envelope density from Class 0 to 
flat-spectrum protostars; to achieve such a decrease with a constant infall rate
of $2.5 \times 10^{-5}$ \Msun\ yr$^{-1}$, the stellar mass would have to increase
by a factor of 2500. Thus, within the context of our model fits, we can conclude that,
as envelopes become more tenuous, the infall rates also decrease.

Other trends are also apparent. Class 0 protostars and flat-spectrum sources show 
a relatively flat distribution of cavity opening angles. On the other hand, the best 
fit for a large fraction of Class I protostars (40\%) results in $\theta$=5\degr. 
This could point to a  degeneracy in the models, since protostars with small cavity 
opening angles tend to have lower envelope densities (and also lower total 
luminosities); thus, the smaller cavity partly compensates for the lower opacity 
resulting from the lower envelope density (see also Figure \ref{Rho1000_inc_cavity}). 

Even though our models do not yield reliable disk properties, we can make a
statement about the difference in the best-fit centrifugal radii (or $R_{disk}$),
which are tied to the structure of the rotating envelope given by the model fits.  
It should be noted that the centrifugal radii set a lower limit to the disk radii, 
since disks may spread outward due to viscous accretion.  Most Class I protostars 
and flat-spectrum sources are fit with a centrifugal radius of just 5 AU. Since 
the smallest centrifugal radius in our model grid is 5 AU and the next value is 50 AU, 
we can state that, except for Class 0 protostars, most protostars in our sample 
have $R_{disk} < 50$~AU, and some might even have $R_{disk} <$ 5~AU.

Small disks of those sizes have been observed; radio interferometry  of the 
multiple protostellar system L1551 IRS 5 shows disks whose semi-major axes 
are $\lesssim$~20~AU \citep{rodriguez98,lim06}. However, there is a degeneracy 
between the centrifugal radius and the envelope density; for protostars with low 
envelope densities, the small centrifugal radius can compensate the decrease in 
opacity by concentrating more material closer to the star. As can be seen in Figure 
\ref{Rho1000_inc_Rdisk}, most protostars with $R_{disk}=$ 5 AU also have lower 
envelope densities. Inclination angle also plays a role; protostars seen at 
$i>$ 80\degr\ typically have larger centrifugal radii. In addition, our envelope
models include outflow cavities, which allow some of the mid-IR radiation to
escape. In order to generate model SEDs with silicate absorption features 
and steep mid-IR slopes at low to intermediate inclination angles, a lower 
$R_{disk}$ value is needed.  
We will discuss the potential implications of the small cavity sizes and $R_c$ 
values for Class I protostars and flat-spectrum sources in Section 
\ref{Model_problems}.

\subsubsection{The Most Embedded Protostars}
\label{Class0}

Among the Class 0 protostars, there are protostars in the earliest evolutionary 
stages, when the envelope is massive and the protostar still has to accrete 
most of its mass. 
\citet{stutz13} identified 18 protostars with very red mid- to far-infrared
colors ($\log(\lambda F_{\lambda}(70)/\lambda F_{\lambda}(24))$
$>$ 1.65), of which 11 were newly identified (see Table D\ref{New_proto}). 
\citet{tobin15} added an additional object.
These protostars were named PACS Bright Red sources (PBRs) by \citet{stutz13}; 
they are HOPS 169, 341, 354, 358, 359, 372, 373, 394, 397-405, 407, and 409. 
Based on their steep 24-70 $\mu$m SEDs and large submillimeter luminosities, 
they were interpreted as the youngest protostars in Orion with very dense 
envelopes.

From our best-fit models to the SEDs of the PBRs, we derive a median 
$\rho_{1000}$ value of 1.2 $\times$ 10$^{-17}$ g cm$^{-3}$, 
which is twice as high as the median value of all the Class 0 protostars in 
our sample. These fits also result in a median envelope mass within 2500 AU 
of 0.66 \Msun\ for the PBRs, but the individual objects cover a large 
range, from 0.07 to 1.83 \Msun. The median total luminosity amounts to 
5.6 \Lsun\ (with a range from 0.6 to 71.0 \Lsun), which is very similar
to the median $L_{tot}$ value for the Class 0 protostars in our sample. 
Most PBRs (14 out of 19 protostars) are fit by models with large inclination 
angles ($i \geq$ 70\degr), but, as shown in \citet{stutz13}, high inclination
alone cannot explain the redness of the PBRs. 
Thus, our models confirm the results of \citet{stutz13} that the PBRs
are deeply embedded and thus likely among the youngest protostars
in Orion.

\subsubsection{Flat-spectrum Sources}
\label{flat-spectrum}

A particularly interesting group of protostars that are not easy to categorize
are the flat-spectrum sources. They are thought to include objects in transition 
between Stages I and II, when the envelope is being dispersed \citep{greene94}. 
In particular, those with low envelope densities could be more evolved protostars, 
or they could be protostars that started out with more tenuous envelopes. 
On the other hand, flat-spectrum sources could also be highly inclined 
disk sources \citep[see][]{crapsi08}, or protostars surrounded by dense 
envelopes, but seen close to face-on \citep{calvet94}. This type of 
misclassification could have a large effect on the lifetimes of the earlier 
protostellar stages and thus on the timeline of envelope dispersal. 
Among the 330 HOPS targets in our sample, we identified 102 flat-spectrum 
sources based on their flat ($-0.3$ to $+0.3$) spectral index from 4.5 to 24 
$\mu$m (or 5-25 $\mu$m in a few cases). Thus, they compose a fairly large 
fraction of our protostellar sample. Of these 102 objects, 47 have a negative 
spectral index and 55 have one between 0 and $+0.3$; 41 have a spectral 
index between $-0.1$ and 0.1, which results in a very flat mid-infrared SED. 

Despite a flat SED slope between 4.5 and 24 $\mu$m, many flat-spectrum
sources display a weak silicate emission or absorption feature at 10 $\mu$m,
which may indicate the presence of a very tenuous infalling envelope or
may be the result of the viewing geometry. Some SEDs are very flat out to 
100 $\mu$m, others have negative spectral slopes beyond 40 $\mu$m, 
and again others a rising SED from the mid- to the far-IR.
There are also objects with more pronounced absorption features due to 
not only silicates but also ices, as are typically found in Class 0 and I protostars,
but also edge-on disks (see HOPS 82, 85, 89, 90, 92, 129, 150, 200, 210, 211, 
281, 304, 331, and 363). Only two flat-spectrum sources have prominent silicate 
emission features, and their SEDs are reminiscent of protoplanetary disks 
(see HOPS 187 and 199). Thus, flat-spectrum sources likely include objects of a 
variety of evolutionary stages.

\begin{figure}[!t]
\centering
\includegraphics[scale=0.39, angle=90]{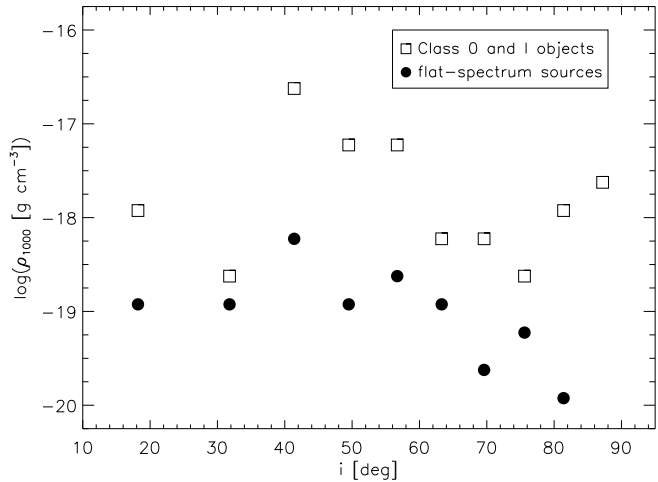}
\caption{Median best-fit $\rho_{1000}$ values at each inclination angle 
for the Class 0 and I protostars ({\it squares}) and the flat-spectrum sources
({\it circles}) in our sample. 
\label{Rho_inc_flat-spectrum}}
\end{figure}

The latter conclusion can also be drawn when analyzing the distribution
of envelope reference densities and inclination angles for flat-spectrum
sources. In Figure \ref{Rho_inc_class}, we showed that flat-spectrum
sources typically have intermediate inclination angles and lower envelope
densities. To compare their properties more directly to Class 0 and I protostars,
in Figure \ref{Rho_inc_flat-spectrum} we show the median best-fit 
$\rho_{1000}$ value at each best-fit inclination angle; it is larger for Class 0
and I protostars than for flat-spectrum sources at all inclination angles.
For Class 0 and I protostars, the median $\rho_{1000}$ value is highest
at intermediate inclination angles, decreases at larger inclination angles,
and then increases again for $i>$ 80\degr. For flat-spectrum sources, the 
median $\rho_{1000}$ value is relatively flat over the 18\degr--63\degr\
region but has its peak value at $i=$41\degr; it decreases for larger 
inclination angles. The only flat-spectrum source with a best-fit inclination 
angle of 81\degr, HOPS 357, has a very low envelope density (the lowest 
value for this parameter in the model grid), and its spectrum displays a 
deep silicate absorption feature.

Overall, this shows that, while a range of envelope densities and inclination 
angles can explain flat-spectrum sources, their envelope densities are typically
lower than for Class 0 and I protostars. The higher-density objects are seen at low
to intermediate inclination angles, while only the lowest-density objects
are seen closer to edge-on. Some of the high-density flat-spectrum sources
could actually be more embedded protostars (Stage 0 objects) seen
face-on (which would be classified as Class 0 objects if seen at larger
inclination angles). Thus, in terms of envelope evolution, they include a
diverse group of objects. 

\begin{figure*}[!t]
\centering
\includegraphics[scale=0.71, angle=90]{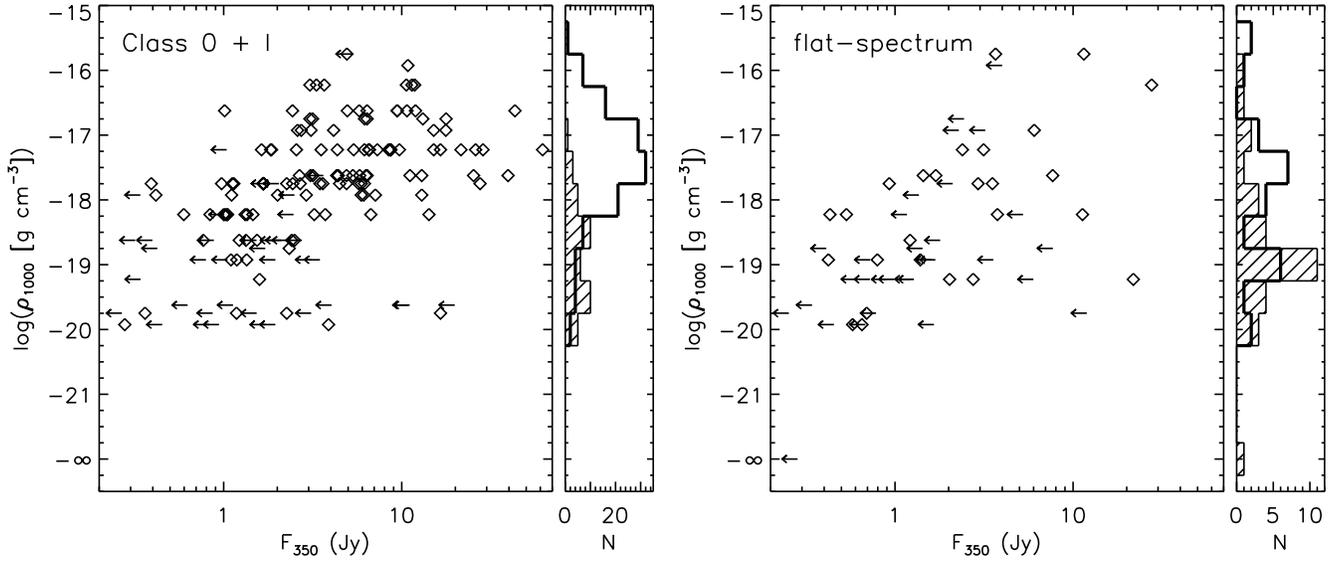}
\caption{Best-fit $\rho_{1000}$ values versus the 350 $\mu$m fluxes
for the Class 0 and I protostars ({\it left}) and the flat-spectrum sources
({\it right}) in our sample. Detections at 350 $\mu$m are shown with 
diamonds, while upper limits are shown with arrows. The histograms 
show the distribution of best-fit $\rho_{1000}$ values for sources with 
a 350 $\mu$m flux measurement ({\it thick solid line}) and with 350 
$\mu$m upper limits ({\it shaded area}). 
\label{Rho_F350}}
\end{figure*}

\begin{figure*}[!t]
\centering
\includegraphics[scale=0.71, angle=90]{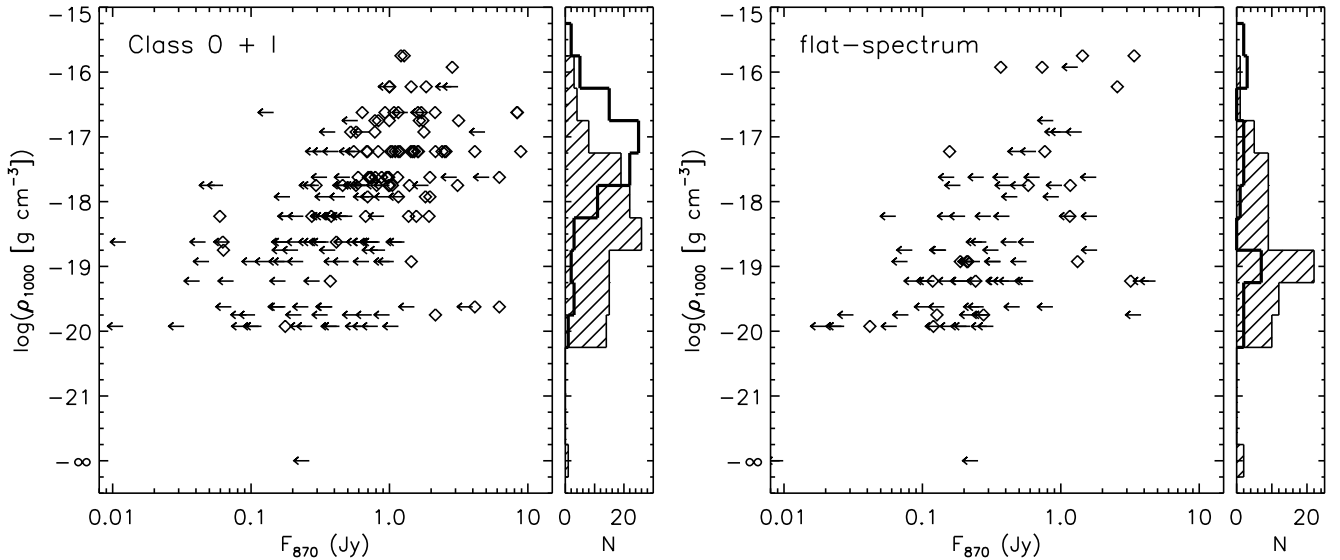}
\caption{Similar to Figure \ref{Rho_F350}, but for the 870 $\mu$m fluxes.
\label{Rho_F870}}
\end{figure*}

We note that even though we find that flat-spectrum sources have in general
lower envelope densities than Class 0 and Class I objects, their best fit does
include an envelope in almost all cases; just 3 of the 102 flat-spectrum 
sources are best fit without an envelope. This seems to contradict recent
findings by \citet{heiderman15}, who found that only about 50\% of flat-spectrum
sources were actually protostars surrounded by envelopes. This could be partly
explained by different criteria used to select flat-spectrum sources; in the 
\citet{heiderman15} sample, flat-spectrum sources are selected by their
extinction-corrected 2-24 $\mu$m spectral index (see also \citealt{evans09,
dunham13}), while our sample uses a flat 4.5-24 $\mu$m spectral index.
Moreover, in their study \citet{heiderman15} detected the presence of an 
envelope via HCO$^+$ emission, and they found that almost all sources 
detected in the sub-mm are also detected in HCO$^+$ (but the opposite 
does not always hold). For our sample of Orion protostars, we find that 
75\% of Class 0+I protostars observed with SABOCA (350 $\mu$m) are 
detected, while only 47\% of flat-spectrum sources have detections. For 
LABOCA observations (870 $\mu$m), these two fractions amount to 
41\% and 21\%, respectively. Thus, we find that flat-spectrum sources 
have a $\sim$ 50\% lower sub-mm detection rate than Class 0+I protostars. 
Flat-spectrum sources without sub-mm detections would likely also not 
display HCO$^+$ emission and thus would be considered as protostars 
without envelopes by \citet{heiderman15}.

To compare how our submillimeter detections correlate with the presence
of an envelope, in Figures \ref{Rho_F350} and \ref{Rho_F870} we show the
derived best-fit reference envelope densities as a function of 350 or 870 $\mu$m
fluxes for the combined Class 0+I sample and the flat-spectrum sources. We
also differentiate the distribution of envelope densities between measured
flux values and upper limits; at 870 $\mu$m, the upper limits are often cases
where the sources are not detected due to confusion with bright, spatially
varying emission. We find that even protostars with upper limits at 350 and 
870 $\mu$m are best fit with an envelope; however, the envelope density 
is lower for objects with upper limits in the sub-mm. This is especially evident 
for Class 0+I protostars; for flat-spectrum sources, the distributions of envelope 
densities for sub-mm detections and upper limits show  significant overlap. Four 
times as many flat-spectrum sources have upper limits instead of detections 
at 870 $\mu$m, but their derived $\rho_{1000}$ values span almost the full 
range of values. Furthermore, the median $\rho_{1000}$ value of 1.19 $\times 
10^{-19}$ g cm$^{-3}$ for sources without detections is relatively close to the 
median value of 5.95 $\times 10^{-19}$ g cm$^{-3}$ for the sources with 
870 $\mu$m detections. Thus, our model fits do not rely on sub-mm detections 
to yield a best fit with an envelope; in most cases the near- to far-IR SED is 
sufficient to constrain the properties of the envelope.

\subsubsection{Sources without an Envelope and Class II Objects}
\label{disk_sources}

Among the six objects whose best-fit SED required no envelope ($\rho_{1000}$
value of 0), three are flat-spectrum sources (HOPS 47, 187, 265), two are Class 
II pre-main-sequence stars (HOPS 113, 293), and one is a Class I protostar (HOPS 232).  
The low 70 $\mu$m fluxes of HOPS 47 and 265 constrained the best model to 
one without an envelope. The SED of HOPS 187 looks like that of a transitional 
disk, which are disks with gaps or holes in their inner regions (see \citealt{espaillat14} 
and references therein). If HOPS 187 were a transitional disk, it would not have 
an envelope. HOPS 232 has a rising SED over the mid-IR spectral range; its best 
fit requires no envelope, but an edge-on disk with a high accretion luminosity.

It would be expected that the SEDs of Class II objects can be best fit by a 
model that does not include an envelope. This is the case for HOPS 113 and 
293. Of the nine remaining Class II objects in our sample, four have very 
low envelope densities ($\rho_{1000} \sim (1-2.5) \times 10^{-20}$ g cm$^{-3}$; 
HOPS 22, 26, 98, 283), while five have $\rho_{1000}$ between $6.0 \times 
10^{-20}$ and $1.8 \times 10^{-19}$ g cm$^{-3}$ (HOPS 184, 201, 222, 272, 277).
The SEDs of HOPS 22, 184, and 201 are similar to those of transitional disks,
with some silicate emission at 10 $\mu$m and a rising SED between about
13 and 20 $\mu$m. The best-fit models require some envelope emission
to fit the long-wavelength data. 
HOPS 222, 272, and 277 lie close to the border between a Class II 
pre-main-sequence star and a flat-spectrum source based on their 
4.5-24 $\mu$m spectral index, and therefore they could have some 
envelope material left, despite being classified as Class II objects.

Overall, of the 330 YSOs in our sample, 319 were classified as
either Class 0, Class I, or flat-spectrum protostars based on their SEDs. 
However, four of them are best fit without an envelope. Conversely, of the
11 Class II objects in our sample, nine are best fit with an envelope; however,
three of these might be transitional disks. Thus, based on our model fits and
SEDs, 321 of our 330 YSOs are protostars with envelopes, and nine are likely
pre-main-sequence stars with disks.

\clearpage

\subsection{The Total Luminosities of Protostars}

The luminosity distribution of protostars is a significant constraint on 
protostellar evolution, and it is important to understand the effect of the 
envelope on the observed luminosity \citep[e.g.,][]{offner11}. The bolometric 
luminosity distribution of the HOPS protostars is very similar to that determined 
for the {\it Spitzer}-identified protostars by \citet{kryukova12} with a 
peak near 1~\Lsun\ (Fig.\ \ref{HOPS_n_Tbol_Lbol_histo}). In contrast, the 
distribution of the total luminosities from the models shows a peak near 
2.5~\Lsun\ (Fig.\ \ref{Ltot_class_histo}), indicating that the luminosities of 
protostars may be systematically underestimated by the bolometric 
luminosities, which do not take into account the inclination angle (and thus
beaming of the radiation along the outflow cavities) as well as foreground
extinction (see Fig.\ \ref{Ltot_Lbol} in section \ref{results_overview}).

Higher intrinsic luminosities for protostars could help address the 
``luminosity problem'' first pointed out by \citet{kenyon90}, who found 
that the luminosities of protostars are lower by about an order of magnitude 
than a simple estimate of the expected accretion luminosity. However, 
an increase in the luminosity by a factor of 2.5-3 would not solve the 
problem; solutions proposed by other authors, such as mass-dependent
accretion rates \citep{offner11} or episodic accretion events \citep{dunham12}, 
are still needed.

Our best-fit models also suggest that Class 0 protostars have a 
different distribution of $L_{tot}$ values compared to Class I protostars or 
flat-spectrum sources. Their median total luminosity is higher, which could 
be an indication of larger accretion luminosities for younger protostars. 
We must bear in mind the caveats and degeneracies mentioned above; 
in particular, in some cases the higher luminosity could be related to the 
adoption of an overly large inclination angle, which results in most 
of the emitted radiation not reaching the observer. Nevertheless, these 
differences have potentially important implications for protostellar evolution, 
which will be discussed in a future publication (W. Fischer et al. 2016, in 
preparation).

\subsection{Potential Problems with TSC Models}
\label{Model_problems}

Although the TSC models provide impressive fits to the SEDs, some of the 
observed trends suggest problems with the models.  First, the distribution 
of inclination angles (Fig.\ \ref{Inc_histo}) deviates from what we expect from 
a randomly oriented sample of protostars.  Although this could result from 
unintentional selection biases in our sample of protostars, it may also be the 
effect of applying the wrong envelope model to the data. 

Furthermore, our data show flat distributions in cavity opening angles 
for Class~0 and flat-spectrum sources, but an excess of small cavities for the 
Class~I protostars (Figure \ref{Cav_class_histo}). We also find that protostars 
with large cavities often have high envelope densities (Figure \ref{Rho1000_inc_cavity}).
For example, models with high envelope densities viewed more edge-on require 
large cavity opening angles and high $L_{tot}$ values to 
generate sufficient mid-IR flux; this is the case for a few of our highest-luminosity 
objects (HOPS 87, 108, and 178). These trends do not support the notion of 
increasing cavity size with later evolutionary stage, which would be expected if 
outflows play a major role in dispersing envelopes \citep{arce06}. This may 
suggest that cavity sizes are not growing with time; however, this may also imply  
a deviation from spherical symmetry for the initial configuration of the collapsing 
envelopes.  Such a deviation may result if the envelope collapses from the 
fragmentation of a flattened sheet or elongated filament.  

Finally, we find an excess of small values of $R_{disk}$, and therefore small 
centrifugal radii, for Class I and flat-spectrum protostars (Figure 
\ref{Rdisk_class_histo}). This is contrary to the expectation from the TSC 
model, in which the late stages of protostellar evolution are characterized 
by the infall of high angular momentum material from large radii and hence 
larger values of $R_c$.  This may imply that disks sizes are small, but it 
may also be the result of incorrect assumptions about the distribution of 
angular momentum in the TSC model.  

In total, these ``conundrums'' that arise from our model fits hint that the 
current models do not realistically reproduce the structure of collapsing 
envelopes. Future high-resolution observations at submillimeter and longer
wavelengths that resolve the structure and motions of envelopes may 
provide the means to develop more refined models that can fit the SEDs 
with more realistic envelope configurations.

\vspace{2ex}

\section{Conclusions}

We have presented SEDs and model fits for 330 young stellar objects
in the Orion A and B molecular  clouds. The SEDs include data from 
1.2 to 870 $\mu$m, with near-infrared photometry from 2MASS, mid-infrared 
photometry and spectra from the {\it Spitzer Space Telescope}, far-infrared 
photometry at 70, 100, and 160 $\mu$m from the {\it Herschel Space Observatory}, 
and submillimeter photometry from the APEX telescope. 
We calculated bolometric luminosities ($L_{bol}$), bolometric temperatures
($T_{bol}$), and 4.5-24 $\mu$m spectral indices ($n_{4.5-24}$) for all 330 sources 
in our sample. From the distributions of these three parameters, we find that $L_{bol}$
has a broad peak near 1 \Lsun\ and extends from 0.02 to several hundred \Lsun,
while the distribution of $T_{bol}$ values is broad and flat from about 30 K to 800 K,
with a median value of 146 K. The 4.5-24 $\mu$m spectral indices range from 
-0.75 to 2.6, with a peak near 0.

Based on traditional classification schemes involving $n_{4.5-24}$
and $T_{bol}$, we have identified 92 sources as Class 0 protostars 
($n_{4.5-24} > 0.3$ and $T_{bol} < 70$~K), 125 as Class I protostars 
($n_{4.5-24} > 0.3$ and $T_{bol} > 70$~K), and 102 as flat-spectrum sources 
($-0.3 < n_{4.5-24} < 0.3$). The remaining 11 sources are Class II pre-main-sequence 
stars with $n_{4.5-24} < -0.3$; most of them just missed the flat-spectrum cutoff,
and three have SEDs typical of disks with inner holes. Considering these
transitional disks and YSOs whose best fit does not require an envelope, we
find that 321 of the 330 HOPS targets in our sample are protostars with
envelopes. Class 0 and I protostars often display a deep silicate absorption 
feature at 10 $\mu$m due to the presence of the envelope, while many 
flat-spectrum sources have a weak silicate emission or absorption feature 
at that wavelength. 

We have used a grid of 30,400 protostellar model SEDs, calculated using the
2008 version of the \citet{whitney03a,whitney03b} Monte Carlo radiative 
transfer code, to find the best-fit models for each observed SED. The grid 
is limited to discrete values for protostellar parameters, and their ranges 
were chosen to represent typical protostars. Within the framework of these 
models, we find the following:

\begin{itemize}
\item{About 70\% of Class 0 protostars, 75\% of Class I protostars, and close to 
90\% of flat-spectrum sources have reliable SED fits ($R < 4$, where $R$
is a measure of the average distance between model and data in units
of the fractional uncertainty). Thus, our model grid can reproduce most of the 
observed SEDs of Orion protostars.}
\item{Our results show a clear trend of decreasing envelope densities as we 
progress from Class 0 to Class I and then to flat-spectrum sources: we find that 
the median $\rho_{1000}$ values decrease from 5.9 $\times 10^{-18}$ g cm$^{-3}$ 
to 2.4 $\times 10^{-19}$ g cm$^{-3}$ to 1.2 $\times 10^{-19}$ g cm$^{-3}$.
The decrease in densities implies a decrease in the infall rates of the protostars 
as they evolve.
We find that the PACS Bright Red sources (PBRs) have median $\rho_{1000}$ 
values twice as high as the median value of the Class 0 protostars in our sample, 
supporting the interpretation that they are likely the youngest protostars in Orion.}
\item{There are degeneracies in the parameters for models that reproduce the 
observed SEDs. For example, increasing the mid-IR SED slope and 
deepening the silicate absorption feature at 10 $\mu$m of a model protostar 
can be done by increasing the envelope density or inclination angle, 
decreasing the cavity opening angle or centrifugal radius, or even increasing 
the foreground extinction. Hence, the properties of a specific source may be fit 
by a wide range of parameters. The best-fit model parameters are particularly 
uncertain for objects whose SED is not well constrained by observations. 
Because of these degeneracies, the observed classes contain a mixture 
of evolutionary stages.}
\item{We find that flat-spectrum sources are particularly well fit by our models.
They have, on average, lower envelope densities and intermediate inclination 
angles, so many flat-spectrum sources are likely more evolved protostars, 
but this group also includes protostars with higher envelope densities (and
sometimes larger cavity opening angles) seen at lower inclination angles. 
Flat-spectrum sources seen at $i>$ 65\degr\ have very tenuous envelopes. 
Thus, the sample of flat-spectrum sources includes protostars 
at different stages in their envelope evolution. All but three of the flat-spectrum 
sources in our sample have envelopes in their best-fit models, indicating that, 
with a small number of exceptions, these objects are protostars with infalling 
gas.}
\item{The luminosity function for the model luminosities peaks at a higher 
luminosity than that for the observed bolometric luminosities as a result of 
beaming along the outflow cavities.  Furthermore, the total luminosity 
determined by the models is higher for Class 0 protostars: the median 
total luminosities are 5.5, 2.0, and 3.0 \Lsun\ for Class 0, Class I, and 
flat-spectrum sources, respectively.}
\item{Since heating by external radiation fields is not included 
in our model grid, we assessed its influence by adding an interstellar radiation 
field to a set of models. We find that an ISRF ten times that typical of the solar 
neighborhood can substantially change the SEDs of  sources with internal 
luminosities of 0.1 \Lsun. However, when we incorporate the effect of 
extinction on the external radiation field, the effect on the protostellar
SEDs is smaller; the best-fit luminosities and envelope densities would be 
overestimated by factors of a few for $\sim$ 0.1 \Lsun\ prototars and much
less for higher-luminosity protostars. We estimate that the best-fit parameters
(in particular, $L_{tot}$, $\rho_{1000}$) of $\sim$ 20\% of the HOPS sources 
could be affected by external heating.}
\item{Although the adopted TSC models reproduce the observed SEDs well, 
there are trends that suggest inadequacies with these models.  First, the 
distribution of best-fit inclination angles does not reproduce that expected 
for randomly oriented protostars.  Second, although the distribution of outflow 
cavity sizes for flat-spectrum and Class~0 sources is flat, there is an excess 
of small cavities for Class I sources. This is in contradiction to the typical 
picture that outflow cavities grow as protostars evolve. Finally, the distribution 
of outer disk radii set by the rotation of the envelope is concentrated at small 
values ($<$ 50 AU) for the Class I and flat-spectrum sources but is slightly 
tilted toward large values ($>$ 50 AU) for Class 0 protostars. Again, this trend 
contradicts the expected growth of disks as the infall region in protostellar 
envelopes expands. These findings suggest that either the envelope structure 
of the adopted models is incorrect, or our understanding of the evolution of 
protostars needs to be revised substantially.}
\end{itemize}

Our work provides a large sample of protostars in one molecular cloud
complex for future, more detailed studies of protostellar evolution. For example,
using additional constraints, such as from scattered light imaging, the 
structure of envelope cavities and thus the role of outflows can be better 
understood. In addition, the detailed structure of the envelope and the disk 
embedded within, as well as multiplicity of the central source, can be studied 
with high spatial resolution imaging such as ALMA can provide. With the 
analysis of their SEDs presented in this work, the HOPS protostars constitute 
an ideal sample to derive a better understanding of the early evolution of 
young stars, when the assembly of the stellar mass and the initial 
stages of planet formation likely take place.

\vspace{1ex}

\acknowledgments
Support for this work was provided by NASA through awards issued by
JPL/Caltech.
The work of W.J.F. was supported in part by an appointment to the NASA 
Postdoctoral Program at Goddard Space Flight Center, administered by 
Oak Ridge Associated Universities through a contract with NASA.
J.J.T. acknowledges support provided by NASA through Hubble Fellowship grant 
\#HST-HF-51300.01-A awarded by the Space Telescope Science Institute, which is 
operated by the Association of Universities for Research in Astronomy, Inc., 
for NASA, under contract NAS 5-26555. J.J.T acknowledges further support from 
grant 639.041.439 from the Netherlands Organisation for Scientific Research (NWO).
The work of A.M.S. was supported by the Deutsche Forschungsgemeinschaft
priority program 1573 (``Physics of the Interstellar Medium'').
M.O. acknowledges support from MINECO (Spain) AYA2011-3O228-CO3-01 
and AYA2014-57369-C3-3-P grants (co-funded with FEDER funds). 
We thank Thomas Robitaille for helpful discussions regarding the model grid
and model parameters.
This work is based on observations made with the {\it Spitzer Space Telescope}, 
which is operated by the Jet Propulsion Laboratory (JPL), California Institute of 
Technology (Caltech), under a contract with NASA; it is also based on
observations made with the {\it Herschel Space Observatory}, a European Space
Agency Cornerstone Mission with significant participation by NASA. 
The {\it Herschel} spacecraft was designed, built, tested, and launched under 
a contract to ESA managed by the {\it Herschel/Planck} Project team by an industrial 
consortium under the overall responsibility of the prime contractor Thales Alenia Space 
(Cannes), and including Astrium (Friedrichshafen) responsible for the payload module 
and for system testing at spacecraft level, Thales Alenia Space (Turin) responsible 
for the service module, and Astrium (Toulouse) responsible for the telescope, with 
in excess of a hundred subcontractors.
We also include data from the Atacama Pathfinder Experiment, a collaboration 
between the Max-Planck Institut f\"ur Radioastronomie, the European Southern 
Observatory, and the Onsala Space Observatory. 
This publication makes use of data products from the Two Micron All Sky Survey, 
which is a joint project of the University of Massachusetts and the Infrared Processing 
and Analysis Center/Caltech, funded by NASA and the NSF.

\appendix

\section{Tables and Figures with SEDs and Best Fits}

\begin{deluxetable*}{lrrclllccccccccl}[h] \scriptsize
\tablenum{1} 
\tablecaption{Classification and Best-Fit Model Parameters for the HOPS Sample
\label{bestfit}}
\tablehead{
\colhead{Object} & \colhead{R.A.} & \colhead{Dec. } & 
\colhead{Class} & \colhead{$L_{bol}$} & \colhead{$T_{bol}$} &
\colhead{$n_{4.5-24}$} & 
\colhead{$L_{tot}$} & \colhead{$R_{disk}$} & 
\colhead{$\rho_{1000}$} & \colhead{M$_{env}$} & \colhead{$\theta$} &
\colhead{i} & \colhead{$A_V$} & \colhead{scaling} & \colhead{R} \\
  & \colhead{[$\degr$]} & \colhead{[$\degr$]} & & \colhead{[\Lsun]} &
\colhead{[K]} &  & \colhead{[\Lsun]}  & \colhead{[AU]} & 
\colhead{[g cm$^{-3}$]} & \colhead{[\Msun]} &
\colhead{[$\degr$]} & \colhead{[$\degr$]} & \colhead{[mag]} & \colhead{factor} & \\
\colhead{(1)} & \colhead{(2)} & \colhead{(3)} & \colhead{(4)} &
\colhead{(5)} & \colhead{(6)} & \colhead{(7)} & \colhead{(8)} &
\colhead{(9)} & \colhead{(10)} & \colhead{(11)} & \colhead{(12)} &
\colhead{(13)} & \colhead{(14)} & \colhead{(15)} & \colhead{(16)}}
\startdata
  HOPS 1 & 88.5514 &   1.7099 &    I &     1.517 &    72.6 &   1.469 &   3.0 & 100 &   2.38 $\times 10^{-19}$ &  0.0133 &   5 &  63 &  23.2 &  0.99 &   2.319 \\
  HOPS 2 & 88.5380 &   1.7144 &    I &     0.542 &   356.5 &   0.455 &   1.3 &   5 &   2.38 $\times 10^{-20}$ &  0.0012 &  15 &  32 &  13.1 &  1.30 &   2.476 \\
  HOPS 3 & 88.7374 &   1.7156 & flat &     0.553 &   467.5 &   0.260 &  0.820 &   5 &   1.19 $\times 10^{-20}$ &  0.0007 &   5 &  50 &   3.0 &  0.81 &   3.331 \\
  HOPS 4 & 88.7240 &   1.7861 &    I &     0.422 &   203.3 &   1.243 &  0.600 &   5 &   1.78 $\times 10^{-19}$ &  0.0099 &   5 &  63 &   2.5 &  2.00 &   4.139 \\
  HOPS 5 & 88.6340 &   1.8020 &    I &     0.390 &   187.1 &   0.626 &   1.6 &  50 &   2.38 $\times 10^{-19}$ &  0.0096 &  35 &  63 &  12.4 &  0.52 &   2.459 \\
  HOPS 6 & 88.5767 &   1.8176 &    I &     0.055 &   112.5 &   1.308 &  0.210 &   5 &   1.78 $\times 10^{-20}$ &  0.0010 &   5 &  76 &   8.0 &  2.00 &   4.091 \\
  HOPS 7 & 88.5835 &   1.8452 &    0 &     0.528 &    58.0 &   1.707 &   6.1 & 100 &   1.78 $\times 10^{-20}$ &  0.0010 &  15 &  81 &  18.7 &  2.00 &   2.981 \\
 HOPS 10 & 83.7875 &  -5.9743 &    0 &     3.330 &    46.2 &   0.787 &   5.4 & 500 &   2.38 $\times 10^{-18}$ &  0.135 &   5 &  70 &   0.0 &  1.77 &   3.168 \\
\enddata
\end{deluxetable*}

\begin{deluxetable}{lrrclllccccccccl}\scriptsize
\tablenum{1}\tablecaption{continued.}
\tablehead{
\colhead{Object} & \colhead{R.A.} & \colhead{Dec. } & 
\colhead{Class} & \colhead{$L_{bol}$} & \colhead{$T_{bol}$} &
\colhead{$n_{4.5-24}$} & 
\colhead{$L_{tot}$} & \colhead{$R_{disk}$} & 
\colhead{$\rho_{1000}$} & \colhead{M$_{env}$} & \colhead{$\theta$} &
\colhead{i} & \colhead{$A_V$} & \colhead{scaling} & \colhead{R} \\
  & \colhead{[$\degr$]} & \colhead{[$\degr$]} & & \colhead{[\Lsun]} &
\colhead{[K]} &  & \colhead{[\Lsun]}  & \colhead{[AU]} & 
\colhead{[g cm$^{-3}$]} & \colhead{[\Msun]} &
\colhead{[$\degr$]} & \colhead{[$\degr$]} & \colhead{[mag]} & \colhead{factor} & \\
\colhead{(1)} & \colhead{(2)} & \colhead{(3)} & \colhead{(4)} &
\colhead{(5)} & \colhead{(6)} & \colhead{(7)} & \colhead{(8)} &
\colhead{(9)} & \colhead{(10)} & \colhead{(11)} & \colhead{(12)} &
\colhead{(13)} & \colhead{(14)} & \colhead{(15)} & \colhead{(16)}}
\startdata
 HOPS 80 & 83.8549 &  -5.0860 & flat &     0.079 &   275.3 &   0.016 &  0.330 &  50 &   1.19 $\times 10^{-17}$ &  0.373 &  45 &  57 &   0.9 &  1.08 &   2.328 \\
 HOPS 81 & 83.8665 &  -5.0828 &    0 &     1.238 &    40.1 &   0.851 &   6.1 & 500 &   1.19 $\times 10^{-20}$ &  0.0007 &   5 &  87 &   0.0 &  2.00 &   2.740 \\
 HOPS 82 & 83.8322 &  -5.0818 & flat &     2.404 &   116.4 &   0.180 &  18.7 &   5 &   1.78 $\times 10^{-18}$ &  0.0548 &  45 &  41 &  59.9 &  1.85 &   2.307 \\
 HOPS 84 & 83.8607 &  -5.0653 &    I &    49.112 &    90.8 &   1.492 &  60.4 & 500 &   5.94 $\times 10^{-19}$ &  0.0337 &   5 &  57 &   6.1 &  2.00 &   3.319 \\
 HOPS 85 & 83.8674 &  -5.0614 & flat &    16.277 &   174.2 &   0.253 &  133.1 &  50 &   1.19 $\times 10^{-17}$ &  0.373 &  45 &  50 &  62.3 &  1.32 &   3.129 \\
 HOPS 86 & 83.8485 &  -5.0279 &    I &     3.273 &   112.7 &   1.487 &  46.1 &   5 &   2.38 $\times 10^{-20}$ &  0.0013 &   5 &  70 &  58.1 &  1.53 &   3.704 \\
 HOPS 87 & 83.8478 &  -5.0246 &    0 &    36.488 &    38.1 &   1.915 &  562.9 & 100 &   2.38 $\times 10^{-18}$ &  0.0975 &  35 &  87 &  51.1 &  1.85 &   4.977 \\
 HOPS 88 & 83.8435 &  -5.0206 &    0 &    15.814 &    42.4 &   2.541 &  124.9 & 500 &   1.19 $\times 10^{-17}$ &  0.437 &  45 &  70 &  49.1 &  1.24 &   3.958 \\
\enddata
\end{deluxetable}

\begin{deluxetable}{lrrclllccccccccl}\scriptsize
\tablenum{1}\tablecaption{continued.}
\tablehead{
\colhead{Object} & \colhead{R.A.} & \colhead{Dec. } & 
\colhead{Class} & \colhead{$L_{bol}$} & \colhead{$T_{bol}$} &
\colhead{$n_{4.5-24}$} & 
\colhead{$L_{tot}$} & \colhead{$R_{disk}$} & 
\colhead{$\rho_{1000}$} & \colhead{M$_{env}$} & \colhead{$\theta$} &
\colhead{i} & \colhead{$A_V$} & \colhead{scaling} & \colhead{R} \\
  & \colhead{[$\degr$]} & \colhead{[$\degr$]} & & \colhead{[\Lsun]} &
\colhead{[K]} &  & \colhead{[\Lsun]}  & \colhead{[AU]} & 
\colhead{[g cm$^{-3}$]} & \colhead{[\Msun]} &
\colhead{[$\degr$]} & \colhead{[$\degr$]} & \colhead{[mag]} & \colhead{factor} & \\
\colhead{(1)} & \colhead{(2)} & \colhead{(3)} & \colhead{(4)} &
\colhead{(5)} & \colhead{(6)} & \colhead{(7)} & \colhead{(8)} &
\colhead{(9)} & \colhead{(10)} & \colhead{(11)} & \colhead{(12)} &
\colhead{(13)} & \colhead{(14)} & \colhead{(15)} & \colhead{(16)}}
\startdata
HOPS 150 & 84.5314 &  -7.1414 & flat &     3.769 &   245.2 &   0.139 &   8.3 &  50 &   2.38 $\times 10^{-18}$ &  0.0957 &  35 &  57 &  12.6 &  0.82 &   3.728 \\
HOPS 152 & 84.4948 &  -7.1237 &    0 &     0.700 &    53.8 &   2.204 &   2.0 &   5 &   2.38 $\times 10^{-19}$ &  0.0132 &   5 &  70 &  59.1 &  2.00 &   4.006 \\
HOPS 153 & 84.4875 &  -7.1157 &    0 &     4.433 &    39.4 &   1.695 &  33.2 & 500 &   5.94 $\times 10^{-18}$ &  0.219 &  45 &  87 &   2.7 &  1.10 &   5.470 \\
HOPS 154 & 84.5837 &  -6.9847 &    I &     0.105 &   166.7 &   0.565 &  0.110 &   5 &   1.19 $\times 10^{-19}$ &  0.0066 &   5 &  32 &   0.1 &  1.10 &   2.812 \\
HOPS 156 & 84.5142 &  -6.9711 &    I &     0.266 &    90.1 &   1.139 &  0.320 &  50 &   5.94 $\times 10^{-19}$ &  0.0331 &   5 &  63 &   0.2 &  1.06 &   2.578 \\
HOPS 157 & 84.4857 &  -6.9442 &    I &     3.824 &    77.6 &   1.744 &  22.4 & 100 &   1.19 $\times 10^{-17}$ &  0.381 &  45 &  57 &  43.7 &  0.74 &   3.127 \\
HOPS 158 & 84.3519 &  -6.9758 & flat &     2.672 &   591.6 &  -0.063 &   5.9 &   5 &   2.38 $\times 10^{-20}$ &  0.0012 &  15 &  50 &   9.0 &  1.95 &   1.809 \\
HOPS 159 & 84.4739 &  -6.7880 & flat &     0.449 &   498.4 &   0.265 &   2.8 &   5 &   2.38 $\times 10^{-20}$ &  0.0009 &  35 &  70 &   9.0 &  0.92 &   2.169 \\
\enddata
\end{deluxetable}

\begin{deluxetable}{lrrclllccccccccl}\scriptsize
\tablenum{1}\tablecaption{continued.}
\tablehead{
\colhead{Object} & \colhead{R.A.} & \colhead{Dec. } & 
\colhead{Class} & \colhead{$L_{bol}$} & \colhead{$T_{bol}$} &
\colhead{$n_{4.5-24}$} & 
\colhead{$L_{tot}$} & \colhead{$R_{disk}$} & 
\colhead{$\rho_{1000}$} & \colhead{M$_{env}$} & \colhead{$\theta$} &
\colhead{i} & \colhead{$A_V$} & \colhead{scaling} & \colhead{R} \\
  & \colhead{[$\degr$]} & \colhead{[$\degr$]} & & \colhead{[\Lsun]} &
\colhead{[K]} &  & \colhead{[\Lsun]}  & \colhead{[AU]} & 
\colhead{[g cm$^{-3}$]} & \colhead{[\Msun]} &
\colhead{[$\degr$]} & \colhead{[$\degr$]} & \colhead{[mag]} & \colhead{factor} & \\
\colhead{(1)} & \colhead{(2)} & \colhead{(3)} & \colhead{(4)} &
\colhead{(5)} & \colhead{(6)} & \colhead{(7)} & \colhead{(8)} &
\colhead{(9)} & \colhead{(10)} & \colhead{(11)} & \colhead{(12)} &
\colhead{(13)} & \colhead{(14)} & \colhead{(15)} & \colhead{(16)}}
\startdata
HOPS 220 & 85.3741 &  -8.7128 &    I &     0.267 &   193.6 &   0.829 &  0.600 &   5 &   2.38 $\times 10^{-20}$ &  0.0013 &   5 &  18 &  24.9 &  2.00 &   2.115 \\
HOPS 221 & 85.6961 &  -8.2853 &    I &     8.363 &   172.3 &   0.684 &  14.4 &   5 &   5.94 $\times 10^{-19}$ &  0.0312 &  15 &  63 &   9.0 &  1.42 &   2.935 \\
HOPS 222 & 85.3612 &  -8.7068 &   II &     0.849 &   738.2 &  -0.647 &   2.0 &   5 &   1.19 $\times 10^{-19}$ &  0.0047 &  35 &  63 &   9.0 &  0.66 &   3.237 \\
HOPS 223\tablenotemark{a} & 85.7019 &  -8.2762 &    I &    19.252 &   247.5 &   1.078 &  151.7 &  50 &   2.38 $\times 10^{-18}$ &  0.0746 &  45 &  63 &  14.5 &  0.50 &   8.410 \\
HOPS 224 & 85.3834 &  -8.6694 &    0 &     2.988 &    48.6 &   0.780 &   4.1 & 500 &   2.38 $\times 10^{-18}$ &  0.131 &  15 &  63 &   9.7 &  1.34 &   3.941 \\
HOPS 225 & 85.3764 &  -8.6715 & flat &     0.835 &   432.5 &  -0.118 &   5.4 &   5 &   5.94 $\times 10^{-20}$ &  0.0024 &  35 &  57 &  27.3 &  1.76 &   1.668 \\
HOPS 226 & 85.3753 &  -8.6693 & flat &     0.855 &   350.2 &   0.015 &   4.3 &   5 &   5.94 $\times 10^{-20}$ &  0.0028 &  25 &  50 &  27.4 &  1.40 &   3.390 \\
HOPS 227 & 85.3847 &  -8.6321 & flat &     0.335 &   371.0 &  -0.078 &   1.9 &   5 &   1.19 $\times 10^{-20}$ &  0.0006 &  15 &  70 &  20.4 &  1.93 &   2.576 \\
\enddata
\end{deluxetable}

\begin{deluxetable}{lrrclllccccccccl}\scriptsize
\tablenum{1} \tablecaption{continued.}
\tablehead{
\colhead{Object} & \colhead{R.A.} & \colhead{Dec. } & 
\colhead{Class} & \colhead{$L_{bol}$} & \colhead{$T_{bol}$} &
\colhead{$n_{4.5-24}$} & 
\colhead{$L_{tot}$} & \colhead{$R_{disk}$} & 
\colhead{$\rho_{1000}$} & \colhead{M$_{env}$} & \colhead{$\theta$} &
\colhead{i} & \colhead{$A_V$} & \colhead{scaling} & \colhead{R} \\
  & \colhead{[$\degr$]} & \colhead{[$\degr$]} & & \colhead{[\Lsun]} &
\colhead{[K]} &  & \colhead{[\Lsun]}  & \colhead{[AU]} & 
\colhead{[g cm$^{-3}$]} & \colhead{[\Msun]} &
\colhead{[$\degr$]} & \colhead{[$\degr$]} & \colhead{[mag]} & \colhead{factor} & \\
\colhead{(1)} & \colhead{(2)} & \colhead{(3)} & \colhead{(4)} &
\colhead{(5)} & \colhead{(6)} & \colhead{(7)} & \colhead{(8)} &
\colhead{(9)} & \colhead{(10)} & \colhead{(11)} & \colhead{(12)} &
\colhead{(13)} & \colhead{(14)} & \colhead{(15)} & \colhead{(16)}}
\startdata
HOPS 281 & 85.1026 &  -7.7190 & flat &     1.281 &   189.3 &   0.335 &   3.0 & 500 &   2.38 $\times 10^{-20}$ &  0.0009 &  45 &  41 &  28.7 &  0.99 &   2.204 \\
HOPS 282 & 85.1087 &  -7.6256 &    I &     0.816 &    95.1 &   1.864 &   1.9 &   5 &   2.38 $\times 10^{-19}$ &  0.0132 &   5 &  70 &  25.7 &  1.90 &   2.769 \\
HOPS 283 & 85.1861 &  -7.4985 &   II &     1.115 &   807.9 &  -0.382 &   2.5 &   5 &   2.38 $\times 10^{-20}$ &  0.0011 &  25 &  63 &   6.5 &  0.83 &   1.935 \\
HOPS 284 & 84.7145 &  -8.0243 & flat &     0.699 &   913.9 &  -0.316 &   1.5 &   5 &   1.78 $\times 10^{-20}$ &  0.0008 &  25 &  70 &   1.0 &  1.51 &   1.561 \\
HOPS 285 & 85.0246 &  -7.4925 &    0 &     0.141 &    66.5 &   1.229 &  0.520 &   5 &   1.78 $\times 10^{-19}$ &  0.0084 &  25 &  76 &  22.9 &  0.52 &   4.026 \\
HOPS 286 & 84.9946 &  -7.5200 &    I &     0.586 &   123.7 &   0.442 &  0.930 &   5 &   5.94 $\times 10^{-19}$ &  0.0312 &  15 &  70 &   0.0 &  0.92 &   3.999 \\
HOPS 287 & 85.0366 &  -7.4577 &    I &     0.755 &   117.8 &   0.850 &  0.600 &  50 &   5.94 $\times 10^{-19}$ &  0.0331 &   5 &  41 &   3.7 &  2.00 &   2.477 \\
HOPS 288 & 84.9831 &  -7.5078 &    0 &   135.474 &    48.6 &   2.587 &  606.9 & 500 &   1.78 $\times 10^{-18}$ &  0.0979 &  15 &  76 &  74.6 &  2.00 &   2.981 \\
\enddata
\end{deluxetable}

\begin{deluxetable}{lrrclllccccccccl}\scriptsize
\tablenum{1} \tablecaption{continued.}
\tablehead{
\colhead{Object} & \colhead{R.A.} & \colhead{Dec. } & 
\colhead{Class} & \colhead{$L_{bol}$} & \colhead{$T_{bol}$} &
\colhead{$n_{4.5-24}$} & 
\colhead{$L_{tot}$} & \colhead{$R_{disk}$} & 
\colhead{$\rho_{1000}$} & \colhead{M$_{env}$} & \colhead{$\theta$} &
\colhead{i} & \colhead{$A_V$} & \colhead{scaling} & \colhead{R} \\
  & \colhead{[$\degr$]} & \colhead{[$\degr$]} & & \colhead{[\Lsun]} &
\colhead{[K]} &  & \colhead{[\Lsun]}  & \colhead{[AU]} & 
\colhead{[g cm$^{-3}$]} & \colhead{[\Msun]} &
\colhead{[$\degr$]} & \colhead{[$\degr$]} & \colhead{[mag]} & \colhead{factor} & \\
\colhead{(1)} & \colhead{(2)} & \colhead{(3)} & \colhead{(4)} &
\colhead{(5)} & \colhead{(6)} & \colhead{(7)} & \colhead{(8)} &
\colhead{(9)} & \colhead{(10)} & \colhead{(11)} & \colhead{(12)} &
\colhead{(13)} & \colhead{(14)} & \colhead{(15)} & \colhead{(16)}}
\startdata
HOPS 364 & 86.9024 &   0.3350 &    I &    33.025 &    96.7 &   1.154 &  59.3 & 100 &   2.38 $\times 10^{-18}$ &  0.115 &  25 &  57 &   0.0 &  1.96 &   2.784 \\
HOPS 365 & 86.7942 &   0.3539 &    I &    19.300 &   160.3 &   0.367 &  23.4 &   5 &   1.78 $\times 10^{-18}$ &  0.0937 &  15 &  50 &   1.4 &  0.78 &   3.163 \\
HOPS 366 & 86.7666 &   0.3696 &    I &     6.445 &   292.2 &   0.373 &  151.7 &   5 &   2.38 $\times 10^{-20}$ &  0.0012 &  15 &  76 &  30.0 &  0.50 &   2.955 \\
HOPS 367 & 88.6511 &   1.8983 &    I &     0.046 &   249.4 &   1.946 &  0.08 &   5 &   5.94 $\times 10^{-20}$ &  0.0033 &   5 &  76 &   3.8 &  0.79 &   4.497 \\
HOPS 368 & 83.8530 &  -5.1751 &    I &    68.852 &   137.5 &   0.985 &  48.3 & 100 &   1.19 $\times 10^{-18}$ &  0.0664 &   5 &  18 &  18.9 &  1.60 &   2.025 \\
HOPS 369 & 83.8624 &  -5.1714 & flat &    35.318 &   379.2 &   0.254 &  199.4 &   5 &   1.78 $\times 10^{-20}$ &  0.0010 &   5 &  70 &  15.0 &  1.97 &   1.079 \\
HOPS 370 & 83.8651 &  -5.1593 &    I &   360.859 &    71.5 &   1.389 &  564.8 & 500 &   2.38 $\times 10^{-18}$ &  0.135 &   5 &  87 &   1.9 &  1.86 &   2.793 \\
HOPS 371 & 83.7934 &  -5.9280 &    0 &     0.571 &    31.6 &  \nodata &   1.9 &   5 &   5.95 $\times 10^{-17}$ &  1.83 &  45 &  57 &  16.3 &  1.91 &   0.426 \\
\enddata
\tablecomments{
\parbox{0.9\linewidth}{
Column (1) lists the HOPS name of the object, columns (2) and (3) its J2000 coordinates
in degrees, column (4) the type based on SED classification, column (5) the bolometric
luminosity, column (6) the bolometric temperature, column (7) the 4.5-24 $\mu$m SED slope,
and columns (8) to (16) the best-fit model parameters: the total luminosity, the disk radius
(which is equal to the centrifugal radius), the reference density at 1000 AU $\rho_{1000}$,
the mass of the envelope within 2500 AU, the cavity opening angle, the inclination angle,
the foreground extinction, the scaling factor applied to the best-fitting model from the grid,
and the $R$ value. }}
\tablenotetext{a}{The tabulated properties for HOPS 223 are very unreliable, since its SED is affected
by extreme variability (see text for details).}
\end{deluxetable}

\clearpage

\begin{figure}[h]
\centering
\includegraphics[scale=0.9]{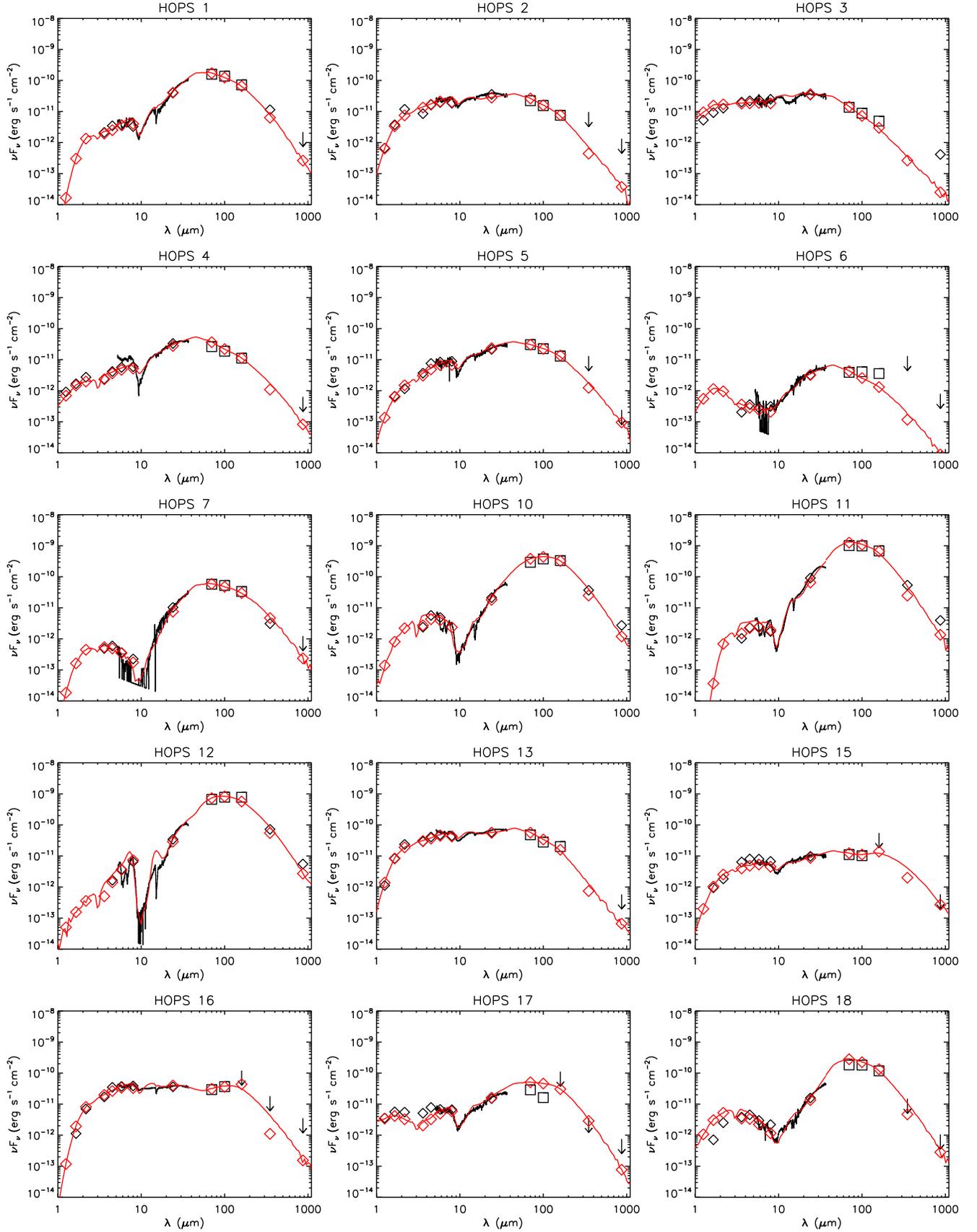}
\caption{\small{SEDs of the HOPS targets modeled in this work ({\it black}; open 
symbols: photometry, arrows: upper limits, line: IRS spectrum). The best-fit model 
for each object is shown as a red line, with fluxes taken from a 4\arcsec\ aperture 
for $\lambda < 8$ $\mu$m, a 5\arcsec\ aperture for $\lambda = 8-37$ $\mu$m, 
and a 10\arcsec\ aperture for $\lambda > 37$ $\mu$m. The red symbols are
the model photometry measured in the same apertures and bandpasses as 
the data (see Section \ref{model_ap} for details)}.
\label{bestSEDs}}
\end{figure}

\begin{figure}[h]
\centering
\includegraphics[scale=0.9]{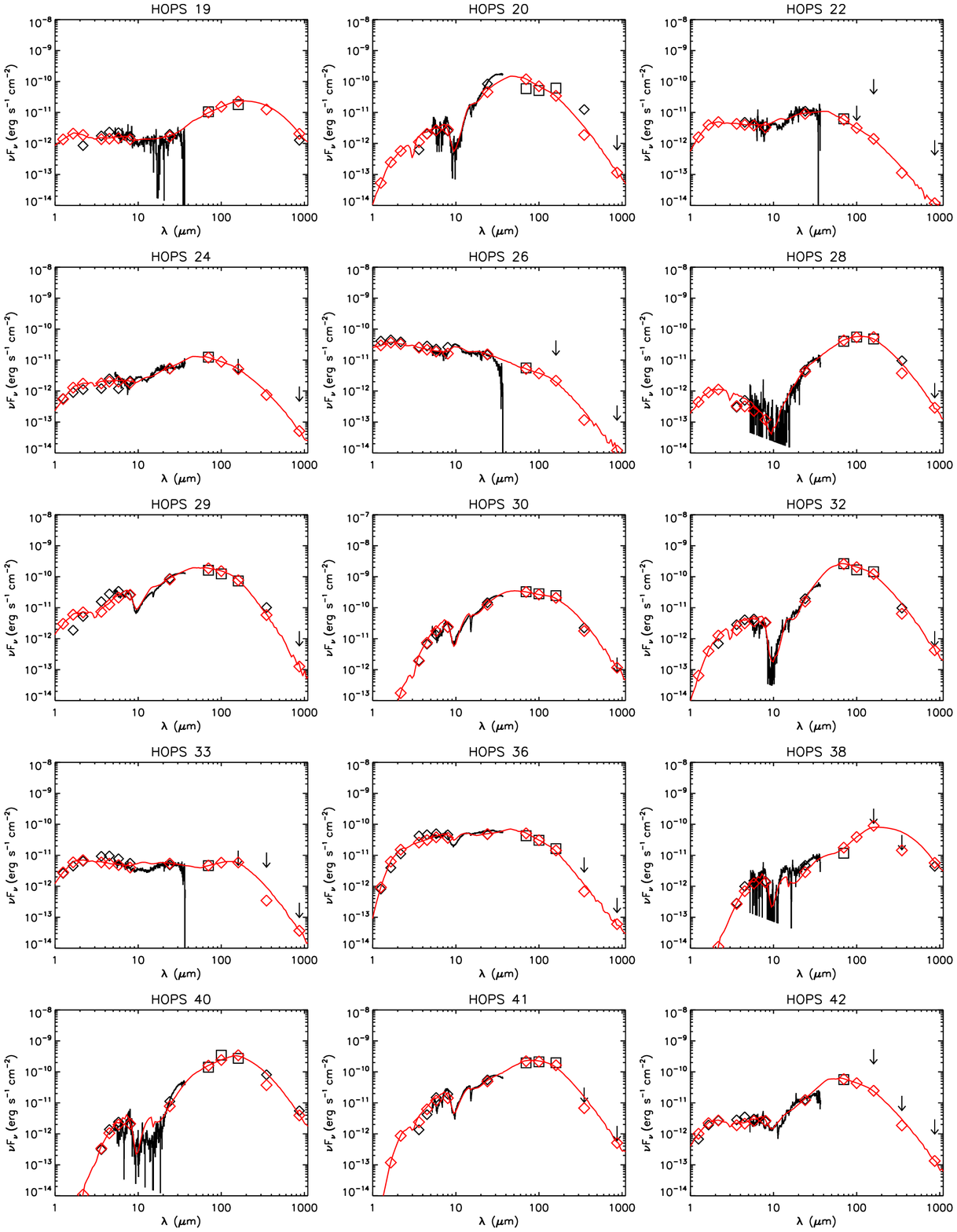}
\figurenum{\ref{bestSEDs}}\caption{continued.}
\end{figure}

\begin{figure}[h]
\centering
\includegraphics[scale=0.9]{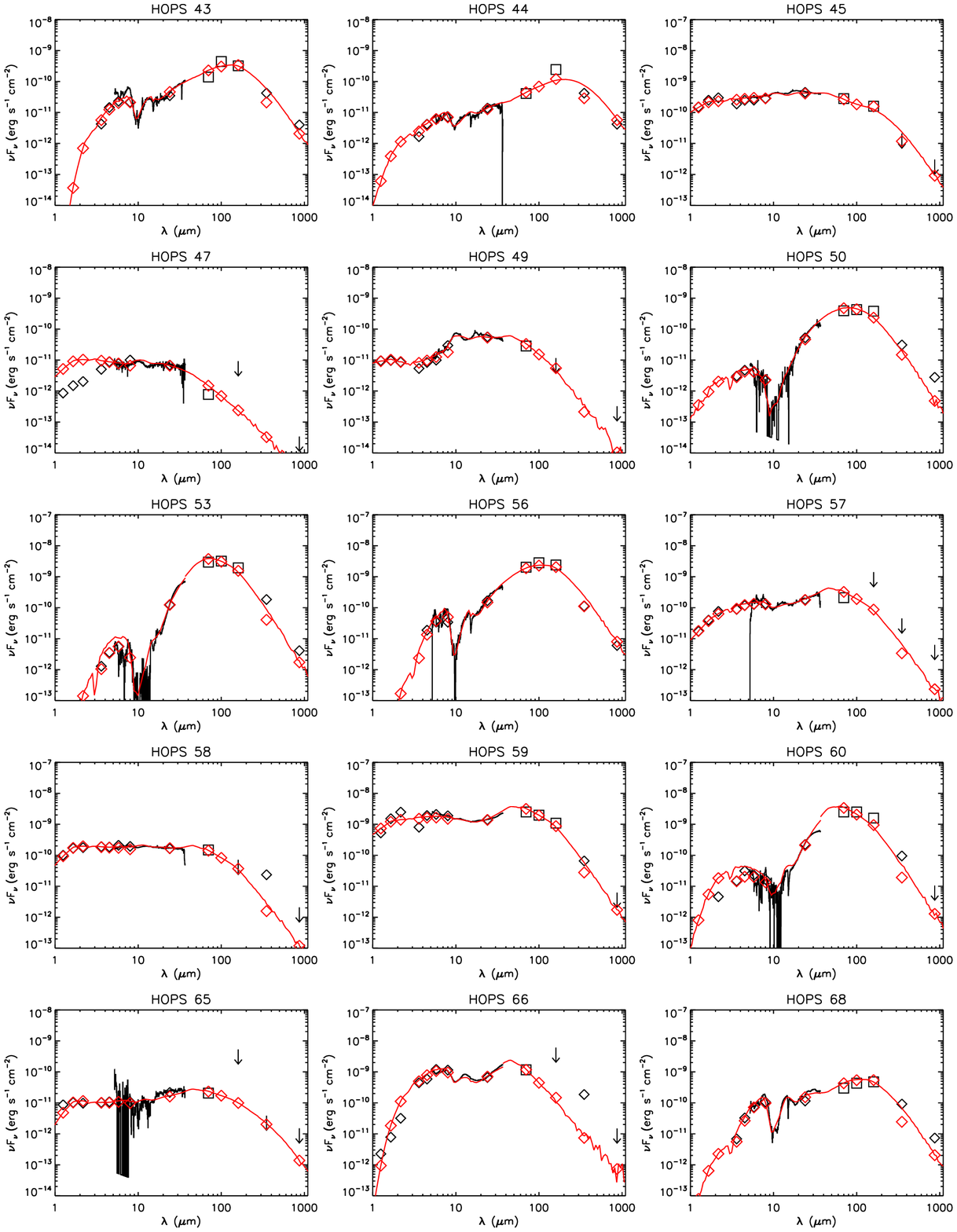}
\figurenum{\ref{bestSEDs}}\caption{continued.}
\end{figure}

\begin{figure}[h]
\centering
\includegraphics[scale=0.9]{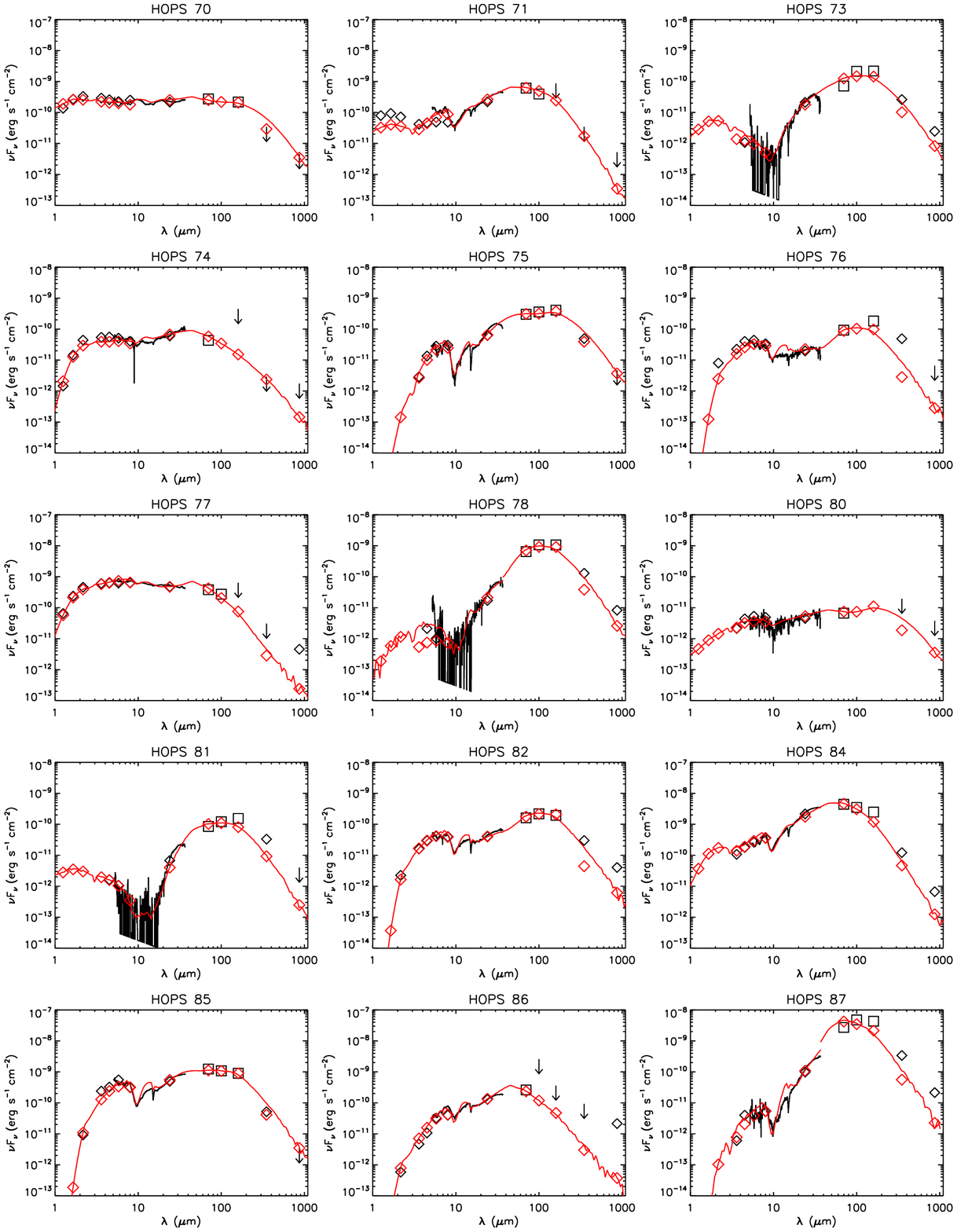}
\figurenum{\ref{bestSEDs}}\caption{continued.}
\end{figure}

\begin{figure}[h]
\centering
\includegraphics[scale=0.9]{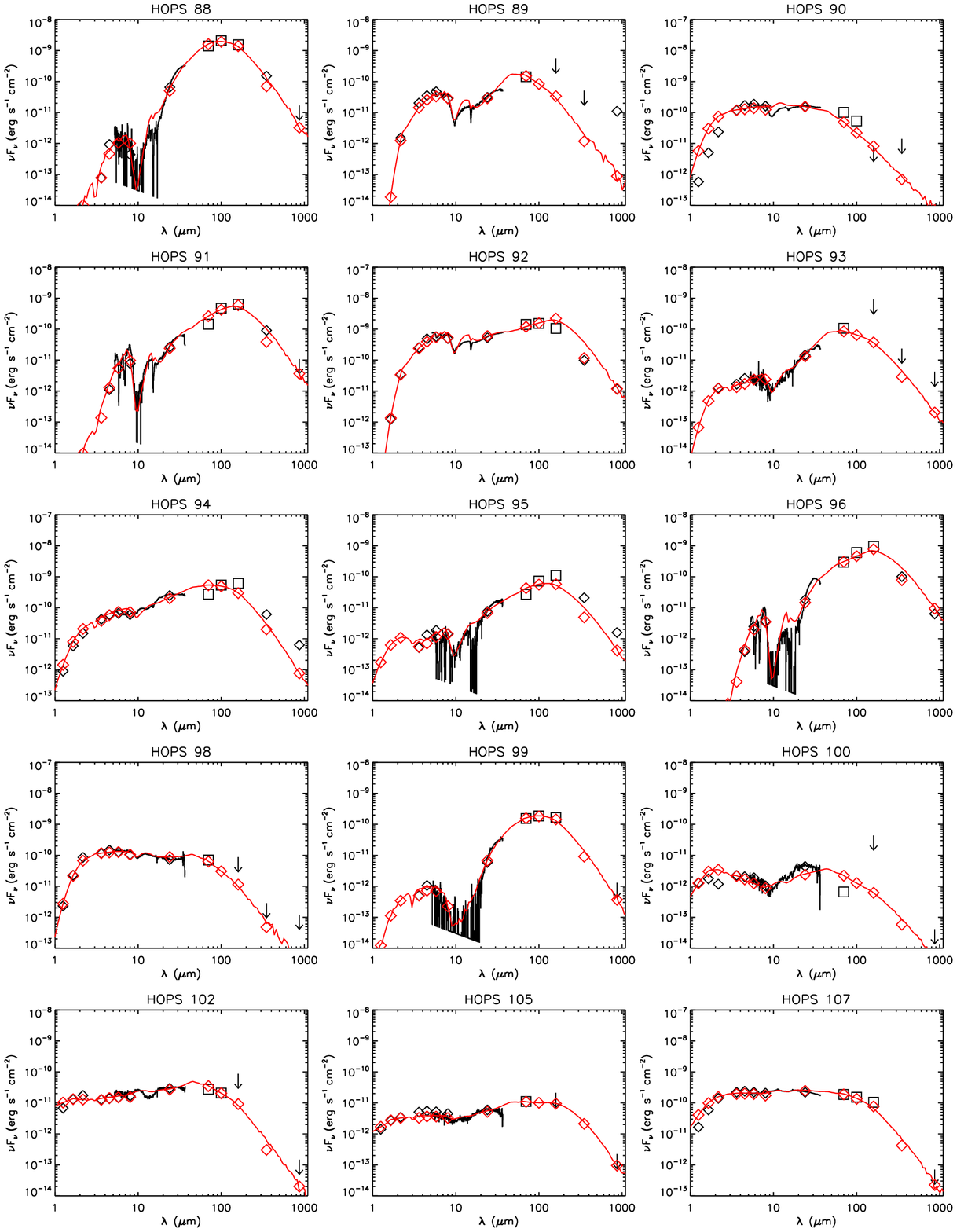}
\figurenum{\ref{bestSEDs}}\caption{continued.}
\end{figure}

\begin{figure}[h]
\centering
\includegraphics[scale=0.9]{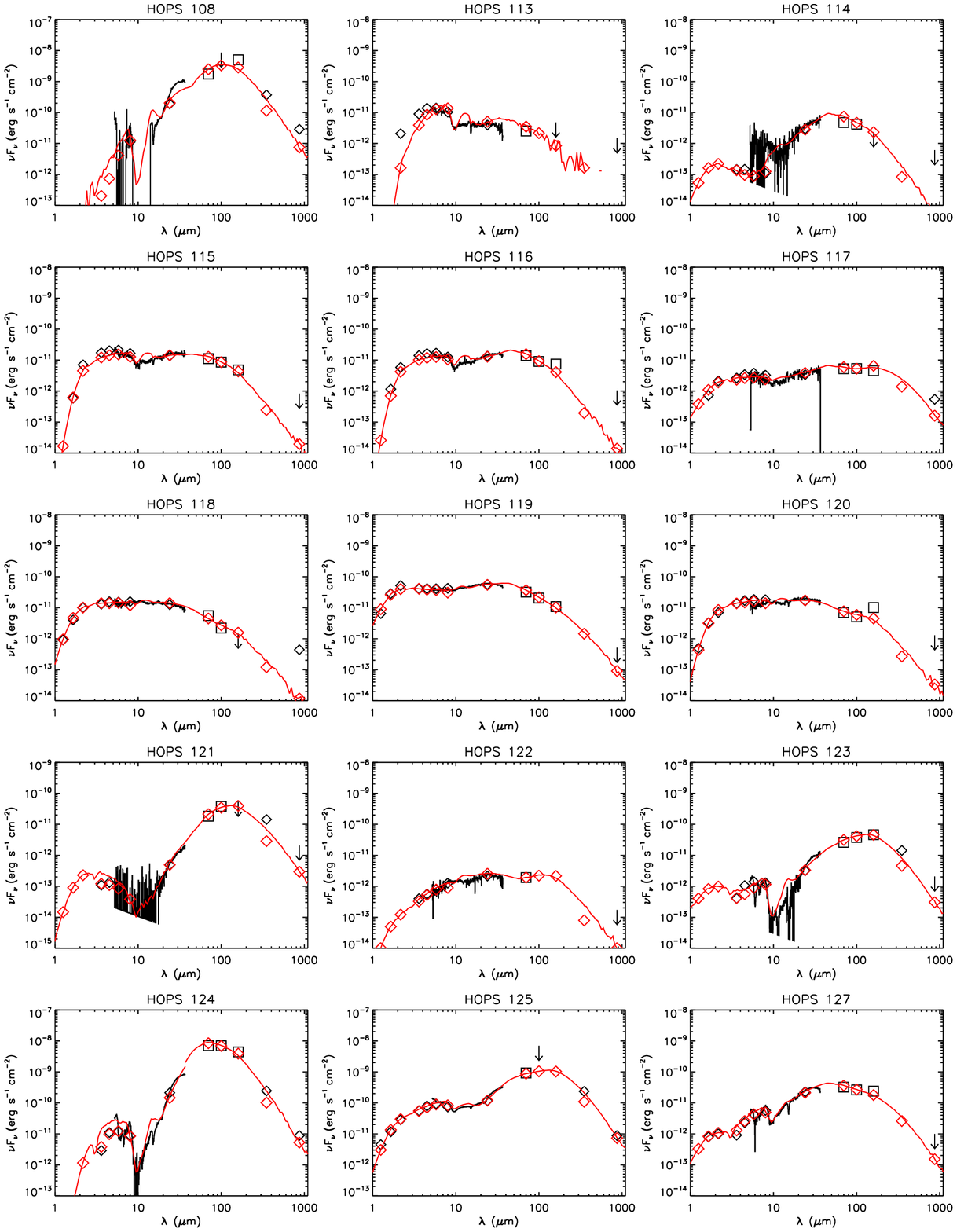}
\figurenum{\ref{bestSEDs}}\caption{continued.}
\end{figure}

\begin{figure}[h]
\centering
\includegraphics[scale=0.9]{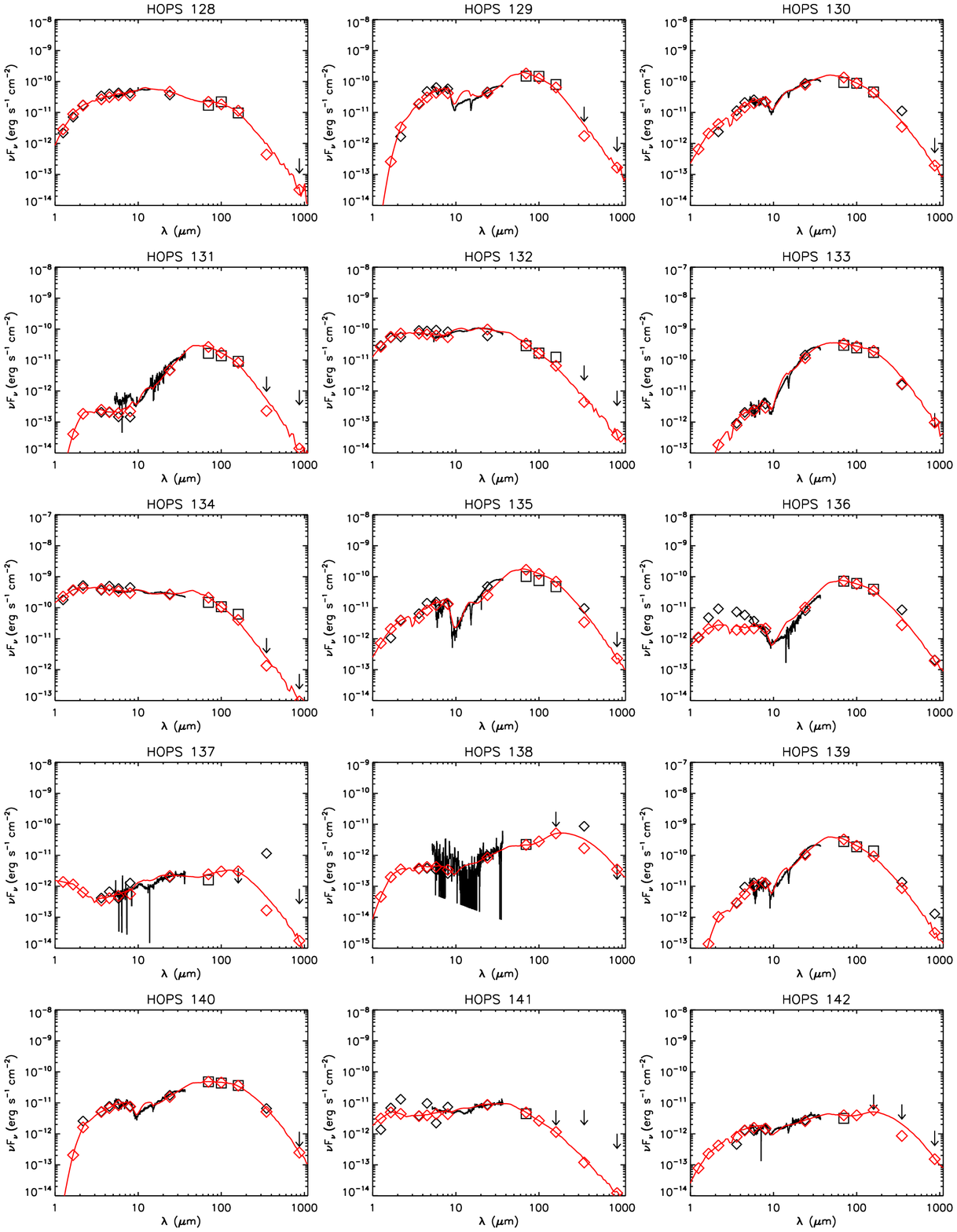}
\figurenum{\ref{bestSEDs}}\caption{continued.}
\end{figure}

\begin{figure}[h]
\centering
\includegraphics[scale=0.9]{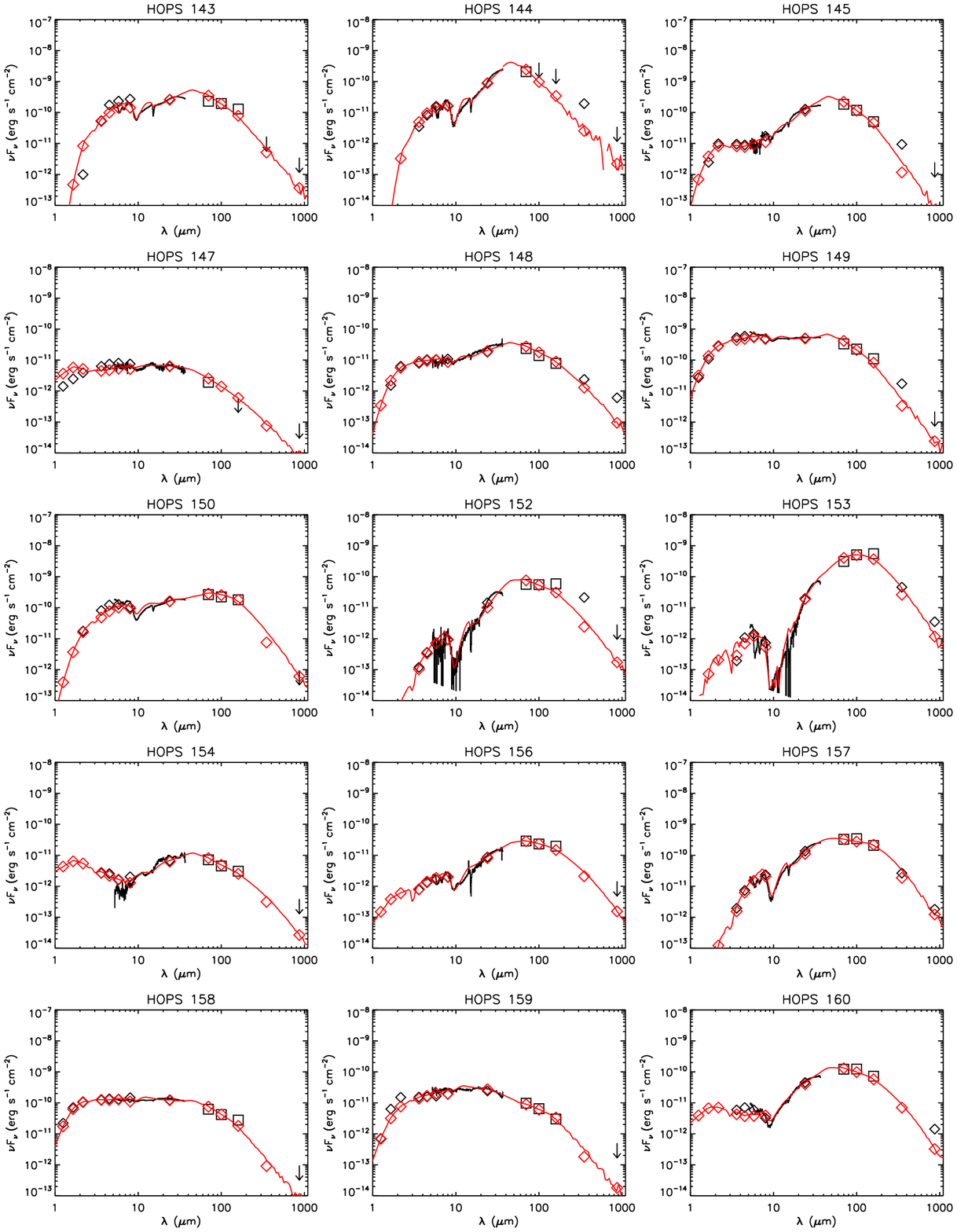}
\figurenum{\ref{bestSEDs}}\caption{continued.}
\end{figure}

\begin{figure}[h]
\centering
\includegraphics[scale=0.9]{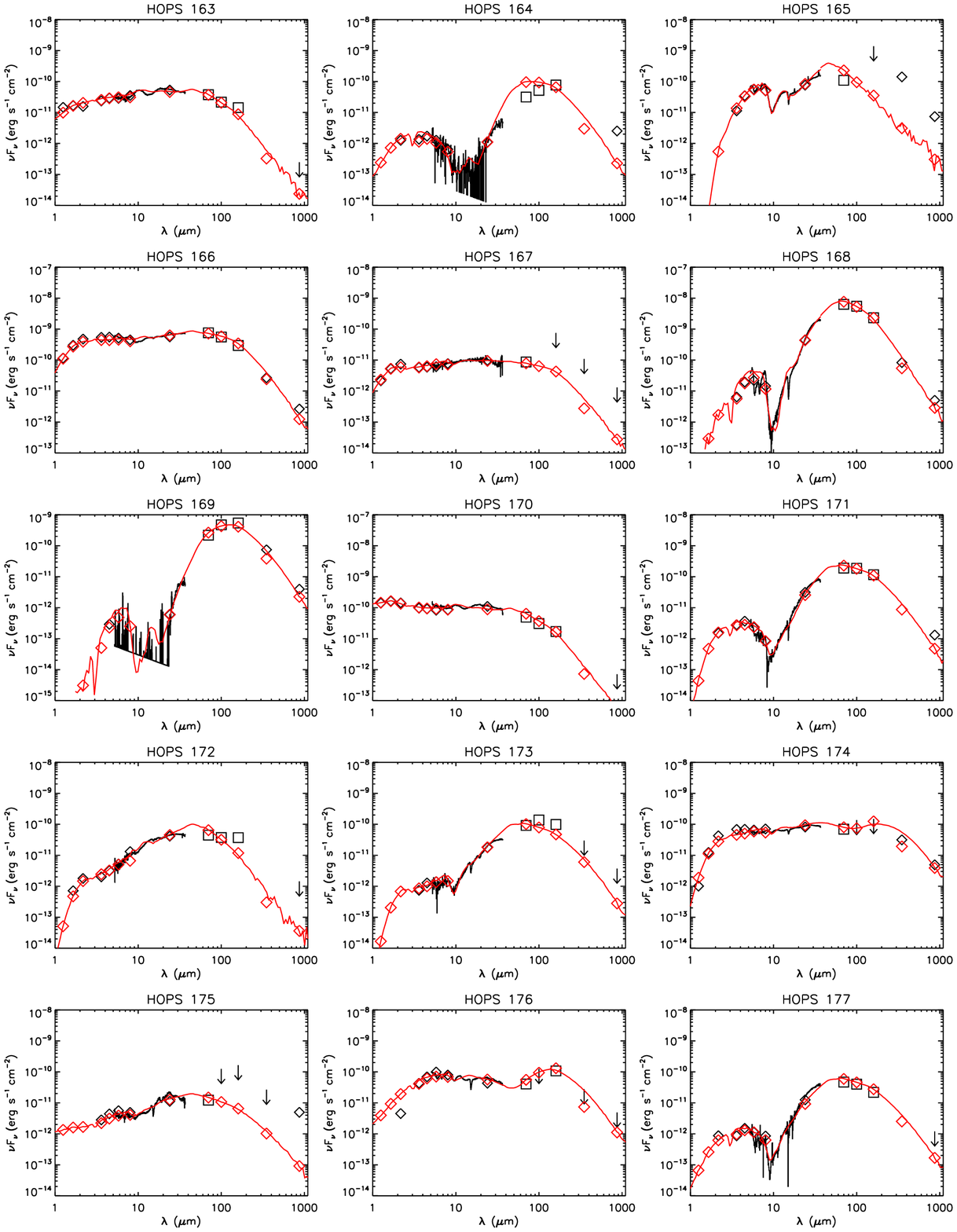}
\figurenum{\ref{bestSEDs}}\caption{continued.}
\end{figure}

\begin{figure}[h]
\centering
\includegraphics[scale=0.9]{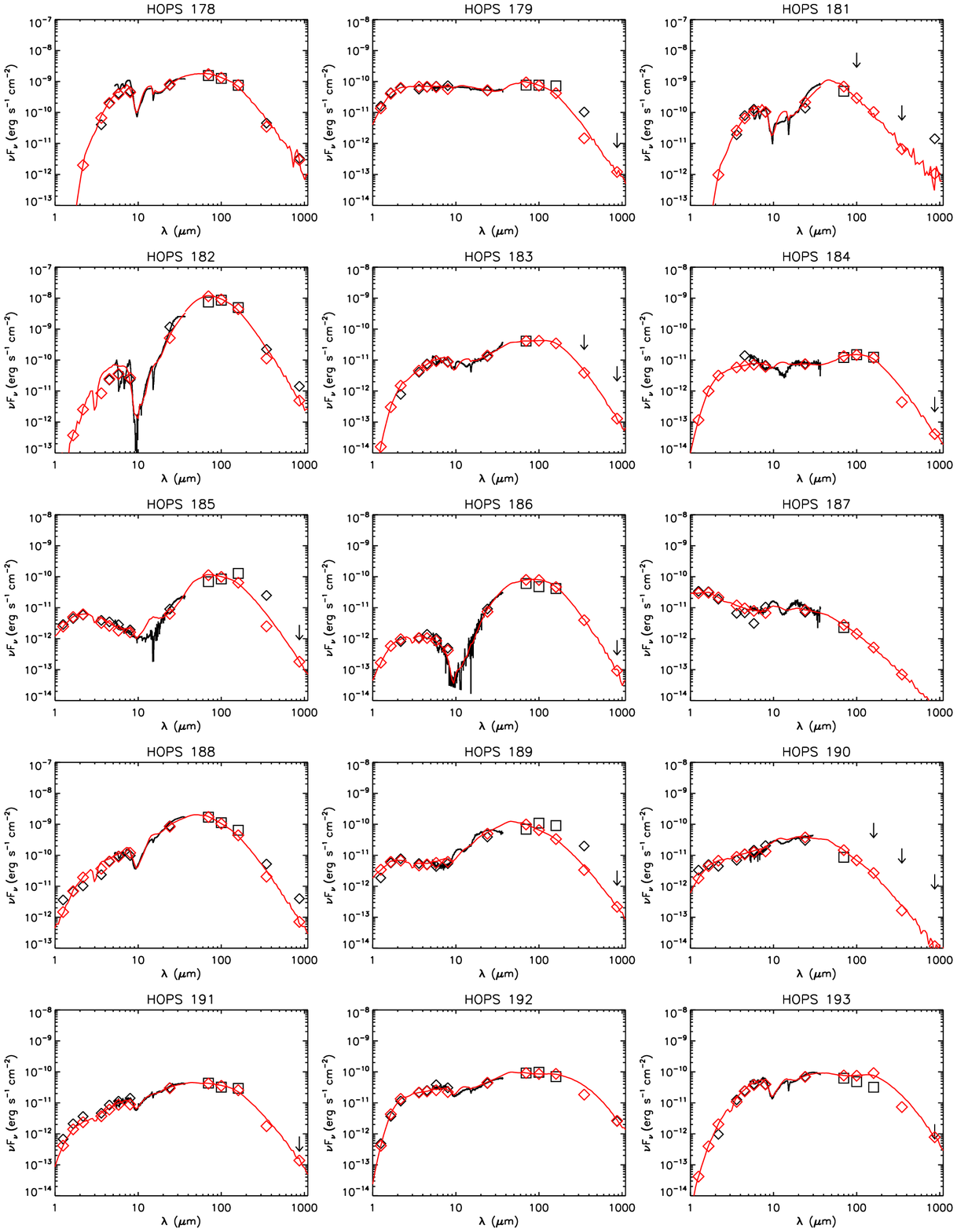}
\figurenum{\ref{bestSEDs}}\caption{continued.}
\end{figure}

\begin{figure}[h]
\centering
\includegraphics[scale=0.9]{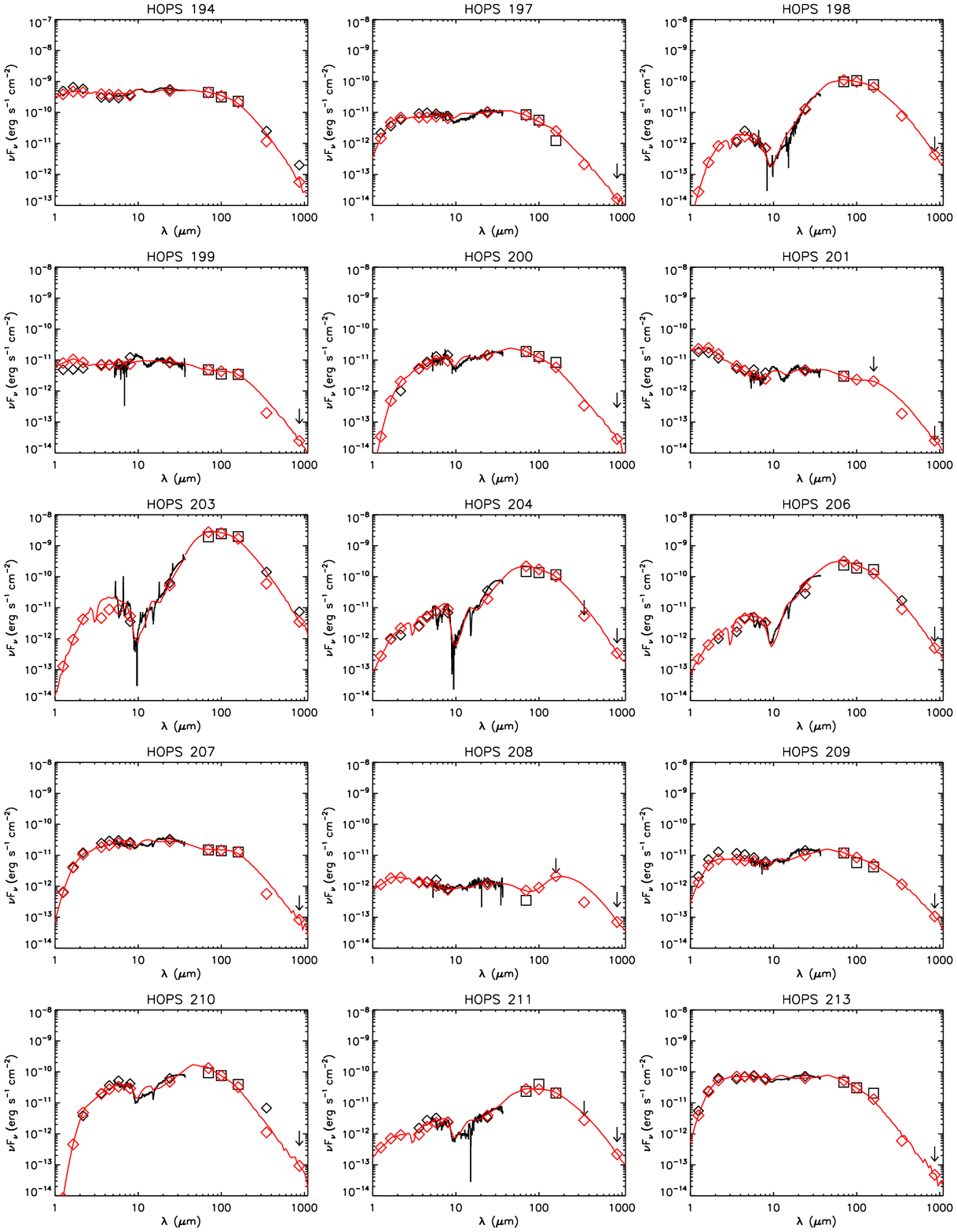}
\figurenum{\ref{bestSEDs}}\caption{continued.}
\end{figure}

\begin{figure}[h]
\centering
\includegraphics[scale=0.9]{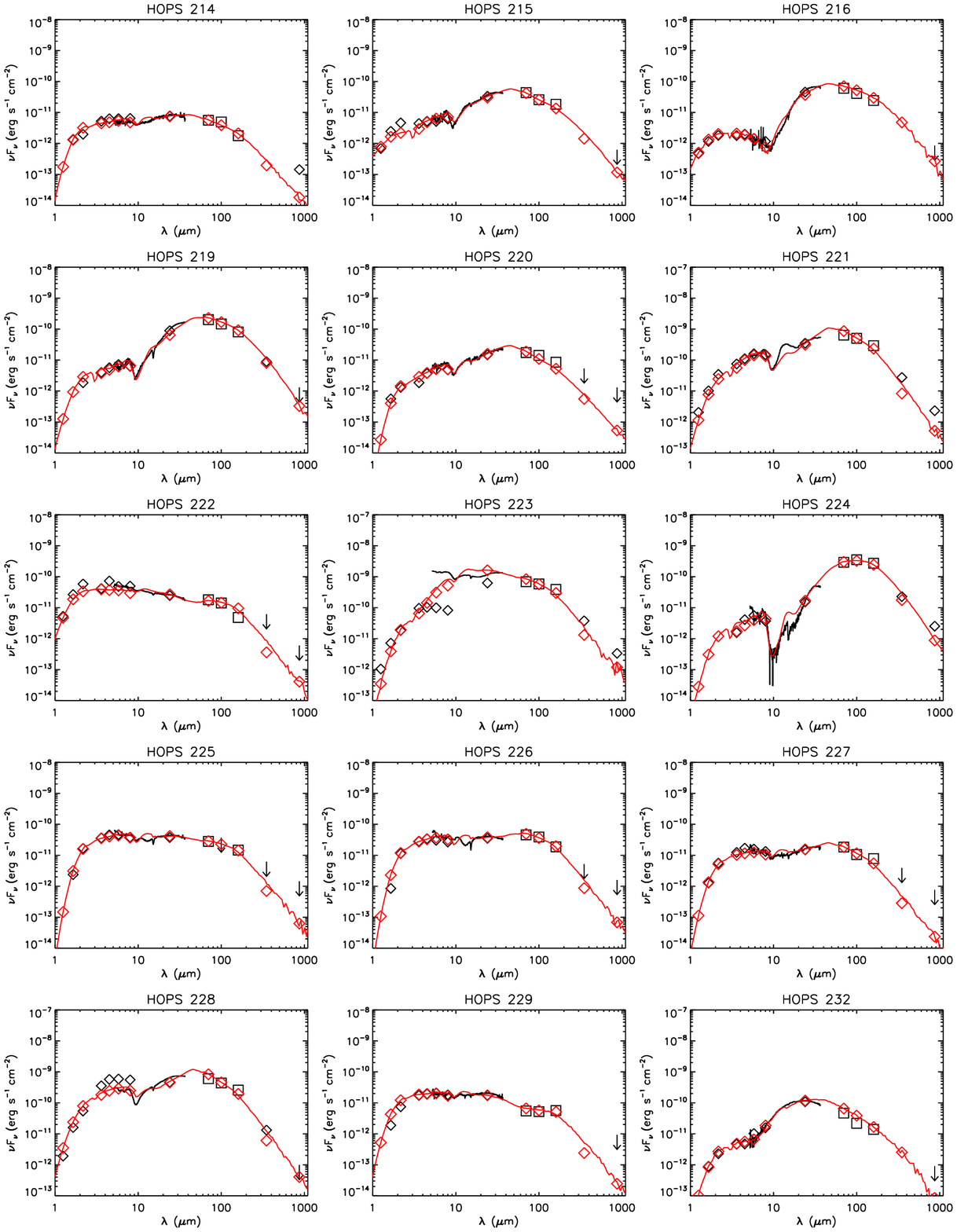}
\figurenum{\ref{bestSEDs}}\caption{continued.}
\end{figure}

\begin{figure}[h]
\centering
\includegraphics[scale=0.9]{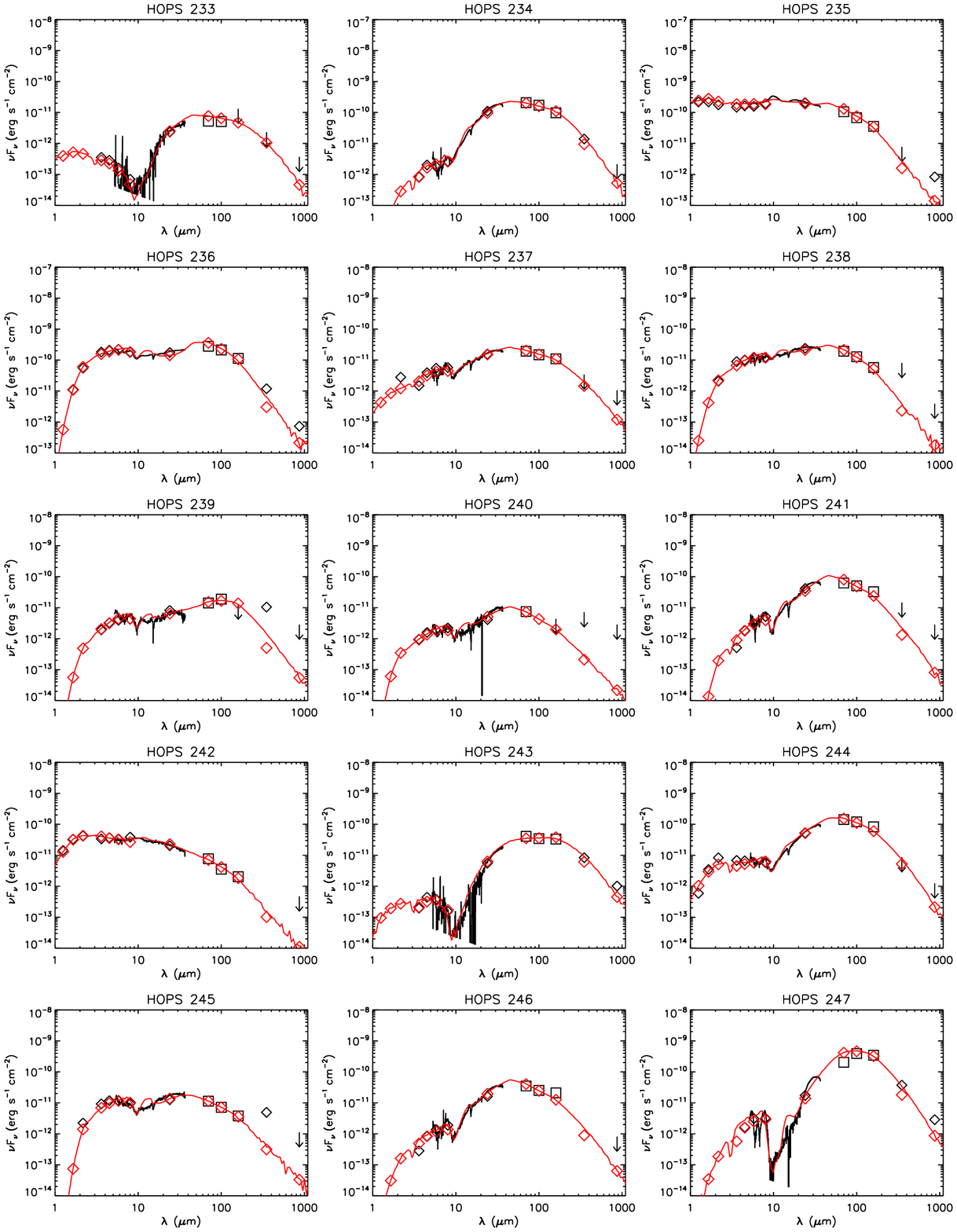}
\figurenum{\ref{bestSEDs}}\caption{continued.}
\end{figure}

\begin{figure}[h]
\centering
\includegraphics[scale=0.9]{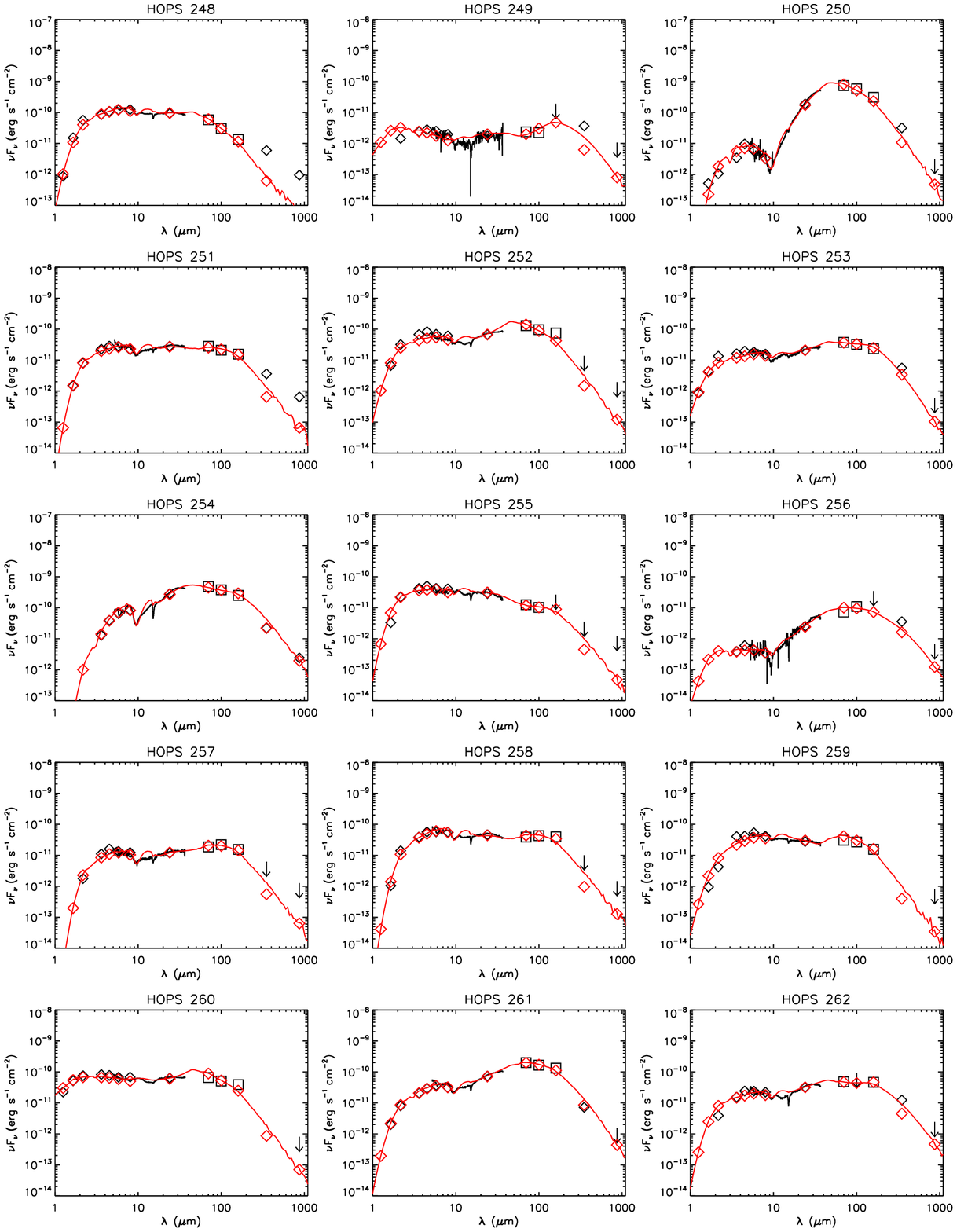}
\figurenum{\ref{bestSEDs}}\caption{continued.}
\end{figure}

\begin{figure}[h]
\centering
\includegraphics[scale=0.9]{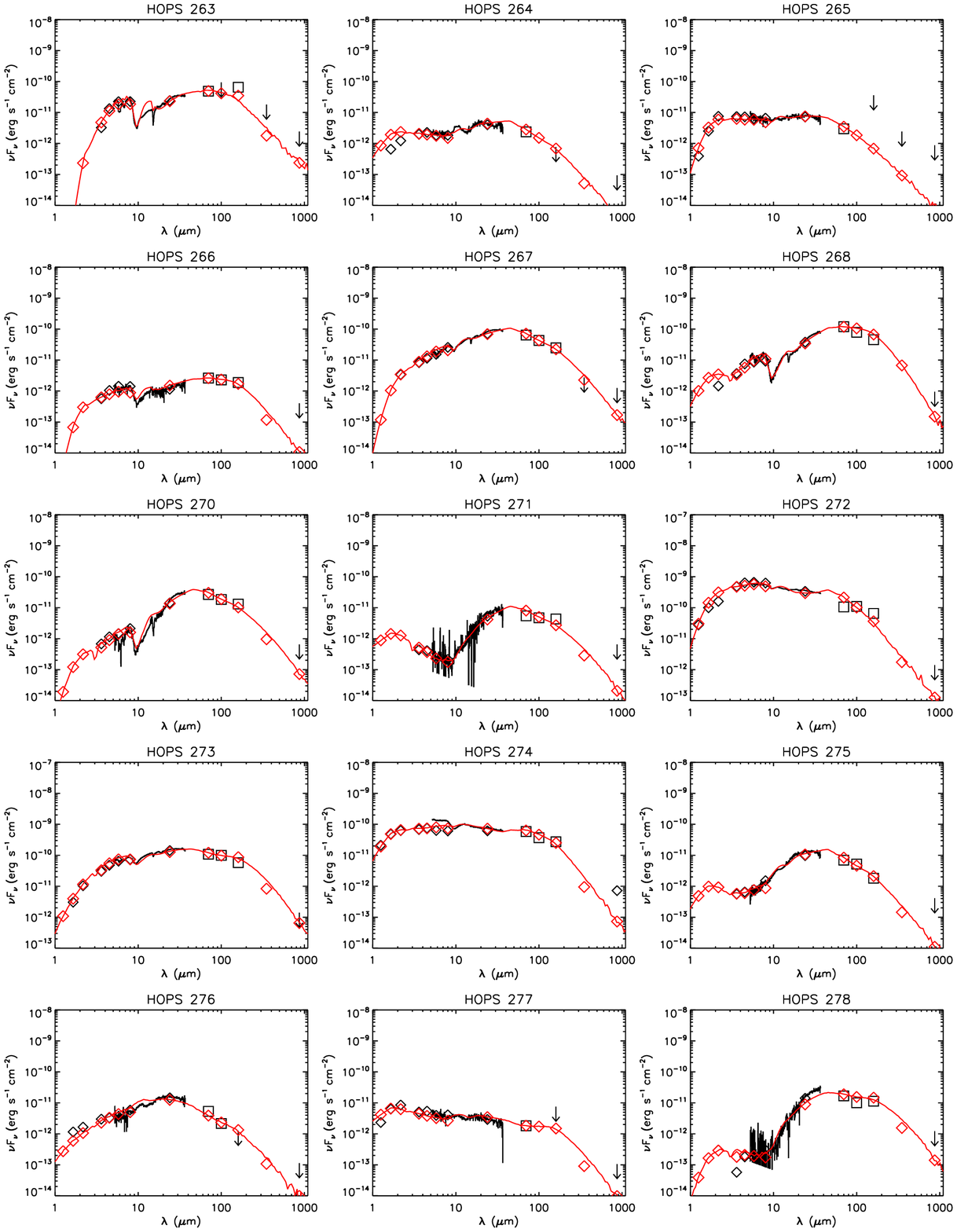}
\figurenum{\ref{bestSEDs}}\caption{continued.}
\end{figure}

\begin{figure}[h]
\centering
\includegraphics[scale=0.9]{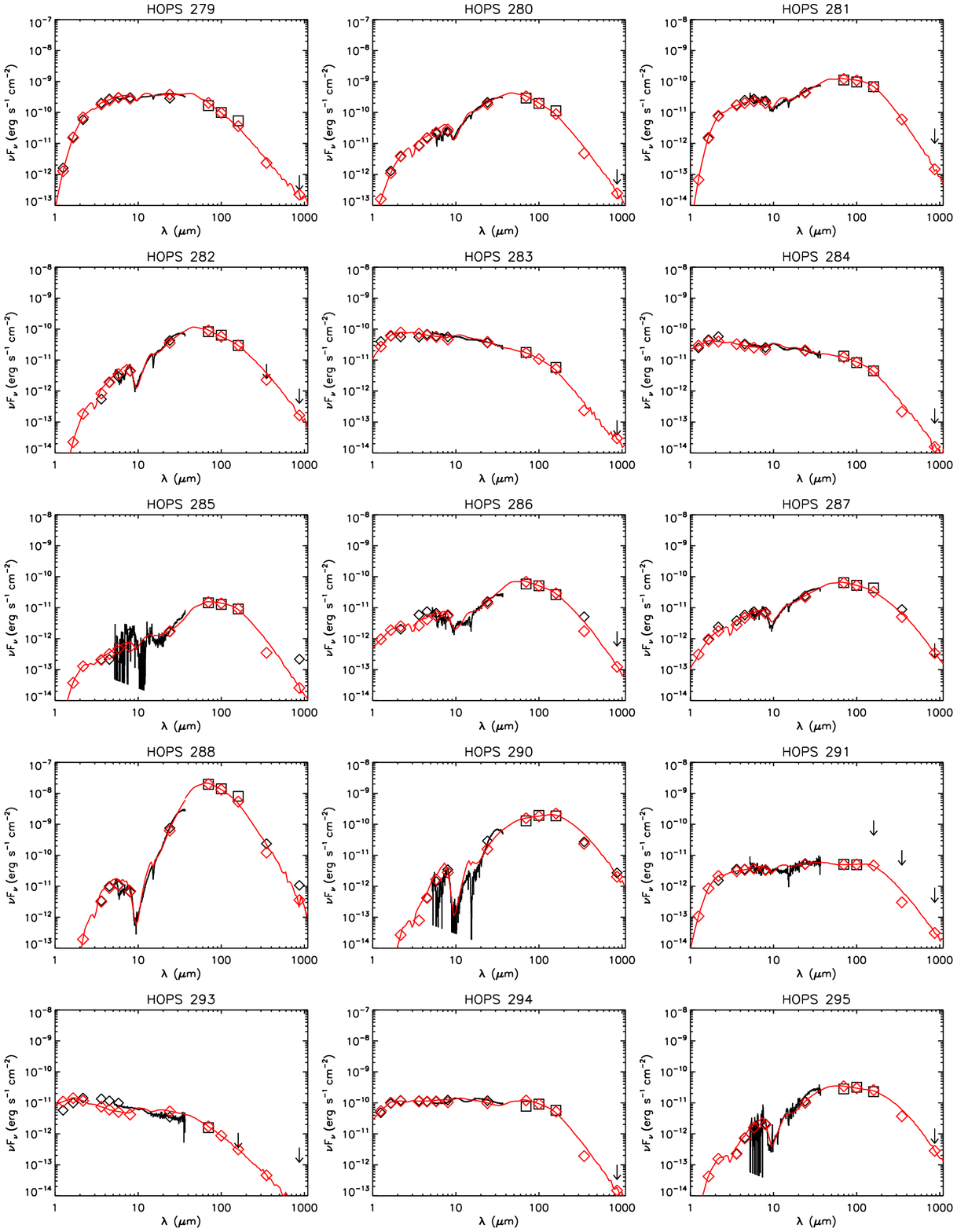}
\figurenum{\ref{bestSEDs}}\caption{continued.}
\end{figure}

\clearpage 

\begin{figure}[h]
\centering
\includegraphics[scale=0.9]{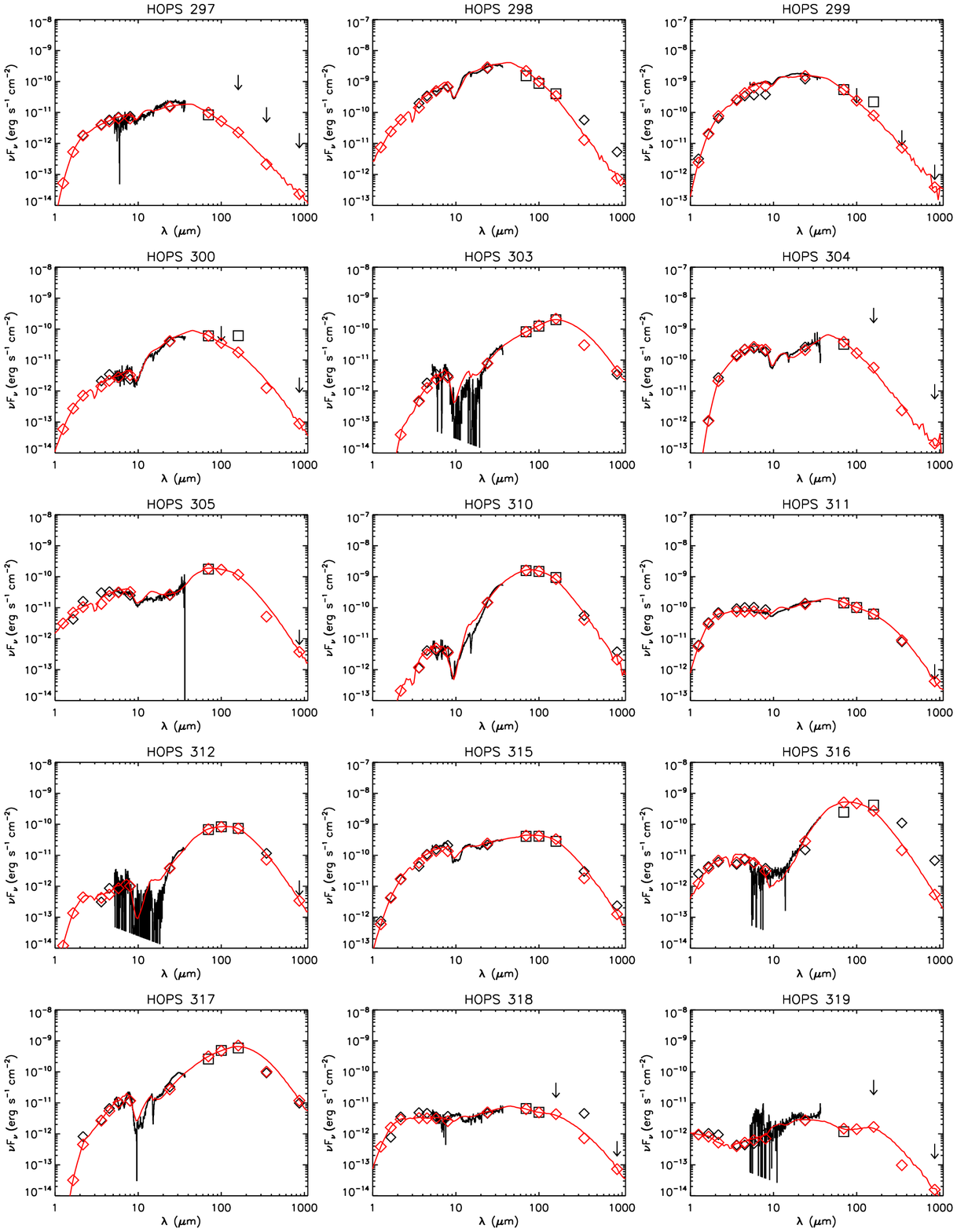}
\figurenum{\ref{bestSEDs}}\caption{continued.}
\end{figure}

\begin{figure}[h]
\centering
\includegraphics[scale=0.9]{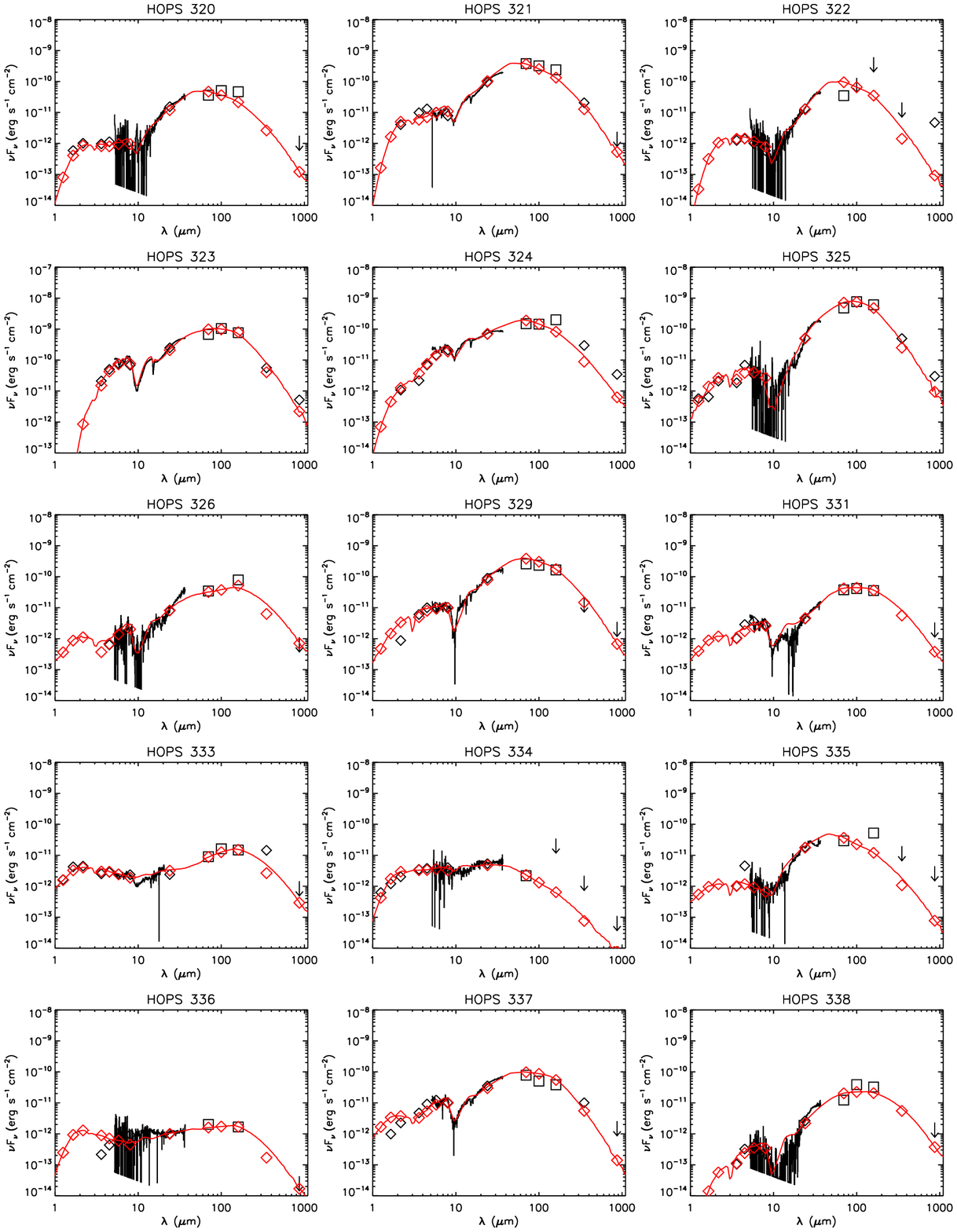}
\figurenum{\ref{bestSEDs}}\caption{continued.}
\end{figure}

\begin{figure}[h]
\centering
\includegraphics[scale=0.9]{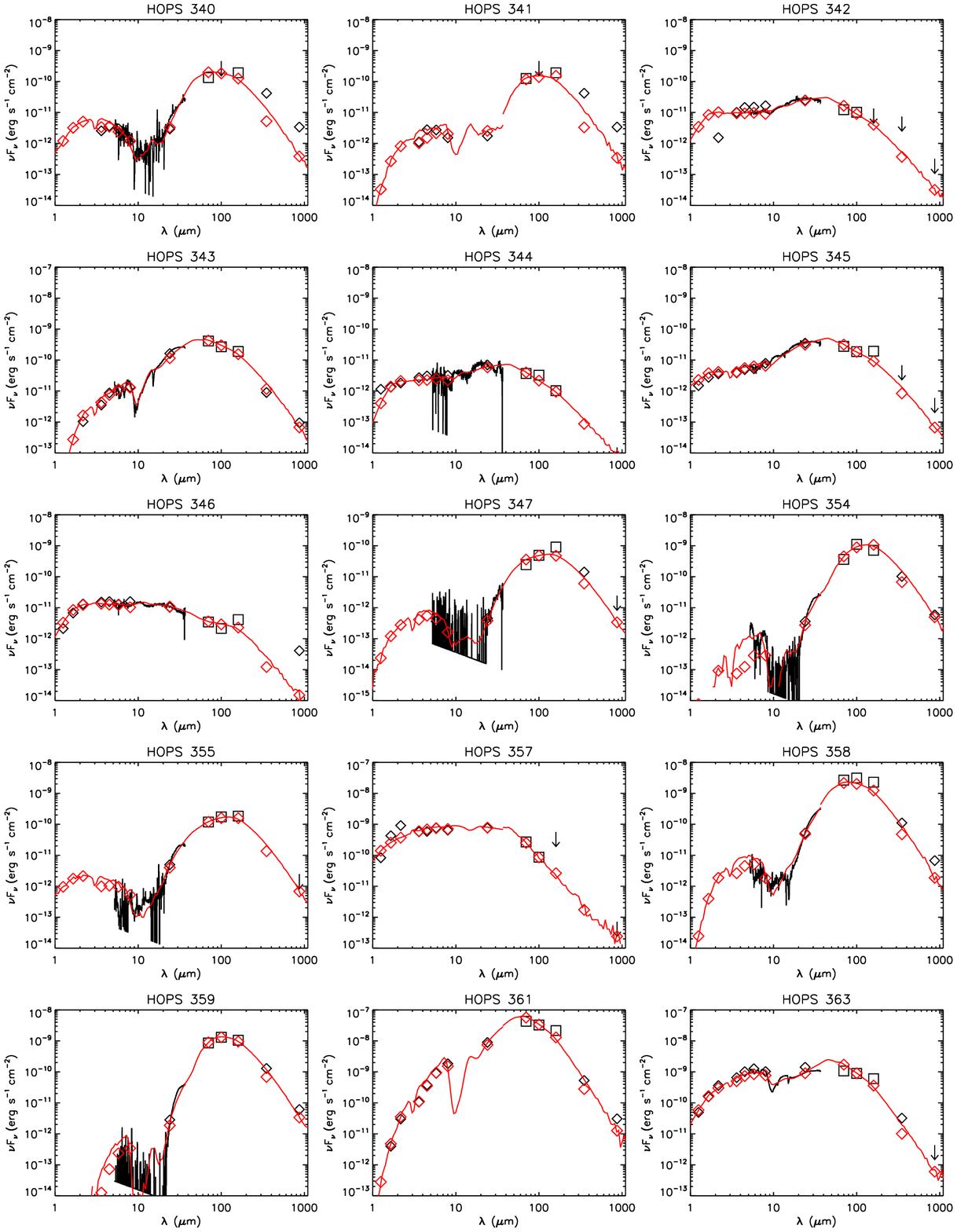}
\figurenum{\ref{bestSEDs}}\caption{continued.}
\end{figure}

\begin{figure}[h]
\centering
\includegraphics[scale=0.9]{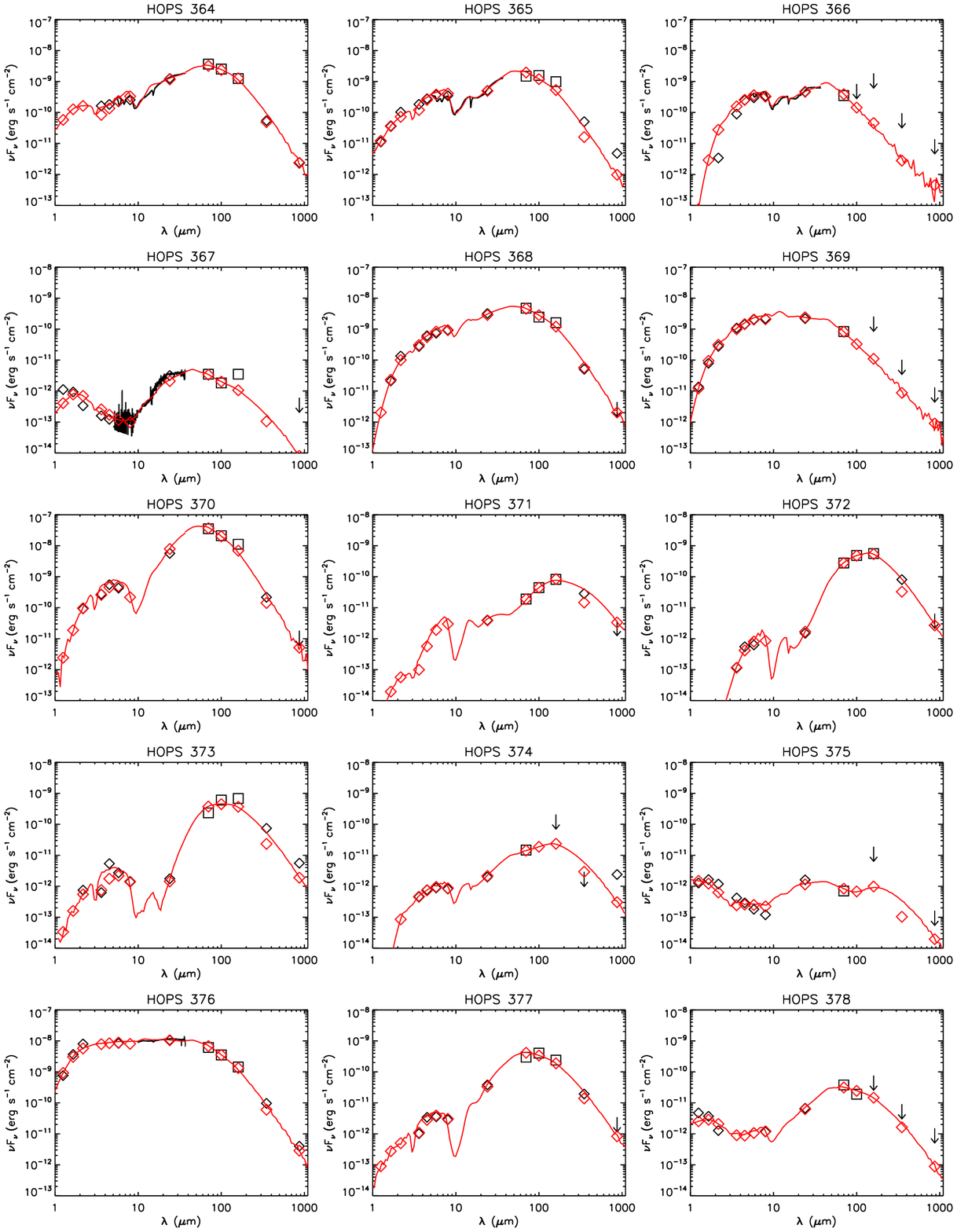}
\figurenum{\ref{bestSEDs}}\caption{continued.}
\end{figure}

\begin{figure}[h]
\centering
\includegraphics[scale=0.9]{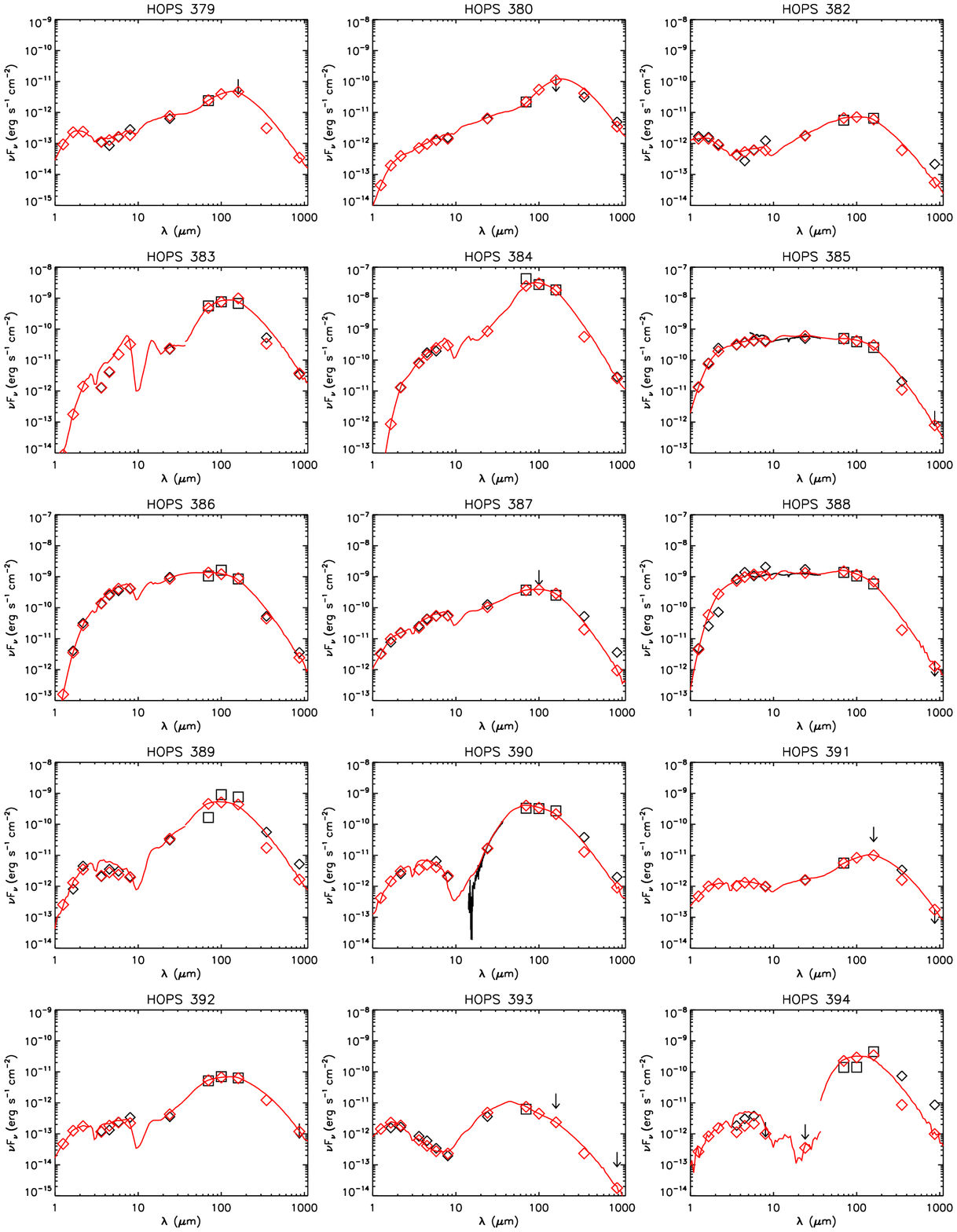}
\figurenum{\ref{bestSEDs}}\caption{continued.}
\end{figure}

\begin{figure}[h]
\centering
\includegraphics[scale=0.9]{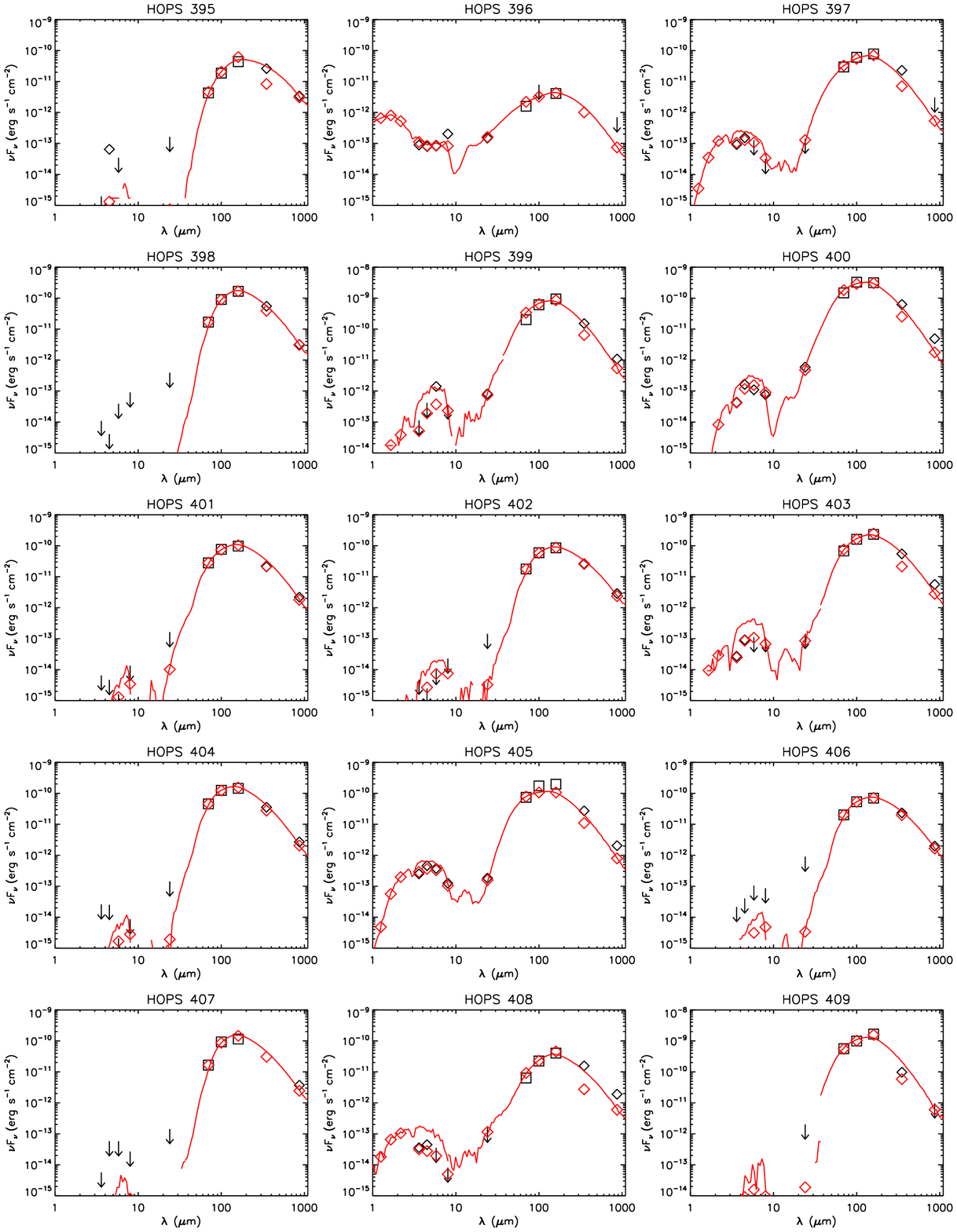}
\figurenum{\ref{bestSEDs}}\caption{continued.}
\end{figure}

\clearpage

\section{Spectral Energy Distributions and Variability}
\label{variability}

The data sets used for the SEDs presented in this work were taken with different 
instruments and telescopes and are not contemporaneous, yet we know that the 
majority of protostars are variable at a $\sim$ 20\% level \citep[e.g.,][]
{morales11,billot12,megeath12}. Therefore, different data sets are snapshots 
of the emission of the protostar at particular times of its unknown duty cycle
of variability. Indeed, in some cases we observe large mismatches between 
different data sets; an extreme example, the outbursting protostar HOPS 
223, was studied by \citet{fischer12}. Another HOPS protostar that recently
experienced an outburst, HOPS 383 \citep{safron15}, does not have a mismatched
SED, since the photometry used here is representative of the post-outburst SED.
In general, variability that is wavelength-dependent or has a long duty cycle 
is more difficult to determine. 

\begin{figure}[!b]
\centering
\includegraphics[scale=0.58,angle=90]{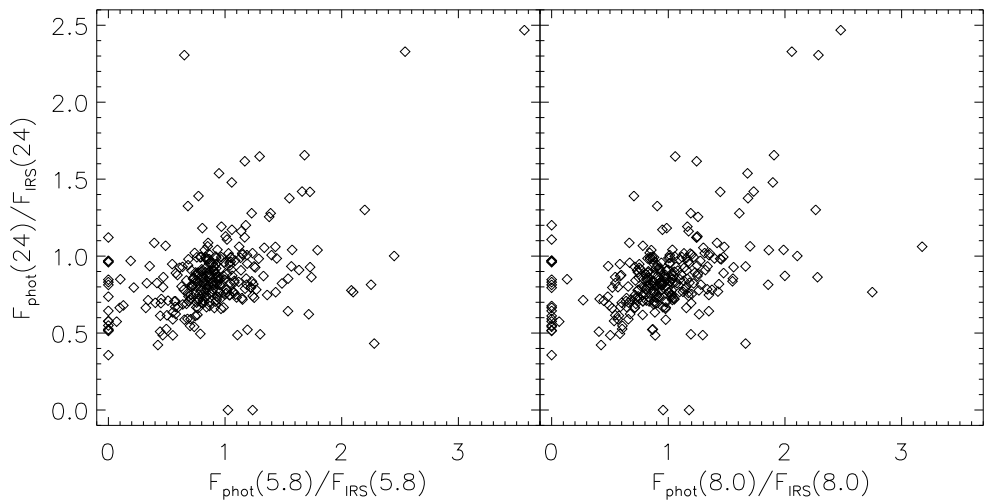}
\caption{Ratios of photometric fluxes in the IRAC and MIPS bands over those 
derived from the IRS spectrum. In the left panel, flux ratios at 24 $\mu$m versus
those at 5.8 $\mu$m are shown, while in the right panel the 24 $\mu$m flux ratios 
are plotted versus the 8.0 $\mu$m flux ratios. A ratio of 0 for a certain band 
means that for that particular object, the IRS spectrum was not available over
the wavelength region of that band.
\label{flux_ratios}}
\end{figure}

Since 290 of the 330 objects in the HOPS sample that were modeled have an 
IRS spectrum and measurements with IRAC or MIPS, we compared fluxes 
measured in the same wave bands, but at different times, to see whether 
there are discrepancies. We used the Spitzer Science Center's 
{\it spitzer\_synthphot} code to calculate IRAC 5.8 and 8.0 $\mu$m and 
MIPS 24 $\mu$m synthetic photometry from the IRS fluxes.
In Figure B\ref{flux_ratios}, we show the flux ratios of IRAC or MIPS photometry
and the synthetic photometry using the IRS spectrum for the protostars
in our HOPS sample. If there were no mismatches, the flux ratios of IRAC 
or MIPS and IRS photometry would be close to one. However, we find that 
they are typically somewhat less than 1; the median ratios at 5.8, 8.0, and 
24 $\mu$m are 0.89, 0.95, and 0.83, respectively.
For small mismatches, calibration uncertainties are a plausible explanation. 
In addition, since we are dealing with objects that are not necessarily point 
sources and often embedded in extended emission, differences in aperture 
sizes between different measurements (IRAC vs.\ MIPS vs.\ IRS) could also 
account for flux mismatches. 

\begin{deluxetable}{lcccc}\scriptsize
\tablewidth{0.7\linewidth}
\tablenum{1}
\tablecaption{Potentially Variable HOPS Targets}
\label{HOPS_variables}
\tablehead{
\colhead{Object} & \colhead{Class} & \colhead{[5.8] Ratio} & 
\colhead{[8.0] Ratio} & \colhead{[24] Ratio}  \\
\colhead{(1)} & \colhead{(2)} & \colhead{(3)} & \colhead{(4)} &
\colhead{(5)}}
\startdata
   HOPS 3 &  flat &  1.681 &  1.906 &  1.656 \\
   HOPS 7 &     0 &  1.121 &  3.176 &  1.061 \\
  HOPS 11 &     0 &  0.444 &  0.405 &  0.510 \\
  HOPS 12 &     0 &  0.553 &  0.888 &  0.485 \\
  HOPS 19 &  flat &  1.296 &  1.059 &  1.648 \\
  HOPS 20 &     I &  2.543 &  2.059 &  2.329 \\
  HOPS 24 &     I &  0.329 &  0.953 &  0.824 \\
  HOPS 32 &     0 &  1.575 &  1.988 &  1.041 \\
  HOPS 38 &     0 &  3.566 &  2.478 &  2.468 \\
  HOPS 41 &     I &  0.788 &  0.586 &  0.495 \\
  HOPS 65 &     I &  0.191 &  1.275 &  0.969 \\
  HOPS 71 &     I &  0.416 &  0.547 &  0.825 \\
  HOPS 78 &     0 &  0.097 &  \nodata &  0.664 \\
  HOPS 85 &  flat &  1.728 &  1.445 &  1.418 \\
  HOPS 91 &     0 &  0.511 &  0.446 &  0.711 \\
  HOPS 95 &     0 &  2.450 &  2.106 &  1.001 \\
 HOPS 108 &     0 &  \nodata &  0.867 &  0.525 \\
 HOPS 114 &     I &  0.102 &  0.132 &  0.850 \\
 HOPS 121 &     0 &  \nodata &  \nodata &  0.357 \\
 HOPS 124 &     0 &  0.402 &  0.601 &  0.697 \\
 HOPS 131 &     I &  0.341 &  0.270 &  0.714 \\
 HOPS 132 &  flat &  1.717 &  1.264 &  0.622 \\
 HOPS 138 &     0 &  0.354 &  0.485 &  0.935 \\
 HOPS 141 &  flat &  0.393 &  1.418 &  1.086 \\
 HOPS 143 &     I &  1.793 &  1.862 &  1.039 \\
 HOPS 154 &     I &  2.252 &  1.856 &  0.814 \\
 HOPS 177 &     I &  1.740 &  2.279 &  0.863 \\
 HOPS 181 &     I &  1.301 &  1.194 &  0.492 \\
 HOPS 182 &     0 &  0.772 &  0.705 &  1.390 \\
 HOPS 183 &  flat &  0.466 &  0.502 &  0.486 \\
 HOPS 186 &     I &  1.062 &  1.999 &  0.872 \\
 HOPS 187 &  flat &  0.321 &  1.155 &  0.664 \\
 HOPS 203 &     0 &  \nodata &  0.481 &  0.737 \\
 HOPS 206 &     0 &  1.106 &  1.295 &  0.487 \\
 HOPS 222 &    II &  0.496 &  0.604 &  0.526 \\
 HOPS 223 &     I &  0.070 &  0.065 &  0.575 \\
 HOPS 228 &     I &  2.098 &  2.749 &  0.766 \\
 HOPS 239 &     I &  0.682 &  0.905 &  1.325 \\
 HOPS 270 &     I &  1.729 &  1.560 &  0.928 \\
 HOPS 271 &     I &  0.948 &  1.679 &  1.538 \\
 HOPS 272 &    II &  1.660 &  1.732 &  1.420 \\
 HOPS 278 &     I &  0.651 &  2.287 &  2.306 \\
 HOPS 290 &     0 &  2.198 &  2.263 &  1.301 \\
 HOPS 297 &     I &  2.083 &  1.263 &  0.777 \\
 HOPS 299 &     I &  0.454 &  0.412 &  0.722 \\
 HOPS 305 &  flat &  0.424 &  0.423 &  0.422 \\
 HOPS 316 &     0 &  2.279 &  1.662 &  0.432 \\
 HOPS 319 &     I &  0.218 &  0.520 &  0.796 \\
 HOPS 321 &     I &  0.769 &  0.518 &  0.662 \\
 HOPS 322 &     I &  0.128 &  0.494 &  0.680 \\
 HOPS 338 &     0 &  1.552 &  1.683 &  1.376 \\
 HOPS 340 &     0 &  1.192 &  0.863 &  0.522 \\
 HOPS 358 &     0 &  \nodata &  \nodata &  0.514 \\
 HOPS 359 &     0 &  \nodata &  \nodata &  0.517 \\
 HOPS 363 &  flat &  1.169 &  1.242 &  1.616 \\
 HOPS 388 &  flat &  1.058 &  1.896 &  1.479 \\
\enddata
\tablecomments{
\parbox{0.63\linewidth}{Column (1) lists the HOPS number of the 
object, column (2) the class based on SED classification, column (3) 
the ratio of the IRAC 5.8 \micron\ flux and the IRS flux over the 
IRAC 5.8 \micron\ band, columns (4) and (5) the ratio of 
photometric and IRS flux for the IRAC 8.0 \micron\ and MIPS 24 
\micron\ band, respectively.}}
\end{deluxetable}

To identify outliers, in Table B\ref{HOPS_variables}, we list those flux ratios 
that lie in the lower or upper 5\% of values. They represent a conservative 
list of potentially variable sources in our sample.
Of the 290 objects for which we calculated flux ratios, 5 have flux 
mismatches larger than a factor of 2 between IRS and both IRAC bands
at 5.8 and 8.0 $\mu$m. Three objects have similarly large mismatches between 
IRS and MIPS. The overlap between these two samples contains two objects, 
HOPS 20 and 38 (the other objects are HOPS 95, 228, 278, and 290).
For most of these objects, the large mismatches can be attributed to noisy 
IRS spectra, especially in the 5-8 $\mu$m region, making the comparison
between IRAC and IRS less reliable.  
Eight objects have IRAC-IRS mismatches smaller than a factor 0.5; one of
these objects and seven different objects have such small mismatches 
between MIPS and IRS (see Table  B\ref{HOPS_variables}). Slightly over 
one-third of these objects have noisy IRS spectra. 
Of the 21 objects that have either large ($>$ factor of 2) or small ($<$
factor of 0.5) mismatches, 9 are Class 0 protostars, 10 are Class I protostars,
and 2 are flat-spectrum sources. 
In cases where the IRS flux is too high relative to the MIPS 24 $\mu$m 
photometry, the mismatch could be due to more extended emission or
flux from a nearby companion being included in the IRS measurement 
(SL and LL slit widths of 3{\farcs}6  and 10{\farcs}5, respectively, versus 
the typical FWHM of the MIPS 24 $\mu$m PSF of $\sim$ 6\arcsec).

For a few sources, the discrepancies between IRS fluxes and IRAC or MIPS 
can be attributed to the scaling factors applied to different parts of the IRS 
spectrum. As mentioned in section \ref{SEDs}, we typically scaled the 
SL spectrum to match the flux of the LL spectrum at 14 $\mu$m, given
that the latter has a larger slit width. However, in the case of HOPS 38,
where the IRAC 5.8 and 8.0 $\mu$m and the MIPS 24 $\mu$m fluxes 
are about a factor of 2.5-3.5 higher than the IRS fluxes, the LL spectrum 
was scaled by 0.4 to match the SL flux at 14 $\mu$m. Given the IRAC and 
MIPS measurements, it would have been more appropriate to scale the 
SL spectrum up. For HOPS 124, the SL spectrum was scaled by 2.5; if
instead the LL spectrum had been scaled down, the discrepancies between
IRS and IRAC and MIPS would be less than 50\%.  

Overall, in cases where the IRS spectrum has sufficient signal-to-noise ratio 
and its fluxes seem lower than the photometric measurements or the 
discrepancies in IRAC-IRS and MIPS-IRS fluxes are quite different, intrinsic 
source variability could be a likely explanation. Among the sample shown 
in Table B\ref{HOPS_variables}, this would apply to HOPS 24, 71, 131,
132, 141, 143, 154, 187, 206, 223, 228, 299, 363, and 388. HOPS 223 is 
indeed variable \citep[see][]{fischer12}, but the other objects still require
confirmation.
Thus, about 5\% of the 290 protostars in our HOPS sample that have
IRS, IRAC, and MIPS data may reveal variability to some degree.
These objects are prime candidates for follow-up observations regarding 
their variability.

\section{Model Parameter Ranges and Degeneracies}
\label{models_unc}

\begin{figure}[!t]
\centering
\includegraphics[scale=0.85]{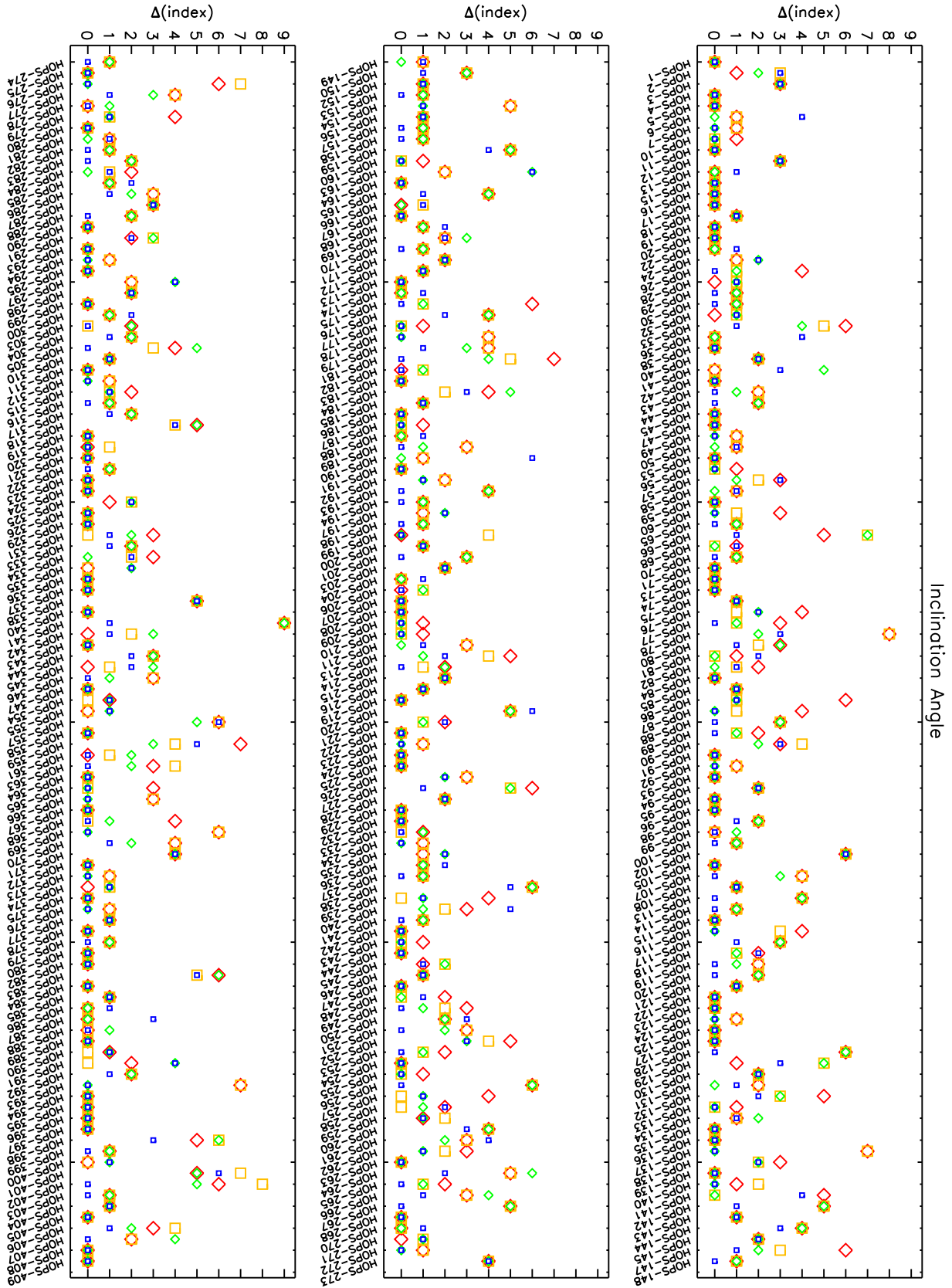}
\caption{For all modeled HOPS targets (see name on x axis), difference 
between the index of the best-fit inclination angle and the index of the mode 
of the inclination angle for models that lie within a difference of 0.5 ({\it small 
blue squares}), 1.0 ({\it small green diamonds}), 1.5 ({\it yellow larger squares}), 
and 2.0 ({\it red larger diamonds}) of the best-fit $R$. {\it The remaining plots 
for the other five model parameters are available in Figure Set C1.}
\label{Inc_modes_bf}}
\end{figure}

In section \ref{deltaR} we discussed how modes can be used to assess
how well model parameters are constrained. Here we analyze the spread
of mode values for individual HOPS targets.
In Figure Set C1 (see Fig.\ C\ref{Inc_modes_bf} for an example) we show the 
difference between the modes and the best-fit values of the major model 
parameters ($i, L_{tot}, \rho_{1000}, \theta, R_{disk}, A_V$) for 
all modeled HOPS targets. As in section \ref{deltaR}, we use models in certain 
$\Delta R$ bins, starting at a range of 0.5 from the best-fit $R$ up to a range 
of 2.0 from the best-fit $R$. For parameters with discrete values, such as the 
inclination and cavity opening angles, we plot the difference between the indices 
of modes and best-fit values. For example, if the best-fit inclination angle has a 
value of 41\degr\ and the mode a value of 57\degr, the difference in indices 
would be 2 (since the discrete values in our model grid are 18\degr, 32\degr, 
41\degr, 50\degr, 57\degr, etc.). Similarly, if the best-fit cavity opening angle 
is 5\degr\ but the mode is 45\degr, the difference in indices would be 4. For 
the total luminosity and foreground extinction, we plotted instead the difference 
between the parameter values of the best fit and the modes.

Objects that are not particularly well fit by their best-fit model from the grid often 
have modes that are quite different from the best-fit value once $\Delta R$ 
reaches 2. For example, for HOPS 181, whose best-fit model has $R$=5.16, 
the mode of the inclination angle for models within $\Delta R$=0.5 (i.e., models
with $R < 5.66$) is the same as the best-fit value, but then the difference increases 
as $\Delta R$ becomes larger. For $\Delta R$=2.0, the mode is seven discrete
values away from the best fit ($i=$18\degr\ for the mode, 76\degr\ for the best fit).
Several other model parameters are also not well constrained. 
There are also objects that have a relatively good fit, but a larger spread in
certain parameters. An example is HOPS 70, whose best-fit model has an $R$ 
value of 2.33; its total luminosity and inclination angle are very well constrained, 
while its reference envelope density is quite uncertain.

Certain protostars are sufficiently well constrained by the available data and 
well fit by our grid of models that their parameters do not change much from 
$\Delta R$=0.5 to $\Delta R$=2.0. For example, the modes of the inclination
angle of HOPS 1 are the same as the best-fit value even for all models within 
$\Delta R$=2, and the other model parameters show a small spread. 
There are 37 protostars with small differences between their best-fit values 
and modes for models within $\Delta R$=2 ($<$ factor of two for
$\rho_{1000}$ and $L_{tot}$, $<$ 50 AU for the disk radius, $<$ 10\degr\ for
the cavity opening angle, $<$ 30\% difference in inclination angle). These
protostars are well characterized by our model fits. The mean and median 
$R$ values for their best-fit models are 3.48 and 3.17, respectively. This 
validates our estimate of $R \sim$ 4 as the boundary between a reliable 
and a less reliable fit.

\figsetstart
\figsetnum{C1}
\figsettitle{Figure set showing the differences between modes and best-fit
parameter values of six model parameters for all modeled HOPS targets.}

\figsetgrpstart
\figsetgrpnum{1.1}
\figsetgrptitle{Difference between modes and best-fit parameter values 
for the inclination angle}
\figsetplot{Inc_modes_best-fit_diff.eps}
\figsetgrpnote{For all modeled HOPS targets (see name on x axis), difference 
between the index of the best-fit inclination angle and the index of the mode 
of the inclination angle for models that lie within a difference of 0.5 ({\it small 
blue squares}), 1.0 ({\it small green diamonds}), 1.5 ({\it yellow larger squares}), 
and 2.0 ({\it red larger diamonds}) of the best-fit $R$.}
\figsetgrpend

\figsetgrpstart
\figsetgrpnum{1.2}
\figsetgrptitle{Difference between modes and best-fit parameter values 
for the total luminosity}
\figsetplot{Ltot_modes_best-fit_diff.eps}
\figsetgrpnote{For all modeled HOPS targets (see name on x axis), difference 
between the best-fit total luminosity and the mode of the total luminosity for 
models that lie within a difference of 0.5 ({\it small blue squares}), 1.0 
({\it small green diamonds}), 1.5 ({\it yellow larger squares}), and 2.0 
({\it red larger diamonds}) of the best-fit $R$.}
\figsetgrpend

\figsetgrpstart
\figsetgrpnum{1.3}
\figsetgrptitle{Difference between modes and best-fit parameter values 
for the reference envelope density}
\figsetplot{Rho1000_modes_best-fit_diff.eps}
\figsetgrpnote{For all modeled HOPS targets (see name on x axis), difference 
between the index of the best-fit reference density $\rho_{1000}$ and the index 
of the mode of $\rho_{1000}$ for models that lie within a difference of 0.5 
({\it small blue squares}), 1.0 ({\it small green diamonds}), 1.5 ({\it yellow 
larger squares}), and 2.0 ({\it red larger diamonds}) of the best-fit $R$.}
\figsetgrpend

\figsetgrpstart
\figsetgrpnum{1.4}
\figsetgrptitle{Difference between modes and best-fit parameter values 
for the cavity opening angle}
\figsetplot{Cavity_modes_best-fit_diff.eps}
\figsetgrpnote{For all modeled HOPS targets (see name on x axis), difference 
between the index of the best-fit cavity opening angle and the index of the mode 
of the cavity opening angle for models that lie within a difference of 0.5 
({\it small blue squares}), 1.0 ({\it small green diamonds}), 1.5 ({\it yellow 
larger squares}), and 2.0 ({\it red larger diamonds}) of the best-fit $R$.}
\figsetgrpend

\figsetgrpstart
\figsetgrpnum{1.5}
\figsetgrptitle{Difference between modes and best-fit parameter values 
for the centrifugal radius}
\figsetplot{Rdisk_modes_best-fit_diff.eps}
\figsetgrpnote{For all modeled HOPS targets (see name on x axis), difference 
between the index of the best-fit centrifugal radius and the index of the mode 
of the centrifugal radius for models that lie within a difference of 0.5 
({\it small blue squares}), 1.0 ({\it small green diamonds}), 1.5 ({\it yellow 
larger squares}), and 2.0 ({\it red larger diamonds}) of the best-fit $R$.}
\figsetgrpend

\figsetgrpstart
\figsetgrpnum{1.6}
\figsetgrptitle{Difference between modes and best-fit parameter values 
for the foreground extinction}
\figsetplot{AV_modes_best-fit_diff.eps}
\figsetgrpnote{For all modeled HOPS targets (see name on x axis), difference 
between the best-fit foreground extinction and the mode of the foreground 
extinction for models that lie within a difference of 0.5 ({\it small blue squares}), 
1.0 ({\it small green diamonds}), 1.5 ({\it yellow larger squares}), and 2.0 
({\it red larger diamonds}) of the best-fit $R$.}
\figsetgrpend

\figsetend

\begin{figure}[h]
\centering
\includegraphics[scale=0.7]{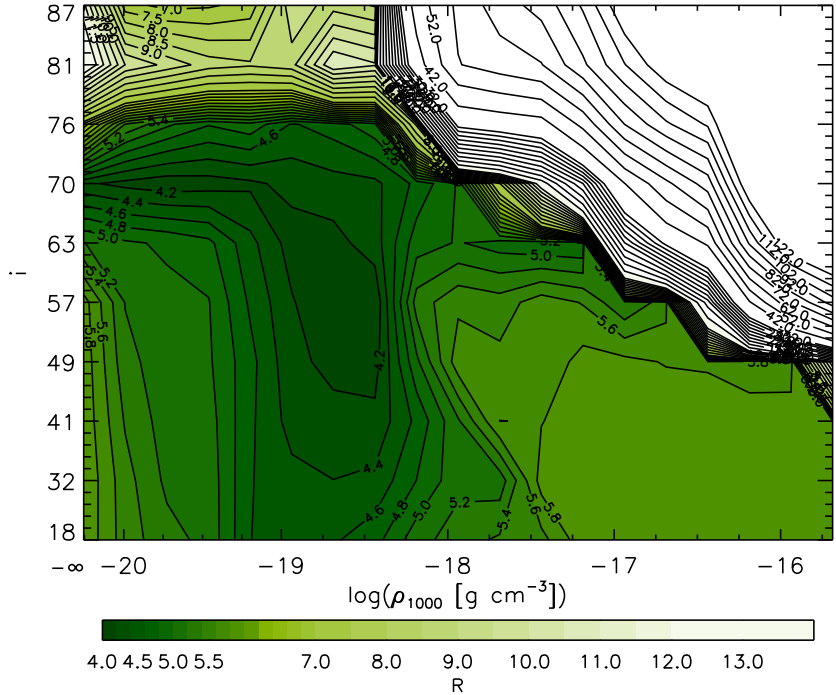}
\caption{$R$ contour plot for the models that fit HOPS 24. For each combination
of inclination angle and reference density, the lowest $R$ values of models with
these two parameter values are shown.  {\it The remaining $R$ contour plots 
are available in Figure Set C2.}
\label{HOPS24_contour1}}
\end{figure}

Part of the parameter uncertainties can be attributed to degeneracies between
model parameters. To illustrate some of these degeneracies, in Figure Set C2
(see Fig.\ C\ref{HOPS24_contour1} for an example) we show contour plots
of $R$ values for sets of two model parameters each (we plot the lowest
$R$ value of models with these two parameter values) resulting from the
model fits of HOPS 24, HOPS 107, and HOPS 149. 
The plots for HOPS 24 show that the inclination angle is somewhat degenerate 
with the envelope density, with higher inclination angles being accommodated by 
lower $\rho_{1000}$ values (Fig.\ C\ref{HOPS24_contour1}). A similar situation 
applies to the disk radius, with larger disk radii requiring higher envelope densities. 
The inclination angle and the cavity opening angle are degenerate, too; for 
higher inclination angles the cavity is larger. 
The $R$ contour plots for HOPS 107 suggest that a certain range of inclination
angles and reference densities can fit the SED, while the disk radius and cavity 
opening angles are not well constrained. However, the plots clearly show that
high-density, high-inclination models fit very poorly.
Finally, we can deduce from the $R$ contour plots for HOPS 149 that also here 
certain parameter values can be excluded; lower inclination angles and 
reference densities in the 10$^{-18}$--10$^{-19}$ g cm$^{-3}$ range yield 
the best fits, with larger densities accompanied by larger disk radii and larger 
cavity opening angles.

\figsetstart
\figsetnum{C2}
\figsettitle{Figure set showing ten examples of $R$ contour plots.}

\figsetgrpstart
\figsetgrpnum{2.1}
\figsetgrptitle{$R$ contour plot for the models that fit HOPS 24, 
$i$ vs.\ $\rho_{1000}$}
\figsetplot{HOPS24_R_contour1.epss}
\figsetgrpnote{$R$ contour plot for the models that fit HOPS 24. For each 
combination of inclination angle and reference density, the lowest $R$ values 
of models with these two parameter values are shown.}
\figsetgrpend

\figsetgrpstart
\figsetgrpnum{2.2}
\figsetgrptitle{$R$ contour plot for the models that fit HOPS 24, 
$R_{disk}$ vs.\ $\rho_{1000}$}
\figsetplot{HOPS24_R_contour2.eps}
\figsetgrpnote{$R$ contour plot for the models that fit HOPS 24. For each 
combination of disk outer radius ($=R_c$) and $\rho_{1000}$, the lowest 
$R$ values of models with these two parameter values are shown.}
\figsetgrpend

\figsetgrpstart
\figsetgrpnum{2.3}
\figsetgrptitle{$R$ contour plot for the models that fit HOPS 24, 
$i$ vs.\ $\theta$}
\figsetplot{HOPS24_R_contour3.eps}
\figsetgrpnote{$R$ contour plot for the models that fit HOPS 24. For each 
combination of inclination angle and cavity opening angle, the lowest 
$R$ values of models with these two parameter values are shown.}
\figsetgrpend

\figsetgrpstart
\figsetgrpnum{2.4}
\figsetgrptitle{$R$ contour plot for the models that fit HOPS 107, 
$i$ vs.\ $\rho_{1000}$}
\figsetplot{HOPS107_R_contour1.eps}
\figsetgrpnote{$R$ contour plot for the models that fit HOPS 107. For 
each combination of inclination angle and reference density, the 
lowest $R$ values of models with these two parameter values are shown.}
\figsetgrpend

\figsetgrpstart
\figsetgrpnum{2.5}
\figsetgrptitle{$R$ contour plot for the models that fit HOPS 107, 
$R_{disk}$ vs.\ $\rho_{1000}$}
\figsetplot{HOPS107_R_contour2.eps}
\figsetgrpnote{$R$ contour plot for the models that fit HOPS 107. For 
each combination of disk outer radius ($=R_c$) and $\rho_{1000}$, the 
lowest $R$ values of models with these two parameter values are shown.}
\figsetgrpend

\figsetgrpstart
\figsetgrpnum{2.6}
\figsetgrptitle{$R$ contour plot for the models that fit HOPS 107, 
$i$ vs.\ $\theta$}
\figsetplot{HOPS107_R_contour3.eps}
\figsetgrpnote{$R$ contour plot for the models that fit HOPS 107. For 
each combination of inclination angle and cavity opening angle, the 
lowest $R$ values of models with these two parameter values are shown.}
\figsetgrpend

\figsetgrpstart
\figsetgrpnum{2.7}
\figsetgrptitle{$R$ contour plot for the models that fit HOPS 107, 
$\theta$ vs.\ $\rho_{1000}$}
\figsetplot{HOPS107_R_contour4.eps}
\figsetgrpnote{$R$ contour plot for the models that fit HOPS 107. For 
each combination of cavity opening angle and $\rho_{1000}$, the 
lowest $R$ values of models with these two parameter values are shown.}
\figsetgrpend

\figsetgrpstart
\figsetgrpnum{2.8}
\figsetgrptitle{$R$ contour plot for the models that fit HOPS 149, 
$i$ vs.\ $\rho_{1000}$}
\figsetplot{HOPS149_R_contour1.eps}
\figsetgrpnote{$R$ contour plot for the models that fit HOPS 149. For 
each combination of inclination angle and reference density, the lowest 
$R$ values of models with these two parameter values are shown.}
\figsetgrpend

\figsetgrpstart
\figsetgrpnum{2.9}
\figsetgrptitle{$R$ contour plot for the models that fit HOPS 149, 
$R_{disk}$ vs.\ $\rho_{1000}$}
\figsetplot{HOPS149_R_contour2.eps}
\figsetgrpnote{$R$ contour plot for the models that fit HOPS 149. For 
each combination of disk outer radius ($=R_c$) and $\rho_{1000}$, the 
lowest $R$ values of models with these two parameter values are shown.}
\figsetgrpend

\figsetgrpstart
\figsetgrpnum{2.10}
\figsetgrptitle{$R$ contour plot for the models that fit HOPS 149, 
$i$ vs.\ $\theta$}
\figsetplot{HOPS149_R_contour3.eps}
\figsetgrpnote{$R$ contour plot for the models that fit HOPS 149. For 
each combination of inclination angle and cavity opening angle, the 
lowest $R$ values of models with these two parameter values are shown.}
\figsetgrpend

\figsetend

\clearpage

\section{Notes on HOPS Targets}

\subsection{HOPS Targets Discovered with {\it Herschel}}

\begin{deluxetable}{llrr}[h]
\tablewidth{0.8\linewidth}
\tablenum{1}
\tablecaption{New Protostars from \citet{stutz13} and \citet{tobin15}
\label{New_proto}}
\tablehead{
\colhead{HOPS Identifier} & \colhead{Original ID} & 
\colhead{R.A.} & \colhead{Dec. } \\
  & & \colhead{[$\degr$]} & \colhead{[$\degr$]}  \\
\colhead{(1)} & \colhead{(2)} & \colhead{(3)} & \colhead{(4)}}
\startdata
HOPS 394 &   019003 &  83.8497 &  -5.1315 \\
HOPS 395 &   026011 &  84.8208 &  -7.4074 \\
HOPS 396 &   029003 &  84.8048 &  -7.2199 \\
HOPS 397 &   061012 &  85.7036 &  -8.2696 \\
HOPS 398 &   082005 &  85.3725 &  -2.3547 \\
HOPS 399 &   082012 &  85.3539 &  -2.3024 \\
HOPS 400 &   090003 &  85.6885 &  -1.2706 \\
HOPS 401 &   091015 &  86.5319 &  -0.2058 \\
HOPS 402 &   091016 &  86.5415 &  -0.2047 \\
HOPS 403 &   093005 &  86.6156 &  -0.0149 \\
HOPS 404 &   097002 &  87.0323 &   0.5641 \\
HOPS 405 &   119019 &  85.2436 &  -8.0934 \\
HOPS 406 &   300001 &  86.9307 &   0.6396 \\
HOPS 407 &   302002 &  86.6177 &   0.3242 \\
HOPS 408 &   313006 &  84.8781 &  -7.3998 \\
HOPS 409 &   135003 &  83.8392 &  -5.2215 \\
\enddata
\tablecomments{
Column (1) lists the HOPS number of the object, column (2) the identifier
of the source from \citet{stutz13} and \citet{tobin15}, and columns (3) 
and (4) its  J2000 coordinates in degrees.}
\end{deluxetable}

\clearpage

\subsection{HOPS Objects not Included in the Modeling Sample}

\subsubsection{Young Stellar Objects}
\label{YSOs_not_modeled}

Among our HOPS sample, there are 41 targets that are likely
YSOs, but they lack PACS measurements at 70 and 160 $\mu$m 
and were therefore not included in the modeling sample. 
There are four additional targets with HOPS numbers that were not 
modeled; they are HOPS 109, 111, 212, and 362, and they are 
duplicates of HOPS 40, 60, 211, and 169, respectively. 
Table D\ref{HOPS_no_model_YSOs} lists the 41 likely YSOs in the 
HOPS catalog that were not part of the modeling sample; their SEDs 
are shown in Figure D\ref{non_model_YSOs}. Among them, 17 were 
not observed by PACS at 70 $\mu$m, while 24 were observed, but 
not detected at 70 $\mu$m. The majority of these targets are Class I 
protostars or flat-spectrum sources; only one is a Class 0 protostar, and 
five are Class II pre-main-sequence stars.

Most of these YSOs have very faint fluxes in the near- to mid-IR.
They could be deeply embedded protostars, like HOPS 307, or just 
very low-mass protostars with weak envelope emission.
Objects with little excess emission out to about 8 $\mu$m, a 
10 $\mu$m silicate emission feature, and a more or less steeply 
rising SED beyond 12 $\mu$m are likely transitional disks 
\citep[see][]{kim13}; good examples are HOPS 51 and 54. 
It is possible that some of the YSOs in this sample are actually 
extragalactic contaminants, in particular objects with flat SEDs 
(see more about this subset of our HOPS sample in the next
subsection).

\begin{deluxetable}{lrrrrrrc}
\tablewidth{\linewidth}
\tablenum{2}
\tablecaption{YSOs in the HOPS Sample with No PACS Data}
\label{HOPS_no_model_YSOs}
\tablehead{
\colhead{Object} & \colhead{R.A.} & \colhead{Dec. } & \colhead{Class} &
\colhead{$L_{bol}$} & \colhead{$T_{bol}$} & \colhead{$n_{4.5-24}$} &
\colhead{PACS Flag} \\
  & \colhead{[$\degr$]} & \colhead{[$\degr$]} & & \colhead{[\Lsun]} &
\colhead{[K]} &  & \\
\colhead{(1)} & \colhead{(2)} & \colhead{(3)} & \colhead{(4)} &
\colhead{(5)} & \colhead{(6)} & \colhead{(7)} & \colhead{(8)}}
\startdata
  HOPS 0 & 88.6171 &   1.6264 &    I &   0.011 &   652.2 & 0.514 & -1 \\
  HOPS 8 & 83.8880 &  -5.9851 &    I &   0.013 &   329.8 & 0.419 &  0 \\
  HOPS 9 & 83.9550 &  -5.9843 &    I &   0.006 &   281.5 & 0.858 & -1 \\
 HOPS 14 & 84.0799 &  -5.9251 & flat &   0.042 &   464.0 & 0.246 &  0 \\
 HOPS 23 & 84.0745 &  -5.7818 &    I &   0.012 &   346.8 & 0.539 &  0 \\
 HOPS 25 & 83.8443 &  -5.7415 & flat &   0.045 &   646.6 & 0.165 &  0 \\
 HOPS 31 & 83.8219 &  -5.6741 & flat &   0.024 &   634.7 & 0.304 &  0 \\
 HOPS 34 & 83.7954 &  -5.6585 &    I &   0.013 &   235.5 & 0.762 &  0 \\
 HOPS 35 & 83.8331 &  -5.6503 &    I &   0.044 &   305.2 & 0.884 &  0 \\
 HOPS 37 & 83.6986 &  -5.6237 & flat &   0.016 &   913.4 & 0.230 &  0 \\
 HOPS 51 & 83.8160 &  -5.5015 &   II &   0.518 &   130.2 & \nodata &  0 \\
 HOPS 52 & 83.8180 &  -5.4924 & flat &   0.641 &   610.8 & -0.163 &  0 \\
 HOPS 54 & 83.3437 &  -5.3841 &   II &   0.097 &  1879.3 & -0.37 & -1 \\
 HOPS 62 & 83.8524 &  -5.1916 & flat &   0.660 &   1154.1 & 0.043 &  0 \\
 HOPS 63 & 83.8538 &  -5.1671 & flat &   0.516 &   544.5 & 0.004 &  0 \\
 HOPS 64 & 83.8625 &  -5.1650 &    I &  15.347 &    29.7 & 0.503 &  0 \\
 HOPS 69 & 83.8551 &  -5.1400 & flat &   2.778 &    31.3 & -0.189 &  0 \\
 HOPS 79 & 83.8662 &  -5.0934 & flat &   0.086 &   666.2 & -0.137 &  0 \\
HOPS 103 & 83.5508 &  -4.8353 & flat &   0.142 &  1484.3 & -0.032 &  0 \\
HOPS 104 & 83.7782 &  -4.8338 &    I &   0.044 &   337.3 & 0.837 &  0 \\
HOPS 110 & 84.0093 &  -5.0472 &    I &   0.014 &   244.0 & \nodata & -1 \\
HOPS 126 & 85.0408 &  -7.1650 & flat &   0.132 &  1865.3 & -0.136 & -1 \\
HOPS 151 & 84.6787 &  -6.9447 &   II &   0.061 &   799.4 & -0.505 &  0 \\
HOPS 155 & 84.3160 &  -7.2972 & flat &   0.013 &   393.8 & 0.133 & -1 \\
HOPS 162 & 84.1291 &  -6.8780 &   II &   0.015 &   909.9 & 0.352 & -1 \\
HOPS 180 & 84.2475 &  -6.1710 &   II &   0.011 &  1493.5 & 0.578 & -1 \\
HOPS 195 & 84.0002 &  -6.1206 & flat &   0.032 &   659.7 & 0.399 &  0 \\
HOPS 217 & 85.7965 &  -8.4056 &    I &   0.008 &   323.8 & 0.773 & -1 \\
HOPS 230 & 85.6283 &  -8.1515 & flat &   0.267 &  1260.2 & -0.104 & -1 \\
HOPS 231 & 85.1189 &  -8.5486 & flat &   0.024 &   386.0 & -0.239 & -1 \\
HOPS 269 & 85.3625 &  -7.7094 & flat &   0.025 &   230.2 & 0.023 & -1 \\
HOPS 289 & 84.9865 &  -7.5017 &    I &   0.095 &   331.1 & 0.868 &  0 \\
HOPS 296 & 85.3215 &  -2.3021 &    I &   0.022 &   326.0 & 0.931 &  0 \\
HOPS 302 & 85.0934 &  -2.2610 & flat &   0.383 &  1367.2 & -0.032 & -1 \\
HOPS 307 & 85.3077 &  -1.7844 &    0 &   0.748 &    57.1 & 1.506 & -1 \\
HOPS 314 & 86.6505 &  -0.3414 &    I &   0.015 &   276.2 & 1.112 & -1 \\
HOPS 327 & 86.6139 &   0.1477 & flat &   0.020 &   991.0 & 0.145 & -1 \\
HOPS 328 & 86.5561 &   0.1759 &    I &   0.012 &   326.3 & 0.868 & -1 \\
HOPS 330 & 86.7140 &   0.3298 & flat &   0.121 &   385.2 & 0.285 &  0 \\
HOPS 332 & 86.8821 &   0.3391 & flat &   0.249 &   145.5 & 0.045 &  0 \\
HOPS 360 & 86.8629 &   0.3425 &    I &   1.017 &    43.2 & \nodata &  0 \\
\enddata
\tablecomments{
Column (1) lists the HOPS number of the object, columns (2) and (3) its 
J2000 coordinates in degrees, column (4) the type based on SED classification,
column (5) the bolometric luminosity, column (6) the bolometric temperature, 
column (7) the 4.5-24 $\mu$m SED slope, and column (8) a flag identifying
whether the object was not observed by PACS (flag value of -1), or not detected
by PACS at 70 $\mu$m (flag value of 0)}
\end{deluxetable}

\begin{figure}[h]
\centering
\includegraphics[scale=0.9]{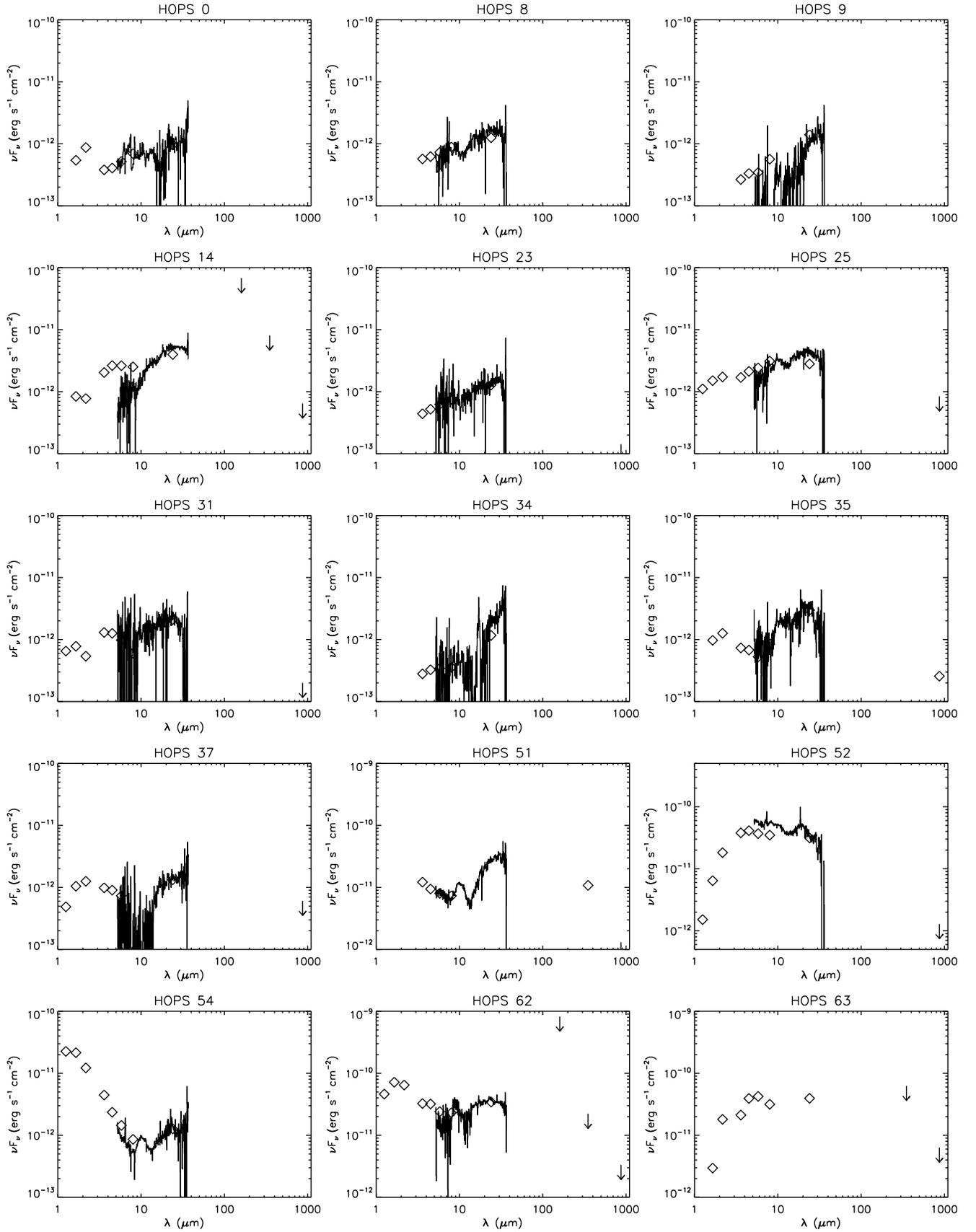}
\caption{SEDs of the HOPS targets not modeled in this work that are
likely YSOs (open symbols: photometry, line: IRS spectrum).
\label{non_model_YSOs}}
\end{figure}

\begin{figure}[h]
\centering
\includegraphics[scale=0.9]{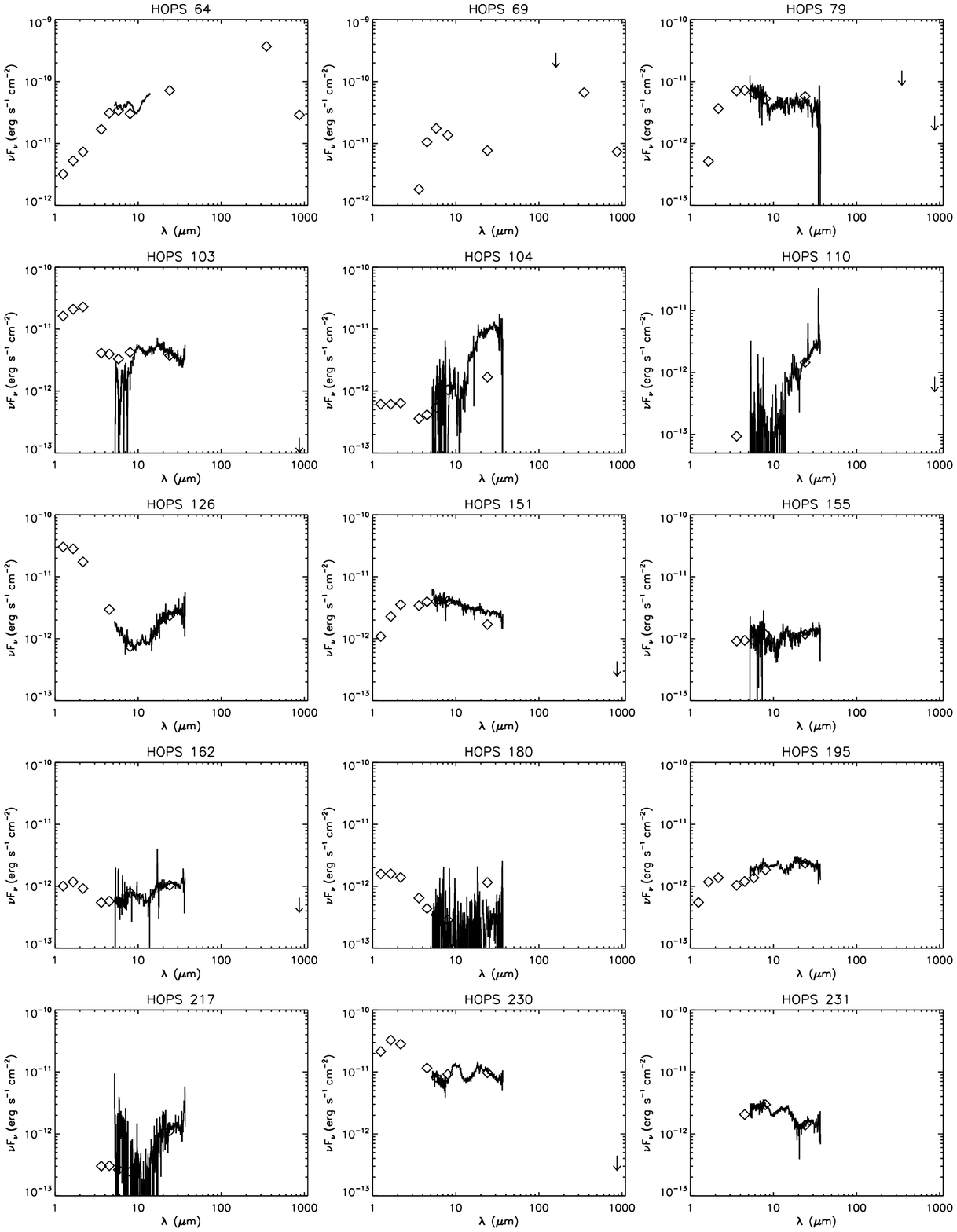}
\figurenum{\ref{non_model_YSOs}}\caption{continued.}
\end{figure}

\begin{figure}[h]
\centering
\includegraphics[scale=0.9]{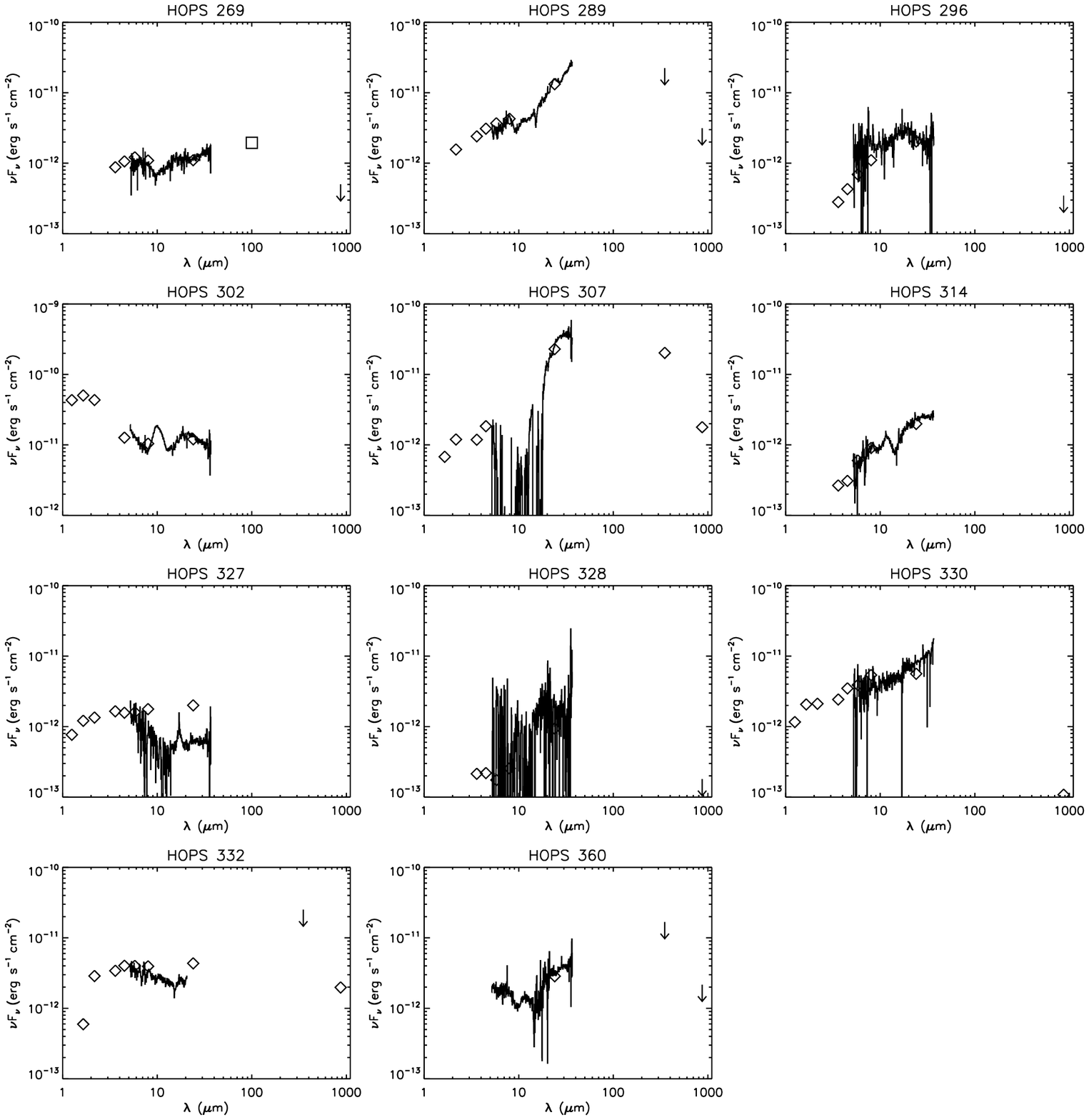}
\figurenum{\ref{non_model_YSOs}}\caption{continued.}
\end{figure}

\clearpage

\subsubsection{Contaminants}
\label{exgal_not_modeled}

Our HOPS sample contains 29 targets that turned out to be likely extragalactic
contaminants. These objects are listed in Table D\ref{HOPS_contaminants},
and their SEDs are shown in Figure D\ref{non_model_exgal}. Most galaxies 
were identified based on the presence of PAH features or emission lines in 
their IRS spectra (in particular, the 5--15 $\mu$m region), the absence of 
a silicate absorption feature at 10 $\mu$m, and an overall flat or slightly
rising mid-infrared continuum. Clear examples are HOPS 21, 27, 46, 48, 55, 
61, 72, 106, 161, and 301. 

Two objects, HOPS 202 and HOPS 205, have a tentative 10 $\mu$m 
silicate emission feature and a steep rise of their SED beyond 12 $\mu$m,
but also PAH emission features in their IRS spectrum; they could be
transitional disks and not galaxies.

In some cases, e.g., HOPS 308, targets classified as extragalactic contaminants 
are very faint in the near- to mid-infrared; instead of galaxies, they could be very
low-mass or deeply embedded protostars. 

The mid-IR SED of HOPS 339 is mostly flat but displays a sharp 10 $\mu$m
absorption feature; based on the SED alone, it would not necessarily be
classified as a galaxy, but high-resolution near-IR {\it HST} images resolve its
extended emission and reveal a spiral galaxy (J. Booker et al. 2016, 
in preparation).

Finally, HOPS 349, 350, 352, 353, 356, and 381 have poorly sampled SEDs
(just one or two flux measurements), so their nature is quite uncertain. We list
their coordinates in Table D\ref{HOPS_uncertain_obj}.

\begin{deluxetable}{lrrrrr}
\tablewidth{\linewidth}
\tablenum{3}
\tablecaption{Likely Extragalactic Contaminants in the HOPS Sample}
\label{HOPS_contaminants}
\tablehead{
\colhead{Object} & \colhead{R.A.} & \colhead{Dec. } & 
\colhead{$L_{bol}$} & \colhead{$T_{bol}$} &
\colhead{$n_{4.5-24}$}  \\
  & \colhead{[$\degr$]} & \colhead{[$\degr$]} & \colhead{[\Lsun]} &
\colhead{[K]} &  \\
\colhead{(1)} & \colhead{(2)} & \colhead{(3)} & \colhead{(4)} &
\colhead{(5)} & \colhead{(6)}}
\startdata
 HOPS 21 & 84.0421 &  -5.8356 &   0.074 &   584.5 & 0.824 \\
 HOPS 27 & 84.0905 &  -5.6995 &   0.102 &   118.2 & 0.458 \\
 HOPS 39 & 84.0934 &  -5.6069 &   0.100 &   159.4 & 1.010 \\
 HOPS 46 & 83.6758 &  -5.5509 &   0.141 &  1081.9 & -0.083 \\
 HOPS 48 & 83.7773 &  -5.5477 &   0.608 &   611.0 & 0.252 \\
 HOPS 55 & 83.4754 &  -5.3638 &   0.316 &   101.5 & 1.732 \\
 HOPS 61 & 83.3579 &  -5.2007 &   0.045 &   721.4 & -0.031 \\
 HOPS 67 & 83.8445 &  -5.1428 &   0.044 &   278.7 & 0.900 \\
 HOPS 72 & 83.8571 &  -5.1296 &   0.545 &   693.0 & 0.068 \\
 HOPS 83 & 83.9822 &  -5.0771 &   0.131 &   293.9 & -0.119 \\
 HOPS 97 & 83.8704 &  -4.9608 &   0.190 &   403.8 & 0.125 \\
HOPS 101 & 83.7843 &  -4.9027 &   3.355 &   481.2 & -0.207 \\
HOPS 106 & 84.0518 &  -4.7544 &   0.016 &   359.7 & 0.943 \\
HOPS 112 & 85.1833 &  -7.3786 &   0.014 &   390.2 & 0.093 \\
HOPS 146 & 84.6840 &  -7.0112 &   0.053 &   519.7 & -0.178 \\
HOPS 161 & 84.1448 &  -7.1871 &   0.062 &   179.1 & 0.458 \\
HOPS 196 & 83.8371 &  -6.3062 &   0.065 &   165.5 & -0.042 \\
HOPS 202 & 83.4330 &  -6.2295 &   0.018 &   736.3 & 0.534 \\
HOPS 205 & 85.7620 &  -8.7971 &   0.067 &   427.8 & 0.549 \\
HOPS 218 & 85.7912 &  -8.2232 &   0.007 &   332.5 & 0.653 \\
HOPS 292 & 84.4787 &  -7.6890 &   0.068 &   280.5 & 0.421 \\
HOPS 301 & 85.4366 &  -2.2654 &   2.955 &   518.8 & 0.271 \\
HOPS 306 & 85.7630 &  -1.8013 &   0.038 &   310.8 & 0.060 \\
HOPS 308 & 85.8082 &  -1.7195 &   0.042 &   156.0 & 0.791 \\
HOPS 309 & 85.6973 &  -1.4131 &   0.041 &   111.9 & -0.168 \\
HOPS 313 & 85.2532 &  -1.1529 &   0.030 &   154.1 & 1.039 \\
HOPS 339 & 86.4733 &   0.4243 &   0.128 &   398.3 & 0.246 \\
HOPS 348 & 86.7511 &   0.3438 &   0.286 &    84.0 & \nodata \\
HOPS 351 & 83.8809 &  -5.0797 &   0.016 &   217.1 & \nodata \\
\enddata
\tablecomments{
\parbox{0.6\linewidth}{
Column (1) lists the HOPS number of the object, columns (2) and (3) its 
J2000 coordinates in degrees, column (4) the bolometric luminosity, 
column (5) the bolometric temperature, and column (6) the 4.5-24 
$\mu$m SED slope.}}
\end{deluxetable}

\begin{deluxetable}{lrr}[!h]
\tablewidth{0.75\linewidth}
\tablenum{4}
\tablecaption{Targets in the HOPS Sample with Uncertain Nature}
\label{HOPS_uncertain_obj}
\tablehead{
\colhead{Object} & \colhead{R.A.} & \colhead{Dec. }  \\
\colhead{(1)} & \colhead{(2)} & \colhead{(3)}}
\startdata
HOPS 349 & 83.8592 &  -5.1426 \\
HOPS 350 & 83.8758 &  -5.1386  \\
HOPS 352 & 83.8617 &  -5.0675  \\
HOPS 353 & 88.5556 &   1.7175  \\
HOPS 356 & 85.5341 &  -1.4438 \\
HOPS 381 & 83.7816 &  -5.6986 \\
\enddata
\tablecomments{
\parbox{0.45\linewidth}{
Column (1) lists the HOPS number of the object, columns (2) and (3) its 
J2000 coordinates in degrees.}}
\end{deluxetable}

\begin{figure}[h]
\centering
\includegraphics[scale=0.9]{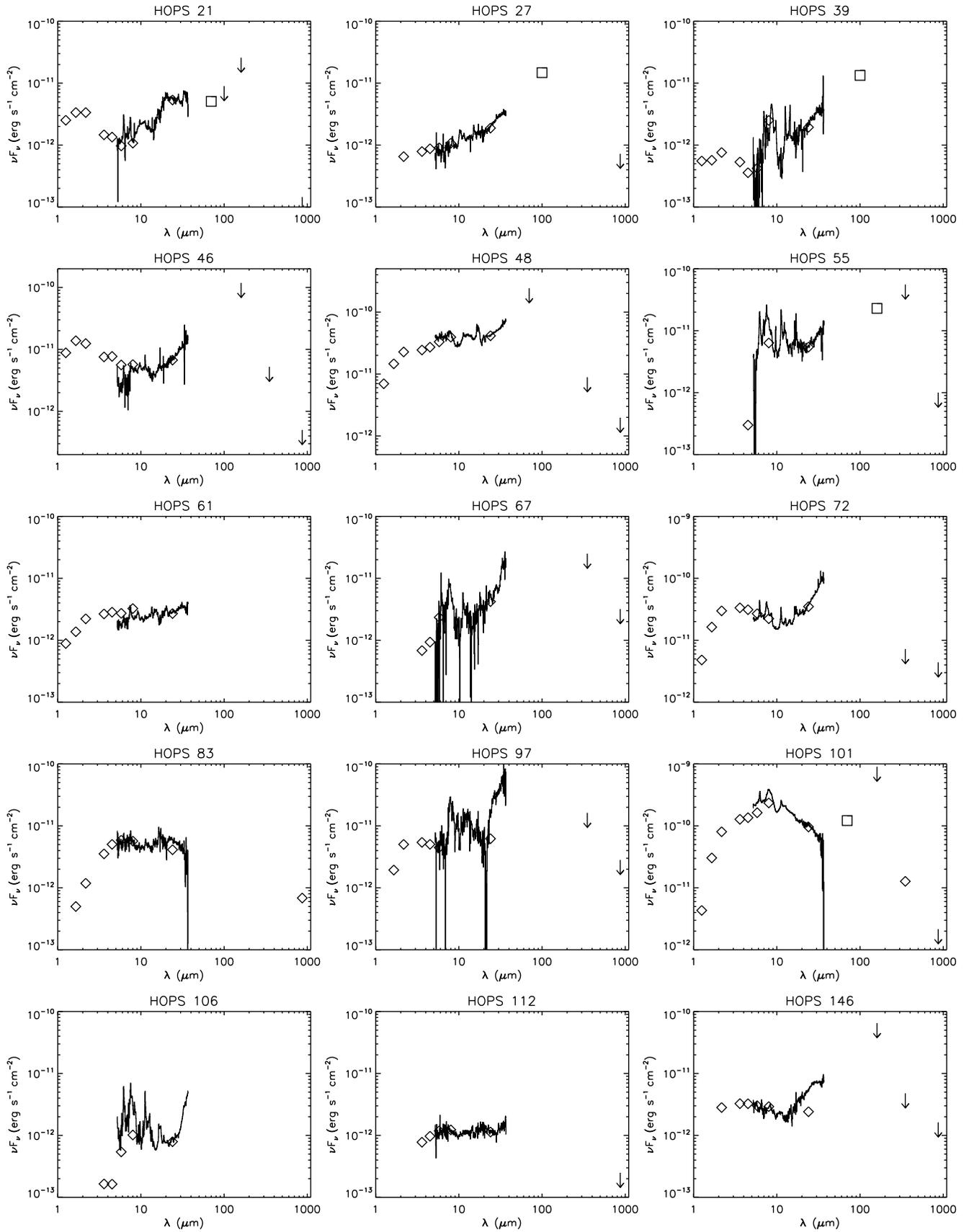}
\caption{SEDs of the HOPS targets not modeled in this work that are
likely extragalactic contaminants (open symbols: photometry, line: IRS spectrum).
\label{non_model_exgal}}
\end{figure}

\begin{figure}[h]
\centering
\includegraphics[scale=0.9]{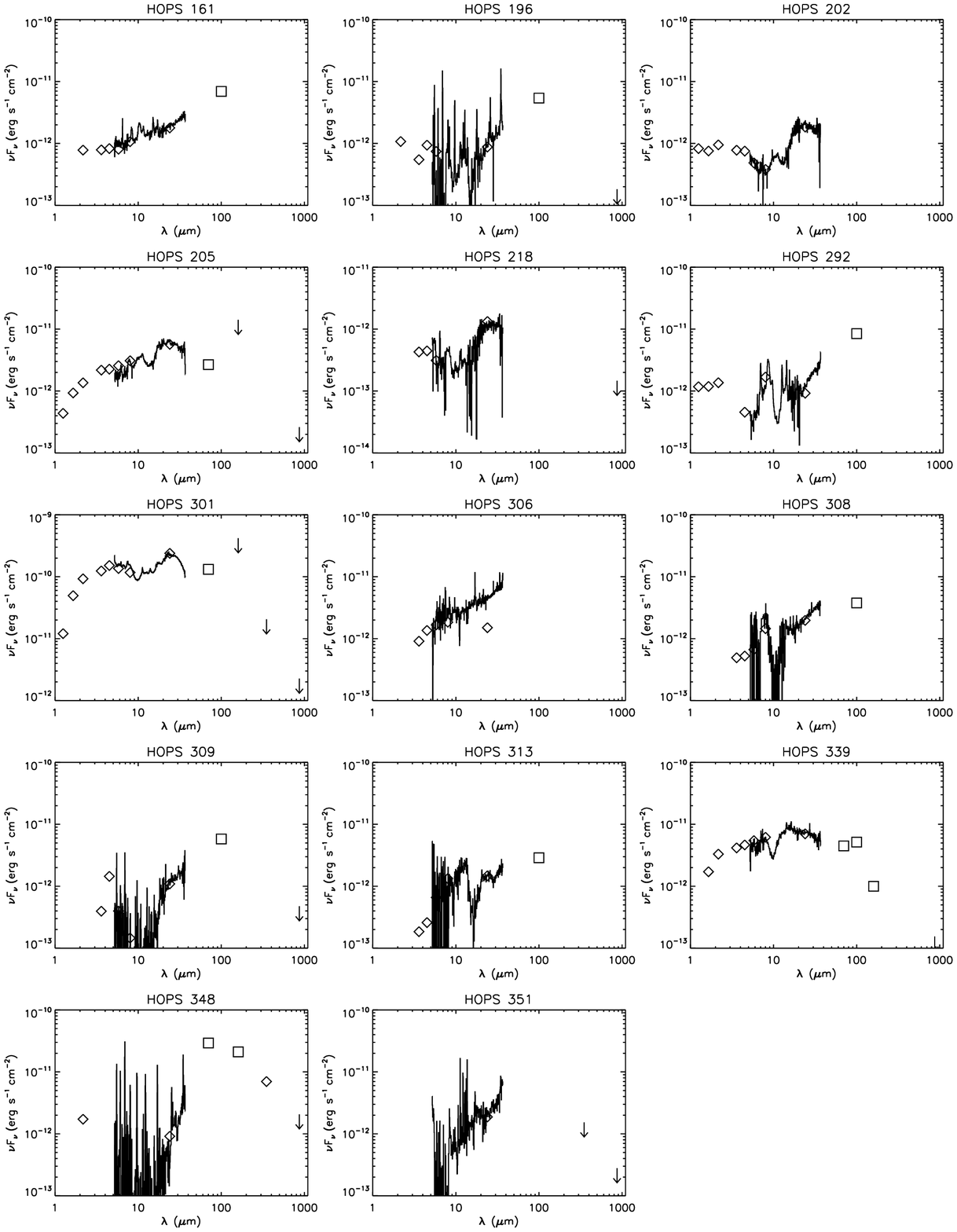}
\figurenum{\ref{non_model_exgal}}\caption{continued.}
\end{figure}


\vspace{2ex}

\begin{thebibliography}{}
\bibitem[Adams \& Shu(1986)]{adams86} Adams, F. C., \& Shu, F. H. 1986, \apj,
308, 836
\bibitem[Adams, Lada, \& Shu(1987)]{adams87} Adams, F. C., Lada, Ch. J., \& Shu, F. H. 
1987, \apj, 312, 788
\bibitem[Adams et al.(2012)]{adams12} Adams, J. D., Herter, T. L., Osorio, M., et al.
2012, ApJL, 749, L24
\bibitem[Ali et al.(2010)]{ali10} Ali, B., Tobin, J. J., Fischer, W. J., et al.  
2010, \aap, 518, L119
\bibitem[Andr\'e et al.(1993)]{andre93} Andr\'e, P., Ward-Thompson, D., \& Barsony, M.
1993, \apj, 406, 122
\bibitem[Andr\'e \& Montmerle(1994)]{andre94} Andr\'e, P., \& Montmerle, T.
1994, \apj, 420, 837
\bibitem[Andr\'e et al.(2010)]{andre10} Andr\'e, P., Men'shchikov, A.,
Bontemps, S., et al.  2010, \aap, 518, L102
\bibitem[Arce \& Sargent(2006)]{arce06} Arce, H. G., \& Sargent, A. I.  2006,
\apj, 646, 1070
\bibitem[Billot et al.(2012)]{billot12} Billot, N., Morales-Calder\'on, M., Stauffer, 
J. R., et al.  2012, ApJL, 753, L35
\bibitem[Boogert et al.(2008)]{boogert08} Boogert, A. C. A., Pontoppidan,
K. M., Knez, C., et al.  2008, \apj, 678, 985
\bibitem[Calvet et al.(1994)]{calvet94} Calvet, N., Hartmann, L., Kenyon, S. J.,
\& Whitney, B. A.  1994, \apj, 434, 330
\bibitem[Caratti o Garatti(2011)]{caratti11} Caratti o Garatti, A., Garcia Lopez, R.,
Scholz, A., et al.  2011, \aap, 526, L1
\bibitem[Cassen \& Moosman(1981)]{cassen81} Cassen, P., \& Moosman, A.
1981, \icarus, 48, 353
\bibitem[Chen et al.(1995)]{chen95} Chen, H., Myers, P. C., Ladd, E. F., \& Wood, 
D. O. S. 1995, \apj, 445, 377
\bibitem[Cody et al.(2014)]{cody14} Cody, A. M., Stauffer, J., Baglin, A., et al.
2014, \aj, 147, 82
\bibitem[Cohen et al.(2003)]{cohen03} Cohen, M., Wheaton, W. A., \&
Megeath, S. T.  2003, \aj, 126, 1090
\bibitem[Crapsi et al.(2008)]{crapsi08} Crapsi, A., van Dishoeck, E. F., Hogerheijde, 
M. R., et al.  2008, \aap, 486, 245
\bibitem[di Francesco et al.(2007)]{diFrancesco07} di Francesco, J., Evans, N. J., II,
Caselli, P., et al.  2007, in Protostars \& Planets V, Reipurth B., Jewitt D., Keil K. 
(eds.), Univ. Arizona, Tucson, p. 17
\bibitem[Draine(2003)]{draine03} Draine, B. T.  2003, \araa, 41,241
\bibitem[Dunham et al.(2008)]{dunham08} Dunham, M. M., Crapsi, A., Evans, N. J., II,
et al.  2008, \apjs, 179, 249
\bibitem[Dunham et al.(2010)]{dunham10} Dunham, M. M., Evans, N. J., II, Terebey, S.,
et al.  2010, \apj, 710, 470
\bibitem[Dunham \& Vorobyov(2012)]{dunham12} Dunham, M. M., \& Vorobyov, E. I.
2012, \apj, 747, 52
\bibitem[Dunham et al.(2013)]{dunham13} Dunham, M. M., Arce, H. G., Allen, L. E.,
et al.  2013, \aj, 145, 94
\bibitem[Dunham et al.(2014)]{dunham14} Dunham, M. M., Stutz, A. M.,
Allen, L. E., et al.  2014, in Protostars \& Planets VI, Beuther H., Klessen R. S., 
Dullemond C. P., Henning T. (eds.), Univ. Arizona, Tucson, p. 195
\bibitem[Dunham et al.(2015)]{dunham15} Dunham, M. M., Allen, L. E., Evans, 
N. J., II, et al.  2015, \apjs, 220, 11
\bibitem[Engelbracht et al.(2007)]{engelbracht07} Engelbracht, C. W.,
Blaylock, M., Su, K. Y. L., et al.  2007, \pasp, 119, 994
\bibitem[Enoch et al.(2009)]{enoch09} Enoch, M., Evans, N. J., II, 
Sargent, A. I., \& Glenn, J.  2009, \apj, 692, 973
\bibitem[Espaillat et al.(2014)]{espaillat14} Espaillat, C., Muzerolle, J., Najita, J.,
et al.  2014, in Protostars \& Planets VI, Beuther H., Klessen R. S., Dullemond
C. P., Henning T (eds.), Univ. Arizona, Tucson, p. 497
\bibitem[Evans et al.(2001)]{evans01} Evans, N. J., II, Rawlings, J. M. C.,
Shirley, Y. L., \& Mundy, L. G.  2001, \apj, 557, 193
\bibitem[Evans et al.(2009)]{evans09} Evans, N. J., II, Dunham, M. M.,
J{\o}rgensen, J. K., et al.  2009, \apjs, 181, 321
\bibitem[Fazio et al.(2004)]{fazio04} Fazio, G. G., Hora, J. L., Allen, L. E.,
et al.  2004, \apjs,154, 10
\bibitem[Fischer et al.(2010)]{fischer10} Fischer, W. J., Megeath, S. T.,
Ali, B., et al.  2010, \aap, 518, L122
\bibitem[Fischer et al.(2012)]{fischer12} Fischer, W. J., Megeath, S. T.,
Tobin, J. J., et al.  2012, \apj, 756, 99
\bibitem[Fischer et al.(2013)]{fischer13} Fischer, W. J., Megeath, S. T.,
Stutz, A. M., et al.  2013, AN, 334, 53
\bibitem[Fischer et al.(2014)]{fischer14} Fischer, W. J., Megeath, S. T.,
Tobin, J. J., et al. 2014, \apj, 781, 123
\bibitem[Furlan et al.(2008)]{furlan08} Furlan, E., McClure, M., Calvet, N.,
et al.  2008, \apjs, 176, 184
\bibitem[Furlan et al.(2014)]{furlan14} Furlan, E., Megeath, S. T., Osorio, M.,
et al. 2014, \apj, 786, 26
\bibitem[Greene et al.(1994)]{greene94} Greene, T. P., Wilking, B. A.,
Andr\'e, P., et al.  1994, \apj, 434, 614
\bibitem[G\"unther et al.(2014)]{guenther14} G\"unther, H. M., Cody, A. M.,
Covey, K. R., et al.  2014, \aj, 148, 122
\bibitem[Hartmann et al.(1996)]{hartmann96} Hartmann, L., Calvet, N., \& 
Boss, A., 1996, \apj, 464, 387
\bibitem[Heiderman \& Evans(2015)]{heiderman15} Heiderman, A., \& Evans,
N. J., II  2015, \apj, 806, 231
\bibitem[Houck et al.(2004)]{houck04} Houck, J. R., Roellig, T. L.,
van Cleve, J., et al.  2004, \apjs, 154, 18
\bibitem[Kenyon et al.(1990)]{kenyon90} Kenyon, S. J., Hartmann, L. W., 
Strom, K. M., \& Strom, S. E.  1990, \aj, 99, 869
\bibitem[Kenyon et al.(1993)]{kenyon93} Kenyon, S. J., Calvet, N., 
\& Hartmann, L. 1993, \apj, 414, 676
\bibitem[Kim et al.(2008)]{kim08} Kim, M. K., Hirota, T., Honma, M., et al.
2008, \pasj, 60, 991
\bibitem[Kim et al.(2013)]{kim13} Kim, K. H., Watson, D. M., Manoj, P.,
et al.  2013, \apj, 769, 149
\bibitem[Kim et al.(2016)]{kim16} Kim, K. H., Watson, D. M., Manoj, P.,
et al.  2016, \apjs, submitted
\bibitem[Kounkel et al.(2016)]{kounkel16} Kounkel, M., Megeath, S. T., 
Poteet, C. A., et al.  2016, \apj, in press, arXiv:1602.07635
\bibitem[Kryukova et al.(2012)]{kryukova12} Kryukova, E., Megeath, S. T.,
Gutermuth, R. A., et al.  2012, \aj, 144, 31
\bibitem[Kryukova et al.(2014)]{kryukova14} Kryukova, E., Megeath, S. T., 
Hora, J. L., et al.  2014, \aj, 148, 11
\bibitem[Lada(1987)]{lada87} Lada, Ch. J. 1987, in Star Forming Regions, 
proceedings of the IAU Symposium No. 115, ed. M. Peimbert \& J. Jugaku, 
Dordrecht:Reidel, 1
\bibitem[Launhardt et al.(2013)]{launhardt13} Launhardt, R., Stutz, A. M.,
Schmiedeke, A., et al. 2013, \aap, 551, A98
\bibitem[Lim \& Takakuwa(2006)]{lim06} Lim, J., \& Takakuwa, S.  2006,
\apj, 653, 425
\bibitem[Manoj et al.(2013)]{manoj13} Manoj, P., Watson, D. M., Neufeld, D. A.,
et al.  2013, \apj, 763, 83
\bibitem[Mathis et al.(1983)]{mathis83} Mathis, J. S., Mezger, P. G., \&
Panagia, N.  1983, \aap, 128, 212
\bibitem[Mathis(1990)]{mathis90} Mathis, J. S. 1990, \araa, 28, 37
\bibitem[Maury et al.(2011)]{maury11} Maury, A. J., Andr\'e, P., Men'shchikov, A.,
et al.  2011, \aap, 535, A77
\bibitem[McClure(2009)]{mcclure09} McClure, M.  2009, ApJL, 693, L81
\bibitem[Megeath et al.(2012)]{megeath12} Megeath, S. T., Gutermuth, R.,
Muzerolle, J., et al.  2012, \aj, 144, 192 
\bibitem[Menten et al.(2007)]{menten07} Menten, K. M., Reid, M. J., Forbrich, J.,
\& Brunthaler, A.  2007, \aap, 474, 515
\bibitem[Morales-Calder\'on et al.(2011)]{morales11} Morales-Calder\'on, M.,
Stauffer, J. R., Hillenbrand, L. A., et al.  2011, \apj, 733, 50
\bibitem[Myers \& Ladd(1993)]{myers93} Myers, P. C., \& Ladd, E. F.  1993,
ApJL, 413, L47
\bibitem[Offner \& McKee(2011)]{offner11} Offner, S. S. R., \& McKee, C. F.
2011, \apj, 736, 53
\bibitem[Ormel et al.(2011)]{ormel11} Ormel, C. W., Min, M., Tielens, A. G. G. M.,
Dominik, C., \& Paszun, D.  2011, \aap, 532, A43
\bibitem[Ossenkopf \& Henning(1994)]{ossenkopf94} Ossenkopf, V., \& Henning, T.
1994, \aap, 291, 943
\bibitem[Pilbratt et al.(2010)]{pilbratt10} Pilbratt, G. L., Riedinger, J. R., 
Passvogel, T., et al. 2010, \aap, 518, L1
\bibitem[Pillitteri et al.(2013)]{pillitteri13} Pillitteri, I., Wolk, S. J., Megeath, S. T.,
et al. 2013, \apj, 768, 99
\bibitem[Poglitsch et al.(2010)]{poglitsch10} Poglitsch, A., Waelkens, C.,
Geis, N., et al. 2010, \aap, 518, L2
\bibitem[Pontoppidan et al.(2008)]{pontoppidan08} Pontoppidan, K. M.,
Boogert, A. C. A., Fraser, H. J., et al. 2008, \apj, 678, 1005 
\bibitem[Poppenhaeger et al.(2015)]{poppenhaeger15} Poppenhaeger, K.,
Cody, A. M., Covey, K. R.  2015, \aj, 150, 118
\bibitem[Poteet et al.(2011)]{poteet11} Poteet, C. A., Megeath, S. T., Watson, D. M.,
et al.  2011, ApJL, 733, L32
\bibitem[Reach et al.(2005)]{reach05} Reach, W. T., Megeath, S. T.,
Cohen, M., et al.  2005, \pasp, 117, 978
\bibitem[Rebull et al.(2014)]{rebull14} Rebull, L. M., Cody, A. M., Covey, K. R.,
et al.  2014, \aj, 148, 92
\bibitem[Rebull et al.(2015)]{rebull15} Rebull, L. M., Stauffer, J. R., Cody, A. M.,
et al.  2015, \aj, 150, 175
\bibitem[Rieke et al.(2004)]{rieke04} Rieke, G. H., Young, E. T., 
Engelbracht, C. W., et al.  2004, \apjs, 154, 25
\bibitem[Robitaille et al.(2006)]{robitaille06} Robitaille, T. P., Whitney, B. A.,
Indebetouw, R., et al.  2006, \apjs, 167, 256
\bibitem[Rodr{\'i}guez et al.(1998)]{rodriguez98} Rodr{\'i}guez, L. F., D'Alessio, P.,
Wilner, D. J., et al.  1998, Nature, 395, 355 
\bibitem[Sadavoy et al.(2014)]{sadavoy14} Sadavoy, S. I., Di Francesco, J., 
Andr\'e, P., et al.  2014, ApJL, 787, L18
\bibitem[Safron et al.(2015)]{safron15} Safron, E. J., Fischer, W. J., Megeath, S. T.,
et al. 2015, ApJL, 800, L5
\bibitem[Shirley et al.(2002)]{shirley02} Shirley, Y. L., Evans, N. J., II, \& Rawlings, 
J. M. C.  2002, \apj, 575, 337
\bibitem[Siringo et al.(2010)]{siringo10} Siringo, G., Kreysa, E., De Breuck, C., 
et al. 2010, The Messenger, 139, 20
\bibitem[Siringo et al.(2009)]{siringo09} Siringo, G., Kreysa, E., Kov\'acs, A., 
et al. 2009, \aap, 497, 945
\bibitem[Skrutskie et al.(2006)]{skrutskie06} Skrutskie, M. F., Cutri, R. M.,
Stiening, R., et al.  2006, \aj, 131, 1163
\bibitem[Stanke et al.(2010)]{stanke10} Stanke, T., Stutz, A. M., Tobin,
J. J., et al.  2010, \aap, 518, L94
\bibitem[Stutz et al.(2010)]{stutz10} Stutz, A. M., Launhardt, R., Linz, H.,
et al. 2010, \aap, 518 L87
\bibitem[Stutz et al.(2013)]{stutz13} Stutz, A. M., Tobin, J. J., Stanke, T.,
et al. 2013, \apj, 767, 36
\bibitem[Stutz \& Kainulainen(2015)]{stutz15} Stutz, A. M., \& Kainulainen
2015, \aap, 577, L6
\bibitem[Terebey et al.(1984)]{terebey84} Terebey, S., Shu, F. H., \&
Cassen, P. 1984, \apj, 286, 529 
\bibitem[Tobin et al.(2008)]{tobin08} Tobin, J. J., Hartmann, L., Calvet, N., \&
D'Alessio, P.  2008, \apj, 679, 1364
\bibitem[Tobin et al.(2012)]{tobin12} Tobin, J. J., Hartmann, L., Bergin, E., 
et al.  2012, \apj, 748, 16
\bibitem[Tobin et al.(2015)]{tobin15} Tobin, J. J., Stutz, A. M., Megeath, S. T.,
et al. 2015, \apj, 798, 128
\bibitem[Ulrich(1976)]{ulrich76} Ulrich, R. K. 1976, \apj, 210, 377
\bibitem[Werner et al.(2004)]{werner04} Werner, M. W., Roellig, T. L.,
Low, F. J., et al.  2004, \apjs, 154, 1
\bibitem[Whitney et al.(2003a)]{whitney03a} Whitney, B. A., Wood, K.,
Bjorkman, J. E., \& Wolff, M. J.  2003a, \apj, 591, 1049
\bibitem[Whitney et al.(2003b)]{whitney03b} Whitney, B. A., Wood, K.,
Bjorkman, J. E., \& Cohen, M.  2003b, \apj, 598, 1079
\bibitem[Whitney et al.(2013)]{whitney13} Whitney, B. A., Robitaille, T. P.,
Bjorkman, J. E., et al.  2013, \apjs, 207, 30
\bibitem[Winston et al.(2010)]{winston10} Winston, E., Megeath, S. T.,
Wolk, S. J., et al. 2010, \aj, 140, 266
\bibitem[Wolk et al.(2015)]{wolk15} Wolk, S. J., G\"unther, H. M., 
Poppenhaeger, K., et al.  2015, \aj, 150, 145
\bibitem[Young et al.(2003)]{young03} Young, C. H., Shirley, Y. L., 
Evans, N. J., II, \& Rawlings, J. M. C.  2003, \apjs, 145, 111
\bibitem[Young \& Evans(2005)]{young05} Young, C. H., \& Evans, N. J., II
2005, \apj, 627, 293
\end{thebibliography}
\end{document}